\documentclass[12pt, oneside]{book}
\usepackage[latin1]{inputenc}                 
\usepackage{amsmath, amssymb, bm}
\usepackage{graphicx}
\usepackage{natbib}

\usepackage[letterpaper,top=3.6cm,left=4cm,bottom=2.9cm,right=2.5cm,ignorehead,footskip=1cm]{geometry}
\usepackage{fancyhdr}
\usepackage{wrapfig}
\usepackage{setspace}
    \fancypagestyle{plain}{
      \fancyhf{}
      \fancyfoot[R]{\thepage}
      
    }

\usepackage{amsmath,longtable}
\usepackage{natbib}
\newcommand{\model}{{\sc m19r8e1.25ni56}}
\newcommand{\remark}[1]{{\small \textbf{[Comment: #1]}}}

\newcommand{\apj}{ApJ}
\newcommand{\apjs}{ApJS}
\newcommand{\apjl}{ApJL}
\newcommand{\pasp}{PASP} 
\newcommand{\aj}{AJ}
\newcommand{\araa}{ARA\&A}

\newcommand{\mnras}{MNRAS}
\newcommand{\aap}{A\&A}
\newcommand{\nat}{Nature}
\newcommand{\apss}{Ap\&SS}
\newcommand{\physrep}{Phys. Rep.}
\newcommand{\iaucirc}{IAU Circ.}
\newcommand{\aaps}{A\&AS}

\begin{document}
\begin{titlepage}
  \parindent=0em
  \sffamily
  \hspace{-0.3in}  \includegraphics[height=1.2in]{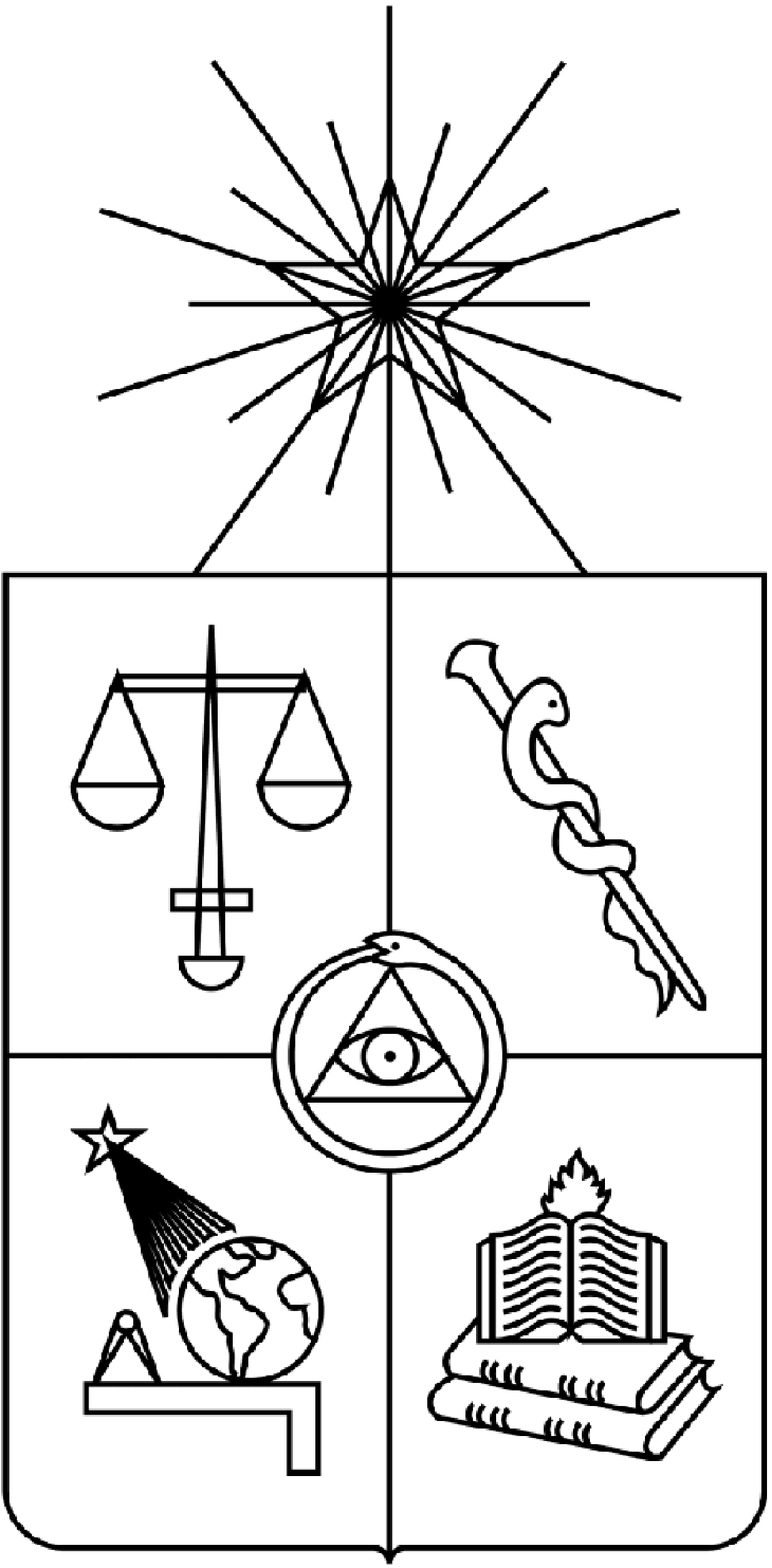}
  \begin{flushleft}

    {\bfseries

      \vspace{-1.15in}

	     {\Large \hspace{1.4cm}UNIVERSIDAD DE CHILE}\\

	     {\large \hspace{1.4cm}FACULTAD DE CIENCIAS FISICAS Y MATEMATICAS\\

	       \hspace{1.4cm}DEPARTAMENTO DE ASTRONOMIA\\}

    }

  \end{flushleft} 
  
  \vspace{\baselineskip}

  \begin{center}
    {\Large \bfseries CONFRONTANDO MODELOS HIDRODINAMICOS
CON OBSERVACIONES DE  SUPERNOVAS DE TIPO  II-PLATEAU \\
    \normalsize \vspace{2\baselineskip}
    TESIS PARA OPTAR AL GRADO DE DOCTOR EN CIENCIAS\\MENCION ASTRONOMIA\\
      \vspace{1cm} \large
      MELINA CECILIA BERSTEN\\}
  \end{center}
  \vspace{2\baselineskip}
  \vspace{-0.5cm}
  \begin{center}
    {\bf PROFESOR GUIA:}\\
    Mario  Hamuy Wackenhut\\
    \vspace{\baselineskip}
	   {\bf MIEMBROS DE LA COMISION:}\\
	   Omar Benvenuto\\
           Alejandro Clocchiatti\\
           Andr\'es Escala Astorquiza\\
	   Jos\'{e} Maza Sancho\\
\vspace{-0.1cm}	   
  \end{center}
  \begin{center}
    \vspace{\baselineskip}
    SANTIAGO DE CHILE\\
    {\small JULIO 2010}
  \end{center}

\newpage
\noindent
Nota: El Prof.\,Omar Benvenuto fue co-director de esta tesis. Por
motivos administrativos, aparece como miembro de la comisi\'on
evaluadora.

\noindent
Note: Prof.\,Omar Benvenuto was co-director of this thesis. For 
administrative reasons, he appears as a member of the evaluation
committee. 

\end{titlepage}

\newpage
\newpage
\frontmatter

\chapter{Resumen}
\label{ch:abs}

El objetivo principal de esta tesis es el desarrollo de
un c\'odigo hidrodin\'amico unidimensional para modelar curvas de luz
bolom\'etricas de supernovas de tipo II plateau (SNs~II-P) con el
prop\'osito final de analizar las propiedades f\'isicas de los
progenitores de estas supernovas comparando nuestros modelos con una
gran base de datos de muy buena calidad y cobertura temporal.\\ 
Para poder comparar nuestro modelo con las observaciones hemos 
derivado calibraciones para la correcci\'on bolom\'etrica (CB) y la temperatura
efectiva basadas en fotometr\'ia $BVI$. La dispersi\'on t\'ipica de la
calibraci\'on encontrada para la CB conduce a una
incerteza de 0.05 dex en la luminosidad.\\
Usando nuestro c\'odido hemos estudiado en detalle el caso de la
SN~1999em, una de las SNs~II-P mejor observadas, y
encontramos un muy buen acuerdo con las observaciones usando los
siguientes par\'ametros f\'isicos: $E=1.25$ foe, $M=
19 \,M_\odot$, $R= 800 \,R_\odot$, and $M_{\mathrm {Ni}}=0.056
\,M_\odot$. En este an\'alisis vemos que es necesario usar una
extensa mezcla de $^{56}$Ni para reproducir un plateau tan chato como
muestran las observaciones.\\
Hemos calculado curvas de luz bolom\'etricas para todas las SNs
de nuestra muestra y definido una serie de par\'ametros que nos
permiten caracterizarlas. Comparando dichos
parametros encontramos un rango
de $1,15$ dex en la luminosidad del plateau, un rango de duraci\'on del
plateau de 64 a 103 d\'ias y masas de niquel inferiores a $0,1$
 M$_\odot$ para todas las SNs en nuestra muestra, con la
 excepci\'on de SN~1992am.\\ 
Finamente, hemos calculado una grilla de 
 modelos hidrodin\'amicos para estudiar la dependencia de diferentes
 observables con los par\'ametros f\'isicos (masas, radio y energ\'ia
de la explosi\'on) y las correlaciones entre distintos
observables. A partir de este an\'alisis vemos que nuestros modelos
reproducen la fuerte correlaci\'on observada entre la luminosidad y la
velocidad de expansi\'on durante el plateau.

\newpage
\section*{Agradecimientos}
Quiero agradecer en primer lugar a Gast\'on Folatelli por su 
apoyo constante a lo largo de esta tesis no s\'olo desde el punto de vista
emocional si no tambi\'en cient\'ifico.
 A mis directores, Omar Benvenuto por su incondicional apoyo y  por
  toda la  experiencia cient\'ifica que me ha transmitido;  y Mario Hamuy por
 haberme impulsado constantemente a la b\'usqueda de un poco m\'as y
 tambi\'en por las oportunidades que me ha brindado.  A ambos les
 agradezco la confianza 
 y la independencia con las que me han dejado trabajar.
 
Quiero agraceder profundamente a Ken'ichi Nomoto por haberme invitado al
IPMU, experiencia que fue fundamental para terminar de
madurar varios de los temas tratados en esta tesis. Un especial
agradecimiento a Sergei Blinnikov  por todas las conversaciones,
sugerencias y por el inter\'es puesto en mi trabajo; y tambi\'en a
Elena Sorokina. Con ambos compart\'i  unos meses muy intensos con
hermosas tardes de t\'e, ciencia y otros temas. Tambi\'en me gustar\'ia
agradecer a Nozomu Tominaga por su ayuda para testear mi modelo. A
Giuliano Pignata por ser el ide\'ologo de mi estad\'ia en 
Jap\'on y a Alejandro Clocchiatti por su apoyo y el permanente
entusiasmo puesto en  mi trabajo.  

No quisiera dejar de agradecer a mi familia, en especial a mi abuela y
a mi hermana Lara. A  mis amigos de La Serena, Julia,
Nidia y Rodolfo, de Santiago en especial a Cristian y Maria Jos\'e y  
por supuesto  tambi\'en a mis amigos de mi querida y extra\~nada La
Plata, a  Andrea F., Andrea T.,  Cecilia, Mariana, Martin, Ver\'onica,
Ricardo  y  a Ramiro  por sus hermosas visitas. Gracias amigos por
aguantar mis descargas y transmitirme siempre alegr\'ia y confianza.

\listoffigures
\listoftables
\tableofcontents
\mainmatter
%
\chapter{Introduction}\label{ch:intro}
\section{Supernova Background}
\label{SNbak}
A supernova ( SN, hereafter) explosion is among the most spectacular
events that can be observed in the Universe.
The term ``supernova'' is somewhat misleading, as such an 
event does not represent  a new star (that is, a ``nova''), but instead the
death of  a star.

The first records on supernovae (SNe) goes back to the second century
A.D., mainly by Chinese astronomers. They observed and recorded 
``guest stars'' , i.e., stars that suddenly appeared in the 
  sky, were visible for a 
certain length of time, and then faded away. The ``guest stars'' that
were visible for 
a year or longer were probably SNe, while the shorter guests were common novae.
Among the most famous historical SNe are the events of  1054 A.D. (the ``Crab
Nebula''),  1572 A.D. (``Tycho's supernova''), and  1604 A.D. (``Kepler's 
supernova'' which appeared only few years  before the invention of the
telescope).
Another SN, known as Cas A (Cas for Cassiopeia), exploded in our
galaxy between 1650 
and 1680. Its remnant is a very strong radio source but, intriguingly,
was not reported by  contemporary observers.

The occurrence of this phenomenon  became more  familiar to astronomers
at the end of the 19th century, with the beginning of the use of 
photographic plates and systematic surveys. Astronomers
designated this objects with the genuine name ``nova''. By 1927
$\sim$100 novae had been discovered in the magnitude range
16--18. However, it was not until the distances to the nearby
galaxies were firmly established through the observation of Cepheid
variables (Hubble, 1925) that the separation of ordinary
novae and the  extraordinary luminous SNe, become clear. 

The first physical explanation for SNe was  proposed by Baade and Zwicky
(1934). They suggested that the SN energy comes from the collapse of
ordinary stars into neutron stars. This idea is a prescient
suggestion for the fate of massive stars, but it is not the correct
explanation for all types of SNe. Nature has more than one way to
explode a star. This was revealed clearly by Minkowski (1941) who
noted that the spectra of SNe indicated at least two different
types of objects. One group, which he named ``Type I'', was
characterized by broad emission 
features and no signs of hydrogen in the spectrum, while the rest, named
``Type II'', showed strong, broad hydrogen lines dominating the
spectrum. Since that time, the 
classification of SNe has evolved parallel to the increasing body of
spectroscopy data and our knowledge about these objects. In the mid-1980s it
was realized that the Type I group could be further differentiated
according to the presence of the Si II $\lambda$6355 line.  If this
line was present in the spectra, the SN was classified as Type Ia; otherwise
it was designated as Type Ib or Ic, depending on the presence or
absence of He lines, respectively. Another critical piece of
information about the nature of these 
objects was provided by the stellar environments where they were
discovered. Supernovae of types II, Ib and Ic are only observed near
star forming regions in late-type galaxies, which indicates their
association with young stellar environments, while Type Ia SNe
(SNe~Ia) occur
both in elliptical and spiral galaxies. Currently, it is accepted that
there are two different 
physical types: the thermonuclear explosion of a white dwarf star in a
binary system, and the gravitational collapse of the core of
a massive star, as suggested by Baade and Zwicky (1934). The latter
are called ``core-collapse SNe'' (CCSNe) and observationally
correspond to SNe~II, Ib and Ic. SNe~Ia correspond to  the first group
of thermonuclear SNe.

\begin{figure}[htbp]
\begin{center}
\includegraphics[scale=.50]{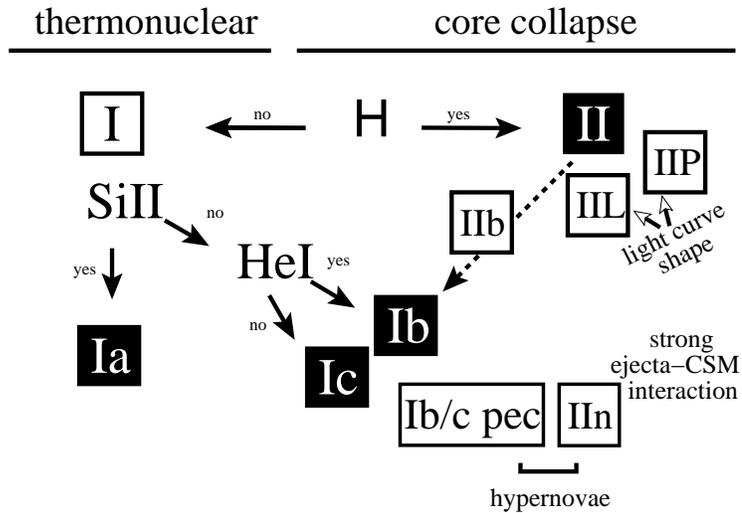}
\caption[SNe classification]{The current classification scheme of
  supernovae. Type Ia SNe are associated with the thermonuclear
  explosion of accreting white dwarfs. Other SN types are associated
  with the core collapse of massive stars. CCSNe
  with explosion energies \\ E $> 10^{52}$erg are often called
  hypernovae. \label{SNclassification}}   
\end{center}
\end{figure}

Unlike   SNe~Ia, CCSNe form a diverse group in terms of their
spectral and photometric properties.  SNe~II have been
subdivided according to their light curve shapes into: II-Plateau
(showing plateau-like light curves), and II-Linear (showing linearly
declining light curves). Further subgroups arise from spectroscopic
properties:  IIn (showing narrow emission lines), and IIb (showing
hydrogen initially but afterward evolving to 
hydrogen-deficient at later
times). There are some SNe~Ib/c associated with long gamma-ray bursts
\citep{2006ARA&A..44..507W} which show much broader lines than
the typical objects of this class. They have 
been referred to as ``hypernovae'' or broad-lined SNe, due to the large
inferred kinetic energies. 
Figure~\ref{SNclassification} shows a diagram of the
current SN classification, as provided by
\citet{2003LNP...598...21T}. A review of SN classification criteria is
given by \citet{1997ARA&A..35..309F}. It should be noted that, because
of its empirical nature, the current classification scheme is 
prone to evolve as more data are obtained. 

A typical SN has a luminosity of $\sim$$10^{10} L_\odot$ during a
period of weeks to months, which is comparable to the luminosity of its host
galaxy. This radiative energy corresponds to only $0.01$\% of their
typical kinetic energy which is $\sim$$10^{51}$ erg (a quantity
commonly known as 1 foe or, more recently, as 1 Bethe). The 
kinetic energy is essentially determined from the observed expansion
velocities, as measured from spectral lines. The characteristic  broad
P-Cygni profiles in SN spectra, with mean expansion velocities
of the order to $ 10^4$ km s$^{-1}$, is an indicator of fast-expanding
atmospheres. In addition, the observed temperatures near maximum brightness
are $\sim$2 $T_\odot$, (where $T_\odot$ stand for the
  effective temperature of the  Sun),  which implies radii $\sim$$10^5$
R$_\odot$, assuming spherical symmetry and that the SN radiates as a
black body. For the case of CCSNe, the kinetic energy represents only 
1\% of the energy released during the explosion. The remaining 99\%
($\sim$$10^{53}$ erg) is carried away by neutrinos created during the
collapse of the core. Such an explosion mechanism was first confirmed
with the detection of neutrinos from SN~1987A. This famous SN
discovered on February 23, 1987 in the Large Magellanic Cloud (LMC)
was the brightest SN detected since the invention of the telescope and
one of the most extensively observed stellar objects in the history of
astronomy. It was 
observed in all wavelengths from gamma rays to radio wavelength, and for the
first time, extra solar neutrinos were detected. It presented a
peculiar light curve shape which we know now that was due to the blue
supergiant nature of the progenitor star, later identified on photographic
plates of the LMC obtained prior to the explosion.

During the SN explosion a fraction of the progenitor's material is
 thermonuclearly  burned into various 
elements. Among these products, unstable isotopes of iron-group elements play an
important role in the subsequent evolution of the object. The most abundant
radioactive isotope produced is $^{56}$Ni which decays  in $^{56}$Co
that in turn decays  to stable $^{56}$Fe. This process provides an
additional source of energy 
which is thermalized in the ejecta and later released during a period
of weeks to 
months.

Given their large intrinsic luminosity, SNe have long been considered
potential probes for extragalactic distance determination and thus for
the measurement of cosmological parameters that describe the expansion
history of the Universe. Historically, SNe~Ia have 
been used for this purpose due to their high degree of
homogeneity. Through an empirical correlation found by
\citet{1993ApJ...413L.105P} between the absolute peak luminosity 
and the width of the light curve, SNe~Ia can be used to determine
distances with a precision of $\sim 7\%$
\citep{1996AJ....112.2398H}. One decade ago, the 
application of this method to high-redshift SNe~Ia in comparison
with the low-redshift sample of the Cal\'an/Tololo survey led to the
remarkable finding of the {\em accelerated\,} expansion of the Universe
\citep{1998AJ....116.1009R,1999ApJ...517..565P}. Although, SNe~II-P
are fainter and less homogeneous than SNe~Ia, they have also been
proposed as good distance indicators with potential 
application to cosmology, in an independent way as
SNe~Ia. The most popular methods in the literature are the  
expanding photosphere method \citep[EPM;][]{E96,D05}, the
spectral-fitting  expanding atmosphere method  \citep[SEAM;][]
{2002ApJ...574..293M,2004ApJ...616L..91B}, and  the
standard-candle method \citep[SCM;][]{HP02}. The former two methods
require detailed atmospheric modeling, while the SCM is a much
simpler approach based on an empirical correlation between luminosity
and expansion velocity at the middle of the plateau phase. In
principle, EPM and SEAM provide distances which are independent of the
standard cosmic 
``distance ladder'', whereas SCM requires an external calibrator in
order to provide absolute distances. By comparing low- and
high-redshift samples, SCM can be
potentially  employed to determine cosmological parameters using
data from upcoming SN surveys.

Recently, a compilation of all SN discoveries within a volume of 28 Mpc
during a period of time of 10.5 yr has provided us the relative rates
of each sub-type of SNe \citep{2009MNRAS.395.1409S}. This study showed
that SNe~Ia represent only 27\% of all SNe. The rest are CCSNe with the 
following rates: 59\% for SNe~II-P, 29\% for SNe~Ib/c, 5\% for
SNe~IIb, 4\% for SNe~IIn and 3\% for SNe~II-L. 

The goal of this thesis is the study of SNe~II-P and their
progenitors. In  \S~\ref{sec:SNII-P} I
review our current
knowledge on these objects focusing on the motivations for this work. In
\S~\ref{sec:goals}, I summarize the main goals of  this thesis and I give a
brief overview of each of its chapters. 

\section{Type II-Plateau SNe}
\label{sec:SNII-P}

\begin{figure}[htbp]
\begin{center}
\includegraphics[scale=.8,height=7cm]{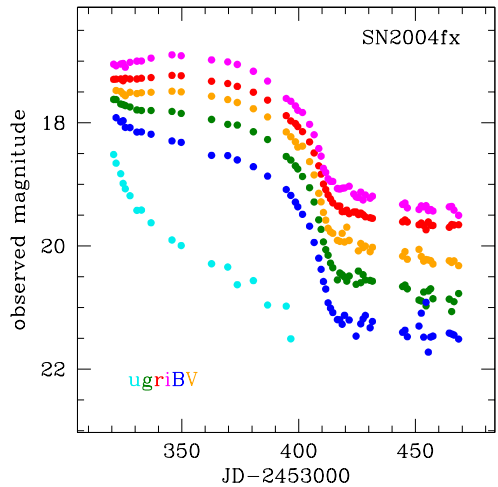}
\caption[LC of SN~II-P]{Light curve SN~2004fx, a typical SN~II-P\label{LCSNIIP}}
\end{center}  
\end{figure}

\begin{figure}[htbp]
\begin{center}
\includegraphics[angle=-90,scale=.8,width=8cm]{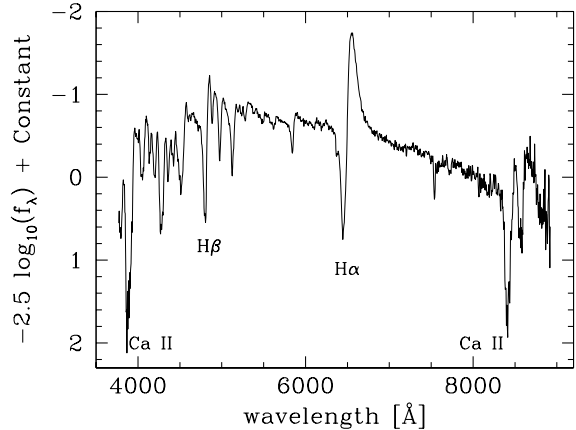}
\caption[Spectra of SN~II-P]{Spectra of SN~2004fx, a typical
  SN~II-P\ \label{ESPSNIIP}}   
\end{center}
\end{figure}

Type~II-Plateau SNe (SNe~II-P) form a well-defined family
characterized by a ``plateau'' in the optical light curve 
\citep{1979A&A....72..287B}, where
the luminosity remains nearly constant for a period of $\sim$100 days
(see Figure~\ref{LCSNIIP}), and the presence of prominent P-Cygni hydrogen
lines in the spectrum (see Figure~\ref{ESPSNIIP}). They are a subclass of
the core-collapse SNe (CCSNe) ---which includes Type Ib,  Type Ic, and other
subclasses of type II SNe--- originated by the violent death of
stars with initial masses greater than 8 $M_\odot$
\citep{2003ApJ...591..288H,2009MNRAS.395.1409S} and sharing, in
general terms, the same explosion mechanism. It has been shown that
SNe~II-P are the most common type of SNe in nature, amounting to 59\%
of all CCSNe. Apart from their astrophysical importance in connection
with stellar evolution and the physics of the interstellar medium,
additional interest on SNe~II-P has recently arisen from the fact that
they have been established as good distance indicators with potential
application to cosmology, independent of Type Ia
SNe (see \S~\ref{SNbak}).

Massive stars may
suffer considerable mass loss during the early phases of their evolution,
due to strong stellar winds or transfer to  binary companions.
Thus, they may lose part or all of their outermost envelope of unprocessed
hydrogen and helium. The vast diversity  
observed among CCSNe is related with the properties of the  
progenitor star. In the
current picture, SNe~II result from progenitors which are able to
retain a significant 
fraction of their external hydrogen layers and possible have the least
massive progenitor of all CCSNe.  This picture is
combined with hydrodynamical 
models of SNe~II-P,  which show that a red 
supergiant progenitor with an extensive H envelope is necessary in
order to reproduce the plateau-shaped light curves 
\citep{1971Ap&SS..10...28G,1977ApJS...33..515F,1976ApJ...207..872C}. Recent
direct detection of the progenitors of several  
SNe~II-P suggest in fact that SNe~II-P arise from red supergiant stars
with $M_{ZAMS} < 18 M_\odot$
(\citet{2003PASP..115.1289V,2004Sci...303..499S} and \citet{2009MNRAS.395.1409S}). 

There is general agreement that the explosion of a massive star is originated
by the collapse of its central parts into a neutron star or black hole
when the iron core is formed at the end of the star evolution, and
further nuclear burning no longer provides thermal pressure to support
the star. During the explosion, the heavy elements (mostly
$\alpha$-nuclei) synthesized inside the SN progenitor over
its life are spread at very high speed into the interstellar
medium. Despite the intensive  theoretical
modeling done during  recent years \citep[see][and references
  therein]{2006NewAR..50..487B,2007PhR...442...38J} the mechanism that
transfers the energy released during the collapse of the core to the
envelope, still remains unknown. The
approach usually adopted to model CCSNe is to decouple the
explosion into two independent parts: a) the core collapse and the
subsequent formation of a shock wave, and b) the ejection of the
envelope. Based on the analysis of 
the propagation of the shock wave through the envelope, independently
of how the shock is formed, it is possible to study the observational
outcome of the explosion such as light curves and spectra. This
approach has been extensively used \citep[][among
  others]{1977ApJS...33..515F,
  1971Ap&SS..10...28G,1988ApJ...330..218W} and has led to the 
  conclusion that 
the main factors influencing the outcome are the explosion
energy and the progenitor properties (mass and radius). 

Observationally, SNe~II-P show a wide range 
of plateau luminosities ($L_p$) and durations ($\Delta t_p$),
expansion velocities ($v_{exp}$), and
nickel masses ($M_{\mathrm Ni}$) \citep{1989ApJ...342L..79Y,H01}. The
morphology of the light curve (LC) of SNe~II-P has been shown to be
connected with physical properties of the progenitor object such as 
ejected mass ($M$), explosion energy ($E$), and 
pre-supernova radius ($R$). The relations
between these physical and observable parameters were first derived
analytically by \citet{1980ApJ...237..541A} and 
then generalized by \citet{1993ApJ...414..712P}. Numerical
calibrations of these relations were then given by
\citet{1983Ap&SS..89...89L,1985SvAL...11..145L} (LN83 and LN85 
hereafter, respectively) based on a grid of hydrodynamical models for
different values of $M$, $R$ and $E$.  
\citet{2003ApJ...582..905H} and \citet{2003MNRAS.346...97N} applied 
such calibrations to a set of $\sim$20 SN~II-P and thereby derived masses,  
radii and explosion energies for their sample. However, these
studies have not been fully satisfactory owing to 1) the lack of
good-quality data, 2) the use of simplified relations between
ill-defined and hard-to-measure photometric and spectroscopic parameters, and 
3) the fact that some of the models are based on simplified physical
assumptions. Some of the weaknesses of the LN83 and LN85 models are that
they did not include the effect of nickel heating in their calculations, the
use of old opacity tables, the neglect of any effect from 
line opacities, and the simplified  initial
pre-supernova models adopted. 

 This thesis was motivated by
the need to remedy such problems in order 
to improve our knowledge of the physical properties of SN~II-P
progenitors. Furthermore, the availability of an enlarged data
set of $BVI$ light curves and spectra of $\sim$30 SNe~II-P
  \citep{H09} justifies a detailed study. To 
accomplish our purpose, we have developed 
our own hydrodynamical model \citep{BBH09} which allows us to perform a
comparison between models and data in a consistent way.

During the development of this thesis there have been important advances in
this field. For example, \citet{2007A&A...461..233U} performed a
detailed analysis of SN~1999em and also studied
how several physical parameters affect the LC. In that work, the
author provided relations between physical and observed parameters for
 SNe~II-P similar  to SN~1999em. More recently,
 \citet{2009ApJ...703.2205K} calculated a 
set of model LC and spectra of SNe~II-P, for different masses,
metallicities, and explosion energies. They used their models to describe 
the dependence of plateau luminosity and duration on explosion
energy and progenitor mass. Nevertheless, the relations they found are
simple and easy to apply {\em only} in the extreme case of no
$^{56}$Ni production. When $^{56}$Ni is considered, the relations
involve more parameters,  
which hinders their applicability to obtain physical parameters from
observations. Other works have also 
explored the effect of several physical parameters on the LC and
other observational properties of SNe~II-P
\citep{2003MNRAS.345..111C,2004ApJ...617.1233Y}, although without
attempting to derive general relations. In  recent years, there has
been a number of works which analyzed the physical properties of
individual SNe~II-P based on hydrodynamical models \citep[][among others]
  {2005AstL...31..429B,2009A&A...506..829U}.    
 
In addition to the comparison with hydrodynamical models,
there is an alternative way to derive progenitor masses of
SNe~II-P, namely, using pre-supernova
images to search for the progenitor star and measuring its brightness
and color to derive a mass in connection with a stellar evolution
model. The last decade has witnessed direct discoveries of several SN
progenitors providing a remarkable and rapid progress in this
field. Currently, there are three SNe~II-P (SN~2003gd, SN~2005cs and
SN~2008bk)  with a firm detection of 
the progenitor  and $\sim$17 others without a positive detection (see
\citet{2009MNRAS.395.1409S} for a review). The latter cases are still 
useful since they provide upper limits to the progenitor masses. These
studies have suggested a minimum of the stellar mass for SNe~II-P of
8$\,\pm$ 1$M_\odot$ and a maximum of 16.5 $\, \pm$
1.5$M_\odot$. Although there seems to be  consensus on the lower mass
limit of mass for SNe~II-P, the upper limit derived by this
method is systematically lower than the  progenitor masses derived
by modelling the light curves, as noted by
\citet{2008A&A...491..507U} and
\citet{2009arXiv0908.0700S}. Specifically, there are three SNe~II-P
with hydrodynamical masses which have 
been studied in pre-supernova imaging (namely, SN~1999em,  
SN~2004et and SN~2005cs) and a few more that have been studied using
semi-analytical models. In all cases the mass estimated by
hydrodynamical models
is higher than the estimate or upper limit given by pre-supernova
imaging. This discrepancy poses a very interesting and unsolved
problem which is necessary to address.   

\section{This thesis}
\label{sec:goals}
Supernovae are very relevant astrophysical objects: they
affect the energetic and chemical evolution of
galaxies, and are also important distance indicators for
cosmology. Determining what ranges of physical parameters (masses, 
radii and energies) give rise to each type of CCSNe is still an
 open question. One way to estimate
such parameters is by comparing observations 
(light curves, colors, spectra) with hydrodynamical models. In addition
to this, the LC provides a useful tool to test 
models of stellar evolution of massive stars and unveil the
internal structure (chemical composition, mixing and mass
distribution) of the progenitor object. Furthermore, models may provide a
theoretical explanation for the correlations that are employed in the
determination of distances to SNe~II-P, and guide future refinements of
this technique.

The main goal of this thesis is the development of a one-dimensional,
Lagrangian hydrodynamic code to model bolometric light curves of
SNe~II-P with the ultimate purpose of analyzing the physical properties of
an available large sample of SNe with highly-precise, well-sampled
observations.

I have organized the thesis as follows. In Chapter~\ref{ch:database},
I give a description of the sample of $\sim$30
SNe~II-P used in this study, the sources of the data, corrections
for extinction and  distances. In Chapter~\ref{ch:code}, I present
our one-dimensional, flux-limited, Lagrangian hydrodynamical code
 for the  modeling of bolometric light curves of SNe~II-P. The
 micro-physics and pre-supernova models used in our calculation are
 also described in  Chapter~\ref{ch:code}. More technical details of
 the calculations are given in appendices~\ref{ch:anexoA} and
 \ref{ch:anexoB}. In Chapter~\ref{ch:model_obs}, I explain the method
 used to compute bolometric corrections (BC) for SNe~II-P which are
 necessary  to derive bolometric light curves from $BVI$ photometry,
this allows us to compare our model with observations. In
Chapter~\ref{sec:sn99em}, I analyze the case of the prototype
SN~1999em,  and discuss  how our model compares with previous
hydrodynamical studies of this object. In
Chapter~\ref{ch:ana}, I derive  bolometric LC for our
sample of SNe and derive physical parameters for this sample. I also
study the correlations between different observables. Finally, in
Chapter~\ref{ch:summ}, I summarize the main 
conclusions of this work.

\chapter{Sample of supernovae}
\label{ch:database}
One of the main motivations for this thesis is the availability of an
unprecedented data set of $BVI$ light curves and spectra of
$\sim$35 SNe which offers the opportunity  
to significantly improve our understanding of SNe~II-P and their
progenitors. This sample of SNe~II-P was observed in the course of
four systematic follow-up programs: the {\em Cerro Tololo Supernova
  Program} (1986-2003),  
the {\em Cal{\'a}n/Tololo Supernova Program} (CT; 1990-1993), 
the {\em Optical and Infrared Supernova Survey} (SOIRS; 1999-2000), 
and the {\em Carnegie Type II Supernova Program} (CATS;
2002-2003). Additionally, I included four SNe  
from the literature: SN 1999gi \citep{2002AJ....124.2490L}, 
SN 2004dj \citep{2006MNRAS.369.1780V}, SN 2004et
\citep{2006MNRAS.372.1315S}, and SN 2005cs
\citep{Pa06,2006A&A...460..769T}. Complementary photometry for
SN~2003gd obtained by \citet{2003PASP..115.1289V} and
\citet{2005MNRAS.359..906H} was also incorporated.  A
complete listing of the SNe used in this work is given in
Table~\ref{sampleSNe} where I present the SN,  host galaxy name, the
SN equatorial coordinates, the heliocentric redshift of the host
galaxy and their sources, the extinction due to our own Galaxy
\citep{SFD98} and the corresponding reference for the photometric and
spectroscopic data of each supernova. 

\begin{table}[htpp]
\begin{center}
{\scshape \caption{List of Type II Supernovae}\label{sampleSNe}}
\scriptsize
\vspace{3mm}
\begin{tabular}{lccccccc}
\hline\hline
SN Name  &  Host Galaxy  &  $\alpha$(J2000)  &   $\delta$(J2000)  &  $z_{\mathrm{host}}^{\mathrm{a}}$   & (s)$^{\mathrm{b}}$ & $E(B-V)_{\mathrm{Gal}}$ & References \\
\hline
1991al   &  LEDA 140858        & 19 42 24.00    & -55 06 23.0 &  0.01525 &  HP02 &   0.051  &   1  \\  
1992af   &  ESO 340-G038       & 20 30 40.20    & -42 18 35.0 &  0.01847 &  NED	 &   0.052  &   1  \\ 
1992am   &  MCG -01-04-039     & 01 25 02.70    & -04 39 01.0 &  0.04773 &  NED  &   0.049  &   1  \\ 
1992ba 	 &  NGC 2082           & 05 41 47.10    & -64 18 01.0 &  0.00395 &  NED	 &   0.058  &   1  \\ 
1993A  	 &  anonymous          & 07 39 17.30    & -62 03 14.0 &  0.02800 &  NED	 &   0.173  &   1  \\  
1999br   &  NGC 4900	       & 13 00 41.80    & +02 29 46.0 &  0.00320 &  NED	 &   0.024  &   2  \\  
1999ca   &  NGC 3120           & 10 05 22.90    & -34 12 41.0 &  0.00931 &  NED  &   0.109  &   2  \\
1999cr   &  ESO 576-G034       & 13 20 18.30    & -20 08 50.0 &  0.02020 &  NED	 &   0.098  &   2  \\  
1999em 	 &  NGC 1637 	       & 04 41 27.04    & -02 51 45.2 &  0.00267 &  NED	 &   0.040  &   2  \\  
1999gi 	 &  NGC 3184 	       & 10 18 17.00    & +41 25 28.0 &  0.00198 &  NED	 &   0.017  &   3  \\  
200210   &  MCG +00-03-054     & 01 01 16.80    & -01 05 52.0 &  0.05140 &  NED  &   0.036  &   4  \\
2002fa   &  NEAT J205221.51    & 20 52 21.80    & +02 08 42.0 &  0.06000 &  NED  &   0.099  &   4  \\
2002gw 	 &  NGC 922 	       & 02 25 02.97    & -24 47 50.6 &  0.01028 &  NED	 &   0.020  &   4  \\  
2002hj   &  NPM1G+04.0097      & 02 58 09.30    & +04 41 04.0 &  0.02360 &  NED  &   0.115  &   4  \\
2002hx 	 &  PGC 23727 	       & 08 27 39.43    & -14 47 15.7 &  0.03099 &  NED	 &   0.054  &   4  \\  
2003B 	 &  NGC 1097 	       & 02 46 13.78    & -30 13 45.1 &  0.00424 &  NED	 &   0.027  &   4  \\  
2003E 	 &  MCG -4-12-004      & 04 39 10.88    & -24 10 36.5 &  0.01490 &  J09  &  0.048   &   4  \\ 
2003T 	 &  UGC 4864 	       & 09 14 11.06    & +16 44 48.0 &  0.02791 &  NED	 &   0.031  &   4  \\  
2003bl 	 &  NGC 5374 	       & 13 57 30.65    & +06 05 36.4 &  0.01459 &  J09  &  0.027   &   4  \\ 
2003bn 	 &  2MASX J10023529    & 10 02 35.51    & -21 10 54.5 &  0.01277 &  NED	 &   0.065  &   4  \\  
2003ci 	 &  UGC 6212 	       & 11 10 23.83    & +04 49 35.9 &  0.03037 &  NED	 &   0.060  &   4  \\  
2003cn 	 &  IC 849 	       & 13 07 37.05    & -00 56 49.9 &  0.01811 &  J09  &  0.021   &   4  \\ 
2003cx 	 &  NEAT J135706.53    & 13 57 06.46    & -17 02 22.6 &  0.03700 &  NED	 &   0.094  &   4  \\  
2003ef 	 &  NGC 4708 	       & 12 49 42.25    & -11 05 29.5 &  0.01480 &  J09  &  0.046   &   4  \\ 
2003fb 	 &  UGC 11522 	       & 20 11 50.33    & +05 45 37.6 &  0.01754 &  J09  &  0.183   &   4  \\ 
2003gd 	 &  M74 	       & 01 36 42.65    & +15 44 20.9 &  0.00219 &  NED	 &   0.069  &   4,5,6  \\  
2003hd 	 &  MCG -04-05-010     & 01 49 46.31    & -21 54 37.8 &  0.03950 &  NED	 &   0.013  &   4  \\  
2003hg 	 &  NGC 7771 	       & 23 51 24.13    & +20 06 38.3 &  0.01427 &  NED	 &   0.074  &   4  \\  
2003hk 	 &  NGC 1085 	       & 02 46 25.76    & +03 36 32.2 &  0.02265 &  NED	 &   0.037  &   4  \\  
2003hl 	 &  NGC 772 	       & 01 59 21.28    & +19 00 14.5 &  0.00825 &  NED	 &   0.073  &   4  \\  
2003hn 	 &  NGC 1448 	       & 03 44 36.10    & -44 37 49.0 &  0.00390 &  NED	 &   0.014  &   4  \\  
2003ho 	 &  ESO 235-G58        & 21 06 30.56    & -48 07 29.9 &  0.01438 &  NED	 &   0.039  &   4  \\  
2003ip   &  UGC 327            & 00 33 15.40    & +07 54 18.0 &  0.01801 &  NED  &  0.066   &   4  \\
2003iq 	 &  NGC 772 	       & 01 59 19.96    & +18 59 42.1 &  0.00825 &  NED	 &   0.073  &   4  \\  
2004dj   &  NGC 2403           & 07 37 17.00    & +65 35 58.1 &  0.00044 &  NED  &  0.040   &   7  \\
2004et   &  NGC 6946           & 20 35 25.30    & +60 07 18.0 &  0.00016 &  NED  &  0.342   &   8  \\
2005cs   &  NGC 5194	       & 13 29 53.40    & +47 10 28.0 &  0.00154 &  NED  &  0.035   &   9,10   \\
\hline
\end{tabular}

\begin{list}{}{}
\item [$^{\mathrm{a}}$] Heliocentric host-galaxy redshifts
\item [$^{\mathrm{b}}$] Sources of host-galaxy redshifts. HP02: \citet{HP02}; J09: \citet{2009ApJ...696.1176J}; NED: NASA/IPAC Extragalactic Database
\end{list}

\begin{list}{}{}
\item References. (1) {\em Cal{\'a}n/Tololo Supernova Program}; (2) SOIRS; (3) \citet{2002AJ....124.2490L}; (4) CATS; (5) \citet{2003PASP..115.1289V}; (6) \citet{2005MNRAS.359..906H}; (7) \citet{2006MNRAS.369.1780V}; (8) \citet{2006MNRAS.372.1315S}; (9)\citet{Pa06}; (10) \citet{2006A&A...460..769T}
 \end{list}
\normalsize
\end{center}
\end{table}

\section{Observations} 
\label{sec:observations}
The observations were made with telescopes from  {\em  Cerro Tololo
  Inter-American Observatory} (CTIO),  {\em Las Campanas Observatory}
 (LCO), the {\em European Southern Observatory} (ESO) in La Silla and  
the {\em Steward Observatory} (SO). Table~\ref{telescopes}  shows
the telescopes and instruments used to obtain the
photometry and spectroscopy of this data set. CCD
detectors and standard Johnson-Kron-Cousins $UBVRIZ$ filters
\citep{1966CoLPL...4...99J,1971ROAn....7.....C,H01} were employed in all
the cases, and for a small subset of SNe, observations in the $JHK$ filters
were also obtained. Additionally, low resolution optical spectra were
taken for each SN at various epochs using CCD detectors in
combination with different gratings/grisms and blocking filters.
Currently, all of the optical data and spectra have been
reduced and they are in course of publication \citep{H09}. The data
reductions were performed using IRAF\footnote{IRAF is distributed by
the National Optical Astronomy Observatories, which are operated by
Association of Universities for Research in Astronomy, Inc., under
cooperative agreement with the the National Science Foundations}
procedures including, among others, the host galaxy subtraction \citep[for
more details, see also][]{H01}.

\begin{table}[htpp]
\begin{center}
\scriptsize
{\caption[Telescopes and instruments]{Telescopes and
    instruments used in the photometric and spectroscopic
    observations}\label{telescopes}} 
\vspace{3mm}
\begin{tabular}{lcc}
\hline\hline
Telescope  &  Instrument  & Spec/Phot$^{\mathrm {a}}$ \\
\hline
CTIO 0.9m    &  CCD      &  P   \\
YALO 1.0m    &  ANDICAM  &  P   \\
YALO 1.0m    &  2DF      &  S   \\
CTIO 1.5m    &  CCD      &  P   \\
CTIO 1.5m    &  CSPEC    &  S   \\
Blanco 4.0m  &  CSPEC    &  S   \\
Blanco 4.0m  &  2DF      &  S   \\
Blanco 4.0m  &  CCD      &  P   \\
Swope 1.0m   &  CCD      &  P   \\
du Pont 2.5m &  WFCCD    &  S/P \\
du Pont 2.5m &  MODSPEC  &  S   \\
du Pont 2.5m &  2DF      &  S   \\
du Pont 2.5m &  CCD      &  P   \\
Baade 6.5m   &  LDSS2    &  S/P \\
Baade 6.5m   &  B\&C     &  S   \\
Clay 6.5m    &  LDSS2    &  S/P \\
ESO 1.52m    &  IDS      &  S   \\
Danish 1.54m &  DFOSC    &  S/P \\
ESO 2.2m     &  EFOSC2   &  S   \\
NTT 3.58m    &  EMMI     &  S   \\
ESO 3.6m     &  EFOSC    &  S   \\
Kuiper 61"   &  CCD      &  P   \\
Bok 90"      &  B\&C     &  S   \\ 
\hline
\end{tabular}

\begin{list}{}{}
\item $^{\mathrm {a}}$ Whether the instrument  was used for photometry (P),
  spectroscopy (S) or both (S/P).
 \end{list}
\normalsize
\end{center}
\end{table}

\section{AKA Corrections and Distances} 
\label{sec:AKA}
In order to study the physical properties of our SNe  and 
to compare them with each other it is necessary to correct the observed
magnitudes by host-galaxy extinction (A$_{host}$), redshift
($K$-correction), and Galactic extinction (A$_G$) dubbed AKA
corrections as described in detail by  \citet{F09}. It is also
important to place all  of the SNe in 
the same distance scale by correcting for systematic differences among 
 the different  distance estimate methods.

The AKA corrections and distances for most  SNe were
calculated by \citet{F09}. While the determination of Galactic
extinction is straightforward using
the IR dust maps of \citet{SFD98} and assuming a standard
reddening law as that given by \citet{Car89}, with $R_V=
3.1$, determining the
extinction due to host-galaxy dust is a more difficult task. \citet{F09}
addressed this issue by assuming that all SNe~II-P should evolve
to a similar asymptotic intrinsic color at the end of the plateau
phase. This is expected because of the hydrogen recombination
  nature of the photosphere. Therefore, color
indices measured at the end of the plateau can be used to determine
reddening by comparing with a sample of unreddened SNe~II-P. 

The distances to this sample are determined using the
Standardized Candle Method (SCM), with
the exception of SN~2003ef and SN~2005cs. For SN~2003ef 
I adopted the EPM distance estimated by \citet{2009ApJ...696.1176J},  
converted to the distance scale of \citet{F09} using 
the conversion coefficients given by the authors. The resulting  
value is $48.55\pm 18.4$ Mpc. For SN~2005cs, I used the distance modulus 
of $\mu= 29.62\pm 0.25$ mag given by Pastorello et al 2006,
which corresponds to $8.4\pm 1$ Mpc. Table~\ref{AD} shows host-galaxy
extinctions and distance moduli for our sample of SNe.

\section{Time Reference Frame} 
\label{sec:time}
An additional important step which is necessary in order to compare
the data of different SNe is to place them in a consistent reference
frame of time. This is achieved by establishing an origin of time for
each SN which places time intervals for the observations of all SNe
on  uniform basis. Time intervals are then corrected for
  time dilation using the redshift of each SN.

A natural choice of the origin of time would be
the moment of the explosion. However, there are only 12 SNe in the
sample with an accurate estimate of the  
explosion time ($t_0$) based on EPM and pre-explosion
imaging \citep{2009ApJ...696.1176J}.  These estimates were
obtained using two independent atmospheric models by \citet{E96} and
\citet{D05} (E96 and D05 
hereafter, respectively), which led to two 
different values of $t_0$ for each
SN. In Table~\ref{AD} I show both values of $t_0$ for those 12
SNe~II-P. I use additional information of the explosion time found in
the literature for SN~1991al, SN~199af, SN~1992am,
SN~1993A, SN~1999ca, and SN~1999cr 
\citep[see Table~2 of][]{2003ApJ...582..905H}. For SN~2003gd
a value of $t_0= 4716.5 \pm 21$ given by
\citet{2004Sci...303..499S}, for SN~2004dj a value of $t_0=5167 \pm
21$ given by 
\citet{2006MNRAS.369.1780V}, for SN~2004et a value of $t_0=5270.5 \pm
2$ \citep{2006MNRAS.372.1315S},
and finally for SN~2005cs a value of $t_0= 5549.9\pm 1$ given by
\citet{Pa06}. All these values of $t_0$ are given respect to
JD $=2448000$.  

An alternative approach was introduced by \citet{F09}. In such work,
the origin of time was defined as the moment at the middle point of
the transition 
between the plateau and the radioactive tail ($t_{PT}$). The values of
$t_{PT}$ for the sample of SNe are given in Table~\ref{AD}.
With this definition, I was able to place all the SNe of our sample in
a common time frame.  Both
origins of time, $t_0$ and $t_{PT}$, are used in this thesis,
depending on the context of the analysis.

\begin{table}[htpp]
\begin{center}
\caption[Host-galaxy extinctions, distances and explosion times for
  the sample of SNe~II-P]{Host-galaxy 
  extinctions, distances and explosion times for the sample of
  SNe~II-P\label{AD}} 
\scriptsize
\vspace{3mm}
\begin{tabular}{lccccc}
\hline\hline
             &                &  & $t_0(\mathrm{E96})^{\mathrm{b}}$ & $t_0(\mathrm{D05})^{\mathrm{b}}$ & $t_{PT}^{\mathrm{c}}$\\ 

SN Name      & $A_V^{\mathrm{a}}$ & $\mu^{\mathrm{a}}$ &
[JD$-2448000$] & [JD$-2448000$] & [JD$-2448000$] \\                  
\hline                                            
1991al        &$-0.17$(21) &34.05(28)                  & $\cdots$    & $\cdots$    & 521.998  \\
1992af        &$-0.37$(21) &34.15(28)                  & $\cdots$    & $\cdots$    & 858.583  \\
1992am        &$ 0.52$(23) &36.57(025)$^{\mathrm{d}}$   & $\cdots$    & $\cdots$    &  947.43   \\
1992ba        &$ 0.30$(21) &31.33(28)                  & 883.9(3)    & 879.8(5.6)  & 1013.714  \\
1993A         &$ 0.06$(25) &35.59(04)$^{\mathrm{d}}$    & $\cdots$    & $\cdots$    & 1106.96   \\
1999br        &$ 0.94$(25) &31.86(31)                  & $\cdots$    & 3275.6(7.7) & 3395.508  \\
1999ca        &$ 0.25$(21) &33.29(11)$^{\mathrm{d}}$    & $\cdots$    & $\cdots$    & 3373.19    \\
1999cr        &$ 0.12$(21) &34.56(28)                  & $\cdots$    & $\cdots$    & 3348.88   \\
1999em        &$ 0.24$(21) &30.00(28)                  & 3476.3(1.1) & 3474.0(2)   & 3590.136  \\
1999gi        &$ 1.02$(21) &30.50(28)                  & 3517.0(1.2) & 3515.6(2.4) & 3647.759  \\
0210          &$ 0.31$(36) &37.31(31)                  & $\cdots$    & $\cdots$    & 4591.917  \\
2002fa        &$-0.35$(26) &37.16(29)                  & $\cdots$    & $\cdots$    & 4576.632  \\
2002gw        &$ 0.18$(22) &33.41(28)                  & 4557.9(2.7) & 4551.7(7.6) & 4662.195  \\
2002hj        &$ 0.24$(22) &34.93(28)                  & $\cdots$    & $\cdots$    & 4660.231  \\
2002hx        &$ 0.38$(22) &35.49(28)                  & $\cdots$    & $\cdots$    & 4657.707  \\
2003B         &$-0.09$(21) &32.18(27)                  & $\cdots$    & $\cdots$    & 4713.878  \\ 
2003E         &$ 0.78$(23) &34.01(28)                  & $\cdots$    & $\cdots$    & 4765.663  \\
2003T         &$ 0.35$(21) &35.17(28)                  & 4654.2(2.7) & 4648.9(3.4) & 4759.381  \\
2003bl        &$ 0.26$(21) &34.07(30)                  & $\cdots$    & 4692.6(2.8) & 4804.609  \\
2003bn        &$-0.04$(21) &33.60(28)                  & 4693.4(2.7) & 4687.0(9)   & 4817.0    \\ 
2003ci        &$ 0.78$(23) &35.30(38)                  & $\cdots$    & $\cdots$    & 4817.720  \\ 
2003cn        &$-0.04$(23) &34.81(28)                  & $\cdots$    & $\cdots$    & 4804.986  \\
2003cx        &$-0.27$(25) &36.20(29)                  & $\cdots$    & $\cdots$    & 4830.927  \\ 
2003ef        &$ 0.98$(21) & 33.43(1.2)$^{\mathrm{e}}$  & 4759.8(4.7) &  748.4(15.6)& 4869.095  \\
2003fb        &$ 1.24$(23) &34.49(29)                  & $\cdots$    & $\cdots$    & 4874.252  \\
2003gd        &$ 0.33$(21) &29.98(28)                  & $\cdots$    & $\cdots$    & 4840.108  \\
2003hd        &$ 0.01$(21) &35.86(28)                  & $\cdots$    & $\cdots$    & 4953.053   \\ 
2003hg        &$ 1.97$(24) &33.31(28)                  & $\cdots$    & $\cdots$    & 4998.604   \\ 
2003hk        &$ 0.44$(25) &34.41(28)                  & $\cdots$    & $\cdots$    & 4957.908  \\ 
2003hl        &$ 1.72$(23) &32.04(28)                  & 4872.3(1.7) & 4865.4(5.9) & 5006.708  \\ 
2003hn        &$ 0.46$(21) &31.13(27)                  & 4859.5(3.8) & 4853.8(9.3) & 4962.898   \\ 
2003ho        &$ 2.19$(21) &34.13(28)                  & $\cdots$    & $\cdots$    & 4919.688   \\ 
2003ip        &$ 0.56$(22) &33.76(28)                  & $\cdots$    & $\cdots$    & 5002.872   \\
2003iq        &$ 0.25$(22) &32.43(28)                  & 4909.6(4.3) & 4905.6(9.5) & 5019.625   \\ 
2004dj        &$-0.09$(23) &28.14(29)                  & $\cdots$    & $\cdots$    & 5286.428   \\
2004et        &$ 0.13$(27) &28.37(31)                  & $\cdots$    & $\cdots$    & 5389.0    \\
2005cs        &$ 0.72$(30) & 29.62(0.25)$^{\mathrm{f}}$ & $\cdots$    & $\cdots$    & 5665.705   \\
\hline
\end{tabular}

\begin{list}{}{}
\item [$^{\mathrm{a}}$] Host-galaxy extinctions and distance moduli calculated by \citet{F09}.
\item [$^{\mathrm{b}}$] Explosion times calculated by
  \citet{2009ApJ...696.1176J} using two atmospheric models (E96 and D05).  
\item [$^{\mathrm{c}}$] Times at the middle point of the transition between the
  plateau and the radioactive tail, as calculated by \citet{F09}.
\item [$^{\mathrm{d}}$] Redshift adopting $H_0= 68$.
\item [$^{\mathrm{e}}$] EPM distance estimated by \citet{2009ApJ...696.1176J},  
converted to the distance scale of \citet{F09}.
\item [$^{\mathrm{f}}$] Distance modulus given by Pastorello et al 2006.
\end{list}

\normalsize
\end{center}
\end{table}

\chapter{Hydrodynamical Code}
\label{ch:code}
The main part of my thesis work is devoted to develop a
one-dimensional hydrodynamic code in order to model bolometric light
curves of SNe~II-P. Due to the difficulty of modelling the core collapse
from first principles, the approach extensively used in the literature,
and followed in this work, is to decouple the SN explosion problem in two 
independent parts: 
a) the core collapse and formation of the shock wave (SW), and b) the
ejection of the envelope. Although the exact nature of the
  mechanism that rips the object in two parts is not clear, the great
  differences in  energetics and time scales (seconds for the collapse
  and days for the ejected envelope) allow the theoretical description
  of a SN~II explosion to be separated into internal and external
  problems. The energy transferred to the envelope (which we
  called explosion energy) plays the role of a coupling parameter
  between the internal and external parts of problem. In addition, the
  processes which control the envelope
  ejection and the supernova radiation do not depend on how the energy
is transferred to the envelope as long as this process occurs in a short enough time.
 Based on the
propagation of the SW through the envelope, independently
of how the shock is formed, it is possible to study the observational
outcome of the explosion, such as light curves and spectra. 

In \S~\ref{sec:method}, I describe the numerical method, equations,
micro-physics and initial models used to compute
bolometric light curves of SNe~II-P. That Section is
complemented by Appendices~\ref{ch:anexoA} and \ref{ch:anexoB}
which provide a technical description of the methods. 
The structure of the code is presented in \S~\ref{sec:structure}. A
discussion of the adopted approximations  is given in
\S~\ref{sec:Approximations}. 

\section{Calculation Method}
\label{sec:method}
Our supernova models are computed by numerical integration of the
hydrodynamical equations assuming spherical symmetry for a
self-gravitating gas. Radiation  
transport is treated in the diffusion approximation 
with the flux-limited prescription of \citet{1981ApJ...248..321L}. The
explosion is simulated by injecting a certain amount of  
energy near the center of the progenitor object during a very short
time as compared with the hydrodynamic time-scale. This energy
induces the formation of a powerful shock wave that propagates through the 
progenitor transforming thermal and kinetic energy of the
matter into energy that can be radiated from the stellar surface. To
calculate shock waves, we include, as is usually
  done, an artificial-viscosity term in
the equations of moment and energy.

The equations and numerical method used are discussed in 
\S~\ref{sec:equations}. In \S~\ref{sec:constitutive} I give a  
description of the constitutive relations included in the code. The energy
deposited by radioactive decay is
discussed in \S~\ref{sec:gamma}. Finally, in \S~\ref{sec:initialmodel}
the initial models are described.

\subsection{Equations for radiation transport and hydrodynamics}
\label{sec:equations}

The differential equations that describe the hydrodynamics and
radiative transfer of the system assuming spherical symmetry
and diffusion approximation in Lagrangian coordinates are: (a) velocity definition, 
 
\begin{equation}
\frac{\partial r}{\partial t}= u, 
\label{sec:equations:eq1} 
\end{equation}
(b) mass conservation,
\begin{equation}
V= \frac{1}{\rho}= \frac{4 \pi }{3} \frac{\partial r^3}{\partial
  m}, 
\label{sec:equations:eq2} 
\end{equation}
(c) momentum conservation,
\begin{equation}
\frac{\partial u}{\partial t}= - 4 \pi r^2  \,\frac{\partial}
         {\partial m}(P+q) - \frac{G m} { r^2},  
\label{sec:equations:eq3}
\end{equation}
(d) energy conservation, 
\begin{equation}
\frac{\partial E}{\partial t}= \epsilon_{\rm Ni} - 
      \frac{\partial L}{\partial m} - (P+q)  \, 
       \frac{\partial V} {\partial t},
  \label{sec:equations:eq4}
\end{equation}
and (e) radiative energy transport,
\begin{equation}
     L = - ( 4 \pi r^2 ) ^2  \, \frac{\lambda ac}
     {3 \kappa} \, \frac{ \partial T^4} {\partial m}, 
  \label{sec:equations:eq5}
\end{equation}

\noindent where $m$ is the Lagrangian mass coordinate chosen as
independent variable in place of the radius ($r$) to describe the
structure of the object, $u$ is the velocity, $V$ is the specific
volume, $P$ is the total pressure
(of gas and radiation), and $q$ is the artificial viscosity which is
included in the equations to spread the
pressure and energy over several mass zones at the shock front. There
are many expressions for the artificial viscosity, 
all dependent on the  velocity gradient, which aim at providing a
convenient interpolation scheme between
unshocked and shocked fluid. We adopt the expression given by
\citet{VN50} (see Appendix~\ref{ch:anexoA}). $E$ is the total internal
energy per unit of  mass, including gas and radiation, $\epsilon_{\rm Ni}$  is
the  energy    
deposited  by the radioactive decay of nickel as I describe in
\S~\ref{sec:gamma}. In equation~(\ref{sec:equations:eq4})
 we do not consider other sources of cooling or  
heating, such as losses due to neutrino
processes or energy released by thermonuclear reactions, because the
energy of the SW overwhelms these other sources. Even if neutrinos are
very important in the   
formation of the shock wave as the explosion depends noticeably on the
efficiency of their energy transfer, most of them are 
emitted before the shock wave reaches the stellar photosphere, so they 
have no effect on later epochs of the SN evolution
\citep[see][]{Hill90,1991supe.conf..393B,2007PhR...442...38J}. The 
energy released by explosive nucleosynthesis is much less than the
energy of the shock wave as has been previously 
established [\citet[see][]{1965SvA.....8..664I,WW90,1996snih.book.....A}].
$T$ is the temperature of both matter and radiation, $\kappa$ is the
Rosseland-mean opacity and $L$ is the luminosity. Finally, $\lambda$
is the so-called  ``flux-limiter'', included in the diffusion
equation to ensure a smooth transition between diffusion and
free-streaming regimes  to assure
causality. The expression adopted for $\lambda$ is,  

\begin{equation}
\lambda = \frac{ 6 + 3 R}{6 + 3 R + R^2}, 
\label{sec:equations:eq6}
\end{equation}

\noindent where  

\begin{equation}
R= \frac{ \mid \nabla T^4 \mid } {\kappa \rho \ T^4}= 
    \frac{4 \pi r^2}{\kappa T^4} \mid  \frac{ \partial T^4} {\partial m} \mid. 
\end{equation}

\noindent Note that when $\kappa \rho \rightarrow \infty$ (short mean
free path), $R \rightarrow 0$ and $\lambda \rightarrow 1 $ as
expected for the diffusion limit. But as $\kappa \rho \rightarrow 0$
(long mean free path),  $R \rightarrow \infty$ and $\lambda
\rightarrow 3/R $. Thus, it can be seen that the flux is limited to the
value $c \, a T^4$, as required physically. However, flux-limited
diffusion is not the correct solution for radiation transfer in
optically thin regions; it is simply an interpolation between the
physically correct optically thick and thin limits. 

The quantities $E$, $P$, and $\kappa$ are functions of $\rho$, $T$,
and chemical composition. Further details on these constitutive  
relations are given in \S ~\ref{sec:constitutive}. 

The overall problem is to solve the complete structure of the
object, i.e., determining the dependent variables $r$, $u$, $V$, $T$,
and $L$ as  a function of $m$ and time. Therefore, the problem
reduces to solving a system of five coupled partial differential
equations~(\ref{sec:equations:eq1})--(\ref{sec:equations:eq5})
together with the constitutive relations (see
\S~\ref{sec:constitutive}) and the expression for $q$, $\lambda$ and
$\epsilon_{\rm Ni}$ (see \S~\ref{sec:gamma}). In addition, it is also
necessary to provide boundary and initial conditions. The boundary
conditions that we have adopted are: $u=0$ near the center 
(at $m=M_{\rm core}$, where we typically set 
$M_{\rm core}= 1.4 \, M_{\odot}$), and $P_{\rm gas}=\rho= 0$ at the
surface ($m=M$).  As  initial condition, we have used different
initial models (see \S~\ref{sec:initialmodel}),  in
hydrostatic equilibrium. As discussed  above, we simulate the
explosion by  artificially adding internal energy almost
instantaneously in the  central region of the core. 

We have developed a code in FORTRAN to  numerically solve this system
of differential equations using the method of finite differences. As a
first step, it is necessary to discretize each equation 
in order to transform the system of differential equations into an algebraic
one. It is important to note that the
discretization process is not trivial as there is no unique way to
discretize a differential equation
\citep{1991nmap.book.....B}. Details of the discretization and  
the scheme of integration are given in
Appendix~\ref{ch:anexoA}. Here, I remark that the code uses a space-centering
discretization with the extensive quantities evaluated  
at the interfaces and the intensive quantities in the midpoints of the
grid zones. Two time steps are adopted in each cycle: 
one to advance the velocity, and the other to advance the material
state variables. The time
step is chosen to comply with stability and accuracy constraints (see
the discussion below). Special care is taken in the centering of the
opacity when discretizing equation~(\ref{sec:equations:eq5}) in order to 
prevent numerical noise to appear due to the propagation of the
radiation flow at the steep front where the opacity changes
significantly \citep{Christy67}.

Depending on how the time derivatives of the differential equations
are numerically evaluated, we have different schemes of integration. For
example, given the equation,  

\begin{equation}
\frac{\partial f(x,t)}{\partial t}= g(f,x,t), 
\end{equation}

\noindent in a single time step the numerical scheme advances the solution from
time $t$ (where we known the solution) to time $t + \Delta t$. The
advanced solution can in general be written as

\begin{equation}
f(t+ \Delta t) \simeq f(t) + \Delta t ( (1- \theta) \, g(t)+ \theta
  \, g(t+ \Delta t) ).
\end{equation}

\noindent If $\theta= 0$  the scheme is called {\em explicit} (or forward time
differencing) and it involves only known values of the functions at time
$t$. In other cases, with $\theta$ between $0.5$ and $1$,
the scheme is called {\em implicit} and the solution involves the
initial unknown function $f(t+ \Delta t)$, requiring an iterative
technique to find the solution. More specifically, when $\theta= 1$
the scheme is called fully implicit (or backward time differencing)
and the case of $\theta= 0.5$ is usually known as centered time
differencing. Although, the {\em explicit} schemes are easier to
solve than the {\em implicit} ones, the former have a stringent
requirement on the time step used in the calculation in order to ensure the
stability of the numerical scheme. This
requirement is known as the {\em Courant condition}, 
which states that $\Delta t < \Delta x / v$ where $\Delta x $ is
the width of the grid zone and $v$ is a characteristic speed which
in hydrodynamics problems is adopted as the sound speed.  
Physically, this condition states that the information may not
spread through the grid at a speed grater than the characteristic
speed defined by the problem.

In our code, we use an explicit scheme for the
integration of the hydrodynamic 
equations, but a semi-implicit scheme for the temperature, similar to the
one used by \citet{1977ApJS...33..515F}. The equations are linearized in  
$\delta T$ and solved iteratively for each time step using the
tridiagonal method (see Appendix~\ref{ch:anexoA}). The discretization
typically uses 300 mesh 
points, with a  finer sampling for the outer layers typically smaller
than $10^{-6} M_\odot$. This value was chosen based on tests performed
with our model and following previous works
\citep{1988ApJ...330..218W,1992ApJ...393..742E} which 
show that the early light curve is sensitive to the mass zoning in the
outer layers  when a coarse grid  is used.

Given that we are using an explicit hydrodynamic scheme, 
the time step should be chosen as a fraction of the minimum of the
Courant condition for all  zones in order to achieve
stability. I note that we have also imposed additional
  conditions on the time step: we have required that changes in temperature,
  density and flux over one time step be less than 5\%.

The formation of the shock wave (SW) following core collapse is simulated by
artificially adding internal energy (``thermal bomb'') almost
instantaneously in the inner region of the core. We usually
set a mass cut of $1.4 \, M_\odot$  which the material assumed
to collapse and form a neutron star or black hole. In order to do
this, we have included an additional term in the energy
equation~\ref{sec:equations:eq4} of the form,  
$dE_{exp}/ dm = C \,exp( -t/ \tau) \, exp(- m/m_0) $. This expression
was arbitrarily chosen but its integral  is exactly the
explosion energy which is free parameter of our model. 
We have also tested explosions generated by 
injecting kinetic energy (in this case it is necessary to include an
additional term in the Euler equation~\ref{sec:equations:eq3}) and we
have obtained similar 
results. The latter method, however, leads to very short 
numerical time steps, and therefore to slower calculations.
We have checked the accuracy of our calculation method by testing the
conservation of  energy. The total amount of energy is
conserved within $0.6$\% but if we consider the conservation of
energy between two consecutive time steps, this is  
is within 4 $\times 10^{-6}$.  We consider it  very
adequate  for our purposes.

\subsection{Input physics}
\label{sec:constitutive}

The equation of  state (EOS) is  calculated using
simple expressions for $T$, $\rho$, composition,  and  for the
ionization degrees of hydrogen and helium which are computed assuming local
thermodynamical equilibrium (LTE). Therefore, the degree of ionization is 
determined by solving the corresponding set of Saha equations for
ionization of hydrogen and the first and second
ionization of helium. Ionization of heavy elements is
neglected in our EOS  (but, of course, it is taken
into account in the calculation of the opacity). In general, our EOS
consists of two terms, the radiation and gas contribution.  The latter
can be also divided into electron and ion terms. Thus, for example  the
total pressure is written as,

\begin{equation}
P= P_{Rad} + P_{Gas}=  P_{Rad} + P_{e-} + P_{ion}
\end{equation}

\noindent with

\begin{equation}
 P_{Rad}= \frac{a}{3} T^4, \hspace{0.5cm} P_{ion}= \frac{R}{\mu} \rho T 
\end{equation}

\noindent where $\mu$ is the molecular weight for the non-ionized material. The
electron pressure, $ P_{e-}$, is a much more complicated function. It is
influenced by the degree of ionization (determined by Saha equation)  and  
degeneracy  effects. We calculate the pressure of degeneracy  using
the method  described  by \citet{KWH68},  although the condition of
degeneracy is rarely reached in our calculation. We tested our results using a
more sophisticated EOS, such as  that of  the Los Alamos Tables
\citep{1996ApJ...456..902R}, and we did not  find any significant 
difference with respect to the results obtained using our simple
EOS. This is expected due to the low densities attained in most of the
layers of the  models.

The Rosseland mean opacity, $\kappa$, is a  function of $T$, $\rho$ (or $P$)
and chemical composition. There are tables available in the literature that
provide the opacity for a wide range of $T$, $\rho$ and chemical composition.
In our calculations  we use the OPAL opacity tables \citep[][and references 
therein]{1996ApJ...464..943I}.  As  these tables are given for $T > 6 \times
10^3$ K, we complement them with the opacity tables provided by 
\citet{1994ApJ...437..879A} for lower temperatures which includes molecular
opacities. The tables are interpolated in order to guarantee a
smooth transition at $T=10^4$ K from one to the other.

These tables allow us to calculate opacities for several
metallicities. Also, for a fixed metallicity, 
different mixtures of H, He, C and O can be used. Although in this 
work we adopt a fixed value of $Z=0.02$, departures from this
value (as expected in the inner regions of the object) are
taken into account as excesses of C and O with respect to the values
of the adopted metallicity, at the expense of He.

\begin{figure}[htbp]
\begin{center}
\includegraphics[scale=.55]{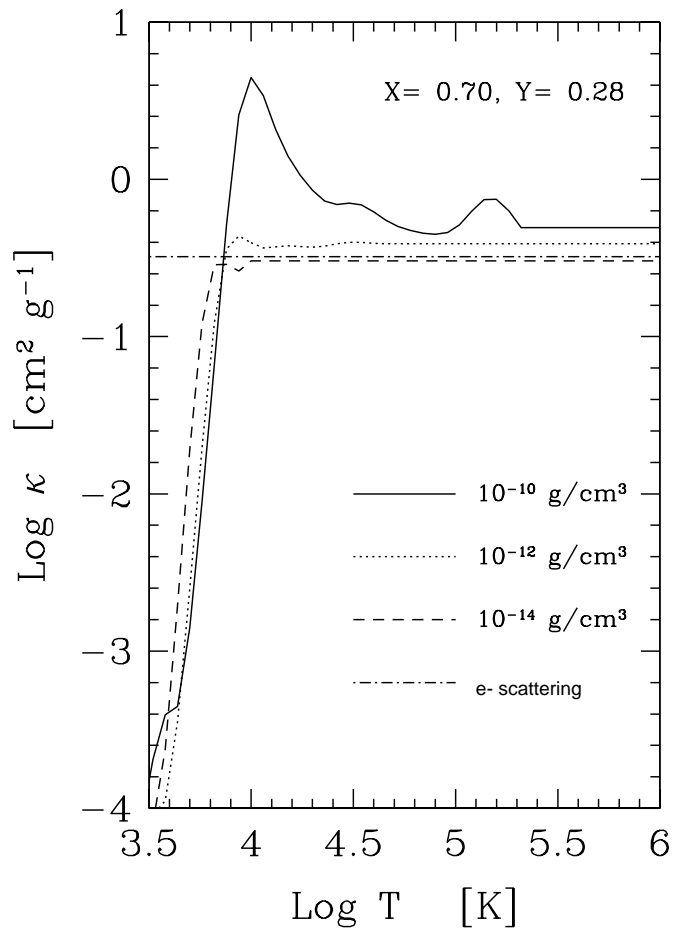}
\caption[Rosseland mean opacity]{The run of the Rosseland opacity on
  temperature ($T$) and density ($\rho$) for solar metallicity
  ($Z=0.02$) as used in our calculation without including any ``opacity
  floor'' (see discussion in Section~\ref{sec:constitutive}). 
  The ranges of $T$ and $\rho$
  shown are typical for SNe~II-P. The hydrogen (X) and  helium (Y)
  mass fractions used are 
  indicated. We also include the electron scattering
   opacity (considering full ionization, which is certainly
   unrealistic for the low temperature sector of 
  this plot) in order to show the dominance of this source for
   $T > 10^4$ K and  $\rho < 10^{-10}$ g cm$^{-3}$. Note that this
  value of the density is 
  reached early-on in the supernovae evolution. For lower 
  temperatures the absorption processes become an important source of
  opacity.\label{fig:opal}} 
\end{center}
\end{figure}

Figure~\ref{fig:opal} shows the opacity given by the tables as a
function of temperature for $Z=0.02$ and different densities. 
The range of temperature shown corresponds to the values reached
during the evolution of a SN~II-P while the values of density are
restricted to those achieved during the plateau phase.
Also shown in the plot is the contribution to the opacity from
electron scattering. Note that electron scattering is the dominant
source of opacity for $T >10^4$ K and $\rho < 10^{-10}$ gr cm$^{-3}$.

The Rosseland mean opacity includes scattering and absorption
processes. Scattering is the dominant process (see Figure
~\ref{fig:opal}) in the supernova 
ejecta until most of the electrons recombine with ions and absorption
processes become an important source of opacity. On the other
  hand,
in rapidly expanding envelopes where large velocity gradients are present,
the Rosseland mean opacity underestimates the true line opacity
\citep{1977ApJ...214..161K}, which hinders the estimation of the 
actual opacity in the outermost (recombined) layers.
Another effect that is not included in the calculation of $\kappa$ is
non-thermal excitation or ionization of atoms/ions which is produced by Compton 
scattering of $\gamma$-rays emitted by radioactive decay of $^{56}$Ni and
$^{56}$Co. The LTE ionization used in the calculation of
$\kappa$ considerably underestimates the true ionization.
The correct way to treat these effects is to calculate the actual
contribution of non-thermal ionization to the opacity and to
include the expansion opacity of lines. However, such treatment is
beyond the scope of this study. We adopt an alternative approach that
has been extensively used in the literature
\citep[e.g.,][]{1990ApJ...360..242S,1990A&A...233..462H,1991ApJ...374..266S,2004ApJ...617.1233Y}     
which consists of using a minimum value of the opacity (or ``opacity
floor''). The minimum opacity values adopted in this work are: $0.01$ cm$^2$ 
g$^{-1}$ for the envelope material, and $0.24$ cm$^2$g$^{-1}$ for the
metal-rich core material. These values were chosen by performing a
comparison with results of the STELLA code. We
found an excellent overall agreement between the light curves given by our
code and  STELLA \citep{1993A&A...273..106B,1998ApJ...496..454B} both
in terms of the 
duration of the plateau and  the morphology of the bolometric
LC) when the previous values of the minimum opacity were adopted (see
\S~\ref{sec:Approximations}.  Note that STELLA is an implicit
hydrodynamic code that incorporates multi-group radiative
transfer, and  additionally uses different opacity tables
and includes the effect of line opacities \citep{{2002nuas.conf...57S}}. 

\subsection{Gamma-ray Deposition}
\label{sec:gamma}

During the explosive nucleosynthesis produced in  supernova
explosions, unstable isotopes of iron elements are formed. The
decay of these isotopes and subsequent thermalization of the decay
products generate extra energy that contributes to power the LC.
The most abundant radioactive isotope produced is $^{56}$Ni which 
decays with a $6.1$ day half-life to $^{56}$Co which in turn decays
with a $77.7$ day half-life to stable $^{56}$Fe. The decays
produce energetic $\gamma$-rays and positrons which 
are  thermalized, providing the thermal luminosity of the
SN.

In order to include this source of energy in the calculations, it is
necessary to determine a local heating rate which  generally is not
the same as the simple radioactive decay rate. Instead,
the decay rate should be modified by the 
probability of thermalization determined by the rate at which
$\gamma$-rays and positrons deposit energy at various points in the
gas as they travel through the ejecta. This
problem is very complicated if tackled from first
principles. A set of $\gamma$-ray transfer equations must
be solved simultaneously with the 
hydrodynamic equations which requires a multiple-energy
group Monte Carlo 
calculation. However, if the  ejecta is optically thick  to
$\gamma$-rays, it is possible to assume that the gamma rays deposit
their energy locally. Several studies of SNe~II-P have previously
assumed this hypothesis considering that the energy contribution of
radioactive material is only relevant toward the end
of the plateau phase. Such approach would be correct only if $^{56}$Ni was
deeply concentrated in the ejecta in which case
gamma photons could hardly diffuse out of the regions where they  form. 
We believe there is no justification to assume this type of $^{56}$Ni
distribution, as  shown in studies of SN~1987A \citep[][among others]
{1988A&A...196..141S,1988ApJ...324..466W,1988ApJ...331..377A,2000ApJ...532.1132B}   
and  the diffusion of gamma rays must be properly calculate from
the location where they are emitted to the outer regions. To calculate
this, we solve the gamma-ray transfer in the gray approximation for any
spherically symmetric distribution of $^{56}$Ni,  assuming that gamma
rays interact with matter 
only through absorption. Details of this calculation are presented in
Appendix~\ref{ch:anexoB}. It has been shown by comparison with Monte
Carlo simulations of Type~Ia supernovae that the complex scattering
process between gamma rays and electrons can be satisfactorily 
approximated as an absorptive process
\citep{1984ApJ...280..282S,1995ApJ...446..766S}. The adopted
value for the gamma-ray opacity  is 
$\kappa_\gamma= 0.06 \, y_e$ cm$^{2}$g$^{-1}$, where $y_e$ is the
number of electrons per baryon. 

\begin{figure}
\begin{center}
\includegraphics[scale=.35]{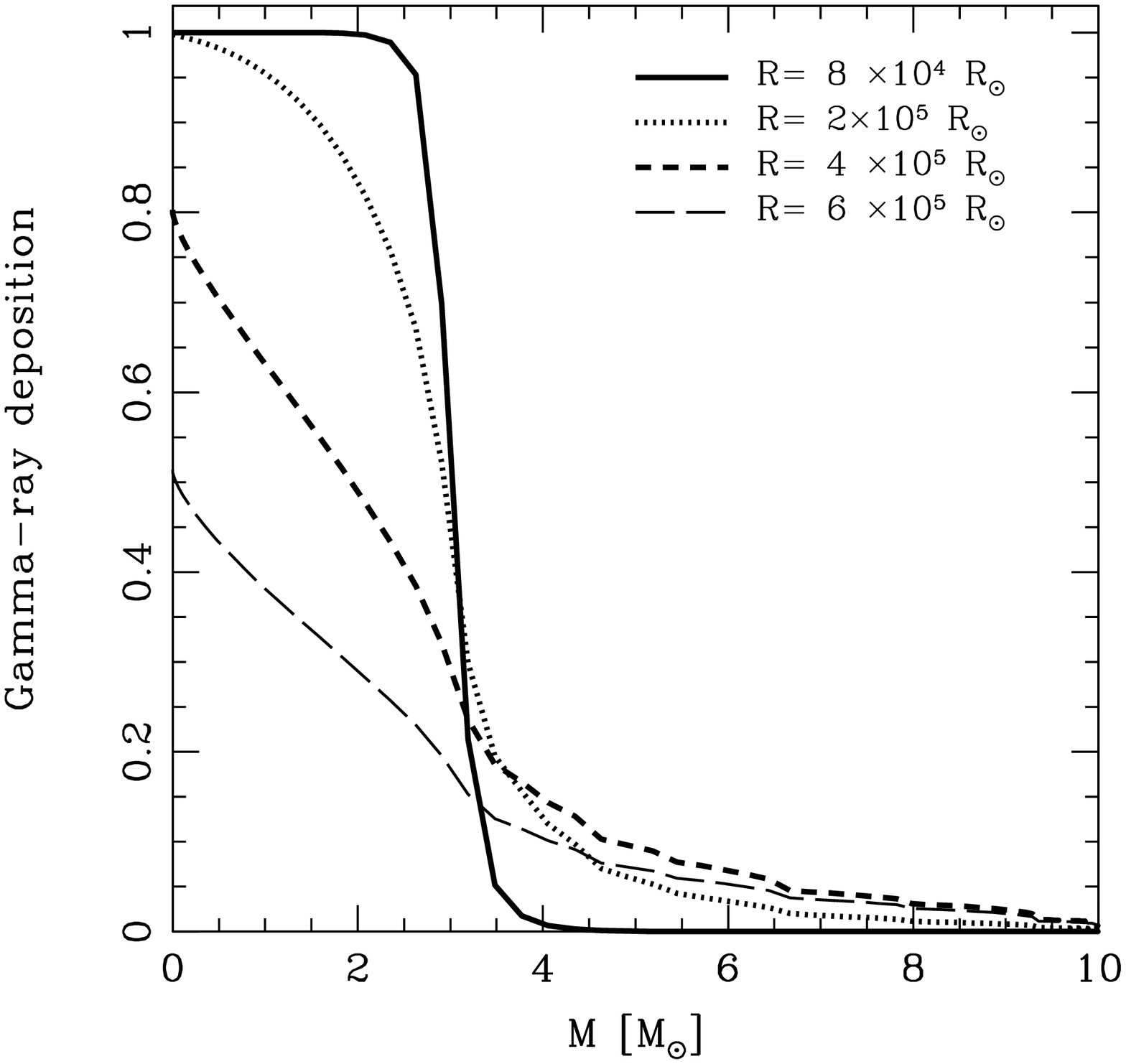}\includegraphics[scale=.35]{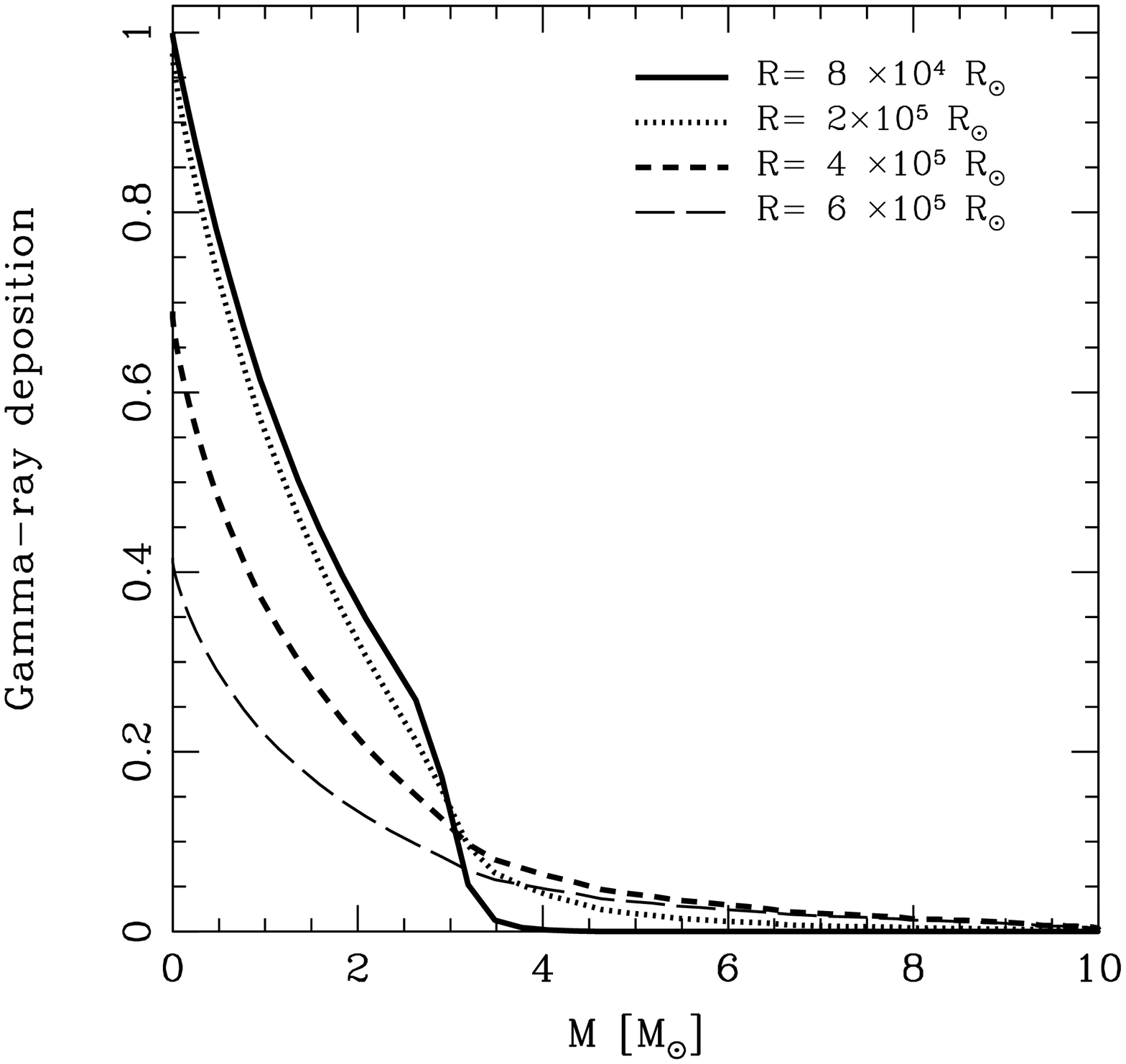}
\caption[Gamma-ray deposition]{Gamma-ray deposition as a function of mass for a
  polytrope with index $n=3$, initial mass  
$10 \, M_\odot$, and different initial radii. {\bf (Left)} for a constant
  distribution of $^{56}$Ni up to $3 M_\odot$; {\bf(right)} for an
exponential distribution of $^{56}$Ni, also up to $3 M_\odot$. The
diffusion of the gamma rays out of the region were they form
becomes more noticeable as the object gets more extended and 
diluted.\label{fig:depo}} 
\end{center}
\end{figure}

The energy rate per gram released by Ni--Co--Fe decay is 

\begin{equation}
\epsilon_{\rm rad} = \, 3.9 \times 10^{10} \,\, \exp(- t/\tau_{\rm Ni}) +
\, 6.78 \times 10^{9} [\exp(- t/\tau_{\rm Co}) -  \exp(- t/\tau_{\rm
    Ni})] \;{\rm erg\; g}^{-1} \,{\rm s}^{-1}, 
\label{code:erad}
\end{equation}

\noindent where $\tau_{\rm Ni}= 8.8$ days, and $\tau_{\rm Co}=113.6$
days are the mean lifetimes of the radioactive isotopes.
The amount of energy deposited at each point is given by the
solution of the gamma-ray transfer multiplied by the previous
expression. In Figure~\ref{fig:depo} we show the gamma-ray
deposition\footnote{The deposition of gamma rays is defined
  by the energy deposited at each point normalized to
  the value corresponding to complete thermalization at the same
  location where the gamma rays are emitted.} profile for the case
of a polytrope with index $n=3$, initial mass 
$10 \, M_\odot$ and different initial radii. The left panel is for a
constant distribution of $^{56}$Ni up to $3 M_\odot$ and the right
panel is for an exponential distribution of $^{56}$Ni, also up to $3
M_\odot$. Note that the diffusion of gamma rays from the region where
they form becomes more noticeable as the object becomes more extended
and diluted.

The possibility of using an arbitrary distribution of $^{56}$Ni allows
us to study different types of mixing and their effect on the
resulting LC. As shown in ~\ref{sec:tail}, we find that
radioactivity becomes an important source of energy, even during
the plateau phase, if we allow an extensive mixing of $^{56}$Ni.

\subsection{Initial models}
\label{sec:initialmodel}

There are two different ways to determine the initial (or pre-SN)
models: those coming from stellar evolution calculations
(``evolutionary'' models), or those from non-evolutionary calculations
where the initial density and chemical composition are parameterized
in a convenient way. Both types of models have been tested with our
code.

\begin{figure}[htbp]
\begin{center}
\includegraphics[angle=-90,width=10cm]{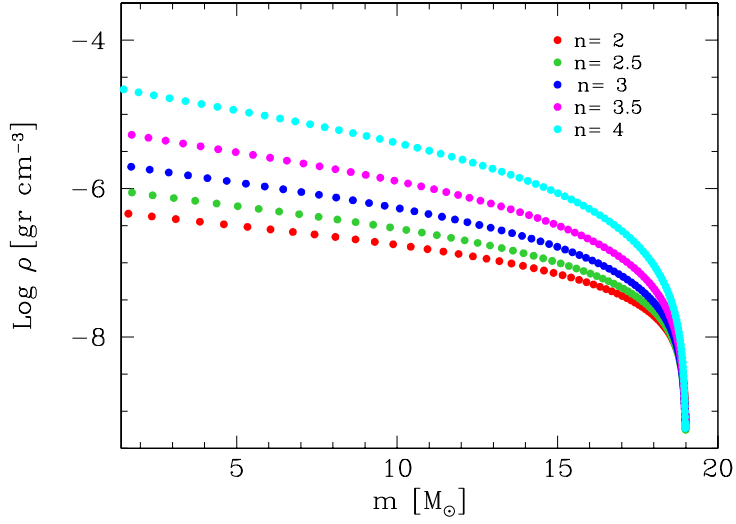}
\caption[Density distribution for single polytropic model]{Density
  distribution as a function of mass  
  for a single polytropic model using different polytropic indices,
  $n$, which are indicated in the plot. The total mass and radius in
  all the models are 19 $M_\odot$ and 800 $R_\odot$. Note that
  higher values of $n$ produce more compact
  configurations.\label{fig:poli}}  
\end{center}
\end{figure}

As non-evolutionary models, we have calculated single and double
polytropes. The single polytropic
models were numerically computed by solving the ``Lane-Emden''
equation for any polytropic index ($n$). This allowed us to study many
different configurations for 
different values of $n$, initial masses and radii, in hydrostatic
equilibrium. Some examples of initial density profiles obtained
using different polytropic indices are shown in Figure~\ref{fig:poli}.

Although a
single polytrope may represent very well the envelope of the real
pre-supernova model, the inner dense 
part which is expected for this type of objects is not well 
reproduced with this simple initial model. One way to 
improve this situation is to consider two polytropes: one  representing the
inner dense core, and the other  accounting for the outer extended
envelope.

 The numerical calculation of double
polytropic models is not trivial because the solution is extremely
sensitive on the adopted initial guess value of the  free
parameters. Specifically, it is necessary to solve the equations of
hydrostatic equilibrium and mass conservation assuming a polytropic
approximation where the  pressure has a dependence on density
 of the form: $P=K \rho^{\gamma}= K \rho^{n / (n+1) }$, where $K$
is a constant and $n$ is the polytropic index. Two different
polytropic relations with different indices ($n_i$ and $n_o$) and
constants ($K_i$ and $K_o$) are adopted to mimic the characteristic
structure of a red supergiant. To calculate the composite polytrope, the
previous equations are numerically integrated outward from the inner
  border (stellar core) to a pre-selected
point (fitting point) using the internal polytropic relation. A
second inward integration  is done  from  the outer border (stellar surface)  
to the fitting point using the external polytropic relation. The
problem is equivalent to a two-point boundary condition for the
case where there are unknown free parameters at both ends of the domain.
In our case, the free parameters are $n_i$, $K_i$ and $K_o$, while the
external index is fixed to $n_o=3$. The inner boundary conditions are
the central density ($\rho_c$) and the core mass ($m_c$), while the
external boundary conditions are the initial mass ($M$) and the
external density ($\rho_o$). The
integrations are performed along the radial variable $r$ from $r_c$
to $R$ (radius of the object). The values of $r_c$, $R$, $M$ and
$\rho_c$ are fixed for
each configuration, while $m_c$, $\rho_o$ are functions of the free
parameters of the problem. Starting from initial guesses, the
values of the free parameters are iteratively sought so that the solution 
joins smoothly at the fitting point. This process is performed using a
shooting method (see Numerical Recipes, chapter 17). Once the values of 
$n_i$, $K_i$ and $K_o$ are found, the density (or pressure) distribution  and
mass distribution can be calculated. The initial temperature
profile is calculated in an iterative fashion using our EOS to ensure
hydrostatic equilibrium. This prevents the formation of a spurious
shock wave.

\begin{figure}[htbp]
\begin{center}
\includegraphics[angle=-90,width=8cm]{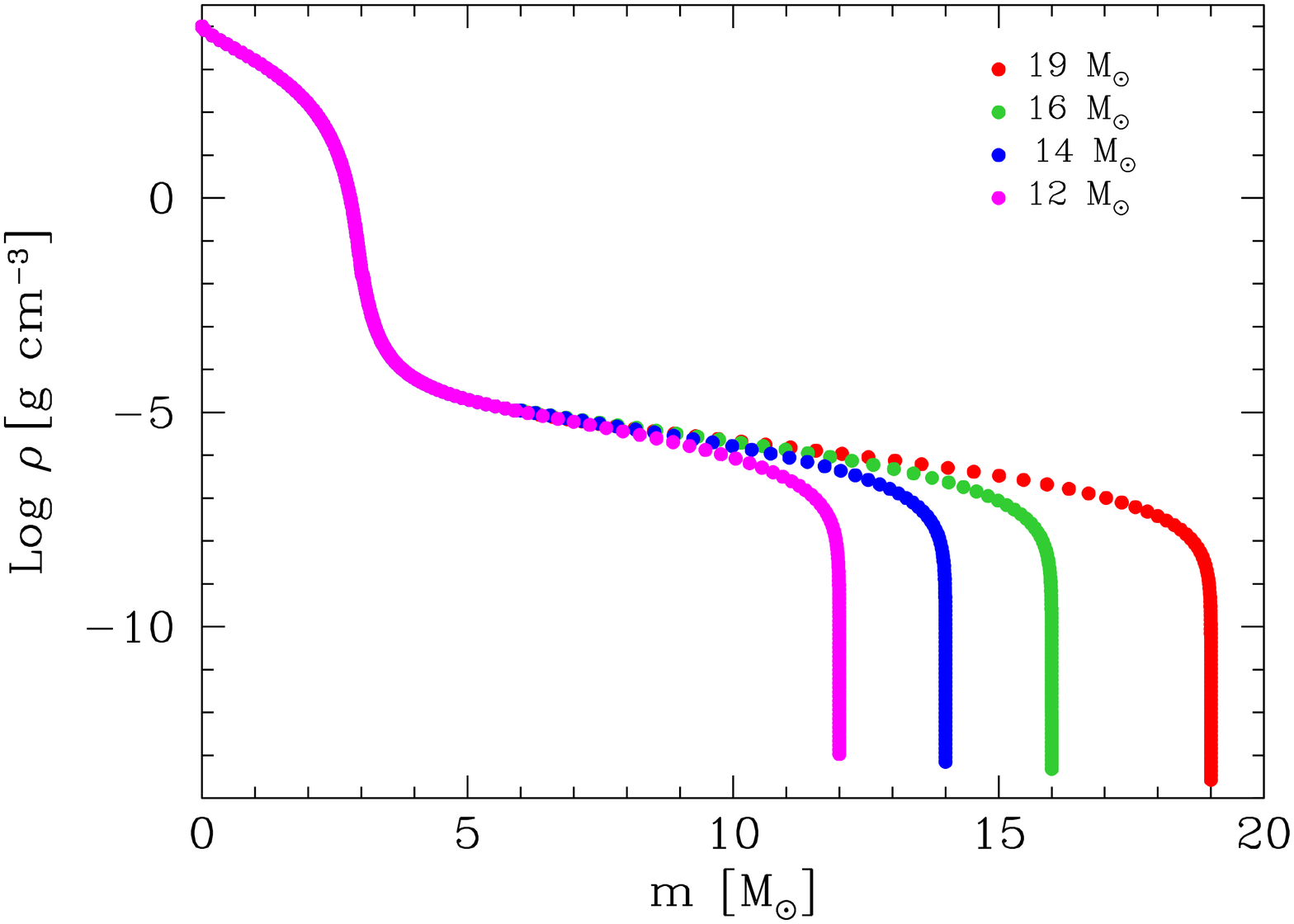}\includegraphics[angle=-90,width=8cm]{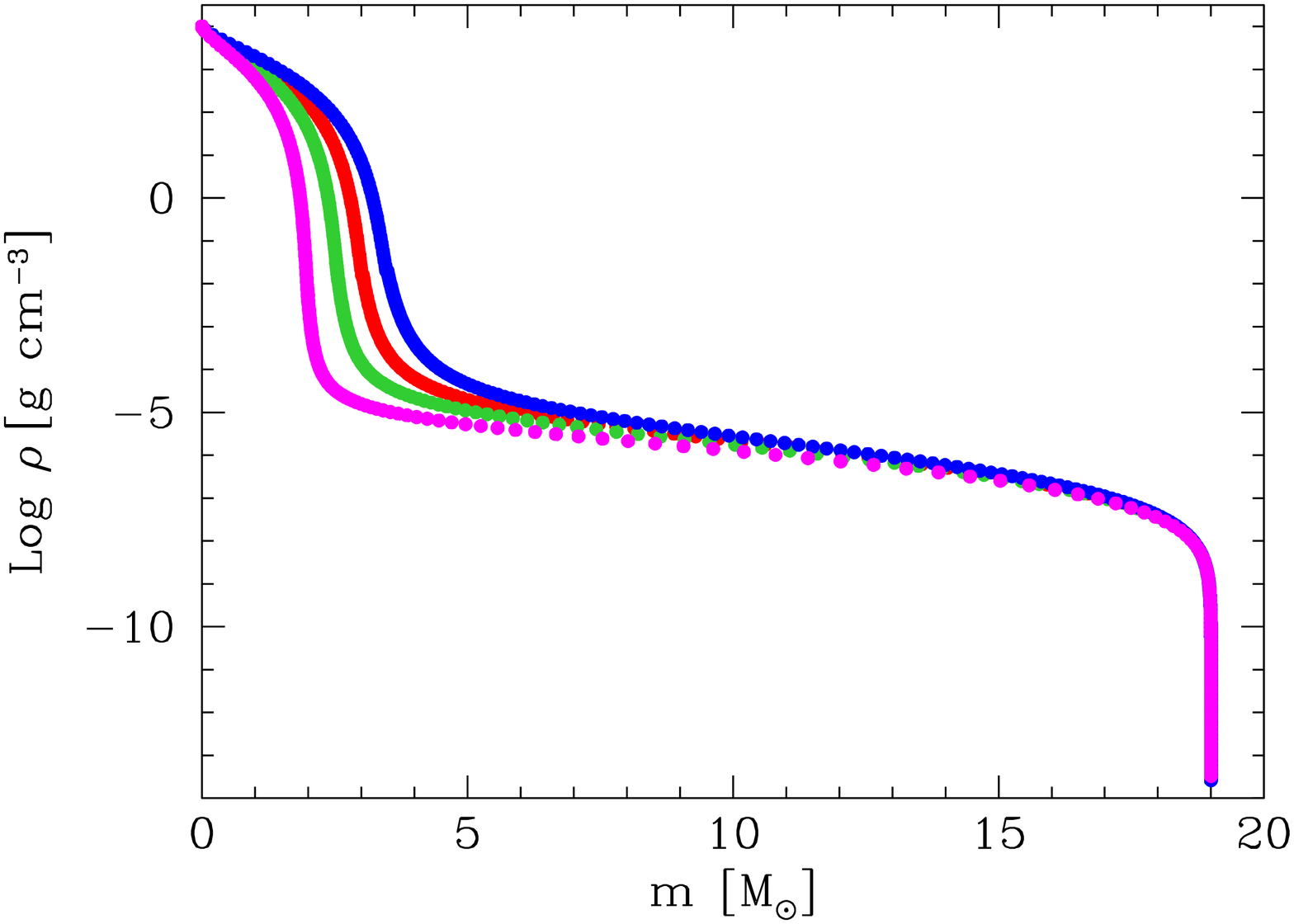}
\caption[Double-polytropic initial models for different masses]{Density
  distribution of double polytropic models as a function of       
  mass for a model with radius of 800 $R_\odot$. {\bf (Left)}
  The effect on the initial density of varying the 
  mass. The values used in the calculation are indicated in the plot.
  {\bf (Right)} The effect on the initial density of varying the fitting point
 for an  mass of 19 $M_\odot$. Note that even if  mass and radius
 are the same, different configurations are obtained by changing the
 fitting point.\label{fig:dospoli1}}  
\end{center}
\end{figure}

\begin{figure}[htbp]
\begin{center}
\includegraphics[angle=-90,width=8cm]{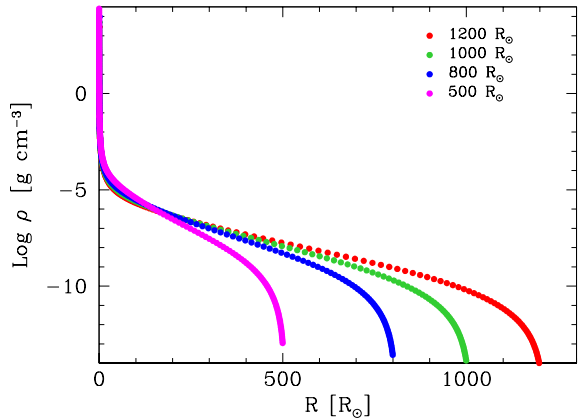}
\caption[Double-polytropic initial models for different radii]{Density
  distribution of double polytropic models as a function of
  radius. The different curves represent different values of the
   radius used in the calculation for a fixed  mass of 19
  $M_\odot$, as indicated in the plot.\label{fig:dospoli2}} 
\end{center}
\end{figure}

Using this method a variety of initial models in hydrostatic
equilibrium 
can be calculated by changing the fitting point, the initial
mass, the radius or the central density as  shown in
Figures~\ref{fig:dospoli1} and \ref{fig:dospoli2}. The effect of
the initial mass and fitting point on the density distribution
is shown in Figure~\ref{fig:dospoli1}. The
effect of the 
initial radius is shown in Figure~\ref{fig:dospoli2}. Note that
different configurations can be obtained by changing 
the fitting point even if the initial mass and radius remain
fixed. The same is valid if one changes the
central density or the external polytropic index. 

\begin{figure}[htbp]
\begin{center}
\includegraphics[angle=-90,width=9cm]{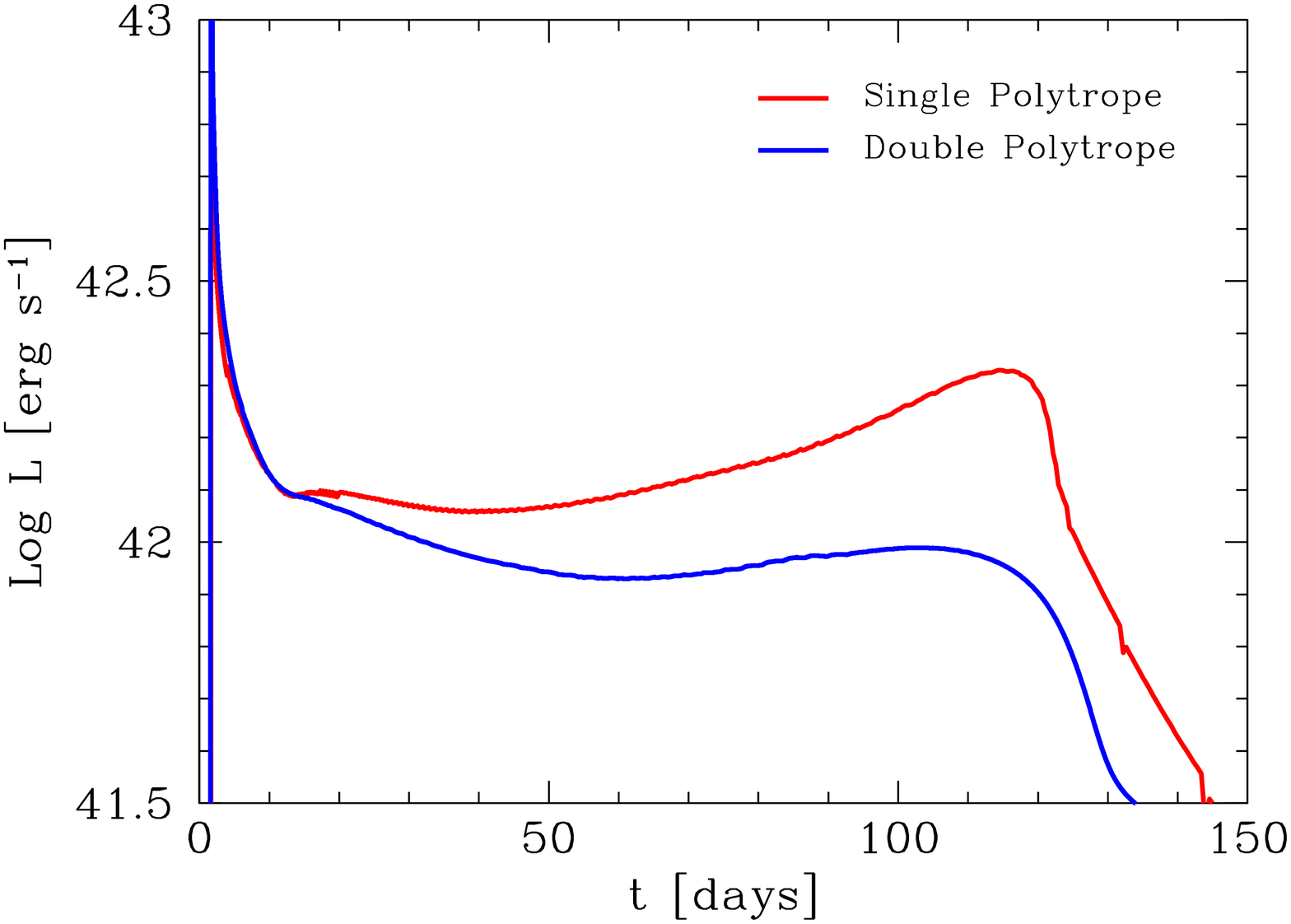}
\caption[LC for single and double polytropic model]{Comparison
  between the LC obtain using a single (red line) and double (blue
  line) polytrope. While the early LCs ($t \lesssim 20$ days) are very
similar,  important differences appear toward the end of the
plateau.\label{fig:comppoli}}  
\end{center}
\end{figure}

These parametric
profiles allow us to study how the internal structure of the progenitor
affects the LC.
Figure~\ref{fig:comppoli} shows a comparison between LCs obtained
using a single and a double polytropic model for an initial mass of 19
$M_\odot$ and an initial radius of 800 $R_\odot$. During the early part
of the LC evolution ($t \lesssim 20$ days) both models are very
similar. Significant differences appear later on. Note
that the single polytropic models produce a prominent bump before the
sharp decrease in luminosity which marks the end of the
  plateau. This bump  is a consequence of the lack of a dense
  core in the initial model  and  is never observed in SNe~II-P.   

\begin{figure}[htbp]
\begin{center}
\includegraphics[angle=-90,width=9cm]{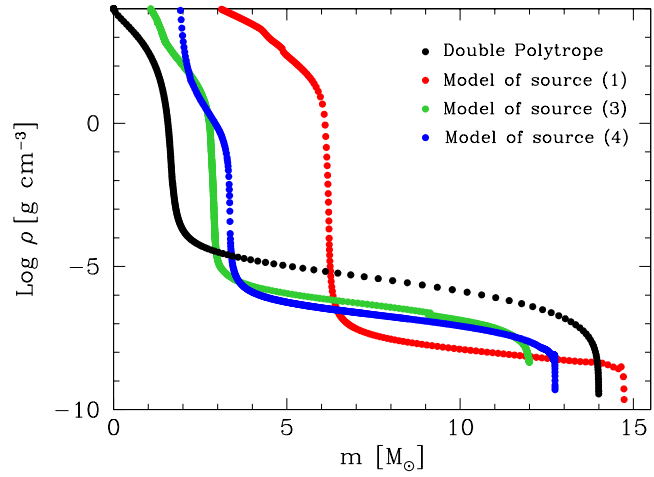}
\caption[Density distribution form stellar
  evolution]{Initial density distributions form
  stellar evolution. The pre-SN models
  come from different sources, as indicated the text. A double
  polytropic model is also included for 
  comparison. \label{fig:severalmodels}} 
\end{center}
\end{figure}

\begin{figure}[htbp]
\begin{center}
\includegraphics[angle=-90,width=9cm]{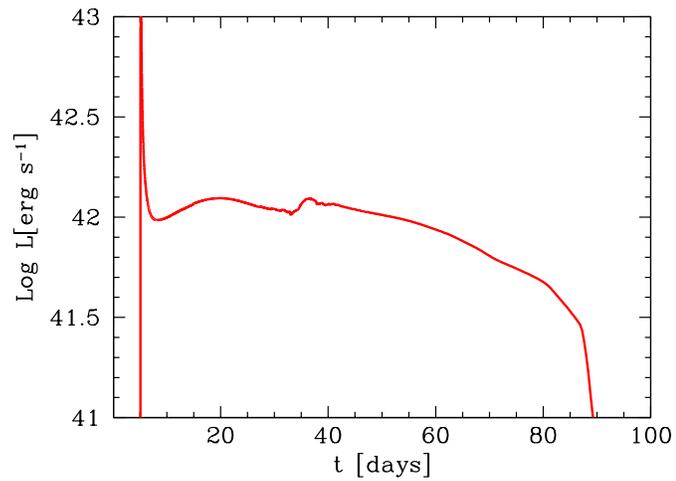}
\caption[LC for stellar evolution progenitor]{LC calculated using the
  initial model from stellar 
  evolution calculations of the ``Supernova Science
  Center-UCSC''. Note that some 
  bumps appear in the LC around $t=$ 35 days which are not present in
  observations. 
\label{fig:clwoosley}} 
\end{center}
\end{figure}

We have also calculated LCs using initial models derived from stellar
evolution calculations. Pre-supernove models of four different
sources were used: (1) ``Supernova Science Center-UCSC''
(http://www.supersci.org/ucsc/), (2) \citet{2000ApJS..129..625L}
(http://web.oa-roma.inaf.it/localinfo/staff/webhost/limongi/data.html),
(3) \remark{Pedir cita a Omar:Benvenutto et al.} and (4) 
\citet{2005ApJ...619..427U}. 
Figure~\ref{fig:severalmodels} shows  
the density distribution for some of these models. Also
shown for  comparison  the density distribution of a double polytropic
model. The LC obtained using a pre-SN model of source (1) is shown in 
Figure~\ref{fig:clwoosley}. Note that some bumps appear around $t= 35$
days in the
LC which are not present in observations. In
\S~\ref{sec:Approximations} a discussion using a pre-SN of source (4)
is presented.

Having tested the code with various
evolutionary initial models, I generally find 
better agreement with observations when using our parameterized double
politropic profiles, in concordance with previous studies 
that employ similar methods to compute the initial structure
\citep[][among
  others]{1993A&A...270..249U,2005AstL...31..429B,2007A&A...461..233U}. 
Thus is the approach adopted throughout the rest of this thesis. 

\newpage

\section{Code structure}
\label{sec:structure}

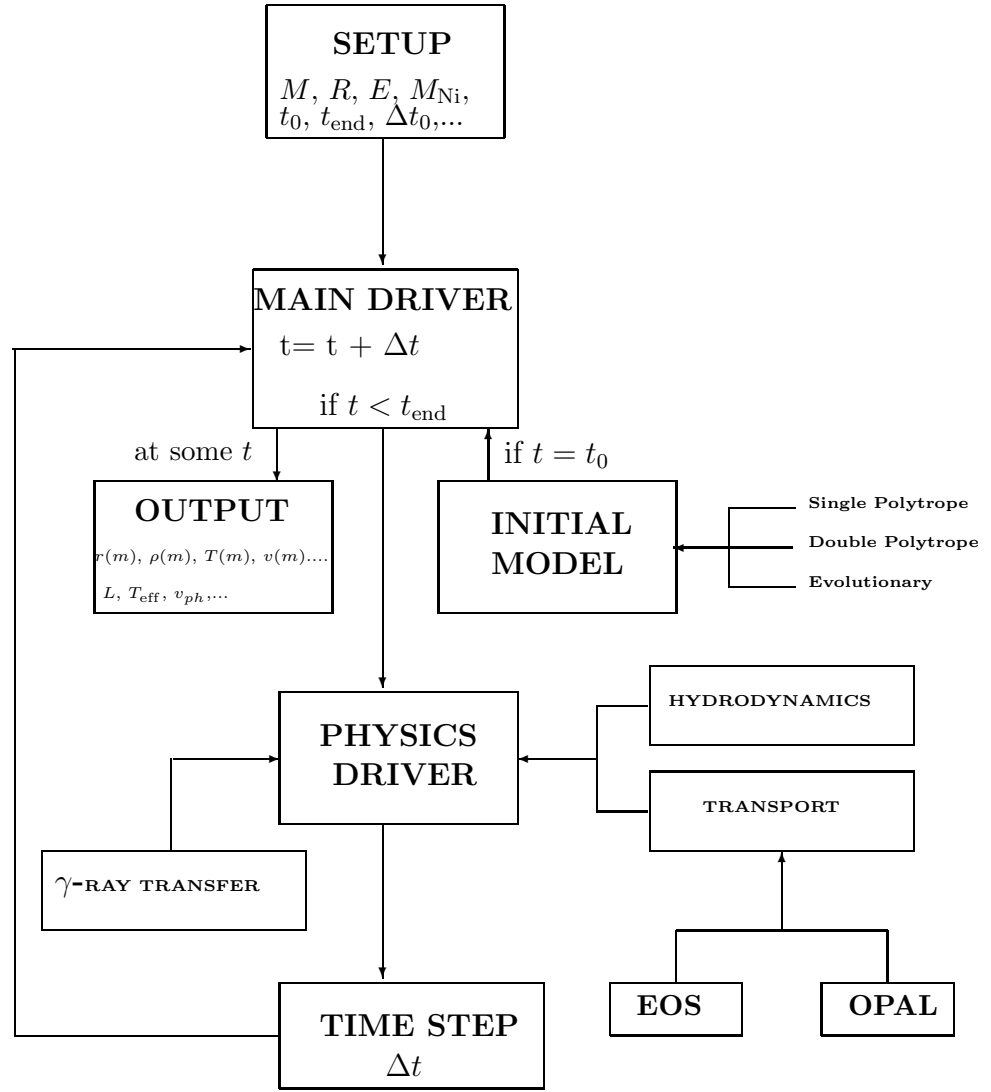
\begin{figure}[htbp]
\begin{center}
\begin{picture}(400,400)
\newsavebox{\foldera}
\savebox{\foldera}
(40,32)[bl]{
\multiput(0,0)(0,50){2}
{\line(1,0){90}}
\multiput(0,0)(90,0){2}
{\line(0,1){50}}}

\newsavebox{\folderb}
\savebox{\folderb}
(40,32)[bl]{
\multiput(0,0)(0,60){2}
{\line(1,0){100}}
\multiput(0,0)(100,0){2}
{\line(0,1){60}}}

\newsavebox{\folderc}
\savebox{\folderc}
(40,32)[bl]{
\multiput(0,0)(0,40){2}
{\line(1,0){100}}
\multiput(0,0)(100,0){2}
{\line(0,1){40}}}

\newsavebox{\folderd}
\savebox{\folderd}
(40,32)[bl]{
\multiput(0,0)(0,30){2}
{\line(1,0){100}}
\multiput(0,0)(100,0){2}
{\line(0,1){30}}}

\newsavebox{\foldere}
\savebox{\foldere}
(40,32)[bl]{
\multiput(0,0)(0,20){2}
{\line(1,0){50}}
\multiput(0,0)(50,0){2}
{\line(0,1){20}}}

\put(155,360){\usebox{\foldera}}
\put(180,392){\bf{SETUP}}
\put(160,375){\small{$M$, $R$, $E$, $M_{\rm Ni}$, }}
 \put(160,365){\small {$t_0$, $t_{\rm end}$, $\Delta t_0$,...}}
\put(195,360){ \vector(0,-1){48}}
\put(150,250){\usebox{\folderb}}
\put(150,295){\bf{MAIN DRIVER}}
\put(160,278){{t= t + $\Delta t$}}
\put(175,255){{if $t < t_{\rm end}$}}
\put(235,230){ \vector(0,1){20}}

\put(241,237){ if $t=t_0$ }
\put(220,180){\usebox{\foldera}}
\put(240,210){\bf{INITIAL }}
\put(240,195){\bf{MODEL }}
\put(350,205){ \vector(-1,0){45}}
\put(330,190){\line(0,1){30}}
\put(326,190){ \line(1,0){25}}
\put(326,220){ \line(1,0){25}}
\put(360,220){\tiny {\bf{Single Polytrope}}}
\put(360,205){\tiny {\bf{Double Polytrope}}}
\put(360,190){\tiny {\bf{Evolutionary}}}
\put(155,250){ \vector(0,-1){20}}

\put(90,180){\usebox{\foldera}}
\put(105,215){\bf{OUTPUT}}
\put(90,200){\tiny {$r(m)$, $\rho(m)$, $T(m)$, $v(m)$....}}
\put(93,186){\tiny {$L$, $T_{\rm eff}$, $v_{ph}$,...}}
\put(105,238){\small{at some $t$}}

\put(195,250){ \vector(0,-1){98}}
\put(160,100){\usebox{\foldera}}
\put(175,130){\bf{PHYSICS}}
\put(180,115){\bf{DRIVER}}

\put(280,125){\vector(-1,0){30}}
\put(280,105){\line(0,1){40}}
\put(276,105){ \line(1,0){19}}
\put(276,145){ \line(1,0){19}}
\put(300,130){\usebox{\folderd}}
\put(307,145){\tiny {\bf{HYDRODYNAMICS}}}
\put(300,90){\usebox{\folderd}}
\put(320,105){\tiny {\bf{TRANSPORT}}}
\put(350,60){\vector(0,1){30}}
\put(310,60){\line(1,0){80}}
\put(306,60){ \line(0,-1){20}}
\put(386,60){ \line(0,-1){20}}
\put(285,20){\usebox{\foldere}}
\put(295,28){\small {\bf{EOS}}}
\put(365,20){\usebox{\foldere}}
\put(375,28){\small {\bf{OPAL}}}

\put(120,125){\vector(1,0){40}}
\put(115,125){ \line(0,-1){35}}
\put(70,60){\usebox{\folderd}}
\put(75,75){\bf{$\gamma$-\tiny{RAY TRANSFER}}}

\put(195,100){ \vector(0,-1){58}}

\put(160,0){\usebox{\folderc}}
\put(60,20){\line(0,1){260}}
\put(60,20){\line(1,0){100}}
\put(175,20){\bf{TIME STEP}}
\put(200,5){$\Delta t$}
\put(55,280){ \vector(1,0){90}}
\end{picture}
\caption[Code Scheme]{Code scheme\label{scheme} }

\end{center}
\end{figure}

The code is organized in several subroutines with different
hierarchy. Figure~\ref{scheme} schematically shows how our
computational program is organized.

\vspace{1cm}

The main routine or ``MAIN DRIVER'' controls the
program during the course of a calculation through  a series of
subroutine calls  which are governed by logical 
sentences based in part on the input information. The
input of the program is represented in the  
``SETUP'' block and includes the information necessary to generate a
specific calculation. There, an input file is read containing the following
information: type of initial model to use in the calculation
(polytrope, double polytrope or evolutionary model), initial mass
($M$), radius ($R$), nickel mass ($M_{\rm Ni}$), distribution of
$^{56}$Ni and explosion energy ($E$). The
initial time step ($\Delta t_0$) and the ending time for the model 
calculations ($t_{end}$) are also included in this file. Other
parameters which control the accuracy of the solution and the number of
iterations allowed for the convergence of the iterative transfer
calculations are also included there. The ``PHYSICS DRIVER'' block 
represents the piece of the code which calls subroutines that
calculate the evolution of the physical quantities. Specifically, the
following subroutines are called from this driver: (1) a
routine to calculate the explicit hydrodynamic equations, (2) a
transport routine which calculates the diffusion radiative transfer
using an iterative method as described in Appendix~\ref{ch:anexoA}, and
(3) the routine for solving the gamma-ray transfer (see
Appendix~\ref{ch:anexoB}). The EOS and OPAL  
subroutines are called from the transport routine and  
represent the calculation of the equation of state and opacity.
The results of the calculations, i.e., the model 
structure, $r(m)$, $P(m)$, $\rho(m)$, $T(m)$, etc., and the evolution of
certain quantities such as bolometric luminosity, $L_{bol}(t)$,
effective temperature, $T_{\rm eff}(t)$,  photospheric velocity,
$v_{ph}(t)$, among others are written in output  
files at specific moments during the evolution. The writing process is 
schematically represented by the  block ``OUTPUT''.  This  subroutine
is called  from the ``Main Driver'' at the specific moments defined 
in the ``SETUP'' routine.  After each step of
calculation, a new step of time, $\Delta t$, is chosen in the ``TIME
STEP'' block depending on stability and accuracy conditions.  
The ``MAIN DRIVER'' routine determines whether a new calculation is
necessary, basically by checking that the new time step, $t= t +
\Delta t$ is less than $t_{end}$. In that case,
the whole process is repeated for time $t + \Delta t$.

\begin{figure}[htbp]
\begin{center}
\includegraphics[angle=-90,width=9cm]{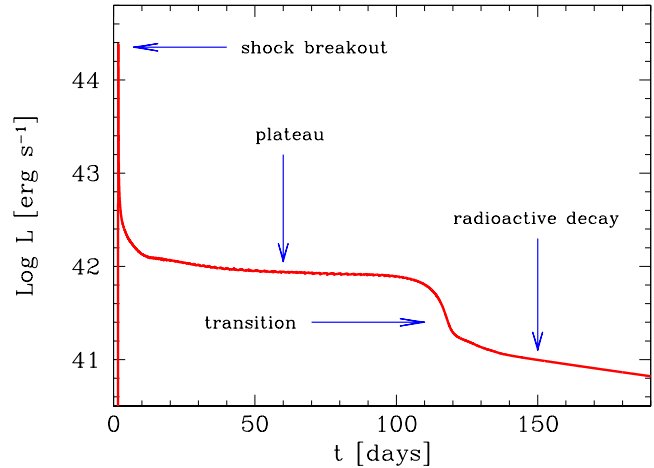}
\caption[Evolutionary phases of a LC]{A typical bolometric LC of
  a SN~II-P. The initial model used in the calculation was a double
polytrope with initial mass 19 $M_\odot$, initial radius  800
$R_\odot$, and injected energy  1 foe. The different phases during
the LC evolution are indicated in the plot. 
\label{fig:LCphases}} 
\end{center}
\end{figure}

Figure~\ref{fig:LCphases} shows a typical bolometric LC obtained with
this code. The initial model used in the calculation is a double
polytrope with initial mass  19 $M_\odot$, initial radius  800
$R_\odot$ and injected energy  1 foe (1 $\times 10^{51}$ erg
s$^{-1}$). From the Figure it is possible to distinguish various
phases: (1) a maximum in luminosity
or `` shock breakout'', (2) a phase where the luminosity is nearly constant or
``plateau'', (3) a transition phase, and (4) a radioactive 
tail. These  phases  are studied in detail in
Chapter~\ref{sec:sn99em}.

\section{Discussion on the approximations}
\label{sec:Approximations}

Several approximations are made in the equations of radiation
hydrodynamics. First, we assume that the fluid motion can be described by 
one-dimensional, radially symmetric flow. The explosion mechanism of 
core-collapse SNe is not well known. It may be an very
asymmetric process. However, for this particular subtype of
supernovae with very extended hydrogen envelopes the asymmetries
expected from the explosion mechanisms itself appears   to be
smoothed. This is supported by recent spectropolarimetric studies
\citep{2005ASPC..342..330L}. 

We use the equilibrium diffusion approximation to describe the 
radiative transfer. This approximation assumes that radiation and matter
are strongly coupled with a single characteristic temperature and a 
spectral energy distribution described by a black body
function (BB hereafter). The approximation breaks down at shock
breakout and at late phases when the ejecta is completely
recombined and the object becomes transparent. Fortunately, during
the plateau phase ---which is our main interest---, this
approximation is fairly adequate. Also note that non-LTE calculations of
SN~II spectra show that deviations from LTE have significant effects
on spectral lines but not on the overall continuum
\citep{1996MNRAS.283..297B,2008MNRAS.383...57D}. At late phases, 
for SNe that experience little interaction with the 
interstellar medium, as is the case of SNe~II-P, the bolometric
luminosity can be simply  approximated as the luminosity deposited by
the radioactive decay of $^{56}$Co, at least during the early part of the
radioactive tail. This is supported by the fact that the luminosity
declines obeying an exponential law with a very similar rate to that
of the decay of $^{56}$Co. The radioactive scenario has been directly
confirmed for SN~1987A through the detection of
$\gamma$-ray lines from $^{56}$Co \citep{1988Natur.331..416M}. No
attempt to apply our model beyond $\sim$180 days is done.

\begin{figure}[htbp]
\begin{center}
\includegraphics[angle=-90,width=12cm]{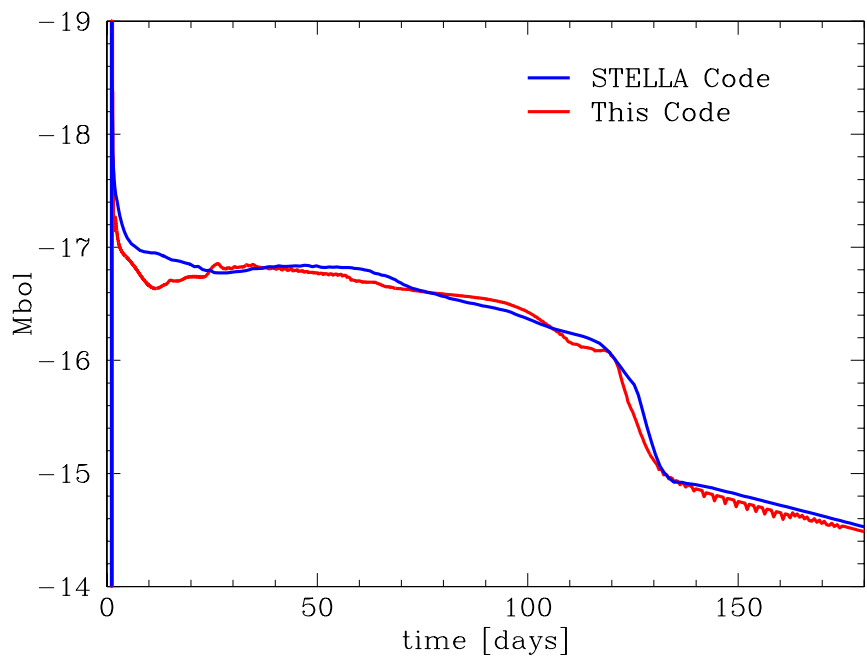}
\caption[Comparison with STELLA Code]{Comparison between the LC obtained
  with our code and the STELLA code. The initial model used was provided
  by \citet{2005ApJ...619..427U}. Note the very good agreement between
  both LCs  despise the simplifications employed here.
\label{fig:comp_stella}} 
\end{center}
\end{figure}

During the transition between plateau and radioactive tail, the
envelope is fully recombined and the notion of a photosphere loses
meaning. Thus, this transition regime is poorly described with our radioactive
transfer prescription and therefore a detailed analysis of this
phase cannot be asessed here. As stated above,
we have performed a comparison of our calculations with those of the
STELLA code using the same initial model provided by
\citet{2005ApJ...619..427U}. We obtained an excellent agreement
between the bolometric luminosities derived by both
methods, as shown in Figure~\ref{fig:comp_stella}.
This is very satisfactory considering that the STELLA code
involves a more sophisticated treatment of the radiative transfer by
solving these equations using a multi-group prescription and including the
effect of line opacities
\citep{1993A&A...273..106B,2002nuas.conf...57S}.

\begin{figure}[htbp]
\begin{center}
\includegraphics[angle=-90,width=12cm]{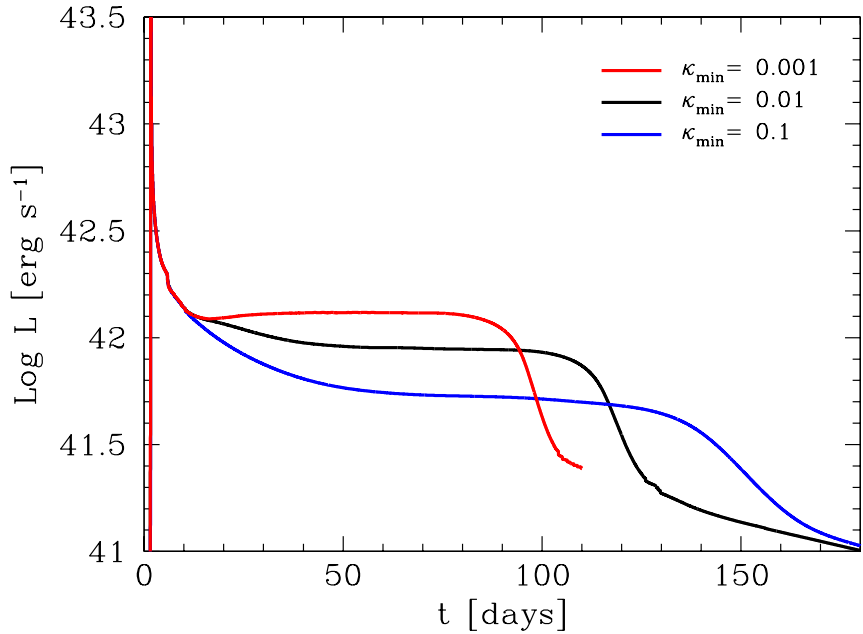}
\caption[LC for different values of the opacity minimum ]{LC for
  three different values of the opacity minimum ($\kappa_{\rm min}$;
  expressed in cm$^2$ g$^{-1}$ in the plot). Note the critical effect
  of the adopted opacity 
  minimum on the LC. The value finally used in our code, 
  $\kappa_{min}= 0.01$ cm$^2$ g$^{-1}$,  was chosen by
  comparison to the STELLA code (see
  Figure~\ref{fig:comp_stella})\label{fig:opacpiso}} 
\end{center}
\end{figure}

Finally, I remark that the very good agreement  obtained between both codes
was reached using the values of opacity minimum ($\kappa_{\rm min}$) mentioned in
\S~\ref{sec:constitutive}, i.e. for the envelope material 
$\kappa_{\rm min}=0.01$ cm$^2$ g$^{-1}$. For other values of the opacity minimum 
commonly used in literature, the comparison with the STELLA code, which
incorporates the effect of expansion opacity in the correct way, was
not satisfactory. In order to show the effect on the LC of the value
adopted for the opacity floor, I calculated models using three 
different  values for the envelope
material: $\kappa_{\rm min}[$cm$^2$ g$^{-1}] = 0.001$, $0.01$ and
$0.1$, as shown in Figure~\ref{fig:opacpiso}. Clearly, from 
this Figure, there is a critical effect of the opacity minimum on the
resulting LC. Hence, it should not be taken as a free parameter.

\chapter{Comparison with observables} 
\label{ch:model_obs} 
Since our code
produces bolometric light curves and effective temperatures (T$_{eff}$) it
proves necessary to calculate these quantities from the observed photometry in
order to compare our model with observations. Although there are many SNe
with optical observations, only a handful objects  are available with
observations 
over UV, optical and IR wavelengths. The purpose of this section is to use those
SNe with observed bolometric light curves and explore the feasibility to derive
a bolometric correction (BC) for other SNe with optical observations alone.\\ 
In \S\ref{sec:ABC}, I present the calibrations derived for BC and T$_{eff}$ from
$BVI$ photometry using data of three well-observed  SNe~II-P, SNe 1987A, 1999em
and 2003hn, and two sets of atmosphere models by Eastman et al. (1996)
and Dessart $\&$ 
Hillier (2005b) (E96 and D05 hereafter, respectively). The typical scatter of
the BC is 0.11 mag and allows us to calculate bolometric luminosities for many
other SNe~II-P having $B V I$ photometry alone, opening thus the possibility for
a statistical analysis of the physical properties of this type of object (see
also \citet{2009ApJ...701..200B}).\\ Additionally, the photospheric velocity is
another useful parameter to compare with  observations.  In
\S~\ref{sec:vphoto}, I describe  the prescription used to compare the observed
velocities with the photospheric velocities provided by our models.

\section{Analysis of BC and T$_{eff}$ -color relations}
\label{sec:ABC}

I begin this section by describing the observational and theoretical
material used to calculate BC and T$_{eff}$. Then, in
\S~\ref{sec:Lbol} I proceed to calculate the bolometric  luminosities for
these data. Finally, in \S~\ref{sec:BCresults} and
\S~\ref{sec:Teffresults} I derive  calibrations for BC and T$_{eff}$
as a function of colors. Along  this section, I 
 refer to the SN evolution in terms of time or color indistinctly.
This is well justified during the plateau phase in which the SN
atmosphere expands, cools and monotonically turns redder.\\

In order to examine if a bolometric correction can be derived from
optical colors, we made use of the three SNe~II that possess the best
wavelength and temporal coverage. Two of these are genuine SNe~II-P,
SN 1999em and SN 2003hn, and the third is the famous SN 1987A which,
except for its peculiar light curve, shares most of the spectroscopic
properties of SNe~II-P. Most of the data of these three SNe were
obtained at {\em Cerro Tololo Inter-American Observatory} (CTIO), {\em
Las Campanas Observatory} (LCO), and the {\em European Southern
Observatory} (ESO) at {\em La Silla}.  For more details see
\citet{H90} and \citet{Bouchet89} for SN 1987A, \citet{H09} for SN
1999em, and 
\citet{Kris08} for SN 2003hn.  The photometric bands used in this
analysis were $U B V R I Z J H K$ for SN 1999em and SN 2003hn, and $U
B V R I Z J H K L M$ for SN 1987A.

We adopt Cepheid distances for SN 1987A and SN 1999em with
corresponding values of 50 kpc \citep{2001ApJ...553...47F} and 11.7
Mpc \citep{Leo03}. For SN 2003hn we used a distance of 16.8 Mpc, as
derived by Olivares et al. 2008 using the Standardized Candle
Method. We correct the photometry for Galactic and host-galaxy
extinction. We perform such corrections using Galactic visual
absorptions of A$_V^{GAL}$=0.249 for SN 1987A, A$_V^{GAL}$=0.13 for SN
1999em, and A$_V^{GAL}$=0.043 for SN 2003hn \citep{SFD98}, assuming a
standard reddening law with R$_V$= 3.1 as given by Cardelli et
al. (1989).  The values used for host-galaxy absorption are,
A$_V^{host}$=0.216 for SN 1987A, A$_V^{host}$=0.18 for SN 1999em
\citep{H01} and A$_V^{host}$=0.56 for SN 2003hn \citep{D08}, again
assuming R$_V$= 3.1.

We also used in this analysis spectral energy distributions (SEDs) from
two sets of SN atmosphere models (E96 and D05). These models depend on
several parameters such as, luminosity, density structure, velocity
and composition.  For more details on the input parameters of such
models the reader is referred to \cite{E96} and \cite{D05}. We use a
total of 61 model spectra from E96, and 107 from D05. We discard 31
spectra from D05 which do not have enough UV coverage. 

\subsection{Bolometric Luminosity Calculations}
\label{sec:Lbol}
By definition, the bolometric luminosity is the integral of the flux
over all frequencies. This integration can be done in a
straight-forward way for the spectral models of E96 and D05 summing
the flux over wavelength.  With the purpose to estimate bolometric
corrections and colors for the models we compute $B V I$ synthetic
magnitudes using the filter transmission functions and zero points
given by \citet{H01}.

For the three well-observed SNe the calculation of bolometric
luminosities is performed from reddening-corrected broadband
magnitudes using the values mentioned in
section~\ref{sec:ABC}. K-corrections are neglected due to the small 
redshifts involved. We began by computing a quasi-bolometric light
curve using all the available broadband data. The magnitudes were
converted to monochromatic fluxes at the specific effective wavelength
of each filter using the transmission functions and zero points of the
photometric system \citep{H01}.  At epochs when a certain filter
observation was not available, we interpolated its magnitude in time
using the closest points.  The total ``quasi-bolometric'' flux,
$F_{qbol}$ was computed using a  trapezium integration over $U-K$ for
SN~1999em and SN~2003hn, and over $U-M$ for SN~1987A.

\begin{figure}
\begin{center}
\includegraphics[scale=.40]{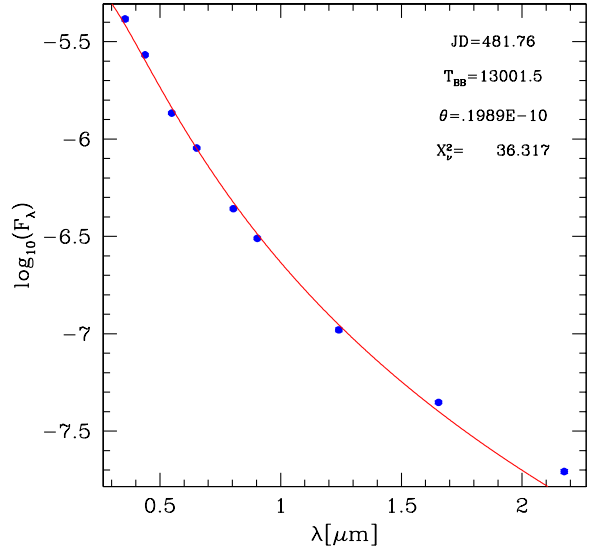}
\caption[Blackbody fit to the monochromatic
fluxes]{Blackbody (BB) fit to the monochromatic
fluxes for the case of SN~1999em at $\sim$ 7 days after 
explosion. The BB function provides a very good representation of the
flux. The T$_{BB}$ and $\theta$ shown in the
inset are the color temperature and angular radius of the SN~1999em
yielded by the BB  fit.  \label{fig:bbfit}}
\end{center}
\end{figure}

To estimate the missing flux in the UV and IR, $F_{UV}$ and $F_{IR}$,
we fit at each epoch a blackbody (BB) function to the monochromatic
fluxes\footnote{These fits are restricted to the plateau phase where
  the envelope of the SN was optically thick, and also to the
  transition to the nebular phase.  On the radioactive tail we did not
  use any UV or IR corrections.} as shown in
  Figure~\ref{fig:bbfit} for the case of SN~1999em at $\sim$ 7 days post
 explosion. At early  epochs the BB model  
provided very good fits to the fluxes in all bands. As the photosphere
became cooler the $U$-band flux started to depart from the BB model in
which cases we exclude this point from the fit.  At later epochs,
subsequently the $B$-band and $V$-band data points show the same
behavior, departing from the BB model. The reason for this is related
to the strong line blanketing that develops with time in that part of
the spectrum.

On the IR side the flux is extrapolated to $\lambda= \infty$ using
the BB fits described above. The integral of that function between the
longest observed effective wavelength and $\lambda= \infty$ is
adopted as the IR correction (F$_{IR}$). This correction increase
with time but always remains below 7$\%$ for the three SNe.
 
\begin{figure}
\begin{center}
\includegraphics[scale=.40]{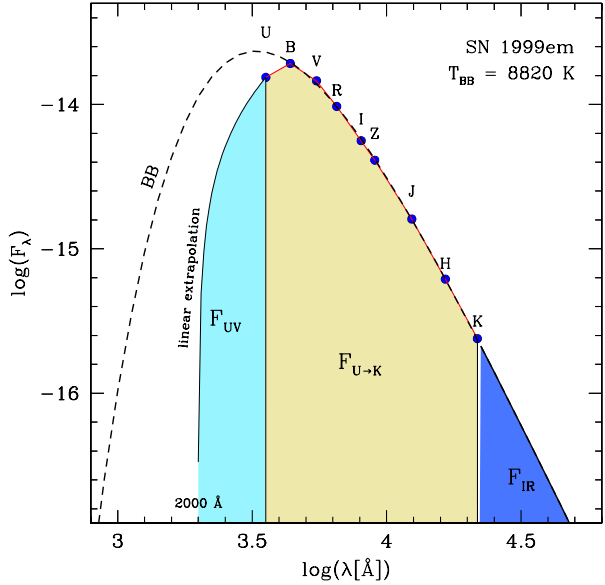}
\caption[Contributions to the bolometric flux]{Contributions
    to the bolometric flux as employed  this work.  F$_{U\rightarrow K}$ is
the  ``quasi-bolometric'' flux computed by integration of  all broad
    band available by each SN. F$_{UV}$ and  F$_{IR}$ represents the
    UV and IR contribution calculated using a  extrapolation of the BB
    fit to the broad band or a linear extrapolation. \label{fig:Flux}} 
\end{center}
\end{figure}

On the UV side, we extrapolate from the effective wavelength of the
$U$ band to $\lambda$=0 using the BB fit on all epochs except when the
$U$-band flux falls below the BB model. In these cases, we extrapolate
the $U$-band flux using a straight line to zero flux at 2000~\AA. Our
choice of $\lambda$= 2000~\AA\ as the wavelength where the flux goes
to zero was based on the behavior of the atmospheric models for which
the flux blueward of 2000~\AA\ is negligible in comparison with the
total flux.  The integrated flux under the Planck function (or
straight line) between the effective wavelength of the $U$ filter and
$\lambda$=0 (or $\lambda$=2000 \AA) is taken as the UV correction
(F$_{UV}$). Figure~\ref{fig:Flux} shows schematically the
  different contributions to the bolometric flux employed  in this
  work.

\begin{figure}
\begin{center}
\includegraphics[scale=.40]{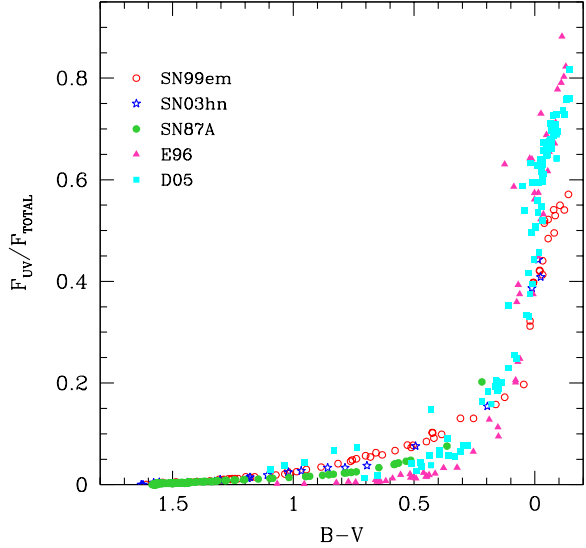}
\caption[UV contribution to the total flux]{UV contribution to the
  total flux as a function of $(B-V)$ 
  for  SN 1987A, SN 1999em and SN 2003hn and for the models of E96
  and D05. At early times, when $(B-V) \lesssim 0.2 $, the UV flux
  represents a significant fraction of the total flux, implying larger
  uncertainties in the UV correction. \label{fig:fuv}}
\end{center}
\end{figure}

 The size of the F$_{UV}$ correction relative to the total flux for
the three SNe and the two sets of atmospheric models is shown in
Figure ~\ref{fig:fuv} as a function of $(B-V)$.  The first thing to
note is the overall good agreement in F$_{UV}$ between the observed
SNe and the atmospheric models.  Second, it is evident that the UV
correction is very important at early epochs and becomes nearly
irrelevant at the latest epochs. Thirdly, note that in the very blue
end, where the UV corrections are of order 50-80\%, there is some
disagreement between the atmosphere models and SN 1999em. The spectral
models suggest larger UV corrections and, as argued below, they are
more trustworthy at these early epochs than the extrapolation of the
broadband magnitudes.  To prove this point, we calculate F$_{UV}$ for
the models using the same technique that we employe for SN 1999em,
i.e. by computing synthetic magnitudes for all passbands, converting
them to monochromatic fluxes, and fitting a BB to the resulting points
(instead of the direct integration of the SED.  In this case the UV
correction proves closer to the UV correction derived from SN
1999em. We conclude that the UV extrapolation using the BB fits to the
broadband magnitudes at very early epochs underestimates somewhat
F$_{UV}$, so we end up using only the atmosphere models at such
epochs.  At later times ($B-V$$> -0.04 $), where the differences
between data and theory become negligible, we adopt both the models
and the observed data. 

\begin{figure}
\begin{center}
\includegraphics[scale=.60]{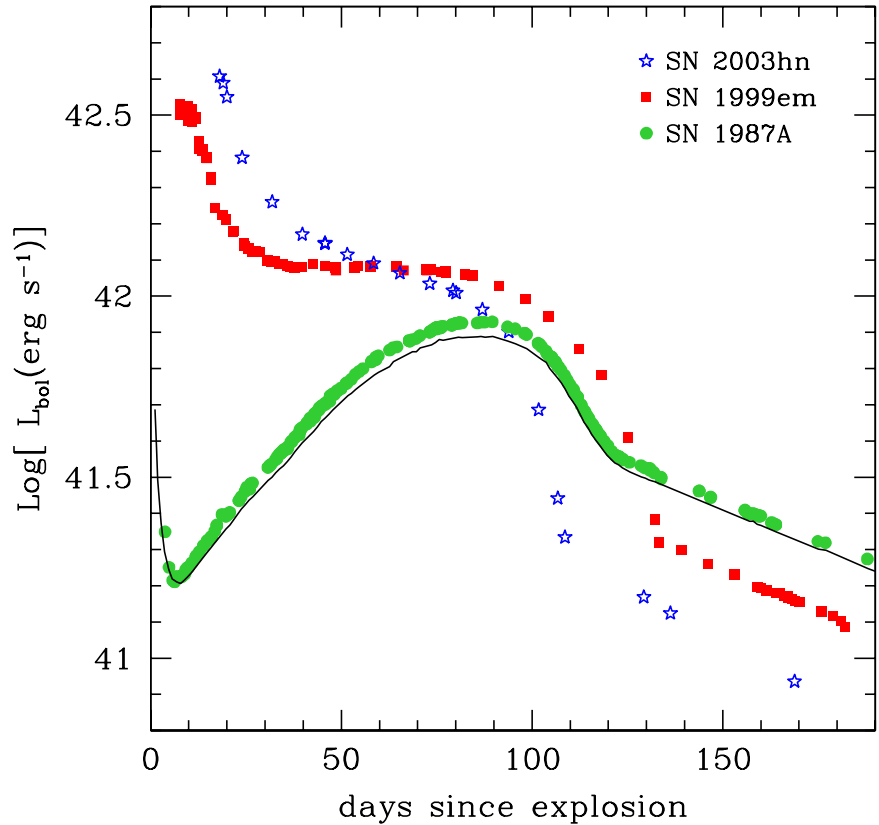}
\caption[Bolometric luminosity of SN 1987A, SN 1999em and SN
  2003hn]{Bolometric luminosity of SN 1987A, SN 1999em and SN 2003hn 
  computed from the integration of broadband optical and near-infrared
  data plus UV and IR contributions as explained in
  section~\ref{sec:Lbol}. For comparison we include the bolometric
  luminosity of SN 1987A obtained by Suntzeff \& Bouchet (1990)(solid
  line).\label{fig:BLSNe}}
\end{center}
\end{figure}

The sum F$_{qbol}$+F$_{UV}$+F$_{IR}$ yields the bolometic flux
F$_{bol}$ (the integration under the  solid curve of
  Figure~\ref{fig:Flux}).  Then we transform flux into 
luminosity using the 
distances given in section \ref{sec:ABC}. The resulting bolometric
luminosities for SN 1987A, SN 1999em and SN 2003hn are shown in Figure
~\ref{fig:BLSNe}. As a comparison, the solid line shows the bolometric
luminosity of the SN 1987A obtained by Suntzeff \& Bouchet (1990). I
find  very good qualitative agreement between both bolometric light
curves for SN 1987A.  There is a systematic difference which remains
smaller than 0.04 dex at all times between both calculations. Such
differences are consistent with the uncertainties arising from  the
use of different photometric  data sets and different integration and
interpolation  scheme.

As can be seen in Figure~\ref{fig:BLSNe}, the morphologies of the
bolometric light curves for the three SNe are very different, especially
that of SN 1987A which shows a broad maximum, not observed in 
classical SNe~II-P, and a less luminous light curve (up to the
transition to the radioactive tail).  The  peculiar light curve of
SN~1987A is  well known and is attributed to the  fact that 
 its progenitor was a compact blue supergiant that lead to a dim initial
plateau and to a light curve promptly powered by radioactivity
\citep{1988ApJ...330..218W,1990ApJ...360..242S} . However, the three
objects showed a similar initial phase of rapid fading and cooling
until the outermost parts of the ejecta reached the
temperature of hydrogen recombination (adiabatic cooling phase). A
second phase can be distinguished for SN 1999em and SN 2003hn which
corresponds to the plateau where the luminosity 
remained nearly constant while hydrogen was recombining. The duration
and the slope of the light 
curve during this phase were different for each supernova. This is
related to the properties of the progenitor object, mainly to the mass
and radius of the hydrogen envelope. The shape of the light curve for
 SN 1987A during this phase was very different as mentioned
above. It showed a broad maximum characterized by a slow rise of $\sim$
90 days followed by a more rapid decline for about 30 days. Finally, I
can distinguish a third phase, the radioactive tail  which has 
similar slope for all three SNe. Here, the luminosity exhibits a
linear decline, and it is dominated by the radioactive decay of
$^{56}$Co into $^{56}$Fe . The luminosity in this part of the light
curve is a direct 
indicator of the amount of $^{56}$Ni synthesized in the explosion
\citep{1989ApJ...346..395W}, the parent product of $^{56}$Co. We
deduce from this that SN 1987A produced 
more $^{56}$Ni than the other two SNe.

\subsection{Bolometric corrections versus Color} \label{sec:BCresults}

There are many SNe that lack IR and UV observations for which it is not 
possible to calculate the bolometric luminosity using 
the method described in the previous section. For these cases it
is necessary to know the bolometric correction required to convert
a $V$-band magnitude into a bolometric flux, i.e.,

\begin{equation}
BC = m_{bol} - [V - A_V],
\label{eq1}
\end{equation}

\noindent
where $A_V$ is the total visual extinction and $m_{bol}$ is the bolometric
magnitude in the Vega system. Note that, since BC is defined as a
magnitude difference, it is independent of the distance assumed for each 
object.

 We calculate BC   at all epochs for each of the calibrating SNe 
and  all of the SN models using the bolometric luminosities computed in
section \ref{sec:Lbol}. The bolometric fluxes were converted into Vega
magnitudes  
in the following manner,

\begin{equation}
m_{bol}=  -2.5 \,  \,log_{10}\, \, F_{bol}\, +  11.64,
\label{eq2}
\end{equation}

\noindent where the zero point is obtained by integrating the SED
of Vega given by Hamuy(2001) and forcing the resulting magnitude
to vanish, i.e., m$_{bol}$(Vega)= 0.

\begin{figure}
\begin{center}
\includegraphics[scale=.40]{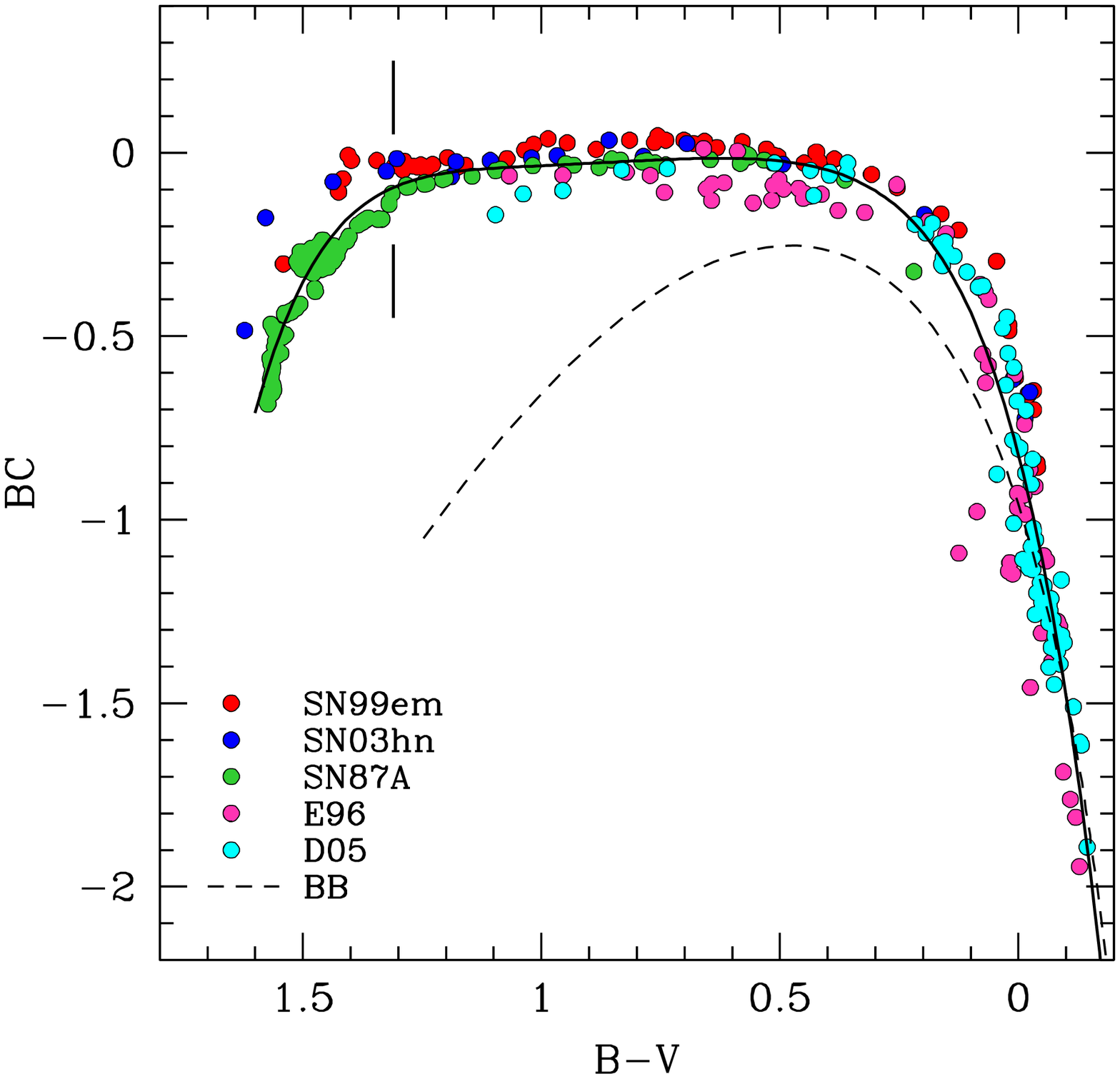}\includegraphics[scale=.40]{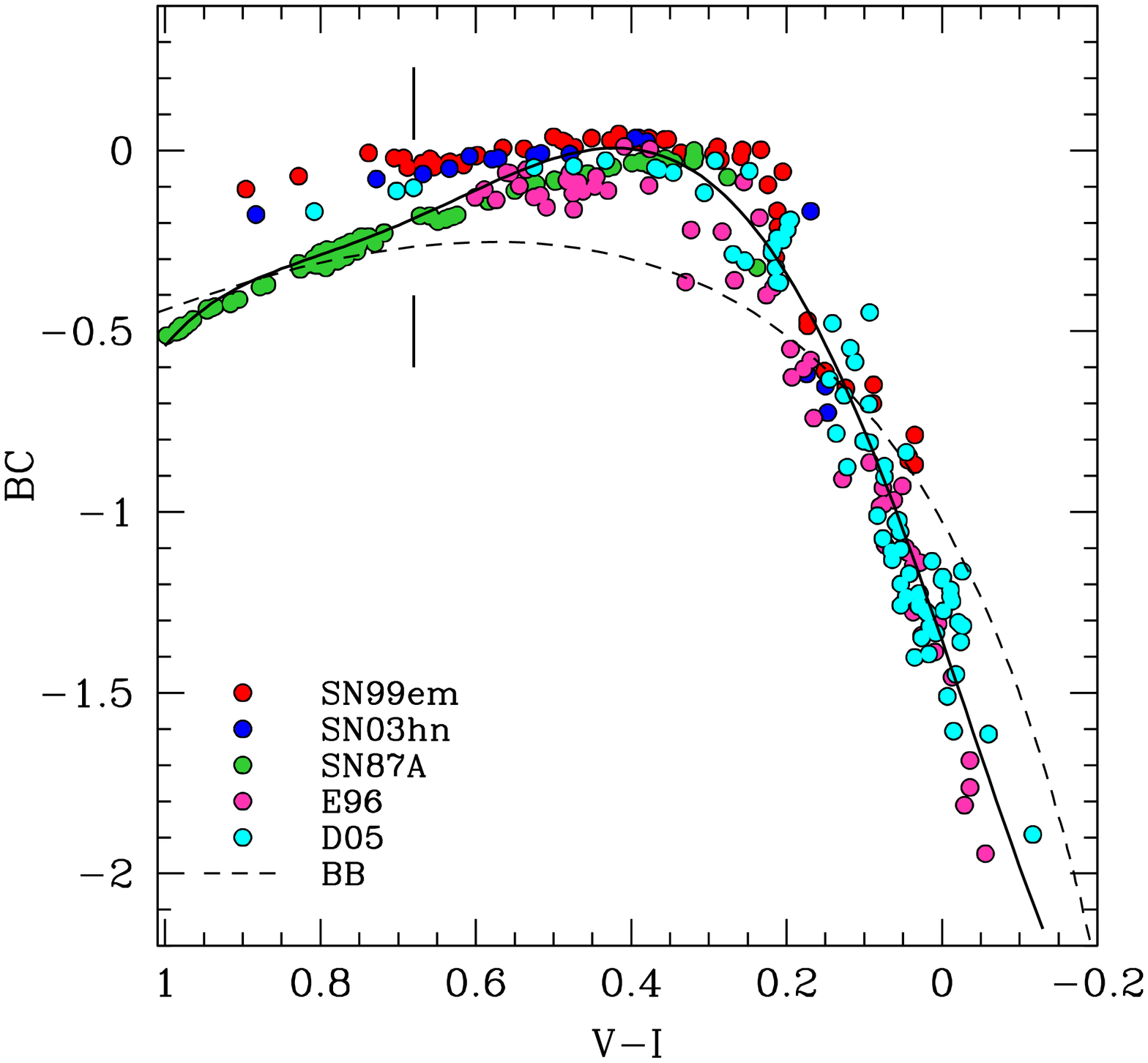} 
\caption[Bolometric corrections versus $(B-V)$ and $(V-I)$]{Bolometric
  corrections versus $(B-V)$ {\bf (left)} and  
$(V-I)$  {\bf  (right)}  for  SN 1999em (open circles),  
SN 2003hn (stars), SN 1987A (filled circles)  and the models of E96
 (triangles) and D05 (squares). The vertical lines indicate 
the color  at  the end of the plateau phase.  The solids line  show   
polynomial fits to the points.  The dashed curves correspond to the 
bolometric corrections of a blackbody spectrum.\label{fig:BC1}}
\end{center}
\end{figure}

\begin{figure}
\begin{center}
\includegraphics[scale=.40]{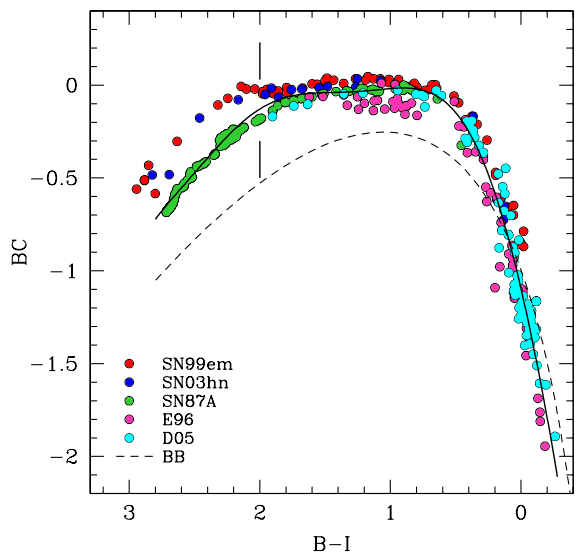}
\caption[Bolometric corrections versus $(B-I)$ ]{Bolometric
  corrections versus $(B-I)$ for  SN 1999em (open circles),  
SN 2003hn (stars), SN 1987A (filled circles)  and the models of E96
 (triangles) and D05 (squares). The vertical line indicates 
the color at the end of the the plateau phase.  The solid line  shows  a 
polynomial fit to the points.  The dashed curve corresponds to the 
bolometric corrections of a blackbody spectrum. \label{fig:BC3}}
\end{center}
\end{figure}

 We  analyzed  the dependence of the BC on color, using
$(B-V)$, $(V-I)$ and $(B-I)$. The main reason why we did not try 
colors involving the $R$ band is because our sample \citep{H09}
has few SNe with $R$-band observations.
Figures \ref{fig:BC1} and  \ref{fig:BC3} show the
resulting corrections  
as a function of the mentioned colors (corrected for dust) 
for SN 1987A, SN 1999em, SN 2003hn, and the models of E96 and D05. 
In each of these plots 
the vertical bar indicates the approximate color corresponding to the end 
of the plateau. 
 To the red of this mark are shown the BC  corresponding to the 
 transition between the plateau and the radioactive tail for the three SNe 
(no atmosphere models cover this phase). We do not include any of the
 nebular data  
in these diagrams, since the BB fits are not appropriate to extrapolate
UV or IR fluxes at these epochs.

These plots reveal a remarkable correlation between BC and intrinsic color, 
both for the objects and the models. It is very satisfactory that, even 
though SN 1987A has a very different light curve compared with normal SNe~II-P,
it matches quite well the behavior of the other two SNe and the models.
At very early times the BCs are quite large owing to the relatively
larger flux contribution of the UV. During most of the plateau 
the BC remains very small around a value of zero, which implies 
that the $V$ magnitude provides a very close proxy bolometric magnitude.
During the transition from the plateau to the radioactive tail (redward from 
the vertical bar) the BC starts to depart from zero due
to the larger flux contribution in the IR. At this phase
SN 1987A shows some discrepancies, at the level of $\sim$0.1-0.2 mag,
with respect to the other two SNe.
Since the bolometric fluxes for SN 1987A comprise two more IR bands, 
$L$ and $M$, we examine the possibility that these discrepancies could be 
due to this fact. For this, we exclude the $L$ and $M$ bands for 
SN 1987A and recompute the bolometric flux in the same manner 
as for the other two SNe, i.e, by calculating $F_{IR}$ as the extrapolation
of a BB fit to $U-K$ photometry.
This exercise shows that, while
the BC corrections over the plateau phase do not change in any significant
way (lending support to the $F_{IR}$ derived from BB fits), by the end of 
the plateau and at later times the new BCs increase and
get closer to the other two SNe. The conclusion is that
the differences observed during the transition are due to an 
inaccurate estimate of $F_{IR}$ from the BB fits restricted to $U-K$ 
photometry. Adding $L$ and $M$ photometry at these late epochs does help
and provides a more accurate estimate of $F_{IR}$. Therefore, during the
transition we decided to exclude the BCs derived from SN 1999em and 2003hn.

A  good representation of the correlation between BC and colors can 
be obtained with polynomial fits of the form,  

\begin{equation}
BC(color)= \sum _{i=0}^{n} a_i \, (color) ^ i,
\end{equation}

\noindent where the order $n$ varies for each color. 
Table ~\ref{clbol:tbl-1} lists the coefficients obtained for the fit of each color, 
their range of validity and the number of  data points used. The fits 
have dispersions (rms) of 0.11 mag for  ($B-V$), 0.11 mag for ($V-I$) 
and 0.09 mag for ($B-I$) in the whole range (plateau plus transition 
to the radioactive tail).

The corresponding polynomials fits are also shown with  solid lines in 
Figure \ref{fig:BC1} and  \ref{fig:BC3}. As  a comparison, 
the dashed lines in these Figures show the BC derived for a blackbody. 
The blackbody models represent well the data at early times (bluest colors),
but evidently differ from the atmosphere models and the observed
SNe at later epochs.

As argued in section ~\ref{sec:Lbol}, we have good reasons to trust more 
the atmosphere models than the early data of SN 1999em, so we decided 
to exclude the latter from our fits. Therefore, our calibration should
be considered more uncertain here. A further complication at early phases
is the steep dependence of the BC on color. This means that a slight 
error in the measurement of the color, such as that due to a poor 
extinction determination, could cause a significant error in the 
determination of the BC.
This problem is less pronounced if we use $(V-I)$  to estimate 
BC at these epochs.  We also test if a bolometric correction with 
respect to the $R$ band instead of $V$ would 
improve the situation, but we do not find any improvements.

\begin{table}
\begin{center}
\caption[Coefficients of the fits to BC]{Coefficients of the fits to
  BC(color)$^{\mathrm{a}}$. \label{clbol:tbl-1}} 
\begin{tabular}{lccc}
\hline
\hline
 a$_i$ &  $B-V$ &  $V-I$ &  $B-I$ \\ 
\hline
a$_0$ &  -0.823 &  -1.355  & -1.096  \\
a$_1$ &   5.027 &   6.262  &  3.038  \\
a$_2$ & -13.409 &  -2.676  & -2.246  \\
a$_3$ &  20.133 & -22.973  & -0.497  \\
a$_4$ & -18.096 &  35.542  &  0.7078 \\
a$_5$ &   9.084 & -15.340  &  0.576  \\
a$_6$ &  -1.950 &  $ \cdots $  & -0.713  \\
a$_7$ &   $ \cdots $ &   $ \cdots $  &  0.239  \\
a$_8$ &   $ \cdots $ &   $ \cdots $  & -0.027  \\
\hline
ranges  & $[-0.2 ,1.65] $  & $ [-0.1, 1] $ & $ [-0.4, 3] $ \\
No. points &  512 &  465 & 512  \\
rms [mag]    &  0.113      &  0.109        &   0.091  \\
\hline
\end{tabular}
\begin{list}{}{}
\item[$^{\mathrm{a}}$]$BC(color)= \sum _{i=0}^{n} a_i \, (color) ^ i$
\end{list}
\end{center}
\end{table}

Using the coefficients given in Table ~\ref{clbol:tbl-1}, it is possible to 
derive a bolometric luminosity for any SN~II-P using only two (or three 
in the case of the $(B-I)$ color) optical filters. If one knows the 
extinction and the distance to the object the bolometric luminosity 
can be computed as follows:

\begin{equation}
\log L[ \mathrm {erg \,~s}^{-1}]=  - 0.4 \,\, [ BC(color) + V - A_{total} (V)  - 11.64  ]  + \log( 4 \, \pi D ^2 ),
\label{eq3}
\end{equation}

\noindent where $D$ is the distance in cm to the SN
and $A_{total}(V)$ is the total, 
host plus Galactic visual extinction. Note that combining this equation 
with equations (\ref{eq1}) and (\ref{eq2}), the luminosity becomes independent
of the arbitrary zero points chosen for the Vega magnitude scale, and it 
only depends on the observed flux density, color, extinction and distance.

\begin{figure}
\begin{center}
\includegraphics[angle=-90,scale=.40]{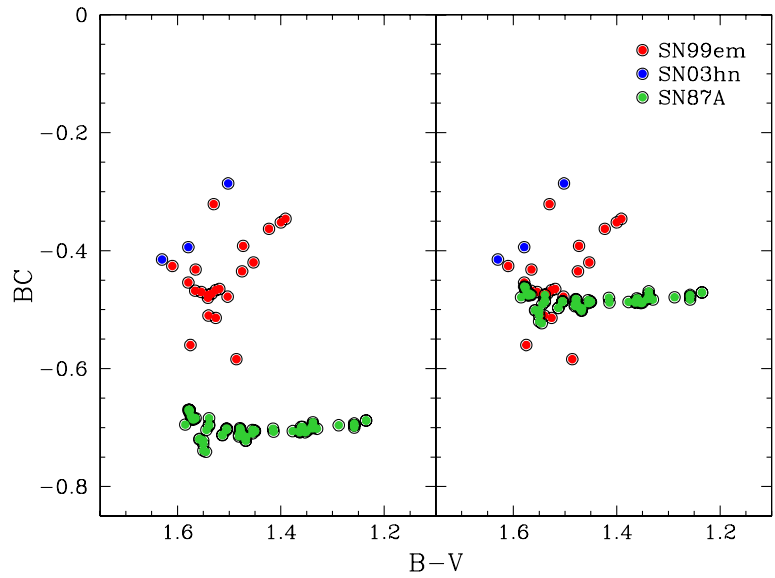}
\caption[Bolometric corrections for radioactive tail phase versus
  $(B-V)$   ]{{\bf( Left)} 
  Bolometric corrections obtained during the 
  radioactive tail phase versus $(B-V)$, as derived using
  $U$ through $M$ photometry for SN 1987A (filled circles), and $U$ through
  $K$ photometry for SN 1999em (open circles) and SN 2003hn (stars).
  {\bf (Right)} Same as before, but excluding the $L$ and $M$ bands 
  for SN 1987A.  \label{fig:BCT}}
\end{center}
\end{figure}

The calibrations of BC versus colors shown above are only valid during 
the optically thick phases since they involve BB fits to the photometry.
In the nebular phase, we calculated BCs for the three SNe using the integrated
flux between the observed bands (i.e. $U-K$ for SN 1999em and SN 2003hn,
and $U-M$ for SN 1987A). We did not attempt to add any flux beyond these
limits since we did not have any physical model to extrapolate. As shown in
 the left panel of Figure \ref{fig:BCT} the BC for SN 1987A is almost
 independent of color,  
with a value  $\sim -0.7$ mag and a scatter of only 0.015 mag.
The other two SNe yield BCs 0.2--0.3 mag higher, with a slight dependence 
on color. 
We investigate whether these differences could be due to the inclusion 
of the two additional bands for SN 1987A: we removed the $L$ 
and $M$ bands from the BC and, not surprisingly, the agreement proved
much better (Right panel of Figure \ref{fig:BCT}). We conclude that the
$L$ and $M$ contributions 
to the bolometric flux is not negligible at the nebular phase. Hence,
we take the value of $-0.70$ derived from SN 1987A as the best estimate 
of the BC at the onset of the nebular phase. 

We conclude this  section with the claim that we have implemented
a robust method to estimate BCs for SNe~II-P which allows one
to derive bolometric luminosities. If we trust the late behavior of SN
1987A as being representative of SNe~II-P in general, our analysis
implies an overall accuracy of 0.05 dex in BCs. Clearly, it would be
interesting to check this result 
using L and M photometric data of other SNe~II-P, but such data are
currently unavailable. Our calibrations have the potential to be applied
to many SNe observed over a limited wavelength range.

\subsection{Effective Temperature-Color Relation}
\label{sec:Teffresults}
Along with bolometric luminosity, 
the effective temperature is a critical parameter in the comparison of
observations with hydrodynamical models. 
Each model spectrum of E96 and D05 has an associated effective temperature
(T$_{eff}$), defined by the relation $L= 4 \pi  R_{ph}^2 \sigma T_{eff}^4$ 
where L is the input luminosity of the atmospheric models and
$R_{ph}$, the photospheric radius, is an output of the models. I
examined the dependence of 
T$_{eff}$ on $(B-V)$ and $(V-I)$ colors derived via synthetic
photometry from the model spectra, as described in section \ref{sec:Lbol}.
The purpose of this analysis was to provide a calibration between
temperature and color which could 
be used to easily derive estimates of T$_{eff}$ from observed colors
for any SN~II-P. Note, however, that T$_{eff}$ does
not have a direct physical meaning for this type of object. It is
simply a convenient contact point between hydrodynamical models and
observations.

\begin{figure}
\begin{center}
\includegraphics[angle=-90,scale=.40 ]{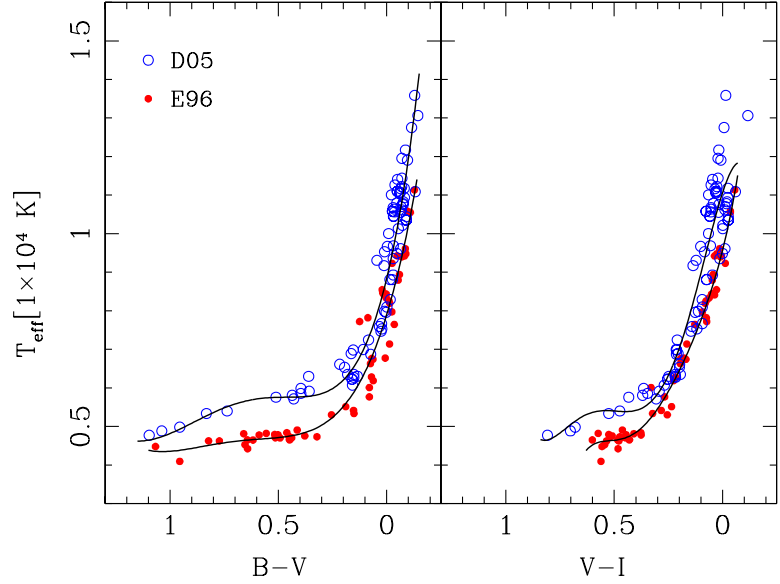}
\caption[Effective  temperature versus $(B-V)$]{ Effective
  temperature versus $(B-V)$ {\bf (left)} and 
  $(V-I)$ {\bf (right)} from the models of E96 
(filled circles) and D05 (open circles). The solid lines show polynomial fits 
for each set of models. \label{fig:Teff}}
\end{center}
\end{figure}

Figure \ref{fig:Teff} shows the effective temperature versus synthetic 
$(B-V)$ and $(V-I)$ colors for E96 and D05 models. 
As expected, there is a tight correlation between these quantities 
for each set of models. At early epochs, when $(B-V) \lesssim 0.2 $ and
$(V-I) \lesssim 0.3$,
both models show consistent values of the effective temperature within
their internal dispersion. Later on, however, when the
plateau phase is well established, there are systematic differences in
the behavior of both sets, with the models of D05 giving   
larger effective temperatures. Similar differences have been reported 
in the literature with regard to the dilution factors calculated from 
both sets of models \citep{D05b,2009ApJ...696.1176J}, but there has
been no clear 
explanation for the discrepancies. Note that
 T$_{eff}$ ( $\propto \sqrt{R_{ph}}$) is an
output of the atmosphere model and depends on complicated 
details of the solution of radiation transport through the envelope,
such as non-LTE treatment of the different species and metal line
opacities.

In order to represent the correlation between  T$_{eff}$  and colors 
shown in Figure \ref{fig:Teff}, we fit polynomial functions 
of the form,

\begin{equation}
T_{eff}(color)[10^4 K]= \sum _{i=0}^{n} a_i \, (color) ^ i.
\end{equation}

\noindent The fits were done for each set of models separately and they 
are shown in Figure \ref{fig:Teff} with solid lines. The coefficients of the 
fits, ranges of validity for $(B-V)$ and $(V-I)$ colors, and dispersions 
are given in Table \ref{clbol:tbl-2}. The fits to the E96 models 
are characterized by a scatter of $\sim$500 K in $ (B-V)$  and $\sim$350 K
in $(V-I)$. For the D05 models the scatter is $\sim $670~K in $(B-V)$ 
and $\sim$800 K in $(V-I)$.  We do not  have any  strong argument to  
rule out either set of models. We therefore keep both results even if they 
show significant systematic differences.  But we notice that
the value of  T$_{eff}$ during the recombination phase (where it is nearly 
constant) appears to be somewhat underestimated  (T$_{eff} \sim 4600$ K) by 
E96. We recall that these calibrations are only valid until the end
of the plateau phase.

\begin{table}
\begin{center}
\caption[Coefficients of the fits to T$_{eff}$]{Coefficients of the
  fits to T$_{eff}$(color)$^{\mathrm{a}}$. \label{clbol:tbl-2}} 
\begin{tabular}{lcccc}
\hline
\hline
   &  $B-V$ & $B-V$&  $V-I$ &  $V-I$ \\
 a$_i$ &  $E96 $  &  $D05 $  &   $ E96 $ &   $ D05 $\\
\hline
a$_0$ &  0.790 &  0.884  &  0.957  &  1.106   \\
a$_1$ & -1.856 & -2.340  & -2.254  & -1.736   \\
a$_2$ &  4.055 &  6.628  &  5.922  & -6.403   \\
a$_3$ & -3.922 & -8.456  & -18.476 &  33.762  \\
a$_4$ &  1.368 &  4.619 &  36.058 & -48.260  \\
a$_5$ &  $\cdots$ & -0.849 & -25.291 &  22.362  \\

\hline
ranges    & $[-0.2,1.15] $  & $[-0.2,1.15]$ & $[-0.1,0.65]$ & $[-0.07,0.83]$  \\
rms [K] &  500 & 670 & 350 &  800 \\
\hline
\end{tabular}
\begin{list}{}{}
\item[$^{\mathrm{a}}$]$T_{eff}(color)[10^4 K]= \sum _{i=0}^{n} a_i \, (color) ^ i$
\end{list}
\end{center}
\end{table} 

\begin{figure}
\begin{center}
\includegraphics[scale=.60 ]{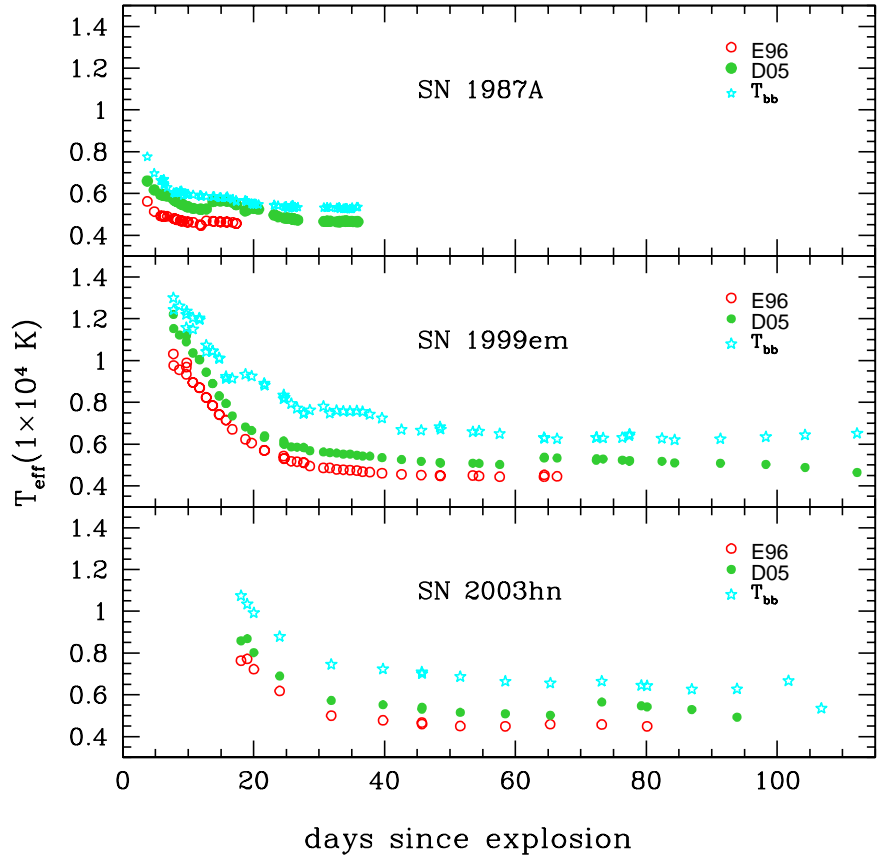}
\caption[Effective temperatures  for SN 1987A, SN 1999em and SN
  2003hn]{Effective temperatures  for SN 1987A ({\bf upper panel}),  
SN 1999em ({\bf middle panel})  and SN 2003hn ({\bf bottom panel }) 
as a function of time since explosion
  calculated using the polynomial fits given in table \ref{clbol:tbl-2} for
  the models of E96 (open circles) and  D05 (filled circles). As a comparison
  we include in these plots the color temperatures obtained from the
  BB fits described in section \ref{sec:Lbol}. 
 \label{fig:Teff2}}
\end{center}
\end{figure}

In Figure \ref{fig:Teff2} we show how these calibrations work for estimating  
T$_{eff}$ for SN 1987A, SN 1999em, and SN 2003hn. 
As a comparison we included in these plots  color temperatures 
obtained from the BB fits described in section \ref{sec:Lbol}. Note that 
for the three SNe the color temperatures are greater than 
$T_{eff}$, which is expected for ``dilute'' atmospheres whose 
continuum opacity is dominated by electron scattering. As  said
above, the usefulness of these fits is that they allow one 
to obtain  T$_{eff}$ for any SN~II-P from their $(B-V)$ or $(V-I)$
colors, and thereby use this quantity  to compare with hydrodynamical
 models. Note that we could equivalently have chosen to calibrate
 R$_{ph}$ vs color. 

\section{Photospheric velocities}
\label{sec:vphoto}

Another important parameter to compare with observations is the
photospheric velocity yielded by our models.  Expansion velocity
estimates from the minimum of several spectral lines (H$_\alpha$, Fe
{\sc ii} $\lambda5169$, H$_\beta$ and H$_\gamma$) of our data set were
given by \citet{2009ApJ...696.1176J}. Since each line forms at a
different shells, it is not straight forward  to compare the observed
velocities with photospheric velocities. \citet{2009ApJ...696.1176J}
got around this problem and derived congruent calibrations between the
velocity derived from the absorption minimum of such lines and the
photospheric velocity using two independent atmosphere models
(E96 and D05). As
noted by \citet{2009ApJ...696.1176J} H$_\beta$ provides a very good
proxy to the photosphere velocity as it is not  highly saturated as 
H$_\alpha$, and is present over most of the SN evolution. Therefore,
in the comparisons with our models, we chose to use the calibration
given by \citet{2009ApJ...696.1176J} for this particular
line. However, in spite of the satisfactory behavior of this
calibration in the high-velocity regime ($v \gtrsim 5000$ km
s$^{-1}$), for lower velocities the atmosphere models fail to
reproduce the behavior shown by the data \citep[see Figure~9
  of][]{2009ApJ...696.1176J}.  Thus, we decided to complement  our
comparison with velocities estimated from the Fe~{\sc ii} $\lambda5169$
line for the case where it is present at later phases of the
  SN evolution and therefore the 
velocities are below the limit imposed by the calibration of hydrogen
lines.

 For consistency with the atmosphere models, we define the photospheric
 position for our hydrodynamical models as the layer where the total
 continuum optical depth is 
 $\tau=2/3$. Note that to calculate $\tau$ we do not include the
 opacity floor (see discussion in section~\ref{sec:constitutive}) because such
 minimum is only included to take into 
 account line effects, and therefore is a bound-bound opacity that
 does not contribute to form the continuum. With this definition, the
 photosphere follows the recombination wave, as we show in section
 ~\ref{sec:CRW},  due to
 the fact that electron scattering is the dominant source of  opacity
 (see Figure ~\ref{fig:opal}). Had we included the opacity floor in
 the definition of $\tau$,  the $\tau=2/3$ surface would be pushed
 well above the recombination front.

 We remark that  with this definition, the photosphere is essentially the
surface of last scattering. However, the surface where the continuum
is actually formed is located in a deeper layer called the
``thermalization depth'' and this is due to the dominance of the
electron scattering over the absorption processes
\citep{1980Afz....16..695S,1991supe.conf..415H,1995ApJ...445..828M}. With
our simple prescription of radiative transfer we cannot accurately
determine the location of the thermalization depth because there
radiation decouples from matter. Note that the color temperature,
determined by a black body fit to the broad-band photometry, is nearly
coincident with the temperature of the thermalization depth but  is
greater than the effective temperature (as also shown  in
  Figure~\ref{fig:Teff2}).

\chapter{Hydro-Model for SN~1999em}\label{sec:sn99em}
In this Chapter, I present  a detailed study of the
prototype SN~1999em using the  hydrodynamical  code described
in Chapter~\ref{ch:code}. The choice of this object was
motivated by the fact that it is one of the best observed SNe of its
type, both in terms of wavelength coverage and  temporal
sampling. In \S~\ref{sec:hydro} I present the bolometric
light curve and photospheric velocities for this object resulting
 from model  calculations. A remarkably 
good agreement with observations was obtained using the following physical
parameters: $E=1.25$ foe, $M= 19 \, 
M_\odot$, $R= 800 \, R_\odot$ and $M_{\mathrm {Ni}}=0.056 \,
M_\odot$. In this analysis, I find that an extensive mixing of
$^{56}$Ni is required in order to reproduce a plateau as flat as that
shown by the observations. In
 \S~\ref{sec:compsn99em} I discuss how our results compare with two
 previous hydrodynamical studies of SN~1999em
 \citep{2005AstL...31..429B,2007A&A...461..233U}. Finally, in
 \S~\ref{sec:lowmass} I study the possibility of fitting the observations
with lower values of the initial mass 
consistently with the upper limits that have been inferred from
pre-supernova imaging of SN~1999em in connection with stellar
evolution models.  I cannot find a set of physical parameters that
reproduce well the  
observations for models with pre-supernova mass of $\leq$ 12
$M_\odot$, although models with 14 $M_\odot$ cannot be fully discarded.

\section{High-Mass Model of SN~1999em}
\label{sec:hydro}

SN~1999em was discovered  shortly after  explosion on 1999 October 29 UT
by the LOSS program \citep{1999IAUC.7294....1L}. Based on EPM study by
\citet{2009ApJ...696.1176J}, we assume SN~1999em exploded
three days before discovery. This is consistent with the
constraint imposed by a negative detection (limiting magnitude 19) on
an image obtained on 1999 October 20 UT. In order to compare with our model,
we correct the times elapsed since explosion  by time
dilation based on the redshift of the host galaxy. We adopted the  
Cepheid distance of 11.7 Mpc given by \citet{Leo03} to compute bolometric
luminosities as explained in \S~\ref{sec:BCresults} using
 photometric data obtained at  
{\em Cerro Tololo Inter-American Observatory} (CTIO),  
{\em Las Campanas Observatory} (LCO), and the 
{\em European Southern Observatory} (ESO) at  {\em La Silla} \citep{H01}.\\  

We use a pre-supernova model with initial parameters
consistent with the optimal hydrodynamical model of
\citet{2007A&A...461..233U}. Specifically, we adopt an initial
mass of 19 $M_\odot$,  radius of  800 $R_{\odot}$, and 
explosion energy of $E=1.25$ foe (1 foe=  $1\times 10^{51}$
erg). This energy was 
released as thermal energy  near  
the core of the object in a very short time scale as compared
with the hydrodynamic time scale of our model. We also assume a
nickel mass of $0.056$ $M_{\odot}$ which was determined by the
luminosity of the radioactive tail. This parameter is quite
different to the nickel mass of $0.036$ $M_{\odot}$ used by
\citet{2007A&A...461..233U} which was determined using the
quasi-bolometric luminosity given by \citet{2003A&A...404.1077E}. Their
luminosity was based on the integration of only $UBVRI$ photometry with
a constant value of $0.19$ dex added to take into account the infrared
luminosity, while our bolometric luminosity is based on a quantitative study
of the bolometric correction for SNe~II-P (see
\S~\ref{ch:model_obs}). Note also that our value for the nickel mass
is closer to the estimate of $0.06$ $M_{\odot}$
given by \citet{2005AstL...31..429B}. In the following analysis, we
denote this model as \model.\\

\begin{figure}[htbp]
\begin{center}
\includegraphics[angle=-90,scale=.30]{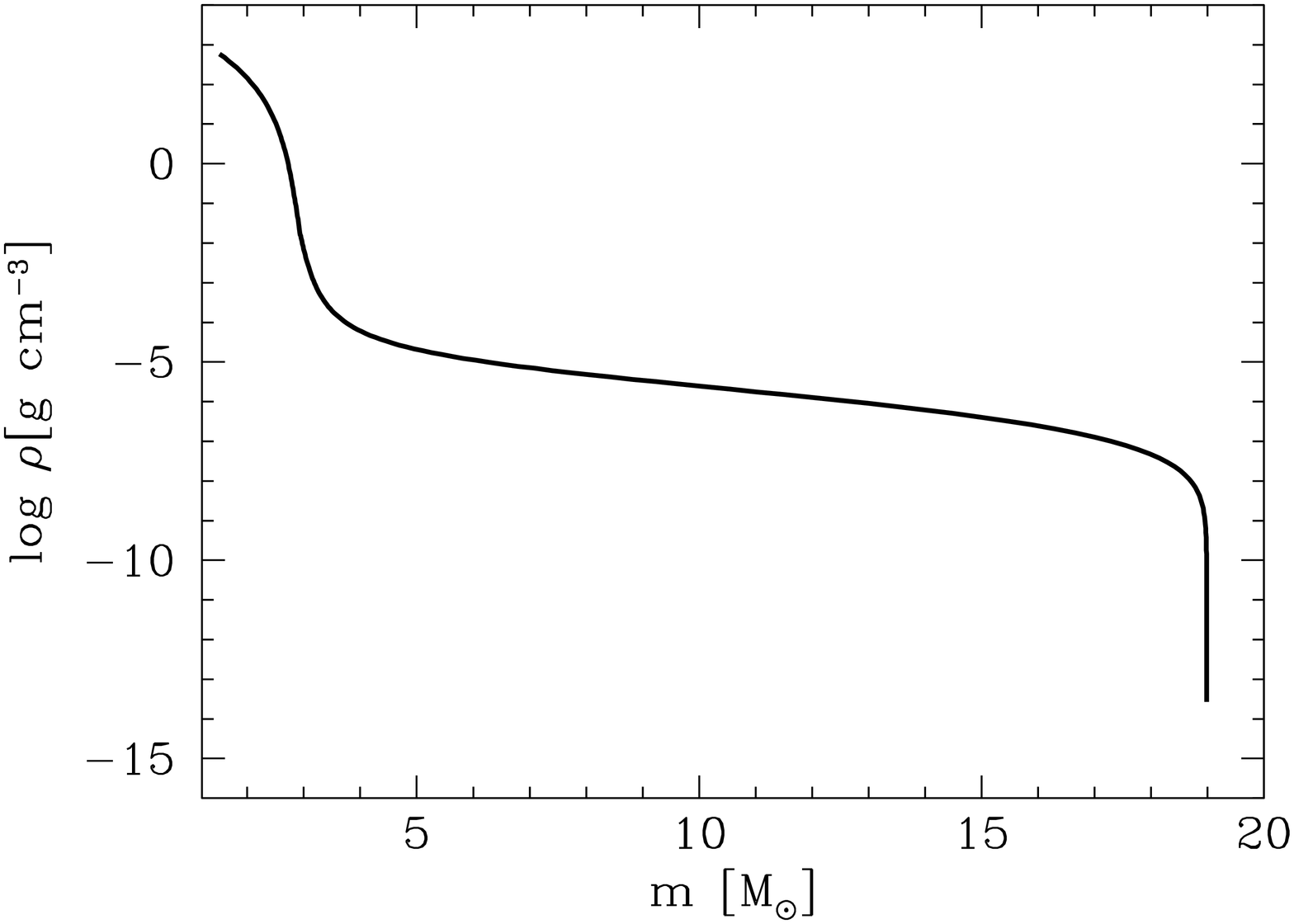}\includegraphics[angle=-90,scale=.30]{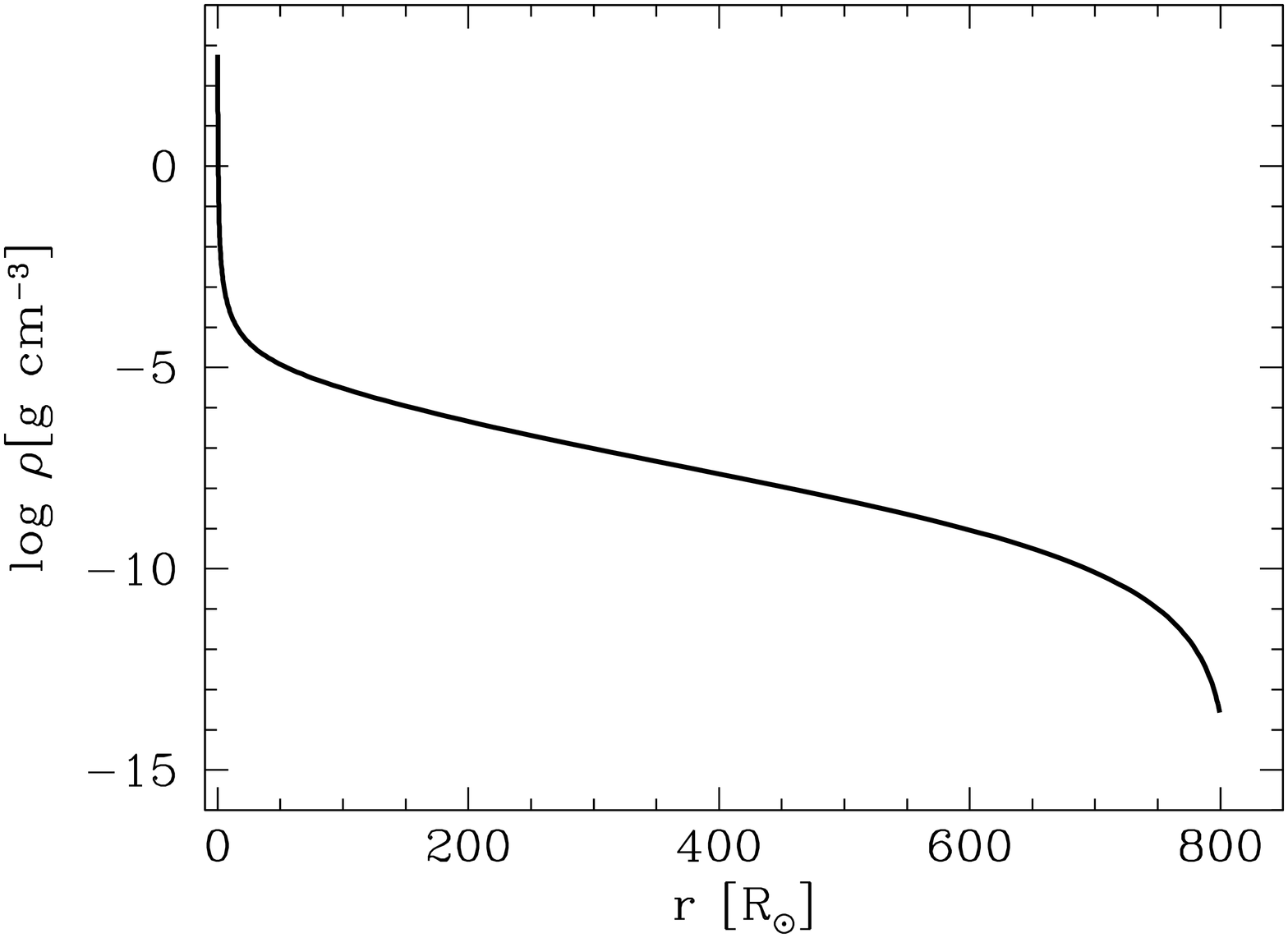} 
\caption[Initial density distribution for model \model]{Initial
  density distribution with respect 
  to interior mass {\bf (left)} and radius {\bf (right)} for
  the pre-supernova model \model.\label{fig:dens}}
\end{center}
\end{figure}

\begin{figure}[htbp]
\begin{center}
\includegraphics[angle=-90,scale=.30]{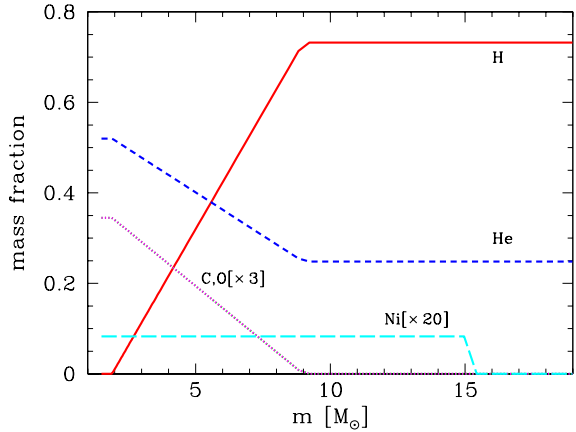} 
\caption [Abundance distribution for model
  \model]{Abundance distribution in the pre-supernova model
  \model\ with respect 
  to Lagrangian mass. For clarity, the abundance of
  carbon (C) and  oxygen (O) were multiplied by 3 and the abundance of
  $^{56}$Ni, by 20. Note that the $^{56}$Ni is uniformly mixed in the
  outer envelope until 15 $M_\odot$. \label{fig:comp}} 
\end{center}
\end{figure}

In our calculations we remove the central $1.4$ $M_\odot$ which is assumed to
form a neutron star. The initial density profile as function of mass
and radius for model \model \  is shown in
Figure~\ref{fig:dens}. Note that the initial 
structure is composed of a dense core
and an extended envelope characteristic of a red supergiant. The
chemical composition profiles are shown in Figure~\ref{fig:comp}. The
outer parts of the envelope, $M > 9 \, M_\odot$, 
have a homogeneous composition with  mass fractions for hydrogen of
$X=0.735$, for helium of $Y=0.251$, and a metallicity of
$Z=0.02$. From there inward, hydrogen and helium are mixed in order to
prevent a sharp boundary between the H-rich and
the He-rich layers. Such sharp boundaries are characteristic of
stellar evolution models but fail to reproduce the observations. Note
that we allow H to mix very deep inside the core and  $^{56}$Ni is mixed 
out in the envelope  up to  $\sim$15 $M_\odot$. This type
of mixing is 
presumably due to Rayleigh-Taylor
instabilities that occur behind the shock front, as  obtained in
multi-dimensional hydrodynamic calculations
\citep{1991A&A...251..505M,2000ApJ...528..989K}
and supported  
 by studies of SN~1987A \citep[][among others]
{1988A&A...196..141S,1988ApJ...324..466W,1988ApJ...331..377A,2000ApJ...532.1132B}. The
presence of H in the core   
leads to a smooth transition between the plateau and the radioactive
tail. The distribution of $^{56}$Ni to external layers helps to
reproduce a plateau as flat as that observed in SN~1999em (see
section~\ref{sec:CRW}).

\begin{figure}[htbp]
\begin{center}
\includegraphics[angle=-90,scale=.35]{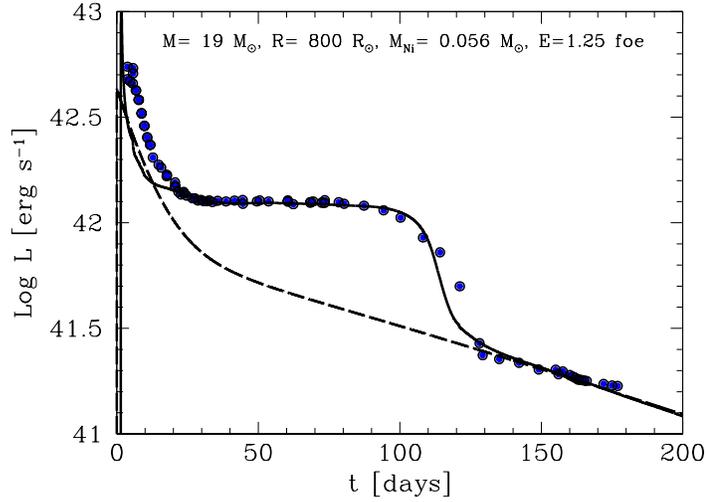}
\caption[The bolometric light curve for  model  \model\ compared with
  the data of SN~1999em]{The bolometric light curve for  model  
 \model\ (solid line) compared with the data of SN~1999em as calculated
 by \citet{2009ApJ...701..200B} (blue dots). The 
  luminosity due to the $^{56}$Ni  $\rightarrow$ $^{56}$Co
  $\rightarrow$  $^{56}$Fe decay is  
 also shown (dashed line). The physical parameters used in the model  
 are indicated. \label{fig:LC}} 
\end{center}
\end{figure}

Figure~\ref{fig:LC} shows a comparison between the bolometric light
curve (solid line) obtained with our code for  model
\model, and the observations of SN~1999em (dots). The luminosity
due to $^{56}$Ni $\rightarrow$ $^{56}$Co $\rightarrow$  $^{56}$Fe
(dashed line) is also shown. Note the very good 
agreement between  model and observations. The largest difference
appears during the earliest phase and  the transition to the
radioactive tail. At earliest epochs our bolometric corrections have
the highest uncertainties, as  noted in \S~\ref{sec:BCresults},
and during the transition 
between the plateau and the radioactive tail, the
diffusion approximation breaks down because
the object is almost completely recombined and the photosphere is not well
defined. During the tail, the bolometric 
luminosity is completely determined by the luminosity of radioactive decay,
which makes our calculations more reliable. Although not shown here,
our code shows that  the shape  
of the light curve at the end of the plateau is very sensitive to the
properties of the core, such as mixing and the form of the 
density transition between the helium-rich core and the hydrogen-rich 
envelope.  In principle, it is possible to find an even better  
fit to the observations but, given the uncertainties introduced by our 
approximations, such a detailed study  should be considered as meaningless.

\begin{figure}[htbp]
\begin{center}
\includegraphics[angle=-90,scale=.3]{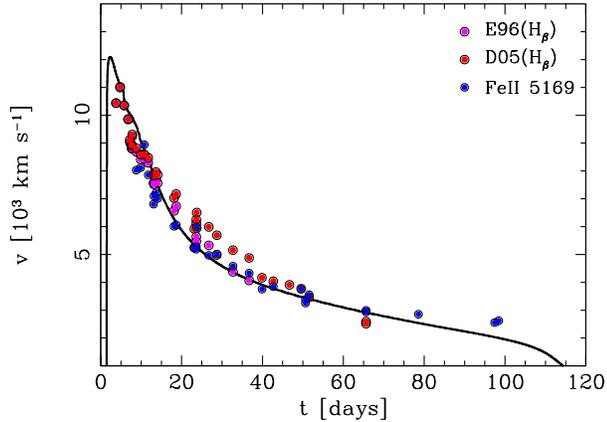}
\caption[The photosphere velocity for model \model\ compared with data
of SN~1999em]{Evolution of the expansion velocity of the photosphere
for model \model\ (solid line) compared with observed  photospheric
velocities  calculated by \citet{2009ApJ...696.1176J} using a
calibration between the velocity derived from the absorption minimum
of H$_\beta$  and two atmosphere models: E96 (filled squares) and D05
(open circles). We also include the velocities calculated from 
Fe~{\sc ii} $\lambda5169$ (filled circles). \label{fig:velo}}
\end{center}
\end{figure}

Figure~\ref{fig:velo} shows the photospheric velocity evolution of our
model compared with observed photospheric velocities as explained in
\S~\ref{sec:vphoto}.  We have also included in the plot the  
spectroscopic velocities measured from the absorption minimum of
the Fe~{\sc ii} $\lambda5169$ line \citep{2009ApJ...696.1176J},
reveling that the velocities
estimated from this line samples quite well the photospheric velocity
at all epochs when it is present. However, it is not present at the
very earliest times ($t< 10$ days). Note the very good overall
agreement between model and observations. 
Fe~{\sc ii} velocities match quite well the model photospheric
velocities except at the latest times. However, it is
important to remark that the photospheric velocity is to
be considered a good discriminator between models only at the early
phases of the evolution while the object is not completely recombined. This  
is because, line velocities become poor photospheric
velocity indicators with time, and the photosphere 
begins to lose its meaning in our models as the ejecta becomes nearly
completely recombined at the end of the plateau.

As shown in Figures~\ref{fig:LC} and \ref{fig:velo} we
obtain a remarkably good agreement with observations (bolometric
light curve and  photospheric velocity evolution) for SN~1999em
despite the simplifications used in our code. These results are very
encouraging and give us confidence in the capability of our code to
infer physical parameters and to study their effect on the observed
quantities with the ultimate aim of understanding the physics of SNe~II-P.

In the following subsections, we describe in some detail
the evolution of the SN for this particular model. At a glance,  the LC is
distinguished by three phases: a) an outburst followed by strong
cooling, b) a plateau, and c) a radioactive tail.  Each phase is
essentially determined by the interplay between the main 
heating and cooling mechanisms. We thus focus our discussion on
these processes along the SN evolution. As a reference,
  Table~\ref{sn99:tbl-1} gives a summary of some
  properties of the model at characteristic times during the
  evolution. The initial 
phase due to shock breakout is discussed in \S~\ref{sec:SW}, the
adiabatic-cooling phase where the homologous expansion is reached is
addressed in \S~\ref{sec:AC}, the cooling and recombination phase is
described in \S ~\ref{sec:CRW}, and finally the radioactive-decay
processes that power the tail are explained in \S~\ref{sec:tail}.

\setlongtables
\begin{table}
\begin{center}
\scriptsize
\caption[Properties of model \model]{Properties of model \model\ at selected time of their evolution.\label{sn99:tbl-1}}
\begin{tabular}{lcccccccccc}
\hline
\hline
          &  $t$ & $\log(L_{\mathrm{bol}})$ & $\log(T_{\mathrm{eff}})$ &
$\log(T_{\mathrm{ph}})$ &  $R_{\mathrm{ph}}$ & $v_{\mathrm{ph}}$  &
$E_{\mathrm{rad}}/E_0^{\mathrm{a}}$& 
 $E_{\mathrm{K}}/E_0^{\mathrm{a}}]$\\
Phase  & [days] & [erg~s$^{-1}$] & [K] & [K]&
[10$^{14}$cm] & [10$^{8}$cm~s$^{-1}$ ] & (\%) &
(\%)\\   
\hline

Peak       &    1.5     &  44.5   &  5.05 & 5.03  &  0.54  & 3.36 & 0.04 & 72  \\
Adiabatic Cooling  &    5  &  42.4   &  4.10 & 4.05  &  3.8   & 11.5 & 0.2  & 93  \\
Plateau    &    50      &  42.1   &  3.73 & 3.77  &  14.7  & 3.47 & 0.65 & 99  \\
Transition &    114     &  41.7   &  3.74 & 3.74  &  10    & 1   & 1.2  & 99.7 \\

\hline
\end{tabular}
\begin{list}{}{}
\item[$^{\mathrm{a}}$] $E_{rad}$ is defined by $\int_0^t L_{bol} \, dt$ and
  $E_K$ is the total kinetic energy in the mass motions. In the table
  we show these quantities normalized to the initial injected  energy
  of  model \model, $E_0=1.25$ foe.
\end{list}
\end{center}
\end{table}

\subsection{Shock wave propagation and the early evolution}
\label{sec:SW}
A powerful shock wave (SW) begins to propagate outward through the envelope when
we artificially inject energy near the center of the star (assumed to occur at
$t=0$). This energy is initially released as internal energy, and
part of it is rapidly transformed into kinetic energy (e.g., by $t=0.5$~days,
the total energy is  approximately equally divided between kinetic
and internal energy). The velocities acquired by matter are so high
that they exceed the local speed of sound, leading to the formation of
a SW. The SW heats and accelerates the matter depositing mechanical
and thermal energy into successive layers of the envelope until it
reaching the surface, where photon diffusion dominates the energy
transfer, and energy begins to be radiated away. 

\begin{figure}[htbp]
\begin{center}
\includegraphics[angle=-90,scale=.30]{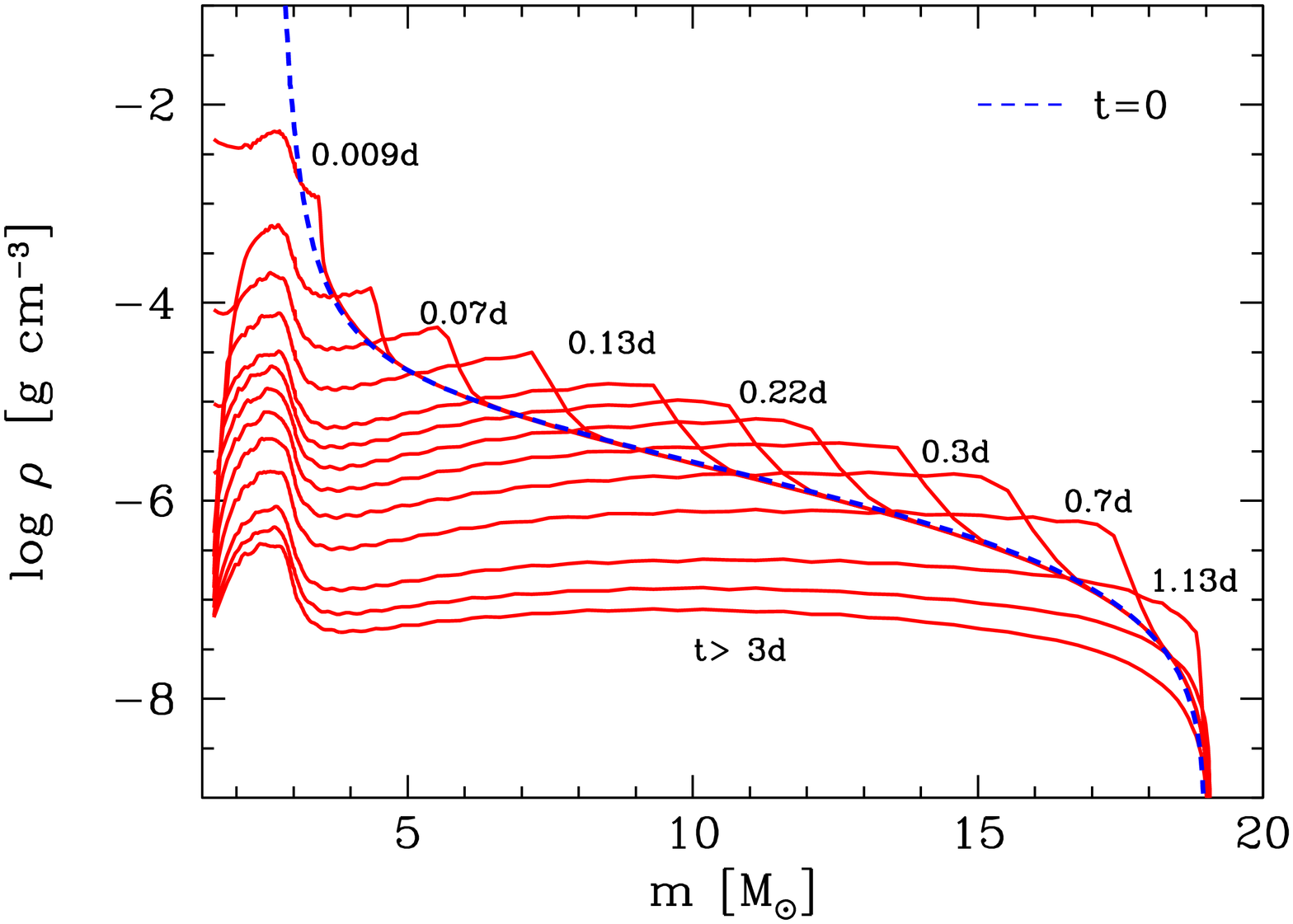}\includegraphics[angle=-90,scale=.30]{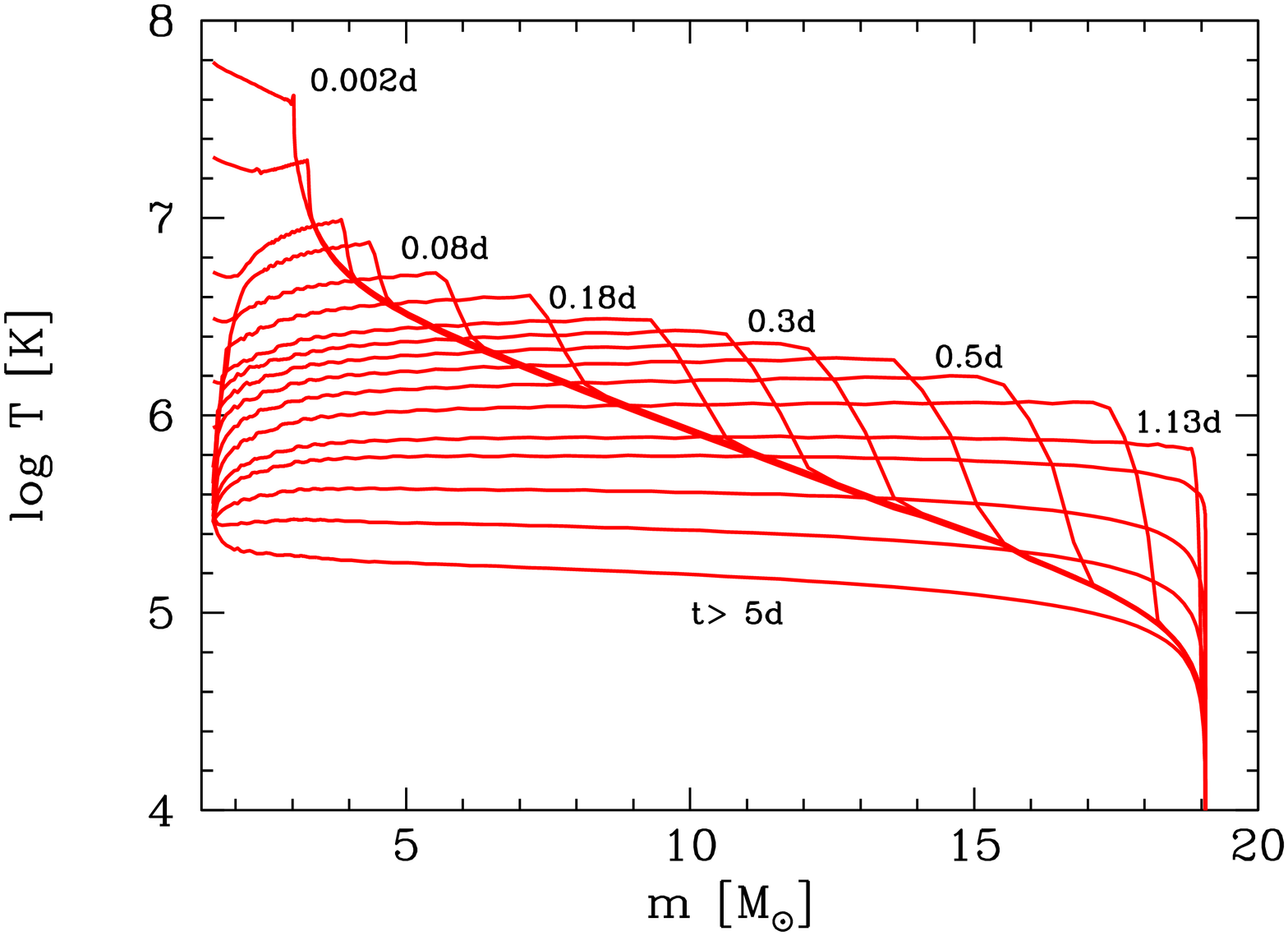} \\ 
\includegraphics[angle=-90,scale=.30]{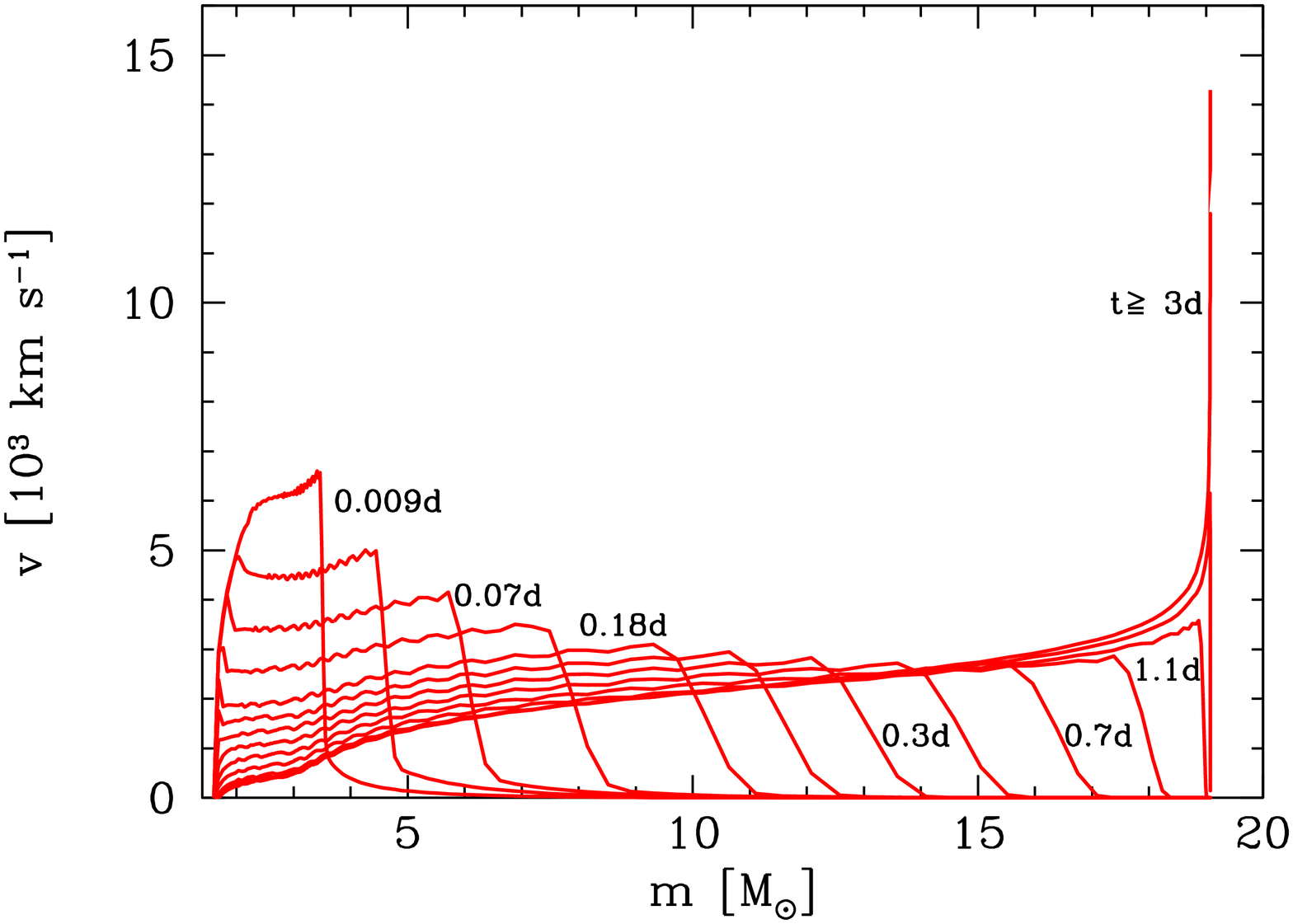}\includegraphics[angle=-90,scale=.30]{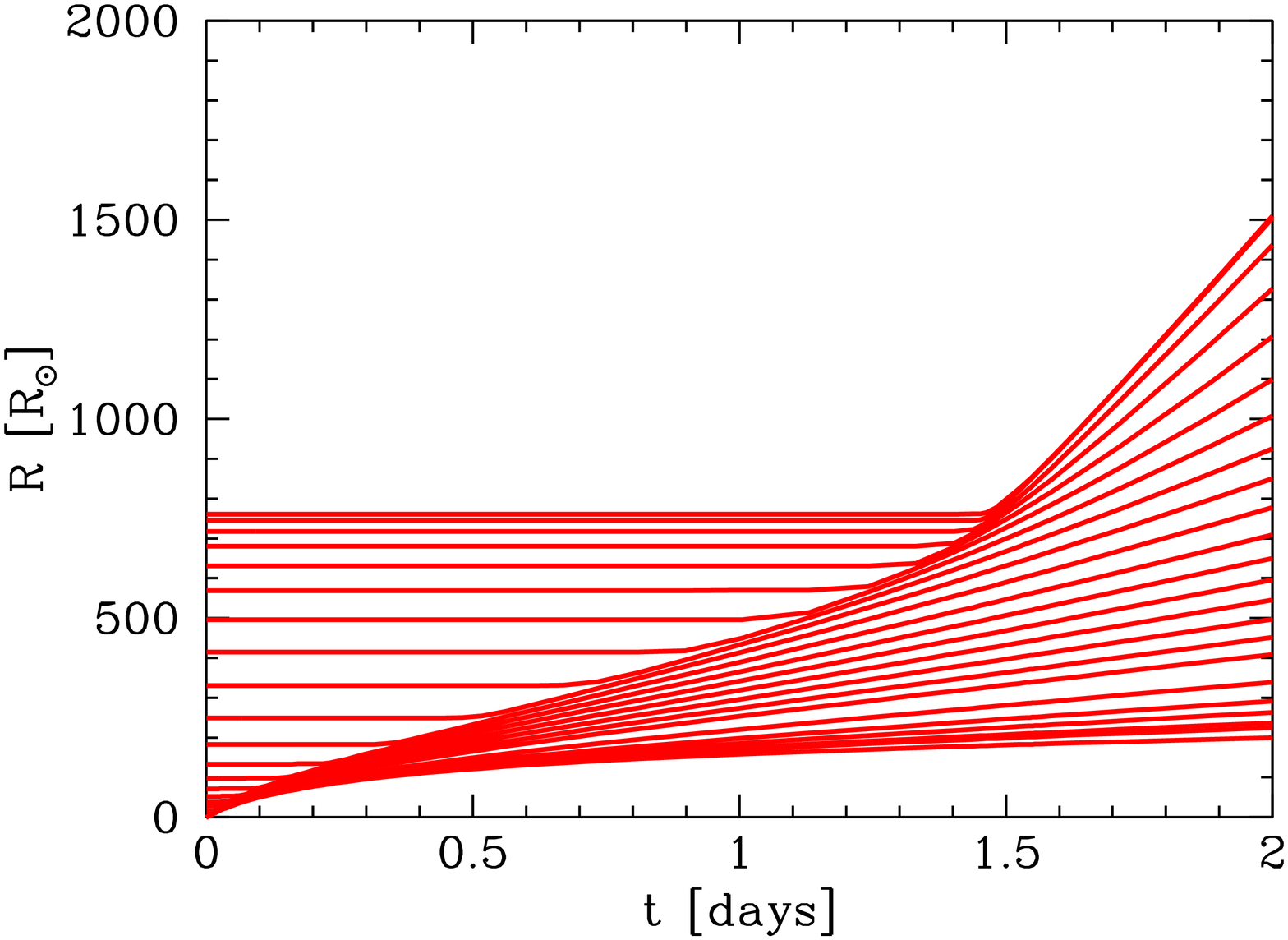} 
\caption[ SW propagation on different physical quantities]{ Changes in  density {\bf  (top left)},
  temperature {\bf  (top right)}, and velocity {\bf  (bottom left)}
  profiles  as a function of interior mass during  shock
  propagation for model \model. Some of the curves are labeled with the
  time elapsed since the energy is injected. The initial density profile 
  is also shown ($t=0$; blue dashed line). Note that a very
  small amount of material near the surface is strongly accelerated as
  the SW passes through the steep density gradients present in the
  outermost layers.   
  {\bf Bottom right:} temporal evolution of
  the radial coordinates corresponding to different interior layers.  
  \label{fig:profiles}}
\end{center}
\end{figure}

Figure~\ref{fig:profiles} shows the effect of the SW propagation on different
physical quantities (velocity, density and temperature) inside the star. The
shock front is evidenced as a sudden change in these quantities. As the shock
moves outward, the material behind is accelerated and
heated. Note that the  outermost layers,  with a sharp
decline in density,  
acquires very high velocities. It represents a small fraction of
the star. 
It is also clear from the velocity profiles that the inner
parts of the object are decelerated. This deceleration is due to the
interaction of the dense core with the extended hydrogen-rich
envelope. The bottom right figure  shows changes in radius for
different shells. From this we can deduce the location of the shock
front at any time as that of the innermost shell which has constant
radius.

\begin{figure}[htbp]
\begin{center}
\includegraphics[angle=-90,scale=.3]{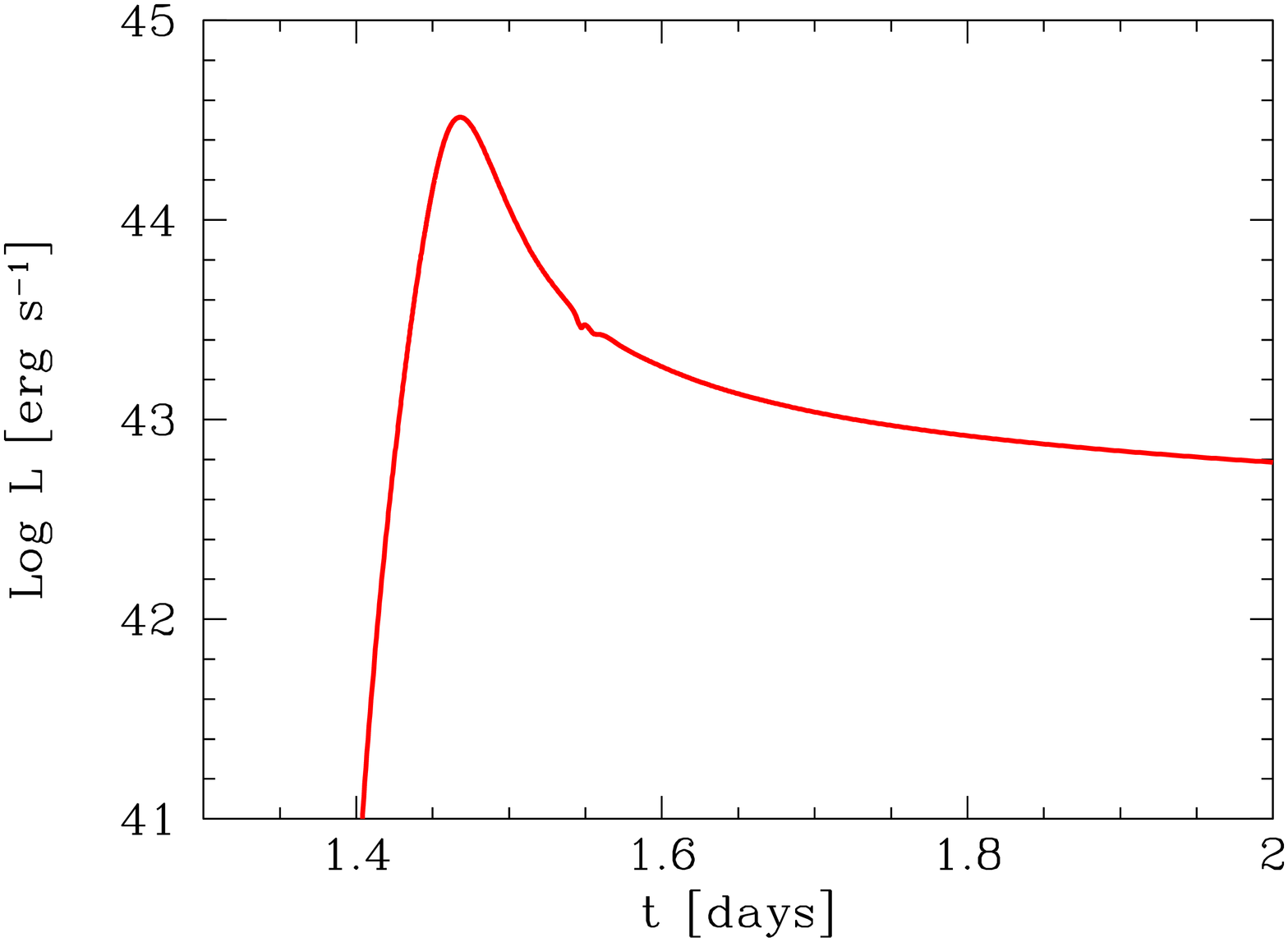}\includegraphics[angle=-90,scale=.3]{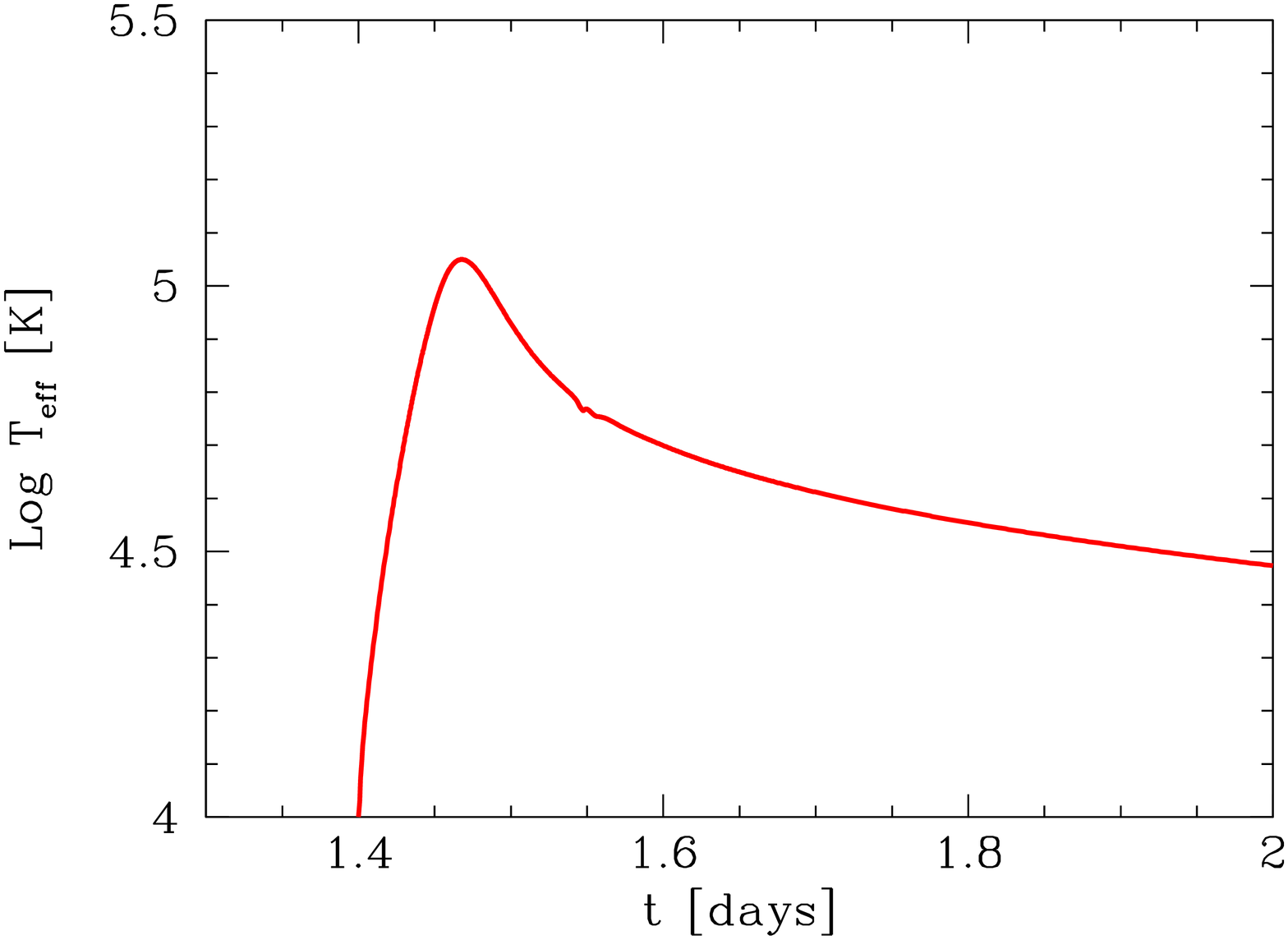}
\caption[Bolometric luminosity and effective temperature during
  breakout ]{Bolometric luminosity ({\bf left}) and effective
  temperature  
  ({\bf right}) during shock breakout \label{fig:LTinc}}
\end{center}
\end{figure}

At $t=1.36$~days, the SW reaches the surface of the object, which
produces the first electromagnetic manifestation of the
explosion (although neutrinos and gravitational waves escape well
before). The effective temperature and bolometric luminosity  suddenly
rise and reach their maximum values a 
few hours after  breakout, specifically, at $t=1.47$~days with
values of $L_{\rm peak}= 3\times 10^{44}$~erg~s$^{-1}$ and
$T_{\rm peak}=1.1\times 10^{5}$~K as shown in Figure~\ref{fig:LTinc}.
At these  temperatures, the peak of the emitted  spectrum
is in the UV range. 

Hereafter, the star begins to expand and cool very
quickly, leading to an increase in photospheric radius and a
decrease in temperature in the external layers. The bolometric
luminosity abruptly decreases but, according to the decrease in
effective temperature and consequent shift of the emission peak
to longer wavelengths, the luminosity in the optical range 
increases. As a result, a sharp peak in bolometric luminosity and
temperature is produced, as shown in Figure ~\ref{fig:LTinc}. 
In our model we obtain a 
  decrease of $1.5$ dex in luminosity only $6.8$ hours after 
  peak brightness.  During this time the total energy radiated is $2.2\times
10^{48}$~erg, emitted essentially as a UV flash. The short
duration of the breakout explains why so few SNe~II-P have been
observed during  this phase.

At temperatures as high as those left by the passage of the
  shock wave, the stellar 
matter is completely ionized  which implies that  the
breakout is accompanied by 
a strong increase in opacity. The position of the photosphere
during the outburst nearly coincides with the outermost shell.
This behavior continues until the onset of recombination. Therefore,
at early stages previous to recombination, the velocity 
of matter in the photospheric position samples the very high velocities of the
outermost layers and reaches values close to $1.2\times
10^4$~km~s$^{-1}$.

 From the energetics  point of view, by $t=0.5$~day the total energy is
approximately divided in equal proportions between internal, dominated
by radiative contribution, 
and kinetic components. Short after breakout the kinetic energy
completely dominates the energetics (the energy radiated away   
at any given time  is less than $1.5$\% of the  total energy).

 It is possible to estimate the average velocity during the SW
propagation: given that the SW takes $1.36$~days to emerge and considering a
radius of $R= 5.5\times 10^{13}$~cm (800~$R_{\odot}$), we obtain an average
speed of $v=4680$~km~s$^{-1}$. It is also
 interesting to calculate the expected  time for breakout using the
 analytic expression given by \citet{1987Natur.328..320S}, 

\begin{equation}
t_{\rm bk} \simeq 1.6 \left( \frac{R_0}{50 R_\odot} \right)  
\times \left[ \left(  \frac{M_{\rm ej}}{10 M_\odot} \right) / \left(\frac{E}{1
    \times 10^{51} \,\,{\rm erg}} \right) \right]^{1/2}  \,\, {\rm hr},
\end{equation}

\noindent where $R_0$ is the initial radius, $M_{\rm ej}$ is the
ejected mass, and $E$ is the explosion energy. Using the values for
our initial model we obtain $t_{\rm bk}=1.26$~days, in very 
good agreement with our numerical calculation.  

\subsection{Adiabatic cooling and homologous expansion}  
\label{sec:AC}
The breakout is followed by a violent expansion, resulting in the cooling of
the outermost layers. During expansion, only a small fraction of the photon
energy can diffuse into the surroundings. Therefore, it is possible to consider
the cooling process to be approximately adiabatic and this approximation remains
valid while the time-scale for radiation diffusion is much longer than the
expansion time-scale\footnote{The expansion time-scale, $\tau_{\rm h}=
R/v$, increases with time while the diffusion time-scale, $\tau_{\rm d}=
\kappa \rho R^2 /c$, decreases because $\rho \propto R^{-3}$. Thus,
there is a time after which the condition for  this approximation
breaks down.}. Note that there are two mechanisms to cool a SN: 
(a) loss of photons or diffusion cooling, and (b) its own expansion or
``adiabatic cooling''. If one of these processes  dominates then we
say that the cooling is 
carried out by such process. In our model, more than 90\% of the
decrease in internal energy happens as adiabatic cooling during the
first 18 days  after  explosion, when the first layers with 
neutral hydrogen appear. After that, diffusion cooling begins to be
significant and the adiabatic cooling phase comes to an
end. 

The internal temperature (and internal energy) decreases almost
adiabatically, i.e.  proportional to $r^{-1}$ due to the dominance of
the radiative term, and quickly reaches a value near 
the  recombination temperature of hydrogen. Quantitatively, the effective
temperature goes from values close to $10^{5}$~K at the time of the
burst when the matter is totally ionized, to  $T_{\rm eff}=
10^{4}$~K five days later. At these temperatures hydrogen begins
to recombine. At the same time, the luminosity reaches 
$L= 1.9 \times 10^{42}$~erg~s$^{-1}$ and the photospheric radius
rapidly increases to $R_{\rm ph}= 5.2 \times
10^{14}$~cm. After that, the luminosity decreases slightly as a result
of the slower decrease in temperature and continuous increase in
radius. By day 18, when the first layers of neutral hydrogen appear,
we have $T_{\rm eff}=6960$~K, $L= 1.2\times 10^{42}$~erg~s$^{-1}$, and
$R_{ph}= 9.1 \times 10^{14}$~cm. Note the slight change in luminosity,
 of only $0.14$ dex, between day 7 and 18 (Figure~\ref{fig:LC}).

\begin{figure}[htbp]
\begin{center}
\includegraphics[angle=-90,scale=.3 ]{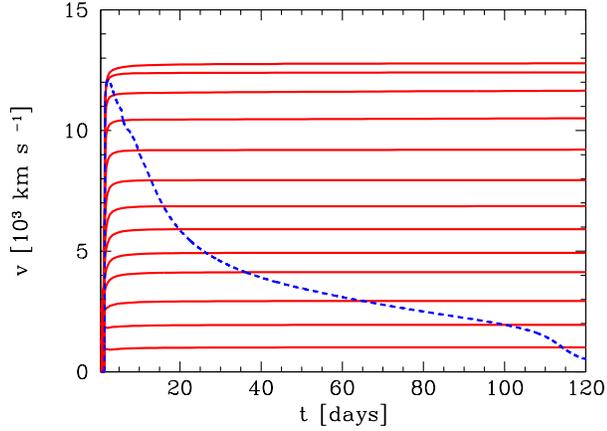}
\caption[Evolution of the velocity for different layers]{Evolution of
  the velocity for different layers 
  (solid lines) and the velocity of material at the photospheric position
  (dashed line). Note that the asymptotic constant velocity is reached at
  latter  times for the more external shells.\label{homovelo}}
\end{center}
\end{figure}

A few days after breakout, the acceleration of the material comes to
an end and the expansion becomes homologous. The
matter reaches velocities of the order of $1.2 \times 10^{4}$~km~s$^{-1}$ in the
outermost layers and the  object enters a state of free expansion where
forces of pressure and gravitation do not have any dynamical effect on the
system. The homologous regime is
characterized by a constant velocity in each layer, a linear growth of
the radial coordinate with time ($r \propto t$) and a
density distribution decreasing with time as $\rho \propto t^{-3}$.  This
behavior is clearly shown in Figure~\ref{fig:profiles} for $t>3$
days and in Figure~\ref{homovelo}. The
nearly constant shape of the density and temperature profiles at 
late times are a result of the expansion being approximately
homologous. This is also evident in the linear behavior of the radial
coordinates for different mass shells. The condition
of constant velocity is very nearly satisfied for each
shell, although the transition between acceleration and homologous
expansion happens at slightly different times for different mass shells
(see Figure~\ref{homovelo})\footnote{After day 8, the
    photosphere remains located in shells where homology has been reached.
   Therefore, our models show that the  expanding photosphere method
   (EPM) which is used to estimate 
  distances assuming homologous expansion, can be 
  safely applied  at epochs later than this.}.

As  mentioned  in \S~\ref{sec:SW}, at the time of breakout the
internal energy of the envelope is small compared to the kinetic
energy. Moreover, the object expands by a factor of $\sim$ 17 during
this phase to a radius of $R_{\rm ph} \sim 9 \times 10 ^{14}$~cm,
before it becomes transparent. Thus, most of the internal energy left
by the passage of the shock has been already degraded by the adiabatic
expansion, and only a small residual will be released as observable
radiation in the following phase.  

\subsection{Cooling and recombination wave}
\label{sec:CRW}

\begin{figure}[htbp]
\begin{center}
\includegraphics[angle=-90,scale=.30]{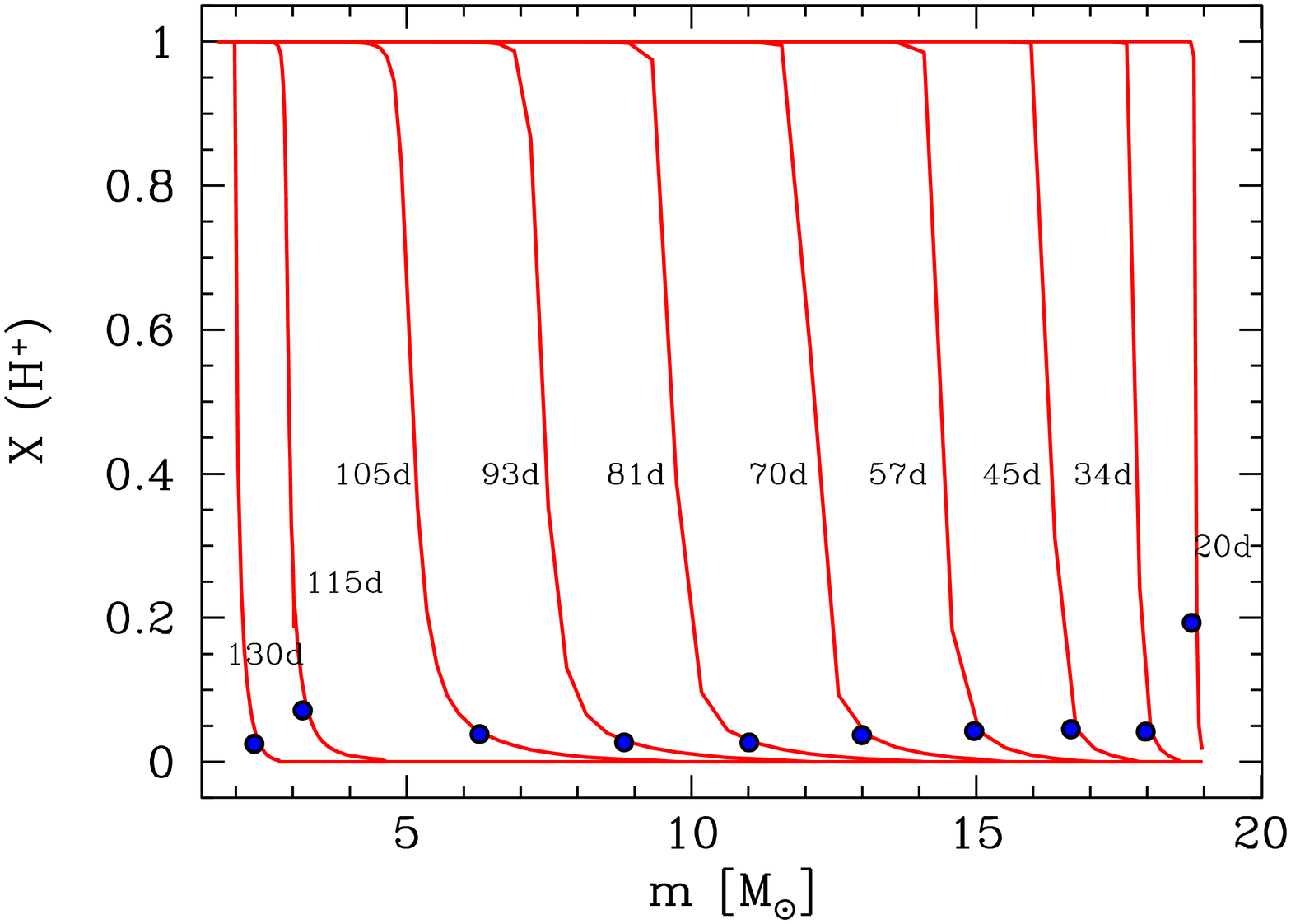}\includegraphics[angle=-90,scale=.30]{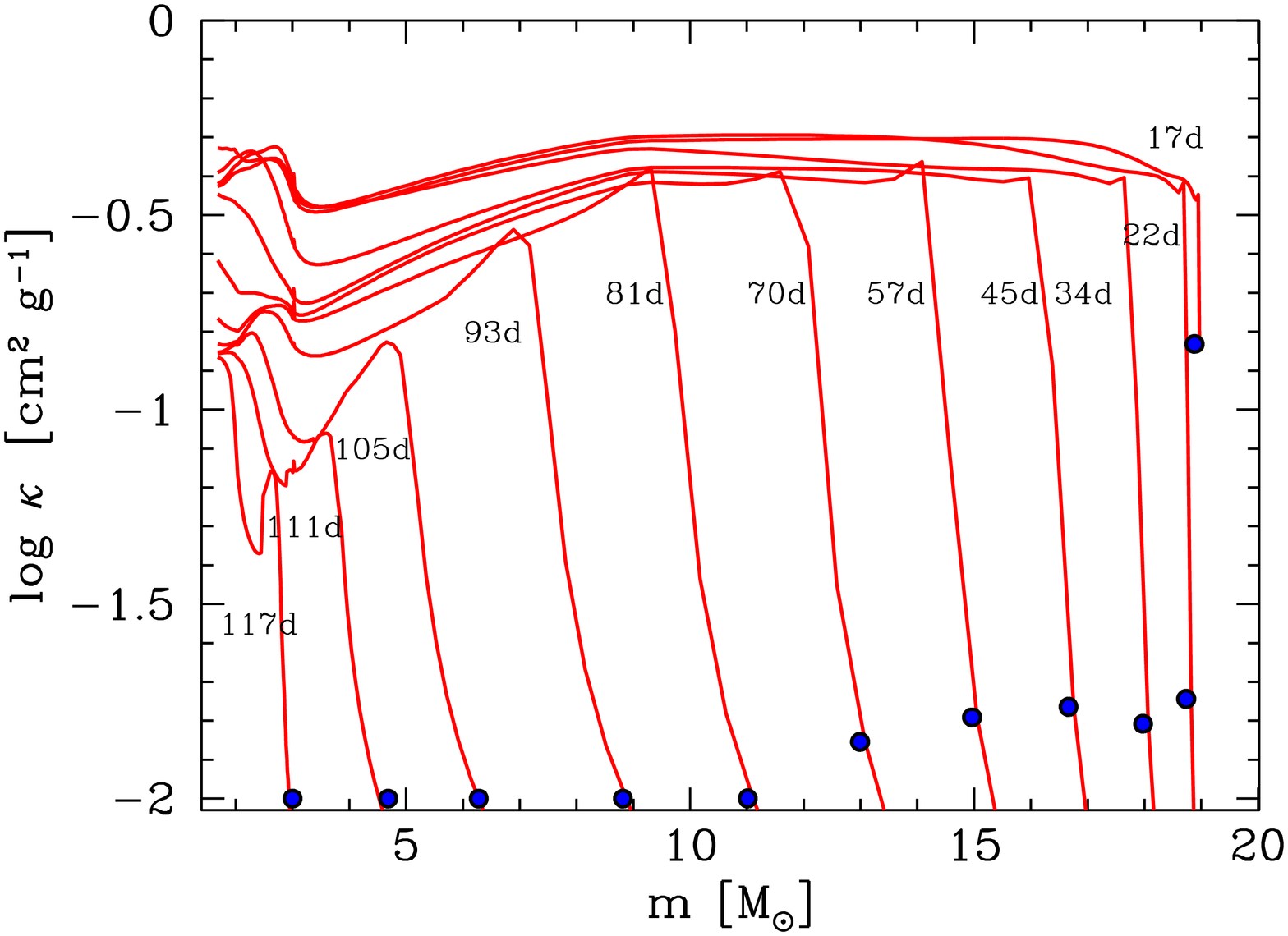} \\ 
\includegraphics[angle=-90,scale=.30]{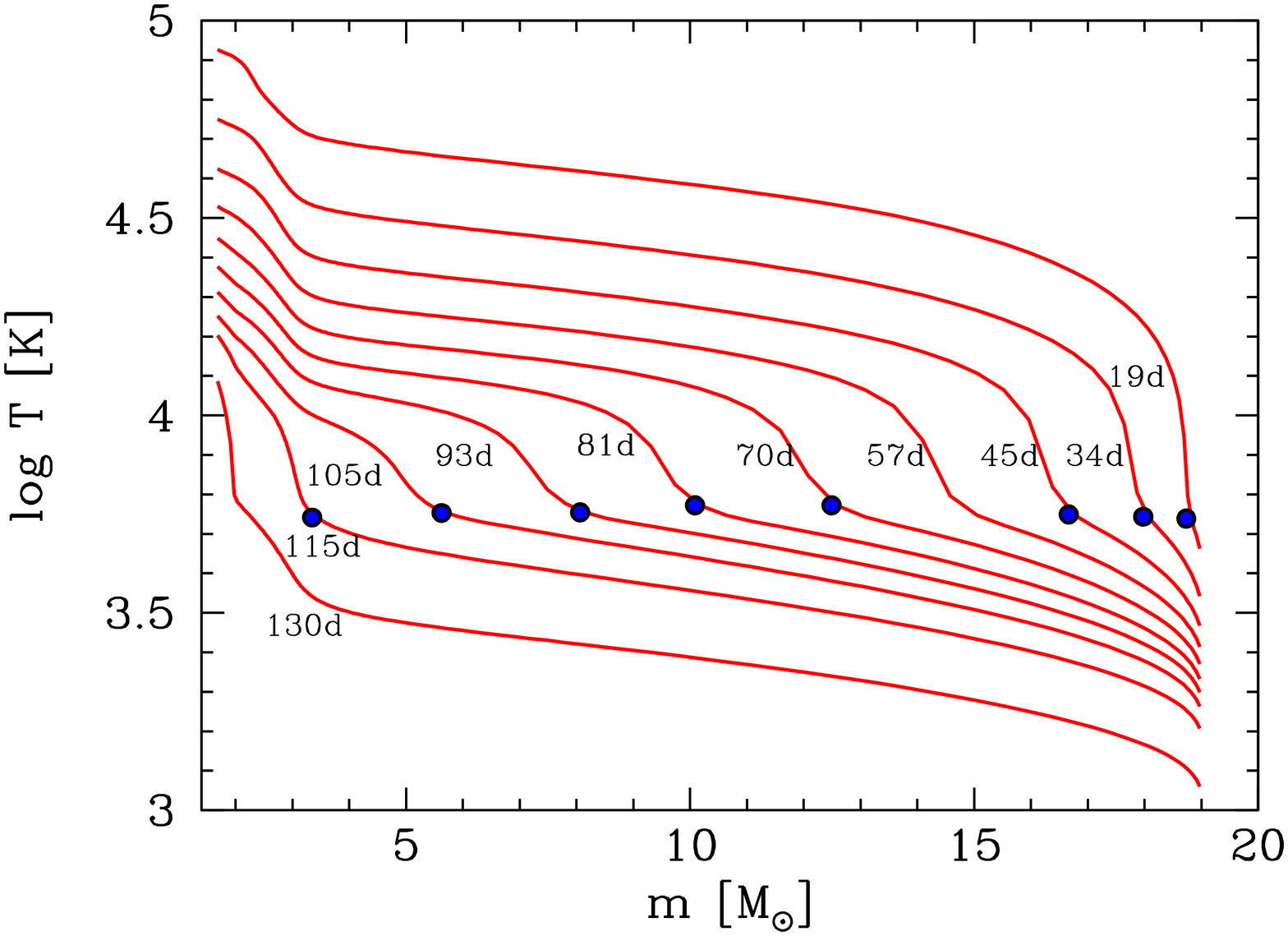}
\caption[Evolution of the fraction of ionized hydrogen, opacity and
  temperature ]{Evolution of the fraction of ionized hydrogen 
{\large\bf(top left)},  opacity {\bf  (top right)}, and temperature
{\bf(bottom left)} 
  as a function of mass. The time elapsed since the energy
  is injected and the photospheric position (blue dot) are indicated
  for each curve. The photosphere  
  is nearly coincident with the outer edge of the CRW which moves
  inward in mass coordinate. Note also that the opacity in the outer
  shells is nearly constant with a value of 0.4 cm$^{2}$ g$^{-1}$,
  close to the electron-scattering opacity for matter
  composed of pure hydrogen, above which it suddenly
  drops.\label{fig:profilesXOT}}  
\end{center}
\end{figure}

The appearance of regions with neutral hydrogen   
determines the onset of a recombination wave (RW). The
duration of the RW is related to the total hydrogen mass and how
deep  hydrogen has been mixed into to the progenitor object. 
As time goes on, hydrogen recombination occurs at different layers
of the object as a wave propagates inwards  (in Lagrangian
coordinate; see 
 Figure~\ref{fig:profilesXOT}, top). Because the opacity is dominated by
electron scattering, it decreases abruptly outwards  at the
recombination front (Figure~\ref{fig:profilesXOT}, top right), increasing the
transparency of these layers and allowing the radiation to easily go
away. Consequently, the internal energy (mostly of the radiation field) is
efficiently radiated away, and the temperature drops sharply 
from $\sim$$10000$~K to $\sim$$5500$~K at the recombination front 
(Figure~\ref{fig:profilesXOT}, bottom). In other words, a cooling
wave associated with the transparency and induced by a recombination
wave propagates inward through the envelope. This is usually called 
``cooling and recombination wave'' (CRW).

In our model the recombination begins approximately at day
  18. From them on, a
CRW develops which moves inward in mass until all the matter is completely
recombined by day 130 when the recombination front reaches the
innermost  H-rich  layers of the envelope (at $m=$2 $M_\odot$ in our
model), and the luminosity suddenly drops, defining the end of this
phase. The propagation of the CRW is clear in
Figure~\ref{fig:profilesXOT} where the evolution of 
the fraction of ionized hydrogen and temperature profiles as a function of mass
for selected times are shown. Note also the behavior of the opacity (see
Figure~\ref{fig:profilesXOT}, top right) which depends strongly on the
ionization state of the matter. In our model, a strong drop in the
opacity of the outer layers is seen at day $\sim$18, which 
leads to a considerable decrease in the 
optical depth of these layers and causes an inward drift in mass of the
photosphere --- defined at a fixed optical depth of $\tau = 2/3$.
The photosphere begins to follow the CRW, as shown in Figure
~\ref{fig:profilesXOT}. It is important to note 
that we define the position of the photosphere using only continuum
opacity sources, i.e. excluding the opacity floor (see
section~\ref{sec:vphoto}).

The photosphere, as defined here, is nearly coincident with the outer
edge of the CRW. Therefore it is just the location of the photosphere with
respect to the CRW what sets the value of the photospheric temperature
close to the temperature of hydrogen recombination ($T \sim
5500$~K). However, note that the effective temperature does not necessarily
have a constant value. Since the effective temperature is defined as
$T^4_{\rm eff}= L/\pi \sigma R_{\rm ph}^2$ and the luminosity remains
nearly constant outside the recombination front (see
Figure~\ref{fig:profilelum}), any changes in effective temperature
are related to changes in  photospheric radius.

The CRW divides the object in two distinct regions: (a) an inner zone
which is hot, optically thick and ionized, and (b) an outer 
zone which is relatively cold, optically thin and completely
recombined. Matter in the inner
  zone is opaque: radiative  transfer is 
too inefficient to produce any appreciable flow of radiative energy
(this would be 
strictly fulfilled if $^{56}$Ni were confined to the innermost layers;
see the discussion below) and the matter cools
down almost adiabatically. On the other hand, the external layers
are transparent and practically do not radiate. Therefore, it is
within the CRW interphase where almost the entire radiant flux is
released.  When 
matter passes through the CRW,
particles begin to be cooled by radiation, emitting more light than
they absorb, and the radiant flux increases. This way, the radiative
flux emerging from the CRW front carries away internal energy of the
matter that is cooled by the wave. More details on the  properties of
the CRW are given by \citet{1971Ap&SS..10...28G}. 

\begin{figure}[htbp]

\begin{center}
\includegraphics[angle=-90,scale=.35]{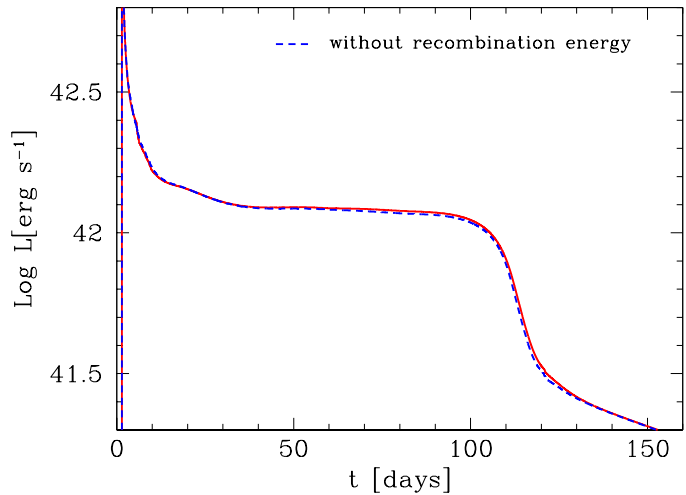}
\caption[Bolometric LC without energy released by
  recombination]{Comparison between bolometric light curves with
  (solid line)   and without (dashed line) 
  energy released by recombination. It is clear that the recombination
  of ions with electrons does not contribute significantly to the
  light curve. The differences between both calculations are
    less than 0.04 dex during whole the evolution. \label{fig:LCSR}}
\end{center}
\end{figure}

\begin{figure}[htbp]
\begin{center}
\includegraphics[angle=-90,scale=.30]{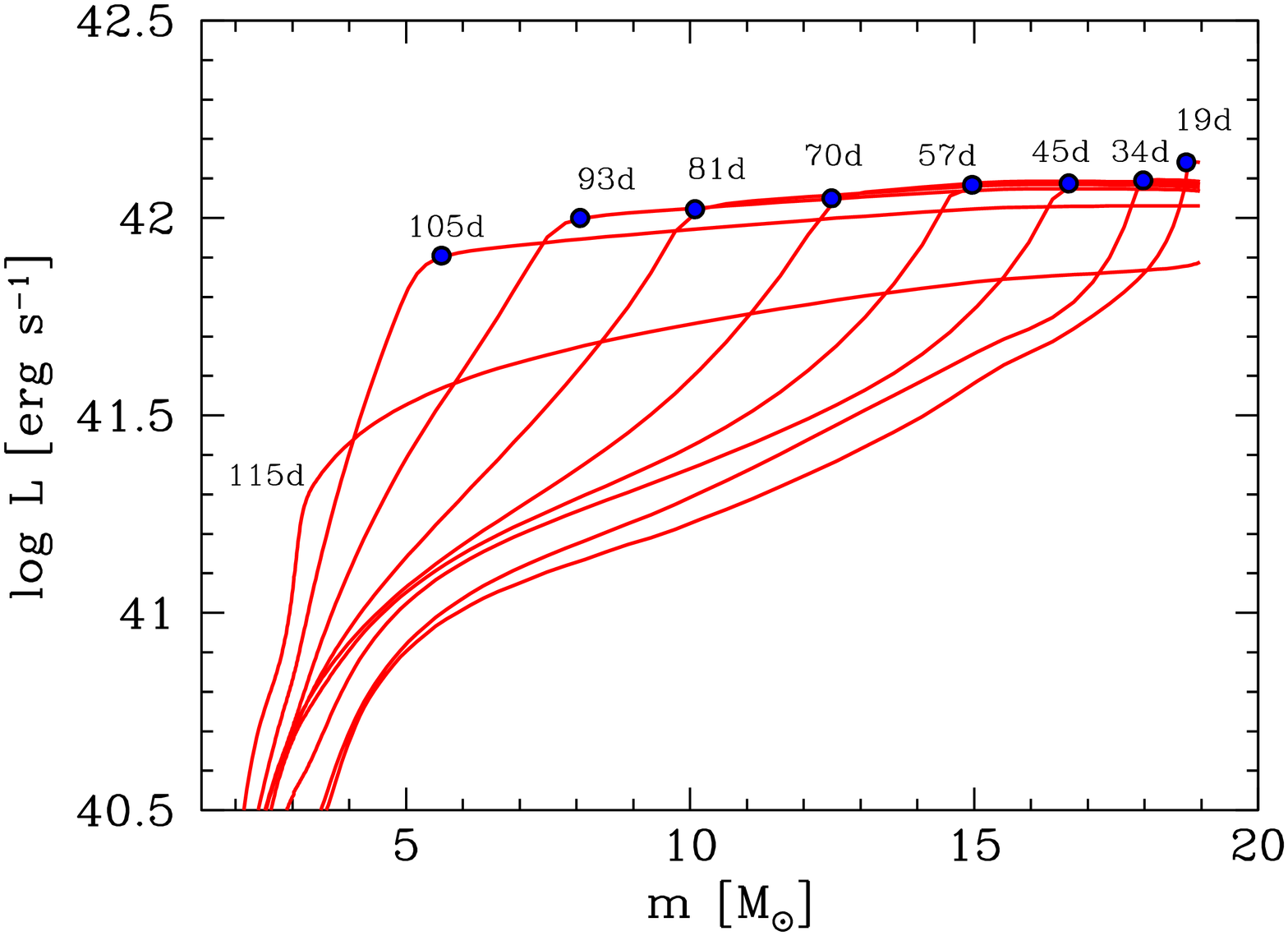}\includegraphics[angle=-90,scale=.30]{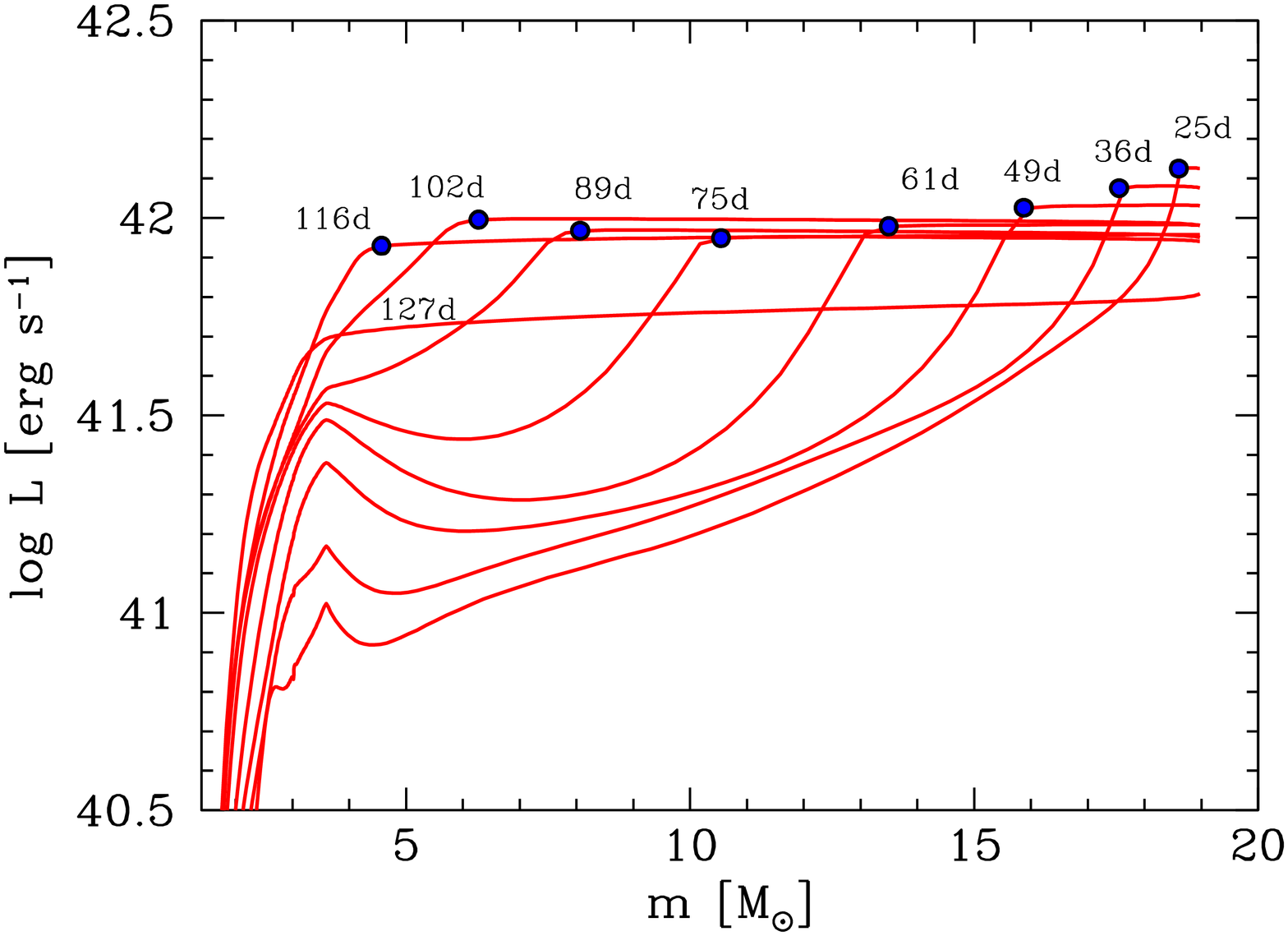} \\ 
\caption[Evolution of the interior luminosity]{Evolution of the
  interior luminosity as a function of 
  mass. The  labels indicate
  the time since explosion. {\bf Left:} $^{56}$Ni is mixed into the
  hydrogen-rich envelope up to 15 $M_\odot$. {\bf Right:} $^{56}$Ni is
  confined to the   
   layers inside 3.5 $M_\odot$. The inward propagation
   of the recombination front is clear in both 
   cases. In the mixed case, the recombination front propagates 
   slower than in the unmixed one because the temperature near the
   front is higher due to radioactive heating. Note that for the
   unmixed case the outward diffusion 
   of the radioactive energy is clear while for the
   mixed case this effect occurs 
   too early to be noticeable. 
 \label{fig:profilelum}}
\end{center}
\end{figure}

Note that the bulk of the radiation does not diffuse to the photosphere;
the photosphere instead moves inward, allowing the 
radiation to escape sooner than it would for a photosphere fixed at
the outer boundary of the ejected mass. That is, the recombination
process is responsible for the energy release during this
phase. {\bf It should be noted that
most of this energy comes from the energy deposited by the SW and not
from the recombination itself}. In order to test the previous
statement, we ran our code  without including the energy released by 
recombination of ions with electrons. The result was that no
appreciable change appeared; quantitatively, the differences between
both calculations are less than 0.04 dex during all the evolution (see
Figure~\ref{fig:LCSR}).

Note, however, that in  model \model\   where we assumed an extended $^{56}$Ni
mixing, the flux of energy inside the CRW is not negligible
(see  left panel of Figure~\ref{fig:profilelum}). 
This is due to the fact that the photosphere meets regions with radioactive
material earlier than in the case where $^{56}$Ni is confined to the innermost
layers. Therefore, the energy deposited by radioactive decay
provides additional power to the LC. A different behavior is seen
when $^{56}$Ni is confined to the innermost layers (inside $2.5$
$M_\odot $; see  right panel of Figure~\ref{fig:profilelum}). In
this case, the statement that the energy flux inside the CRW is very
small is fulfilled at least until day $\sim$120 when
the radiative diffusion of the radioactive decay from the central layers 
begins to dominate. At
earlier times, it is possible to see the outward diffusion of
radioactive energy. Note also that for both models the
luminosity  outside the front is nearly constant.

\begin{figure}[htbp]
\begin{center}
\includegraphics[angle=-90,scale=.40]{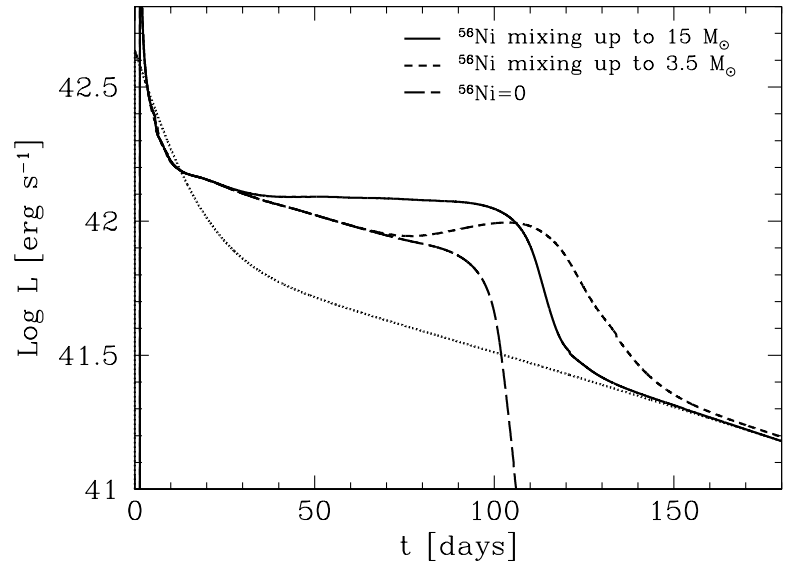}
\caption[Comparison  between bolometric LCs for different  $^{56}$Ni
  mixing ]{Comparison  between bolometric  light curves for extended 
   $^{56}$Ni mixing (solid line) and a model with mixing of  $^{56}$Ni  up
  to 3.5 $M_\odot$ (short-dashed line). We also  include the case
  of a model  without  $^{56}$Ni (long-dashed line) in which
  case  the LC falls abruptly after the plateau phase. The presence of
  $^{56}$Ni extends the plateau and increases the luminosity. This
  is essentially produced when the CRW reaches layers with $^{56}$Ni
  which can thereby power the LC directly. The extended $^{56}$Ni
  mixing reveals this effect earlier in the evolution, which produces
  a flat plateau in concordance with the observations of
  SN~1999em.\label{fig:LCNi}}  
\end{center}
\end{figure}

The assumption of extended mixing was necessary in order to obtain a
plateau as flat as the one observed for
SN~1999em. Figure~\ref{fig:LCNi} shows a comparison of the bolometric
LC for three cases: 1) with extended $^{56}$Ni mixing, 2) no mixing, and
3) without $^{56}$Ni. For case 1), $^{56}$Ni begins to affect the LC
  by day $\sim$35 while for case 2)   
the effect of nickel heating is delayed until day $\sim$75. On the
other hand, in the former case there is a less direct energy
deposition at the  center of the object, and
the plateau declines earlier and steeper to the tail than in case 2).
 Our assumption of mixing of $^{56}$Ni into the hydrogen envelope is not
unreasonable as shown in studies of SN~1987A \citep[][among others]
{1988A&A...196..141S,1988ApJ...324..466W,1988ApJ...331..377A,2000ApJ...532.1132B}. This
type of mixing is   presumably due to Rayleigh-Taylor
instabilities that occur behind the shock front, as  obtained in
multi-dimensional hydrodynamic calculations
\citep{1991A&A...251..505M,2000ApJ...528..989K}. However,
it is important 
to note that \citet{2007A&A...461..233U} found an excellent agreement
with observations of SN~1999em by confining $^{56}$Ni to the innermost
layers and  good agreement
 with  observations of SN~1987A assuming moderate $^{56}$Ni mixing
 \citep{1993A&A...270..249U,2004AstL...30..293U}.

\begin{figure}[htbp]
\begin{center}
\includegraphics[angle=-90,scale=.40]{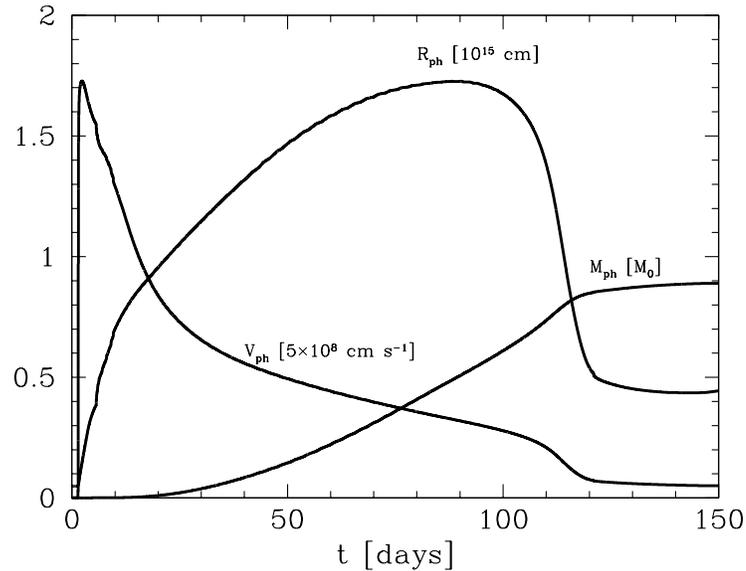}
\caption[Evolution of  photospheric radius, velocity and  mass above
    the photosphere ]{Evolution of  photospheric radius ($R_{ph}$) in
    units of $10^{15}$ cm,  photospheric velocity ($V_{ph}$) in
    units of $5 \times 10^{8}$ cm s$^{-1}$, and  mass above
    the photosphere ($M_{ph}$) in units of $M_0$ ($M_0= 19 \,M_\odot$)
    for   model \model.} \label{fig:VRM}  
\end{center}
\end{figure}

The evolution of the velocity of matter at the photospheric position
$v_{\rm ph}$,  photospheric radius ($R_{\rm ph}$) and  mass above the
photosphere ($M_{\rm ph}$) are shown in Figure~\ref{fig:VRM}. Initially,
the photospheric velocity evolves rapidly, the photospheric position
follows a linear behavior, and there is very little mass above the
photosphere. Later on, when the recombination sets in at $t \sim 18$
day, $R_{\rm ph}$ 
begins to differ noticeably from the linear behavior, the mass above
the photosphere increases, and $v_{\rm ph}$ decreases because it
samples increasingly inner, slower material. Finally, when all the
matter is recombined at $t \sim 110$ days, $R_{\rm ph}$ and $v_{\rm ph}$ 
sharply turn down and the luminosity undergoes a rapid decrease. Note,
however, that the photospheric radius does not drop immediately to zero.
This is due to the heating caused by radioactive decay which produces
some ionization of the gas. On the other hand, the luminosity undergoes
a rapid decrease to values close to the luminosity of radioactive
decays.  If there was no $^{56}$Ni (case 3) the SN luminosity would
abruptly vanish at
this point, as shown with a long-dashed line in Figure~\ref{fig:LCNi}. 

In conclusion, the CRW has a duration of $\sim$$100$~days. The
luminosity experiences  small changes during the CRW propagation.
In order to produce a plateau as flat as that observed in SN~1999em,
we need to invoke significant Ni mixing in the H-rich envelope. {\bf Thus,
the plateau can be seen as a combination of CRW properties plus some
additional energy provided by radioactivity}. Defining, from an
empirical point of view, the plateau phase as the
period of time when the luminosity remains constant within 0.5~mag of
the value at day 50, its duration is $103.5$ days, and the
total energy emitted during this phase is $\sim$$1.12 \times
10^{49}$~erg. Note, that with such definition the plateau
phase includes the final stages of the early adiabatic cooling phase
and not only the recombination phase.

\subsubsection{Radioactive tail}
\label{sec:tail}
The late behavior of the LC (at $t>130$ days) is dominated by the
energy released from radioactive decay. Without radioactive material, the
luminosity would abruptly vanish when  hydrogen gets completely
recombined as shown in Figure~\ref{fig:LCNi}. Instead, the observed LC
decreases to values close to the instantaneous rate of energy
deposition by the radioactive decay,

\begin{equation}
L\sim 1.43 \times 10^{43} M_{\rm Ni}/M_\odot \,\,\exp(-t/111.3). 
\end{equation}

\noindent The decline of the LC at $t>130$ days is nearly
linear in concordance to the exponential decline of the radioactive
decay law. That is, the diffusion time for optical photons becomes so
small that  the decay energy is radiated away nearly instantaneously,
while the gamma-ray optical depth is still sufficiently large in order to
allow a nearly complete local deposition of the decay energy. The
bolometric luminosity in 
this part of the LC is a direct measure of the Ni mass synthesized in the
explosion. 

\section{Comparison with other hydro-models of SN 1999em}
\label{sec:compsn99em}

In this Section, we compare our results with those of
previous hydrodynamical studies of SN~1999em. There are two previous
such studies of SN~1999em: one  by
\citet{2005AstL...31..429B} (BBP05 hereafter) and another by
\citet{2007A&A...461..233U} (U07 hereafter). In Table~\ref{sn99:tbl-2}, we
summarize the physical parameters obtained for SN~1999em in the three
works along with the distance, explosion time, and
the  chemical composition assumed in each model. It is
interesting to note that the three studies employ initial density
distributions and chemical composition distributions of
non-evolutionary models. In our work we  assume H and He profiles
very close to those adopted by U07,  yet a very  different $^{56}$Ni
distribution. While we assume an extended, uniform
mixing of Ni (see Figure~\ref{fig:comp}),  U07 confined 
$^{56}$Ni to the innermost layers (see their Figure~2). In
the case of BBP05, H and He were assumed to be uniformly
mixed throughout the envelope and a radial distribution of $^{56}$Ni up
to $\sim$15 $M_\odot$ was adopted (see their Figure~9). Therefore, our
$^{56}$Ni distribution and the resulting $^{56}$Ni mass are closer
 with those of BBP05 than those of U07. We also note
that the mass of the compact remnant --- which is left aside from the
calculations --- is different in all three works. We  assume a
compact remnant of 1.4 $M_\odot$ while 
U07 assumed 1.58 $M_\odot$ and BBP05 removed  everything  within a
radius of $R_C= 0.1 \, R_\odot$. 

\begin{table}[htbp]
\begin{center}
\scriptsize
\caption[Physical parameters for SN
  1999em from three hydro-codes]{Comparison between physical
  parameters for SN 1999em  
  from three different hydrodynamical codes \label{sn99:tbl-2}}
\begin{tabular}{lccccccccc}
\hline
\hline
  &$D$& $t_{0}$ & &  & $E$ & $M$& $R$ & $M_{\mathrm {Ni}}$ & Ni mixing\\
Code & [Mpc]  & [JD-2451000] & $X_{\mathrm{sup}}$ & $Z$ & [foe] & [$M_\odot$] &
[$R_\odot$]  & [$M_\odot$] & [$M_\odot$] \\  
\hline
This work           &  11.7   &  477.90  & 0.735 &  0.02  & 1.25 & 19    & 800  &
0.056 &  $\sim$15\\ 
BBP05$^{\mathrm{a}}$&  12  & 468.90  &  0.7   &  0.004  & 1   & 18    & 1000 &
0.06 & $\sim$15 \\
U07$^{\mathrm{b}}$ &  11.7  &  476.90 & 0.735 &  0.017 & 1.3$\pm$ 0.1
& 20.58$\pm$1.2 & 500$\pm$ 200  &
0.036$\pm$0.009 &  $\sim$2.5 \\
\hline
\end{tabular}
\begin{list}{}{}
\item[$^{\mathrm{a}}$] \citet{2005AstL...31..429B}
\item[$^{\mathrm{b}}$] \citet{2007A&A...461..233U}
\end{list}
\end{center}
\end{table}

The approaches used in each work are very different. U07 used a 
hydrodynamical code with a one-group approximation for the radiative
transport, including non-LTE treatment on  opacities and thermal
emissivity, non thermal ionization, and expansion
opacity. Their calculations, although involving a more sophisticated
radiative transfer treatment,  yielded bolometric light curves comparable to
those given by our code.  However, the procedure to  produce the
observed bolometric light curve for SN~1999em was quite 
different in both cases. While U07 based their calculations on the
integration of $UBVRI$ photometry with
a constant value of $0.19$ dex added to take into account the infrared
luminosity, we derived $UBVRIJKLM$ bolometric
luminosities from color-dependent bolometric corrections. As noted by
\citet{2009ApJ...701..200B}, the 
  inclusion of the $L$ and $M$ bands in the   
 calculation of the bolometric correction during the tail phase is
 very important and it most likely leads to the 
difference in the nickel mass estimated in this work and in U07.  
BBP05, in turn, used a multi-group hydrodynamical code which allows to
calculate LC in different photometric 
bands. This code also included the expansion opacity
effect. Therefore, BBP05 were able to compare their model with $UBVRI$
LC separately without being affected by the uncertainties in
 the calculation of bolometric luminosities. 

  Table~\ref{sn99:tbl-2}
As shown Table~\ref{sn99:tbl-2} in the parameters yielded by our
calculations are 
intermediate between those estimated by BBP05 and U07. The largest
differences are 2.6 $M_\odot$ in mass, 500 $R_\odot$ in radius, 0.3
foes in energy, 
and $0.045 M_\odot$ in $^{56}$Ni mass. It is quite satisfactory that
our simple 
prescription for the radiative transfer yields physical parameters similar
to those obtained from more sophisticated codes, considering that (a)
the models and  
methods used in each study are quite different, (b) there are differences
in assumed quantities, such as distance, explosion time, mass cut,
photometry, and (c) there may be a degree of degeneracy among explosion
energy, initial radius and initial mass within each model.

\section{Low Mass  Model of SN~1999em}
\label{sec:lowmass}
The mass of the progenitors of SNe~II-P can be derived from hydrodynamical
modeling of light curves and expansion velocities, or from the detection of
the pre-supernova object in archival images of the host galaxy in connection
with stellar evolution models. At the moment three progenitor stars
of SNe~II-P have been firmly detected and there have been negative
detections for 17 other  objects which have led to upper limits of the
progenitor star masses \citep{2009MNRAS.395.1409S}. 
Among these, only three SNe~II-P have masses derived using
hydrodynamical models: SN~1999em, SN~2004et, and SN~2005cs. 
In all of these cases the masses estimated from the hydrodynamical
modeling are   
systematically higher than the values  derived from
the other method. These
discrepancies have been noted previously in the literature
\citep{2008A&A...491..507U,2009arXiv0908.0700S}. For the  particular 
case of SN~1999em, the pre-supernova images give an upper limit of 15
$M_\odot$ for the progenitor star in the ZAMS. It is expected
that the mass of the pre-supernova object should be lower  that this due to 
possible mass loss episodes during the evolution of the star. In this Section we
use our hydrodynamical model to explore a low mass range, consistent
with pre-supernova imaging of SN~1999em, in order to test how well we can  
reproduce the observed properties of this object.

\begin{figure}[htbp]
\begin{center}
\includegraphics[angle=-90,scale=.60]{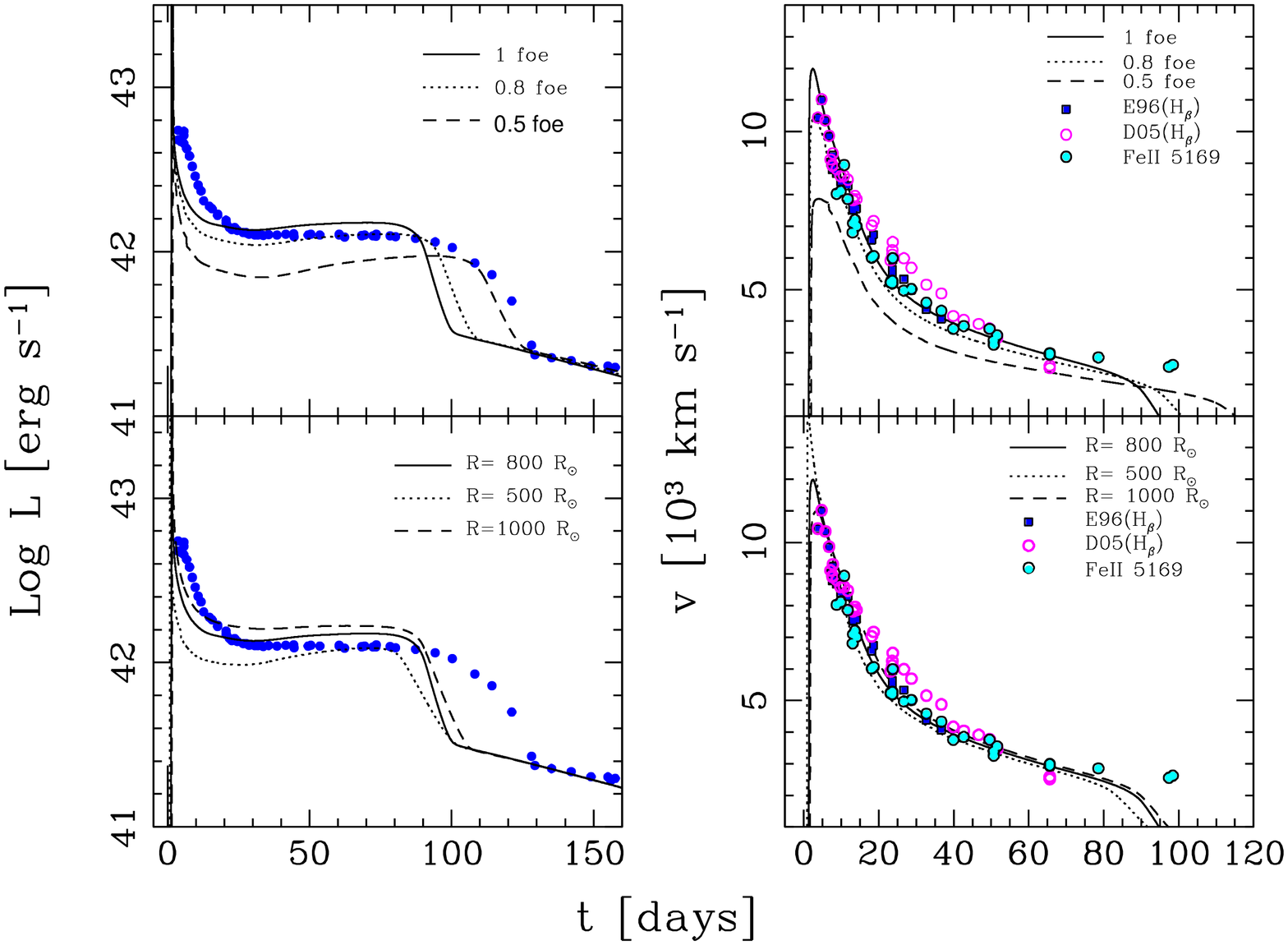}
\caption[Low mass models for SN~1999em]{Comparison between models and observations of SN~1999em for a
  low value of the pre-supernovae mass of $M = 12 \, M_\odot$ and the
  same value of $^{56}$Ni mass and mixing than for the reference model \model. 
  {\bf (Left panels):} bolometric LCs. {\bf (Right
    panels):} photospheric velocity evolution. {\bf (Upper
    panels):} models with different explosion energies as
  indicated in the labels, and a fixed value of 
  $R=800 \,R_\odot$. {\bf (Lower panels):} models with different
  initial radii as indicated in the labels, and a fixed value  of $E =
  1$ foe. \label{fig:M12VER}}   
\end{center}
\end{figure}

\begin{figure}[htbp]
\begin{center}
\includegraphics[angle=-90,scale=.60]{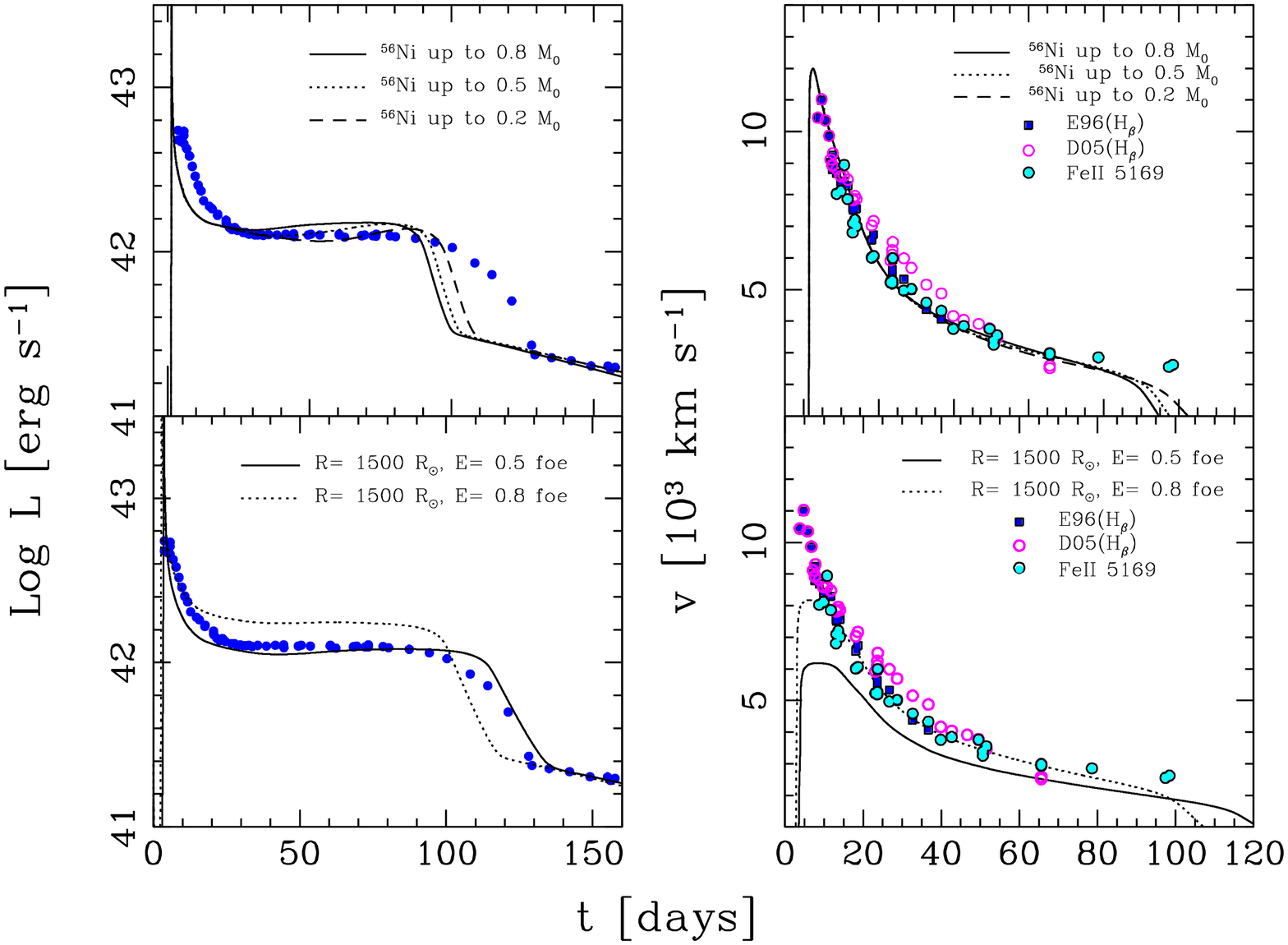}
\caption[Low mass models for SN~1999em (continuation)]{Comparison between models and
  observations of SN~1999em for a 
  low value of the pre-supernovae mass of $M = 12 \,M_\odot$ and the
  same value of $^{56}$Ni mass than for the reference model \model. 
  {\bf (Left panels):} bolometric LCs. {\bf (Right
    panels):} photospheric velocity evolution. {\bf (Upper
    panels):} models with different mixing of
    $^{56}$Ni, $E=1$ foe and $R=800 \,R_\odot$. Note that the degree
  of $^{56}$Ni  is indicated in each figure as a fraction of the
  initial mass of the 
  model. {\bf (Lower panels):} models with an initial
  radius of $R= 1500 \, R_\odot$ and two different values for  the
  explosion energy as indicated (corresponding to models m12r15e08ni56
  and m12r15e05ni56; see Table~\ref{sn99:tbl-3}).  Model m12r15e05ni56
reproduces very well the bolometric light curve of SN~1999em, while it
underestimates the photospheric velocity evolution.\label{fig:M12VNi}} 
\end{center}
\end{figure}

\begin{figure}
\begin{center}
\includegraphics[angle=-90,scale=.60]{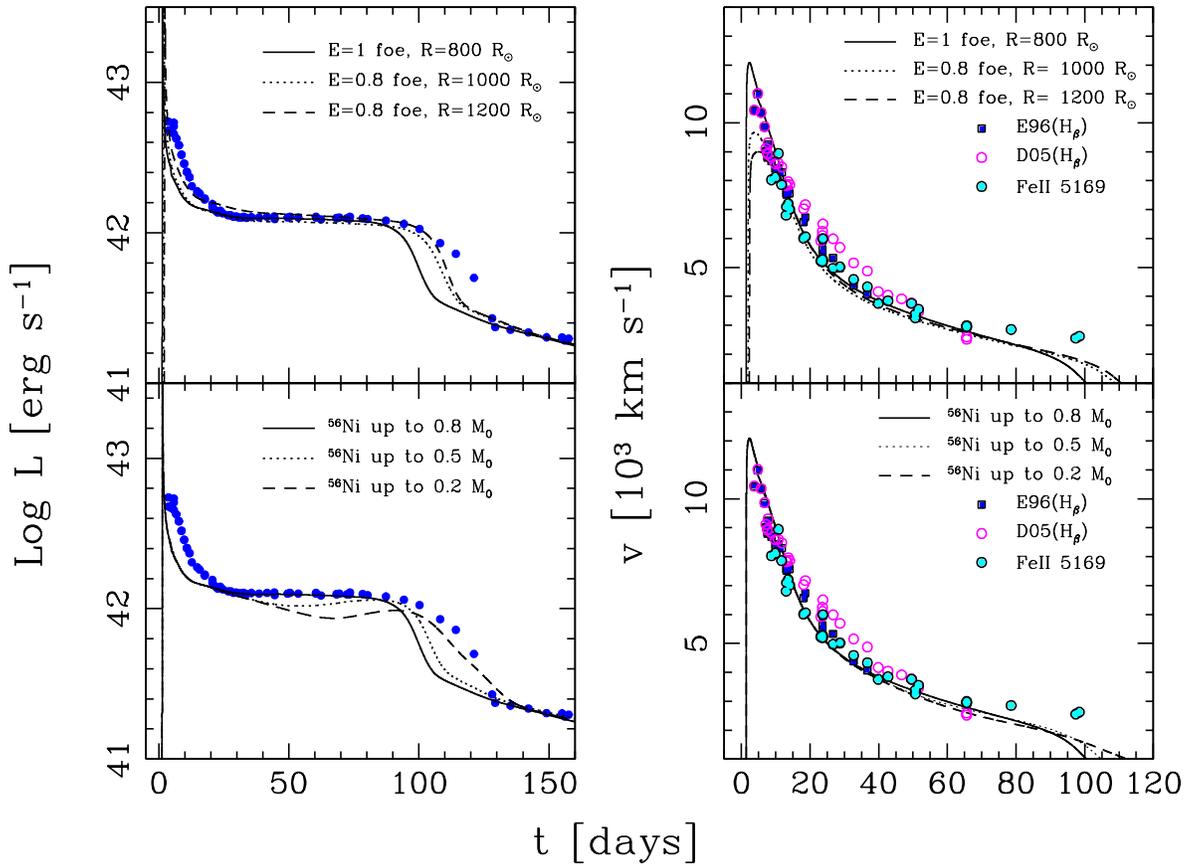}
\caption[Low mass models for SN~1999em (continuation)]{Comparison
  between models (lines) and observations 
    (points) of SN~1999em for a
  low value of the pre-supernova mass of $M = 14 \,M_\odot$ and the
  same mass of $^{56}$Ni as for the reference model \model. 
  {\bf (Left panels):} bolometric LCs. {\bf (Right
    panels):} photospheric velocity evolution. {\bf (Upper
    panels):} models {\sc m14r8e1ni56}, {\sc m14r10e08ni56} and {\sc
    m14r12e08ni56} (see Table~\ref{sn99:tbl-3}). {\bf (Lower 
  panels):} models with different mixing of 
    $^{56}$Ni, $E=1$ foe and $R=800 \,R_\odot$. Note that the degree
  of $^{56}$Ni  is indicated in each panel as a fraction of the
  initial mass of the model. \label{fig:M14VP1}} 
\end{center}
\end{figure}

\begin{figure}
\begin{center}
\includegraphics[angle=-90,scale=.60]{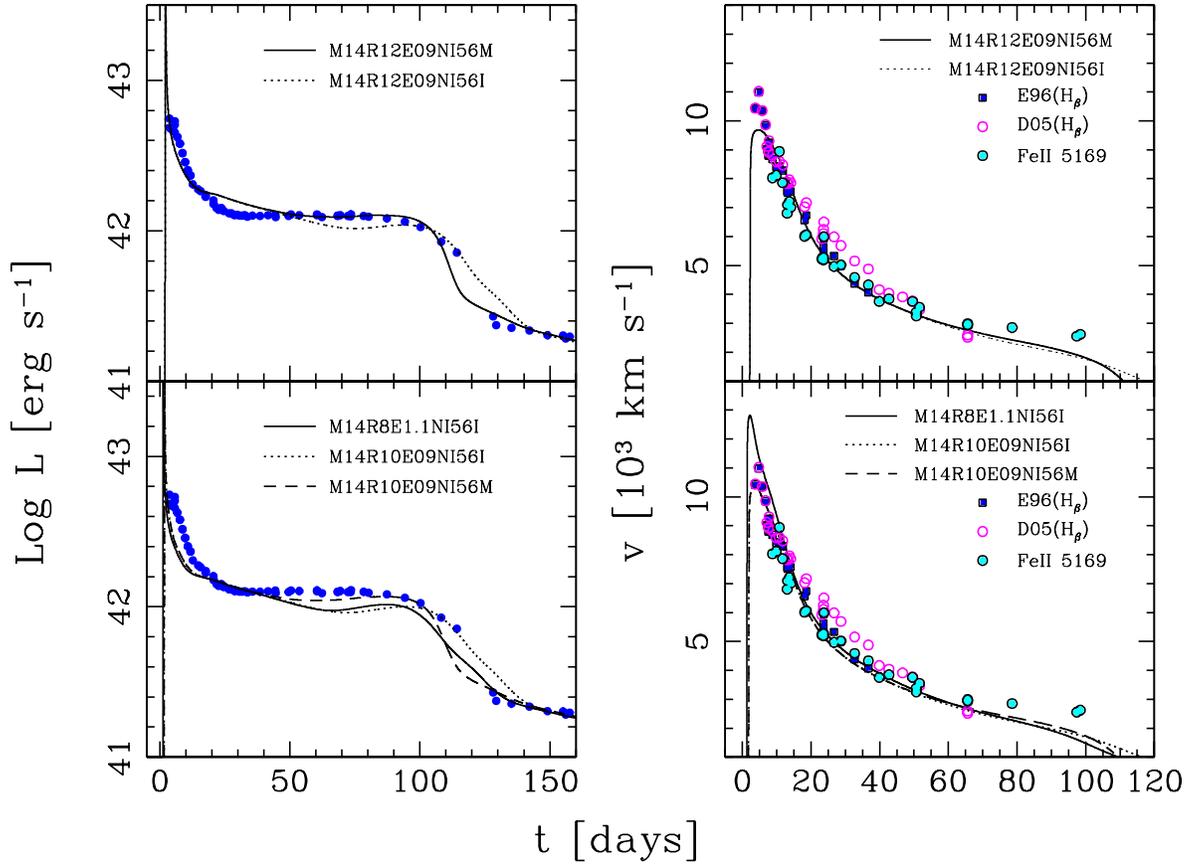}
\caption[Low mass models for SN~1999em (continuation)]{Comparison
  between models (lines) and observations 
    (points) of SN~1999em for a
  low value of the pre-supernova mass of $M = 14 \,M_\odot$ and the
  same mass of $^{56}$Ni as for the reference model \model. 
  {\bf (Left panels):} bolometric LCs. {\bf (Right
    panels):} photospheric velocity evolution. {\bf (Upper
    panels):} models {\sc m14r8e1.1ni56I}, {\sc m14r10e09ni56I} and {\sc
    m14r10e09ni56M}. {\bf (Lower 
  panels):} models {\sc m14r12e09ni56M} and {\sc m14r12e09ni56I} (see
  Table~\ref{sn99:tbl-3}). \label{fig:M14VP2}}  
\end{center}
\end{figure}

We calculate several models using two different values of the
pre-supernova mass: $M= 12 \, M_\odot$ (see Figures~\ref{fig:M12VER} and
\ref{fig:M12VNi}), and $M= 14 \, M_\odot$ (see
Figures~\ref{fig:M14VP1} and \ref{fig:M14VP2}). In all 
Figures, we show the bolometric light curves
and  photospheric velocities ($v_{ph}$)  for
different values of the injected energy ($E$), initial radius ($R$),
and degree of $^{56}$Ni mixing. Table~\ref{sn99:tbl-3} gives a summary of
the parameters used. We remark that in all of these models, we
have assumed the same value for the $^{56}$Ni mass as that of model
\model, i.e. $0.056 M_\odot$,  because this value is required to
  fit  the tail of the LC. Also note that the degree of $^{56}$Ni
  mixing is taken 
as a fraction of the initial mass of the object in order to compare
the different degrees of mixing in a consistent way when models 
with different initial mass are used. For example, a model with a mixing of
$^{56}$Ni up to $0.8 M_0$ has an equivalent degree of $^{56}$Ni mixing
as that of model \model.

For the case of  12 $M_\odot$, we have calculated nine
models. We first analyze the effect on the results of the variation of
one parameter while keeping the other parameters fixed. In the upper panel
of Figure~\ref{fig:M12VER} we show the 
effect on the LC and $v_{ph}$ of the variation of $E$ for a
model with $R = 800 \, R_\odot$ and $^{56}$Ni mixing up to $0.8 \, M_0$. In
the bottom panels of Figure~\ref{fig:M12VER}, we show the effect of varing
$R$, for a model with $E=1$ foe and for the same
$^{56}$Ni mixing. The sensitivity of the LC and $v_{ph}$  on
the extent of  $^{56}$Ni mixing is shown in the upper panel of
Figure~\ref{fig:M12VNi} for a model with $E=1$ foe and $R = 800 \, R_\odot$.

An examination of these Figures yields the following conclusions: 1)
higher injected energy produces higher luminosity and a shorter
plateau, 2) larger initial radius produces higher luminosity and a longer
plateau, 3) more extended $^{56}$Ni mixing produces higher
luminosity and a shorter plateau, 4) $^{56}$Ni mixing and initial
radius have a small 
effect on the $v_{ph}$ evolution as compared with that of the
explosion energy, and 5) as expected, there is no effect of any of
these parameters on the tail luminosity.
Note that the sensitivity of the LC on $E$, $R$, and
$^{56}$Ni mixing is in qualitative  concordance with analytic studies
by \citet{1980ApJ...237..541A}, \citet{1993ApJ...414..712P}, and previous
numerical studies by \citet{1983Ap&SS..89...89L,1985SvAL...11..145L}
and more recently by 
\citet{2004ApJ...617.1233Y} and \citet{2007A&A...461..233U}.

Although we have considered several values of $E$, $R$, and $^{56}$Ni mixing,
neither of the models with  $M= 12 \,  M_\odot$  gives a good
representation of the 
observations of SN~1999em. Note that we have  calculated also  two other
models, {\sc m12r15e05ni56} and {\sc m12r15e08ni56}, adopting even
higher values of the 
initial radius. These models are shown in the bottom panel of
Figure~\ref{fig:M12VNi}. Despite the good representation of the LC
provided by model {\sc m12r15e05ni56}, it 
fails to reproduce the observed  photospheric velocities. We conclude
that it is not possible 
to reproduce the observations of SN~1999em using models with a
  pre-supernova mass of $M = 12 \,  M_\odot$. Clearly, masses lower
  than this value would also fail to match the observations.  

For the case of 14 $M_\odot$, we have calculated ten models for
different combinations of $R$, $E$ and $^{56}$Ni
mixing. Specifically, we used three different values of the initial
radius: $R = 800$, $1000$, and $1200 \,R_\odot$, four values of
explosion energy: $E = 1.1$, $1.0$, $0.9$, and $0.8$ foe, and three
different degrees of $^{56}$Ni mixing: $0.2$, $0.5$, and $0.8 \, M_0$ (see
Table~\ref{sn99:tbl-3}). Figures~\ref{fig:M14VP1} 
and \ref{fig:M14VP2} show the results of these models. At first 
sight we find that these models agree better with the observations
than the 12 $M_\odot$ ones. Therefore the case of 14 $M_\odot$
deserves a more detailed analysis. As described below, we
chose the grid of parameters based on the  
known effects of each physical parameter on the light curves and
velocities, and trying to reach the best possible agreement with the
observations. As noted in Section~\ref{sec:hydro}, in order to assess
the validity  
of models we will focus on how well they reproduce the observed
plateau luminosity and length, and the early photospheric velocities, but we
will disregard differences in the early LC before the plateau phase
where the uncertainties in both models and data are the largest.

Our reference model is shown with a solid line in the upper
  panels of Figure~\ref{fig:M14VP1} and has the following parameters:
  $E = 1$ foe, $R=800\,R_\odot$, and $0.8 \, M_0$ of $^{56}$Ni
  mixing. We note that this reference model has the same initial radius and
  $^{56}$Ni mixing as model \model\ for 19 $M_\odot$. The energy has
  been reduced in order to compensate the effect on the luminosity of
  using a lower mass. As shown in Figure~\ref{fig:M14VP1} the reference model
  produces the correct plateau luminosity and photospheric velocities
  although the plateau duration is too short as compared with the
  data. With the aim of remedying this situation while keeping 
  the mass fixed we invoked lower energies and larger radii (dashed
  and dotted lines). While this served to improve the issue of the
  plateau length, the 
  comparison with photospheric velocities at early times became
  poorer as  expected due to the lower energies. We therefore
   consider these models to be unlikely. In a
  further attempt to find a good 14 $M_\odot$ model, we decided to
  vary the mixing of $^{56}$Ni while keeping the other parameters as in
  the reference model. The results are shown in the lower panels of 
  Figure~\ref{fig:M14VP1}. We now note that while a reduction of the
  $^{56}$Ni mixing serves to increase the length of the plateau, the
  shape of the LC and the plateau luminosity drift away from the
  observations. Thus, we can also discard these models.

Based on the above observations we tested other parameter
combinations, as shown in Figure~\ref{fig:M14VP2}. In the upper panels
we show the tests of models with slightly smaller energies and larger
radii as compared with the reference model, and two different
degrees of $^{56}$Ni mixing. While the match of the LC for these
models is satisfactory, the problem with the early-time velocities
reappears, so we discard these models. 
In the lower panels of Figure~\ref{fig:M14VP2} we show further combinations of
parameters. Here we obtain an improvement in the LC while not
compromising the agreement in the velocities. Among these models the
one called {\sc m14r10e09ni56m} provides the best match  to the LC and
velocity data. We therefore conclude that the 14 $M_\odot$ scenario
for SN~1999em cannot be ruled out, although the comparison with the
data is not as good as that of the 19 $M_\odot$ model (see
Figure~\ref{fig:LC} and \ref{fig:velo}).

\begin{table}
\begin{center}
\caption[Low mass models parameter]{Model Parameters\label{sn99:tbl-3}}
\begin{tabular}{lcccc}
\hline
\hline
    & Mass &  Radius  &  Energy &  Ni  \\
 Model  &  [$M_\odot$]  &  [$R_\odot$]  & [foe]  &
 mixing$^{\mathrm{a}}$\\    
\hline
{\sc m12r8e1ni56}      &  12    &  800   &   1      &   0.8    \\
{\sc m12r8e08ni56}     &  12    &  800   &   0.8    &   0.8    \\
{\sc m12r8e05ni56}     &  12    &  800   &   0.5    &   0.8    \\
{\sc m12r5e1ni56}      &  12    &  500   &   1      &   0.8    \\
{\sc m12r10e1ni56}     &  12    &  1000  &   1      &   0.8    \\
{\sc m12r8e1ni56m}     &  12    &  800   &   1      &   0.5    \\
{\sc m12r8e1ni56in}    &  12    &  800   &   1      &   0.2    \\
{\sc m12r15e05ni56}    &  12    &  1500  &   0.5    &   0.8    \\
{\sc m12r15e08ni56}    &  12    &  1500  &   0.8    &   0.8    \\
{\sc m14r8e1ni56}      &  14    &  800   &   1      &   0.8    \\
{\sc m14r10e08ni56}    &  14    &  1000  &   0.8    &   0.8    \\
{\sc m14r12e08ni56}    &  14    &  1200  &   0.8    &   0.8    \\
{\sc m14r8e1ni56m}     &  14    &  800   &   1      &   0.5    \\
{\sc m14r8e1ni56i}     &  14    &  800   &   1      &   0.2    \\
{\sc m14r8e1.1ni56i}   &  14    &  800   &   1.1    &   0.2    \\
{\sc m14r10e09ni56m}   &  14    &  1000  &   0.9    &   0.5    \\
{\sc m14r10e09ni56i}   &  14    &  1000  &   0.9    &   0.2    \\
{\sc m14r12e09ni56m}   &  14    &  1200  &   0.9    &   0.5    \\
{\sc m14r12e09ni56i}   &  14    &  1200  &   0.9    &   0.2    \\

\hline
\end{tabular}
\begin{list}{}{}
\item[$^{\mathrm{a}}$] The degree of $^{56}$Ni mixing is given as a
    fraction of the initial mass of the model ($M_0$)
\end{list}
\end{center}
\end{table}

\chapter{Observed and Physical Properties }
\label{ch:ana}
In this chapter I use the database presented in chapter~\ref{ch:database}
to examine the observed and physical properties of this sample of
SNe~II-P. I begin by calculating bolometric light curves and
effective temperature evolution for the complete sample of SNe~II-P in
\S~\ref{sec:sample}. Then, I proceed to derive different parameters that
characterize the bolometric LC, and  I find that the SN sample shows a
range of $1.15$ dex in 
plateau luminosity, and plateau durations between 64 and 103 days. 
Comparing the shape of the transition between the plateau and the 
radioactive tail, I find that the size of the drop in
  luminosity is in the range  
 $0.35$ to $1.46$ dex. The radioactive nickel masses
calculated for all the SNe  in 
our sample {\bf except for SN~1992am} are below $0.1$ $M_\odot$.
In section~\ref{sec:grid} I calculate a 
grid of hydrodynamical models for a range of initial masses, radii,
explosion energies, nickel masses, and mixing. From this
set of models and the observations I study correlations between
different observed and physical parameters in \S~\ref{sec:correlatios}.

\section{Application to SN~II-P Data} 
\label{sec:sample}
In this section, we apply the calibrations for 
bolometric corrections and effective temperatures presented in chapter
\S~\ref{ch:model_obs} to our sample of 37 SNe~II-P. Note that in order
to calculate bolometric light curves we just need to apply
equation~(\ref{eq3}) of \S~\ref{sec:BCresults}. Thus, we need to 
have (1) $B V I$ photometry, (2) extinction corrections due to our
own Galaxy, (3) host-galaxy extinction corrections, and (4) distances.     
We obtained Galactic extinction values from \citet{SFD98} (see
Table~\ref{sampleSNe} of chapter~\ref{ch:database}). Host-galaxy
extinctions and distances for most of the present sample were
calculated by \citet{F09} and they are presented in Table~\ref{AD} of
\S~\ref{sec:AKA}. 

\begin{figure}[htbp]
\begin{center}
\includegraphics[scale=0.8]{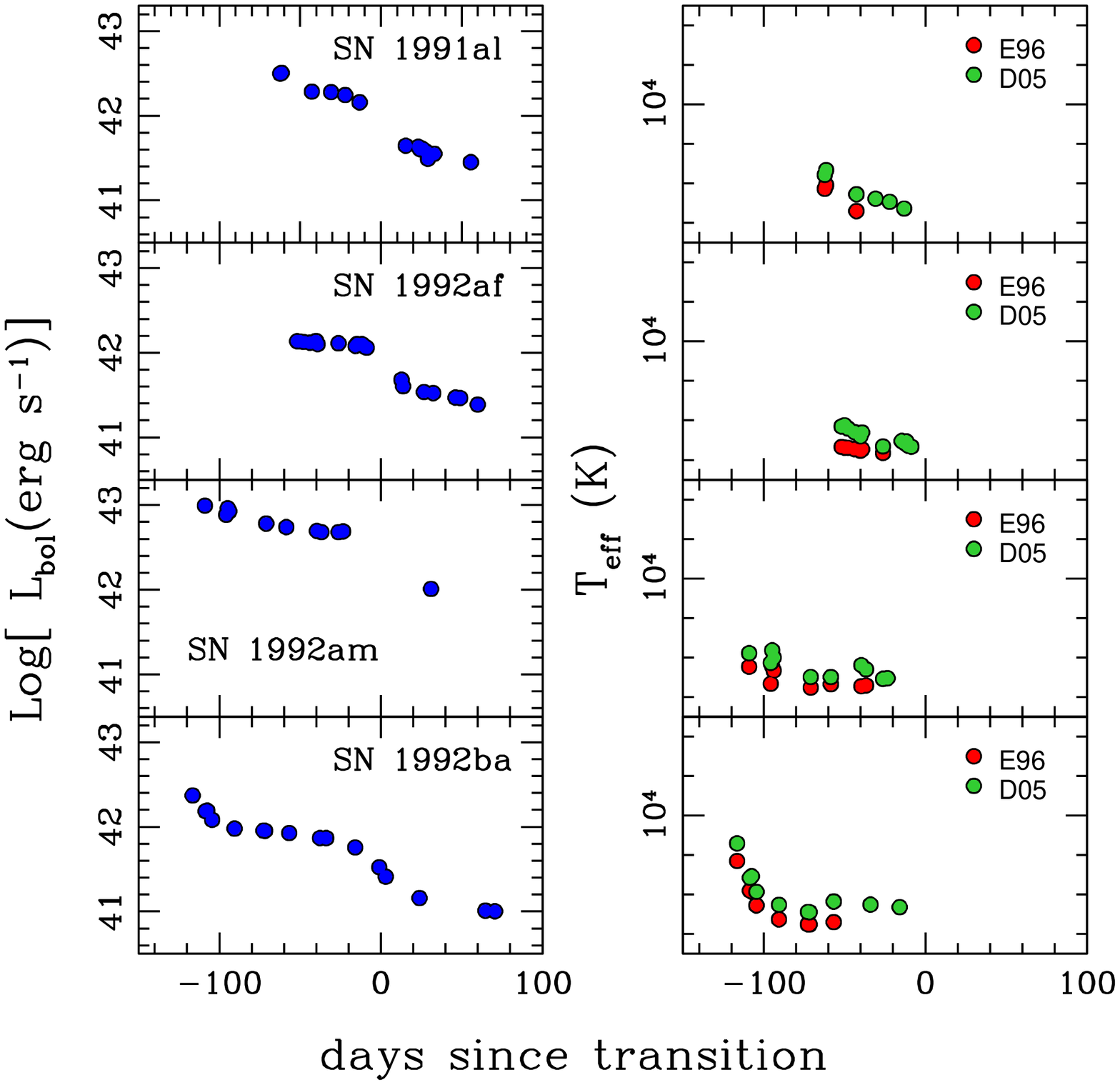}
\caption[Bolometric luminosities and effective temperatures for
  SNe~II-P (part~1)]{Bolometric luminosity {\bf(left)} and
  effective temperatures {\bf(right)} for SNe~II-P as a function of time
  (part~1). The origin of time is the middle point
  between the plateau and the linear tail ($t_{PT}$), as defined by
  \citet{F09} (see Table~\ref{AD} of \S~\ref{sec:time}).}
 \label{fig:Lbol1}
\end{center}
\end{figure}

\begin{figure}[htbp]
\begin{center}
\includegraphics[scale=0.8]{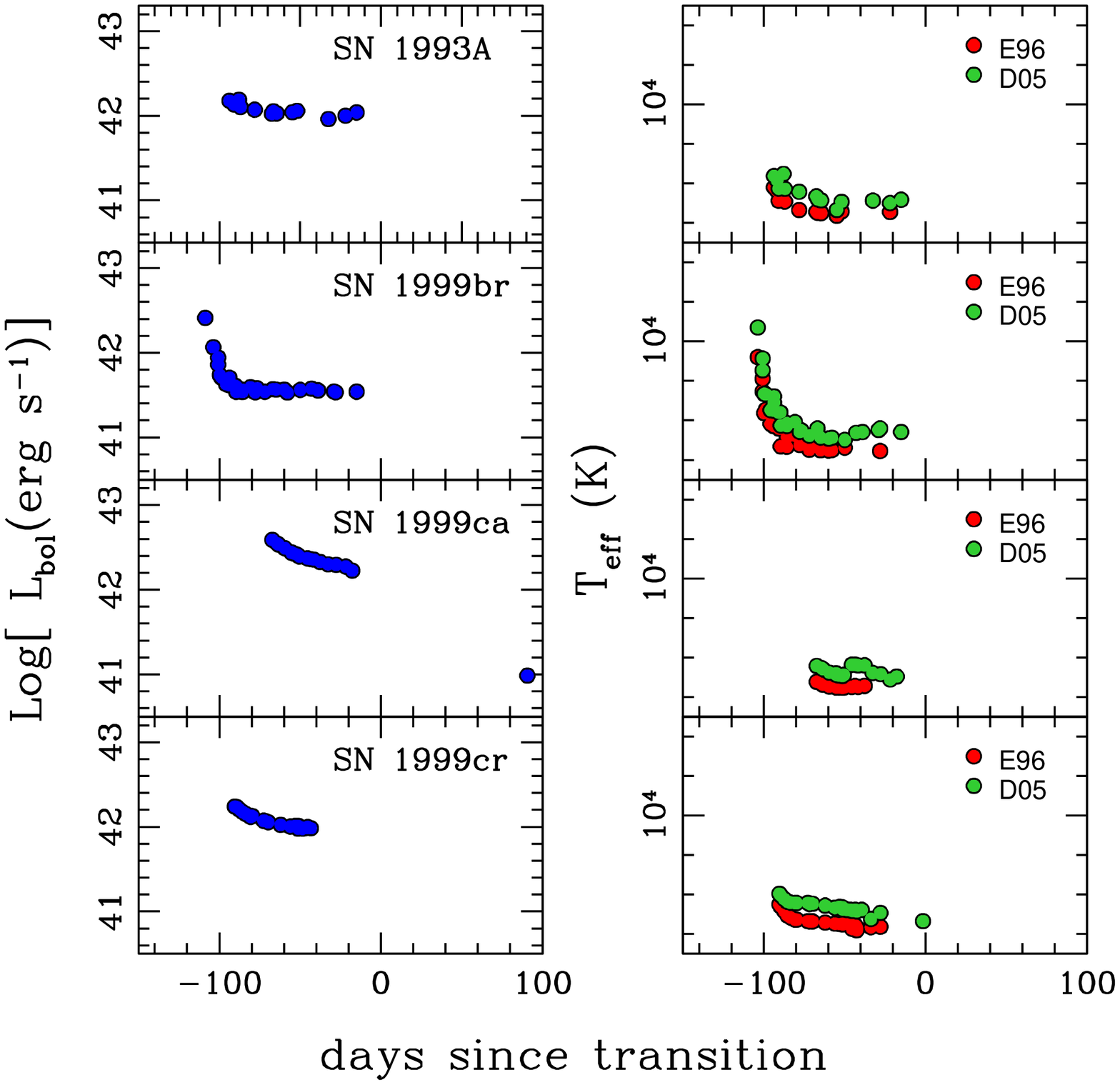}
\caption[Bolometric luminosities and effective temperatures for
  SNe~II-P (part~2)]{Bolometric luminosity {\bf(left)} and
  effective temperatures {\bf(right)} for SNe~II-P as a function of time
  (part~2). The origin of time is the middle point
  between the plateau and the linear tail ($t_{PT}$) as defined by
  \citet{F09} (see Table~\ref{AD} of \S~\ref{sec:time}).}
 \label{fig:Lbol2}
\end{center}
\end{figure}

\begin{figure}[htbp]
\begin{center}
\includegraphics[scale=0.8]{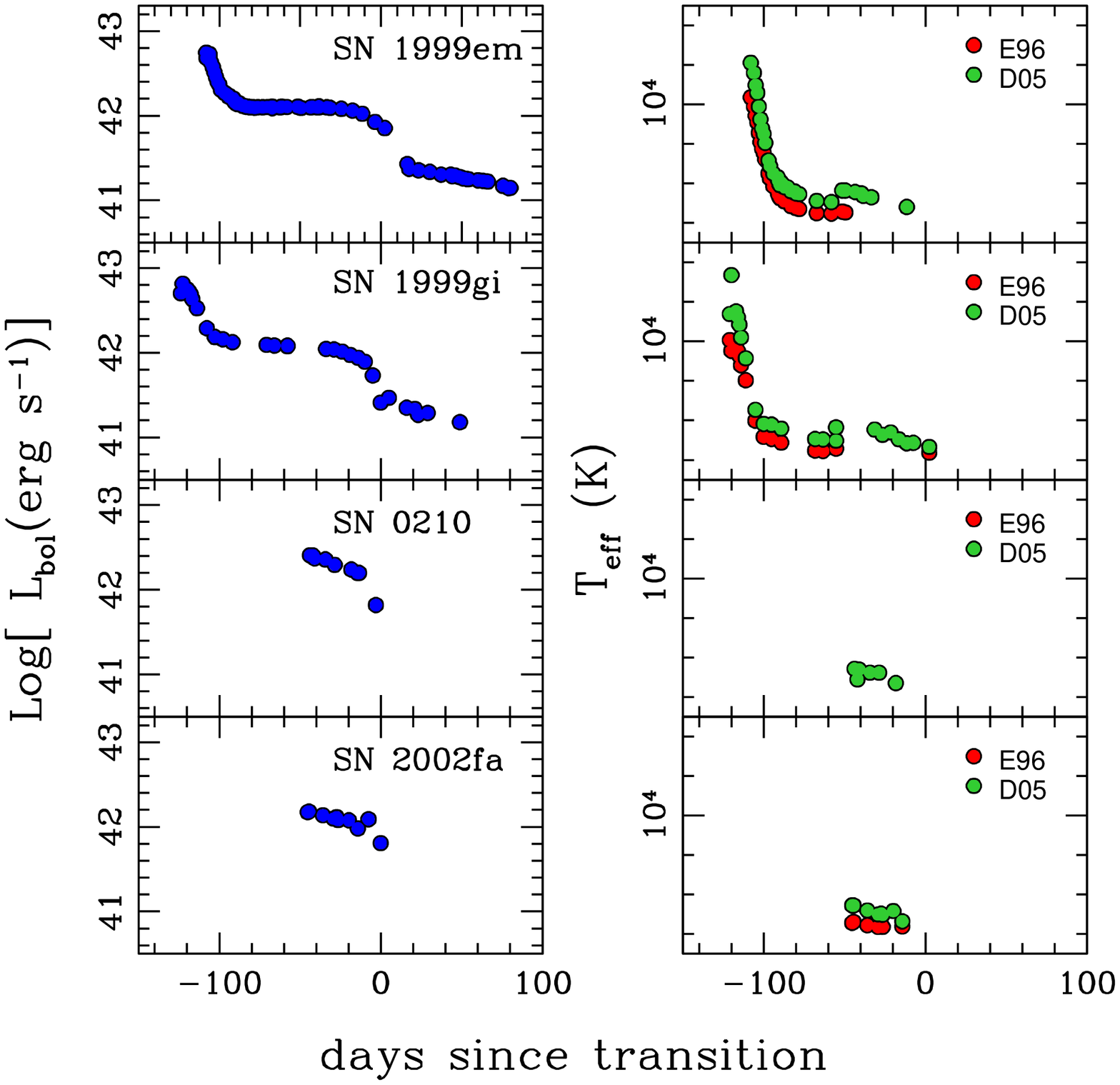}
\caption[Bolometric luminosities and effective temperatures for
  SNe~II-P (part~3)]{Bolometric luminosity {\bf(left)} and
  effective temperatures {\bf(right)} for SNe~II-P as a function of time
  (part~3). The origin of time  is the middle point
  between the plateau and the linear tail ($t_{PT}$) as defined by
  \citet{F09} (see Table~\ref{AD} of \S~\ref{sec:time}).}
 \label{fig:Lbol3}
\end{center}
\end{figure}

\begin{figure}[htbp]
\begin{center}
\includegraphics[scale=0.8]{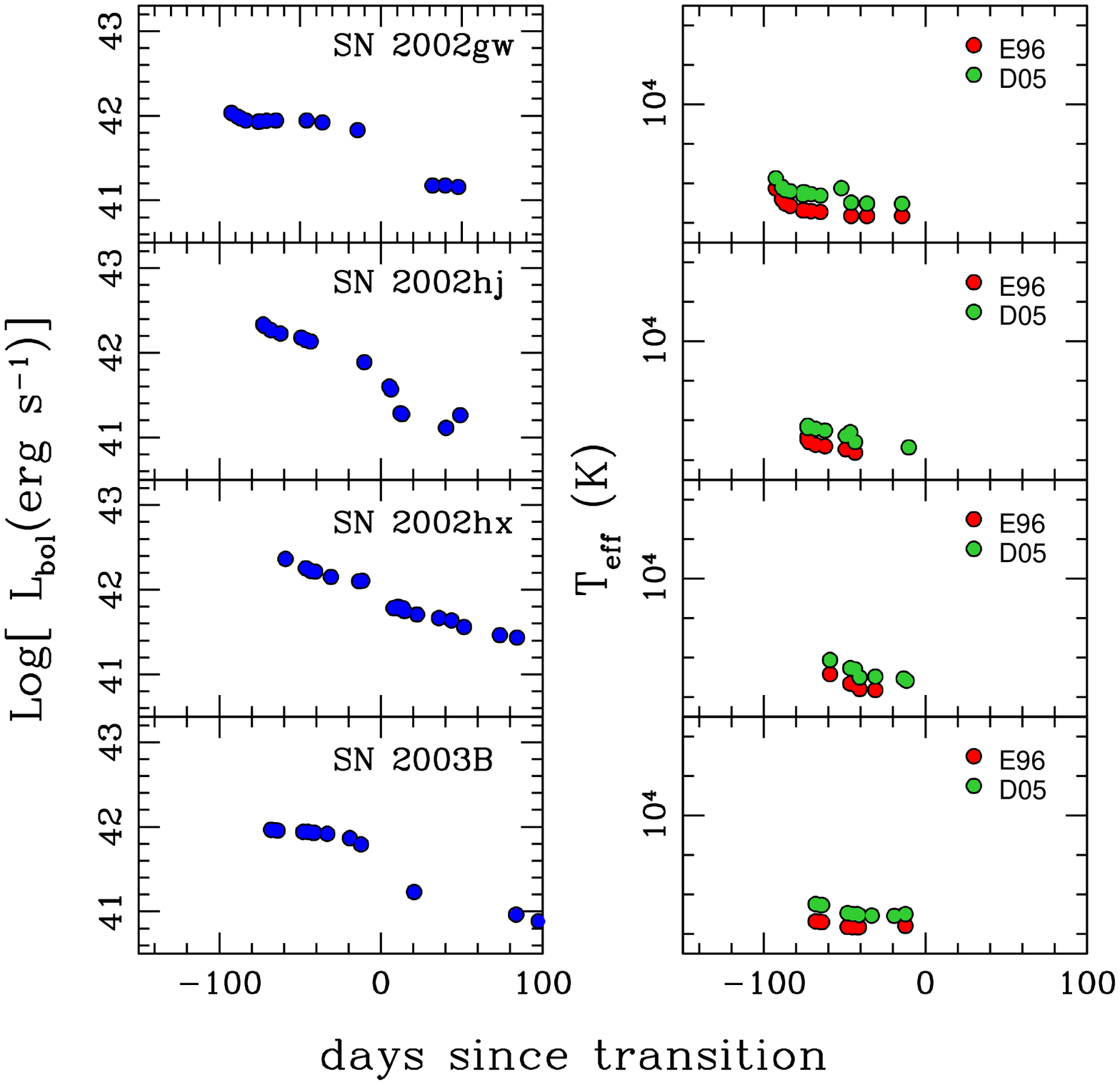}
\caption[Bolometric luminosities and effective temperatures for
  SNe~II-P (part~4)]{Bolometric luminosity {\bf(left)} and
  effective temperatures {\bf(right)} for SNe~II-P as a function of time
  (part~4). The origin of time  is the middle point
  between the plateau and the linear tail ($t_{PT}$) as defined by
  \citet{F09} (see Table~\ref{AD} of \S~\ref{sec:time}).}
 \label{fig:Lbol4}
\end{center}
\end{figure}

\begin{figure}
\begin{center}
\includegraphics[scale=0.8]{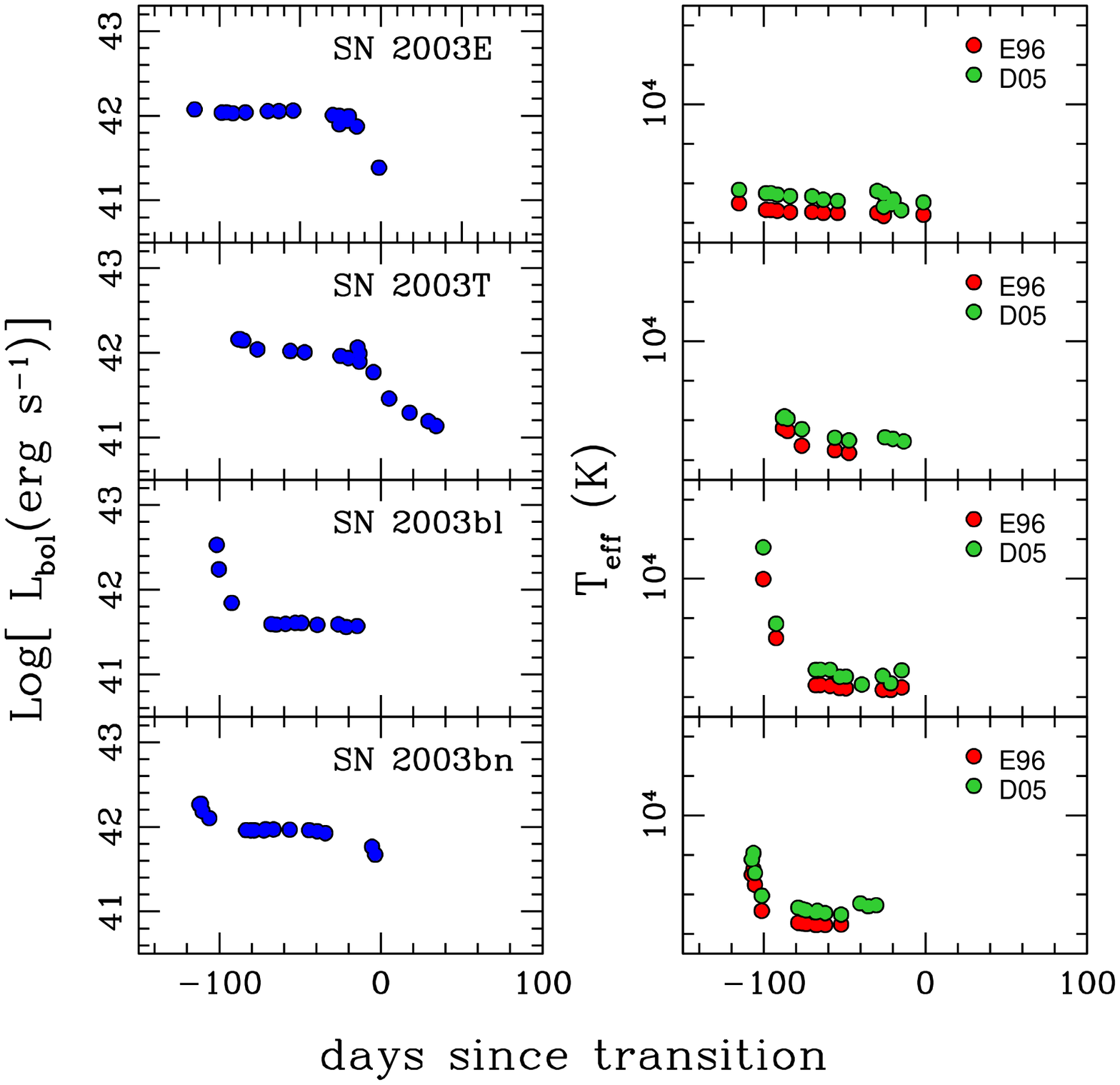}
\caption[Bolometric luminosities and effective temperatures for
  SNe~II-P (part~5)]{Bolometric luminosity {\bf(left)} and
  effective temperatures {\bf(right)} for SNe~II-P as a function of time
  (part~5). The origin of time  is the middle point
  between the plateau and the linear tail ($t_{PT}$) as defined by
  \citet{F09} (see Table~\ref{AD} of \S~\ref{sec:time}).}
 \label{fig:Lbol5}
\end{center}
\end{figure}

\begin{figure}[htbp]
\begin{center}
\includegraphics[scale=0.8]{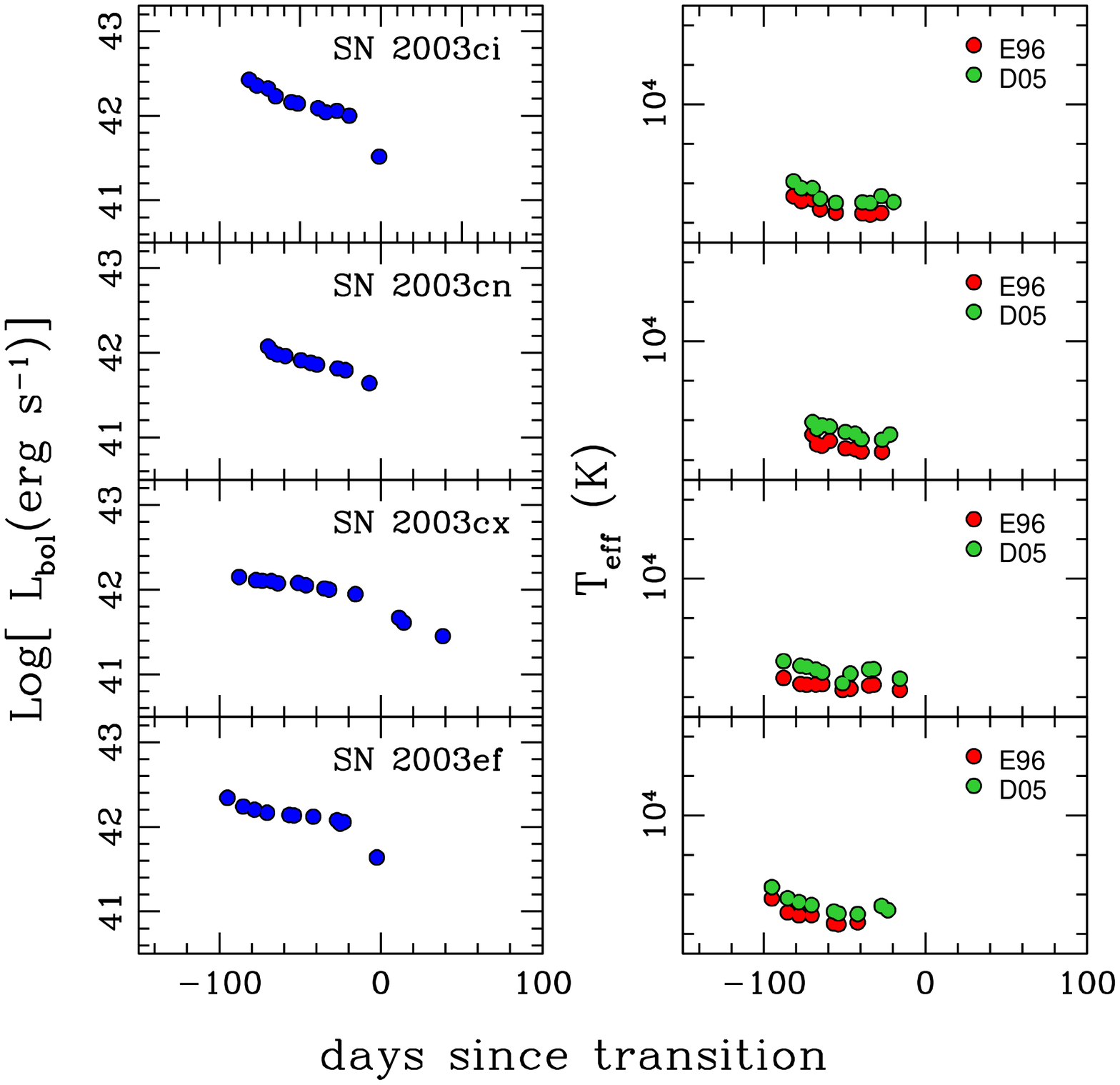}
\caption[Bolometric luminosities and effective temperatures for
  SNe~II-P (part~6)]{Bolometric luminosity {\bf(left)} and
  effective temperatures {\bf(right)} for SNe~II-P as a function of time
  (part~6). The origin of time  is the middle point
  between the plateau and the linear tail ($t_{PT}$) as defined by
  \citet{F09} (see Table~\ref{AD} of \S~\ref{sec:time}).}
 \label{fig:Lbol6}
\end{center}
\end{figure}

\begin{figure}[htbp]
\begin{center}
\includegraphics[scale=0.8]{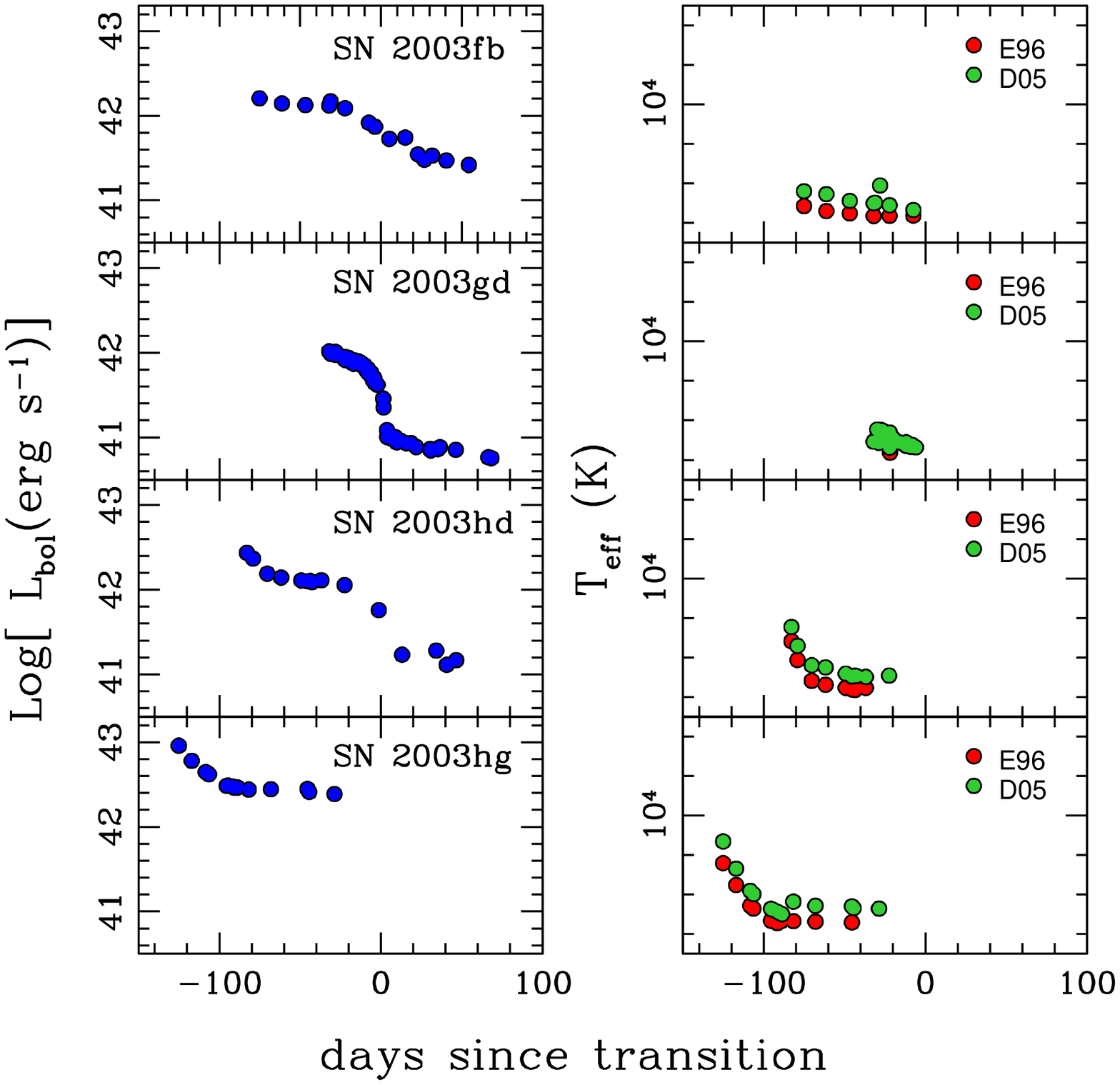}
\caption[Bolometric luminosities and effective temperatures for
  SNe~II-P (part~7)]{Bolometric luminosity {\bf(left)} and
  effective temperatures {\bf(right)} for SNe~II-P as a function of time
  (part~7). The origin of time  is the middle point
  between the plateau and the linear tail ($t_{PT}$) as defined by
  \citet{F09} (see Table~\ref{AD} of \S~\ref{sec:time}).}
 \label{fig:Lbol7}
\end{center}
\end{figure}

\begin{figure}[htbp]
\begin{center}
\includegraphics[scale=0.8]{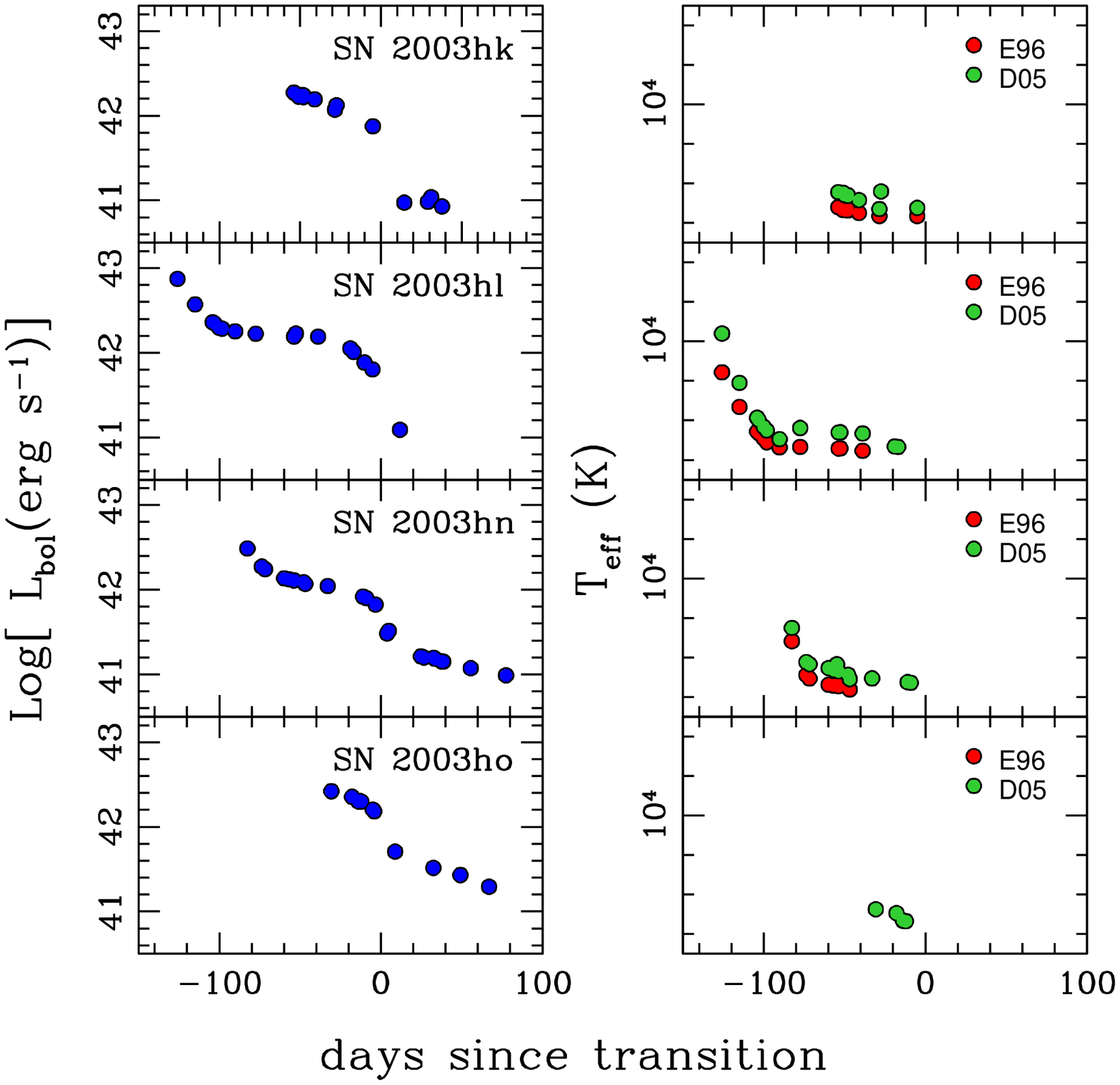}
\caption[Bolometric luminosities and effective temperatures for
  SNe~II-P (part~8)]{Bolometric luminosity {\bf(left)} and
  effective temperatures {\bf(right)} for SNe~II-P as a function of time
  (part~8). The origin of time  is the middle point
  between the plateau and the linear tail ($t_{PT}$) as defined by
  \citet{F09} (see Table~\ref{AD} of \S~\ref{sec:time}).}
 \label{fig:Lbol8}
\end{center}
\end{figure}

\begin{figure}[htbp]
\begin{center}
\includegraphics[scale=0.8]{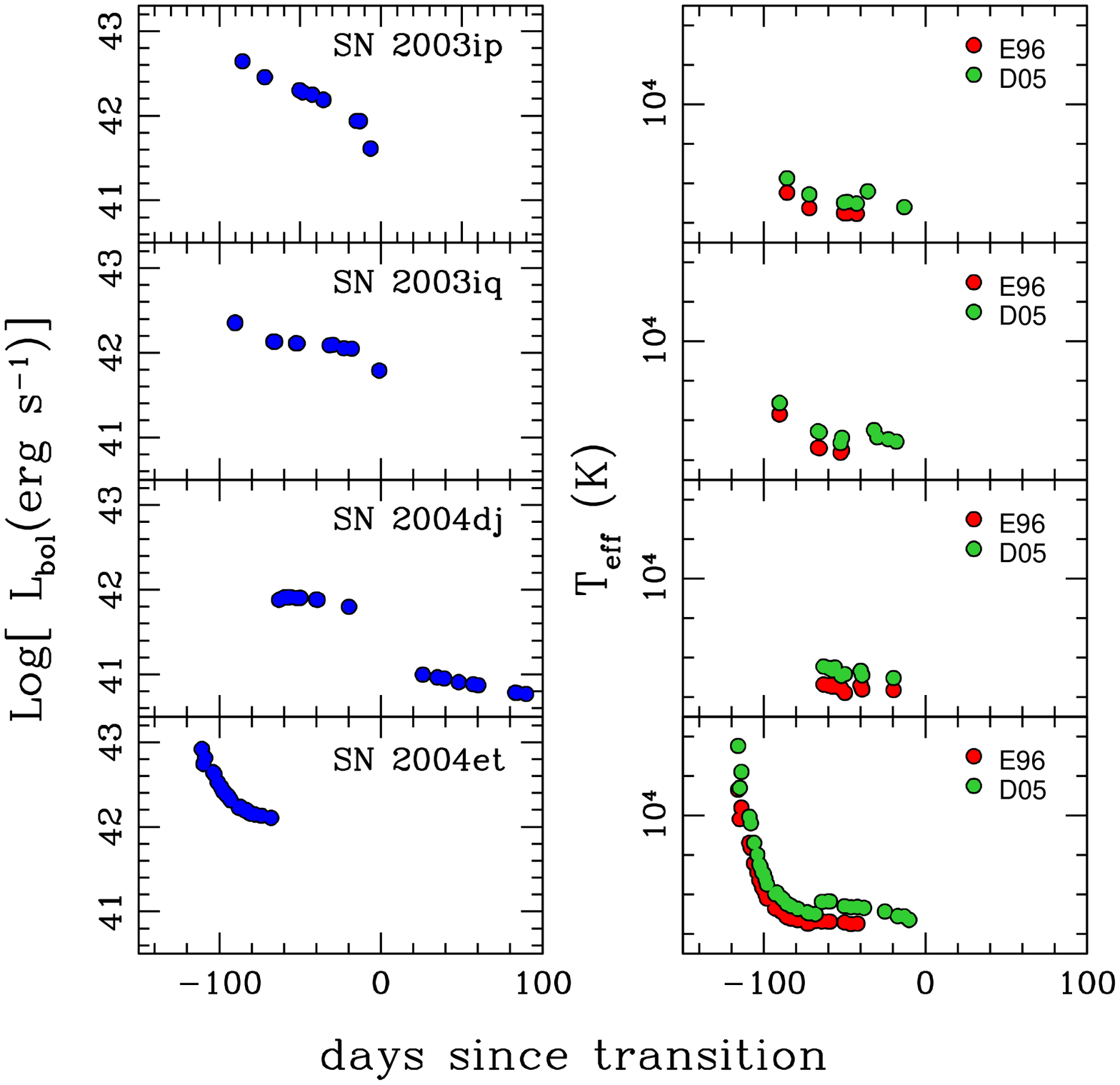}
\caption[Bolometric luminosities and effective temperatures for
  SNe~II-P (part~9)]{Bolometric luminosity {\bf(left)} and
  effective temperatures {\bf(right)} for SNe~II-P as a function of time
  (part~9). The origin of time  is the middle point
  between the plateau and the linear tail ($t_{PT}$) as defined by
  \citet{F09} (see Table~\ref{AD} of \S~\ref{sec:time}).}
 \label{fig:Lbol9}
\end{center}
\end{figure}

\begin{figure}[htbp]
\begin{center}
\includegraphics[angle=-90,scale=.25]{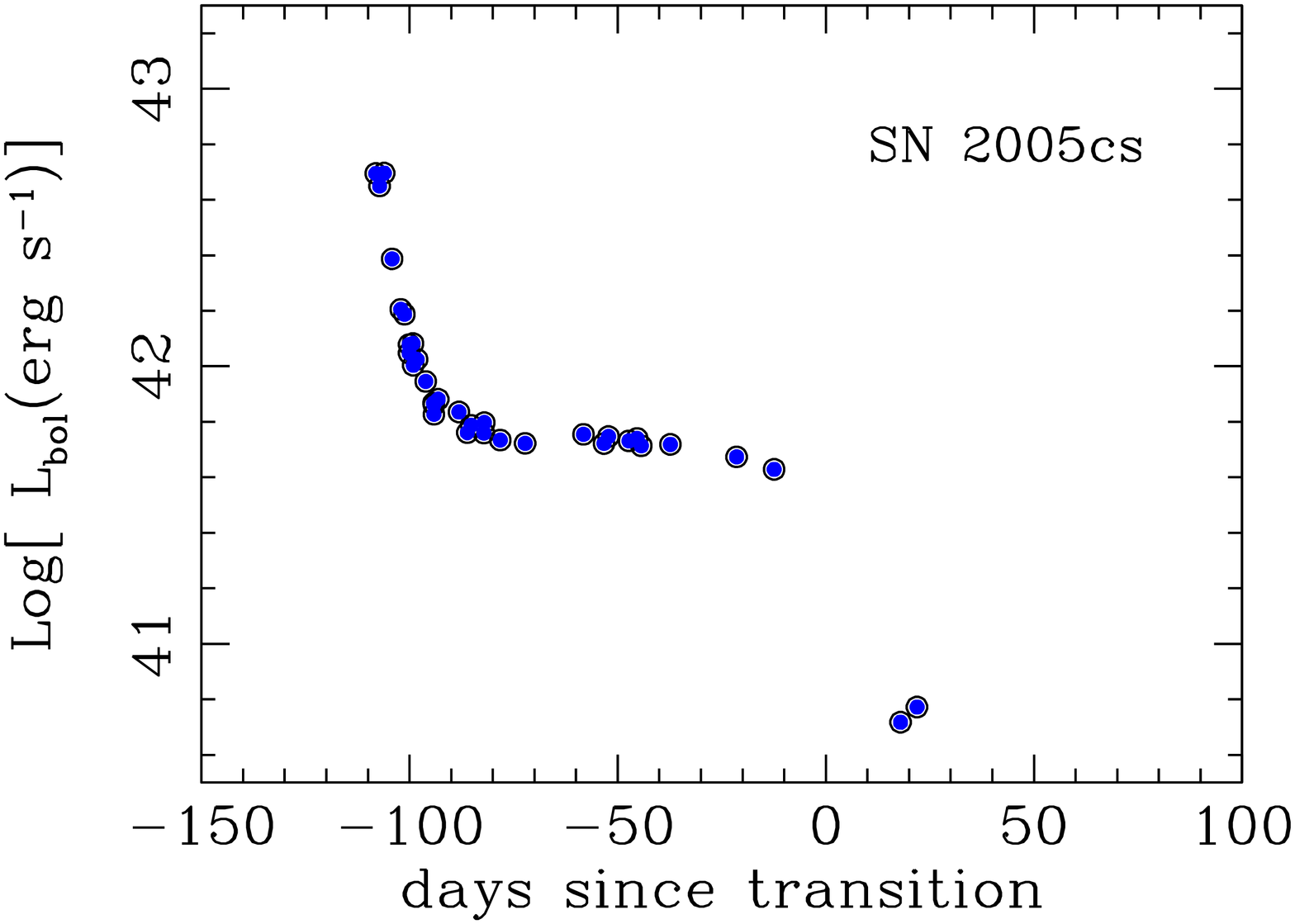}\includegraphics[angle=-90,scale=.25]{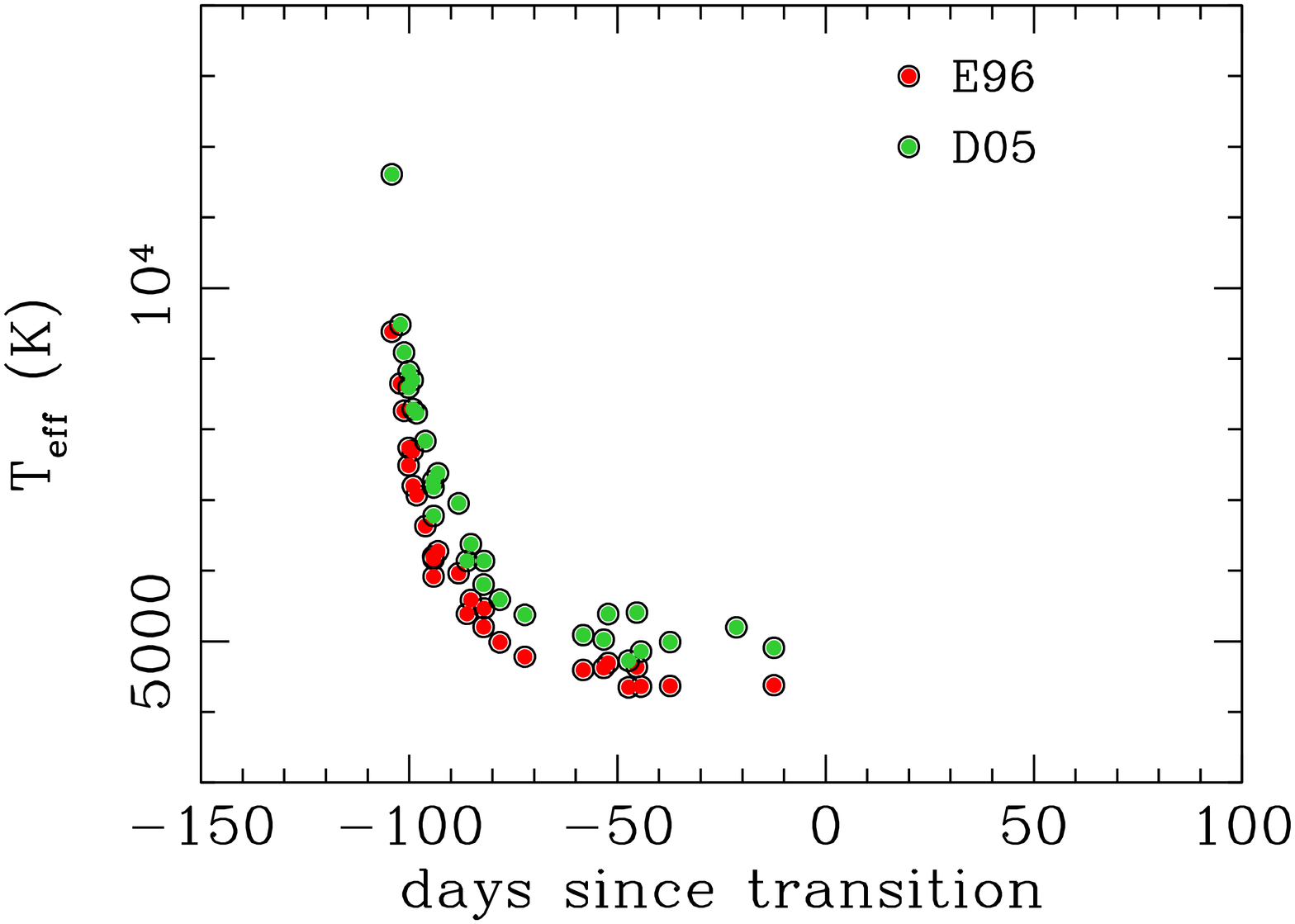} 
\caption[Bolometric luminosities and effective temperatures for
  SNe~II-P (part~10)]{Bolometric luminosity {\bf(left)} and
  effective temperatures {\bf(right)} for SNe~II-P as a function of time
  (part~10). The origin of time  is the middle point
  between the plateau and the linear tail ($t_{PT}$) as defined by
  \citet{F09} (see Table~\ref{AD} of \S~\ref{sec:time}).}
 \label{fig:Lbol10}
\end{center}
\end{figure}

Figures~\ref{fig:Lbol1} -- \ref{fig:Lbol10} present the bolometric light
curves and effective temperatures for the 37 SNe~II-P in our
sample. In order to place all the SNe in the same time scale and
facilitate the comparison among them, the origin of time used is the
middle point of the transition between the plateau and the radioactive
tail ($t_{PT}$; see \S~\ref{sec:time} and Table~\ref{AD}). A simple
inspection of the resulting bolometric light curves reveals a high
degree of heterogeneity and a variety of LC morphologies. 

There are $\sim$16 SNe  which were observed previous to the plateau
phase, i.e. during the adiabatic cooling phase. Among these, there are
three subluminous objects, SN~1999br, SN~2003bl and SN~2005cs which, in
comparison with the rest, appear to have a steeper slope during the
adiabatic cooling phase and a flatter plateau. Note, however, that in
none of the cases we have data covering the maximum produced promptly
after shock breakout, as predicted by the models. This lack of
observation at maximum light is 
consistent with the prediction that such peak has a very short duration
(a few hours), making it very difficult to observe. Only in recent years it
was possible to observe this elusive phase and  to detect the shock
breakout for two SNe~II-P thanks to observations at UV wavelength or to
the grater cadence of the  the Supernova Legacy Survey (SNLS)
\citep{2008ApJ...683L.131G,2009ApJ...705L..10T}.

In all cases where we observe a plateau the bolometric luminosity
remains nearly constant, slowly decreases, but never
increases. \citet{2007A&A...461..233U} has 
predicted, and we have also tested this with our models, that part of the
dense core of the massive star should be 
ejected in order to produce a plateau phase as observed. On the
contrary, if the whole core remains in the compact remnant and the
density profile encountered by the shock wave in the envelope is
relatively flat, the resulting luminosity increases with time (see for
example Figure~\ref{fig:comppoli} of \S~\ref{sec:initialmodel}). Our
data suggest that all SN~II-P ejecta involve part of the dense core of
the progenitor star. Also note that our sample of SNe shows
different slopes ---though never positive--- during the ``plateau''. Some
SNe are more linearly declining instead of showing an obvious plateau in
bolometric luminosity.
Such is the case of SN~1999ca, SN~2002hx, SN~2002hj, SN~2003cn,
SN~2003ci and SN~2003ip. We call these ``intermediate-plateau'' SNe. 
If we consider the definition of Type~II-L SNe
(SNe~II-L) as those objects which show a linear, uninterrupted
luminosity decline until they reach the radioactive tail phase
\citep{2003LNP...598...21T}, then SN~2002hx could perhaps be 
classified as SN~II-L. Interestingly, the existence of this group of
intermediate SNe with tilted ``plateaus'' suggests a possible
continuous connection between SNe~II-P and SNe~II-L, with the
differences probably related to the progenitor mass.
In this scenario, SNe~II-L would have  lower envelope
masses probably due to mass loss during the progenitor evolution,
although there are alternative explanations
(see \citet{1991ApJ...374..266S}). A quantitative
classification based on the average decline rate of the first 100
days of the light curve was given by \citet{1994A&A...282..731P}. In
the following,  we decided to include these intermediate SNe with the aim
of showing how their observable properties compare with the genuine
SNe~II-P and we leave for future research a more precise
definition of the limiting value of the ``plateau'' slope which would
serve to divide both subclasses on the basis of their physical properties. 
It is important to keep in mind, however, that our code may not  be
suitable to  model  SNe~II-L if such objects experience interaction
with the circumstellar medium.

\begin{figure}[htbp]
\begin{center}
\includegraphics[scale=0.35]{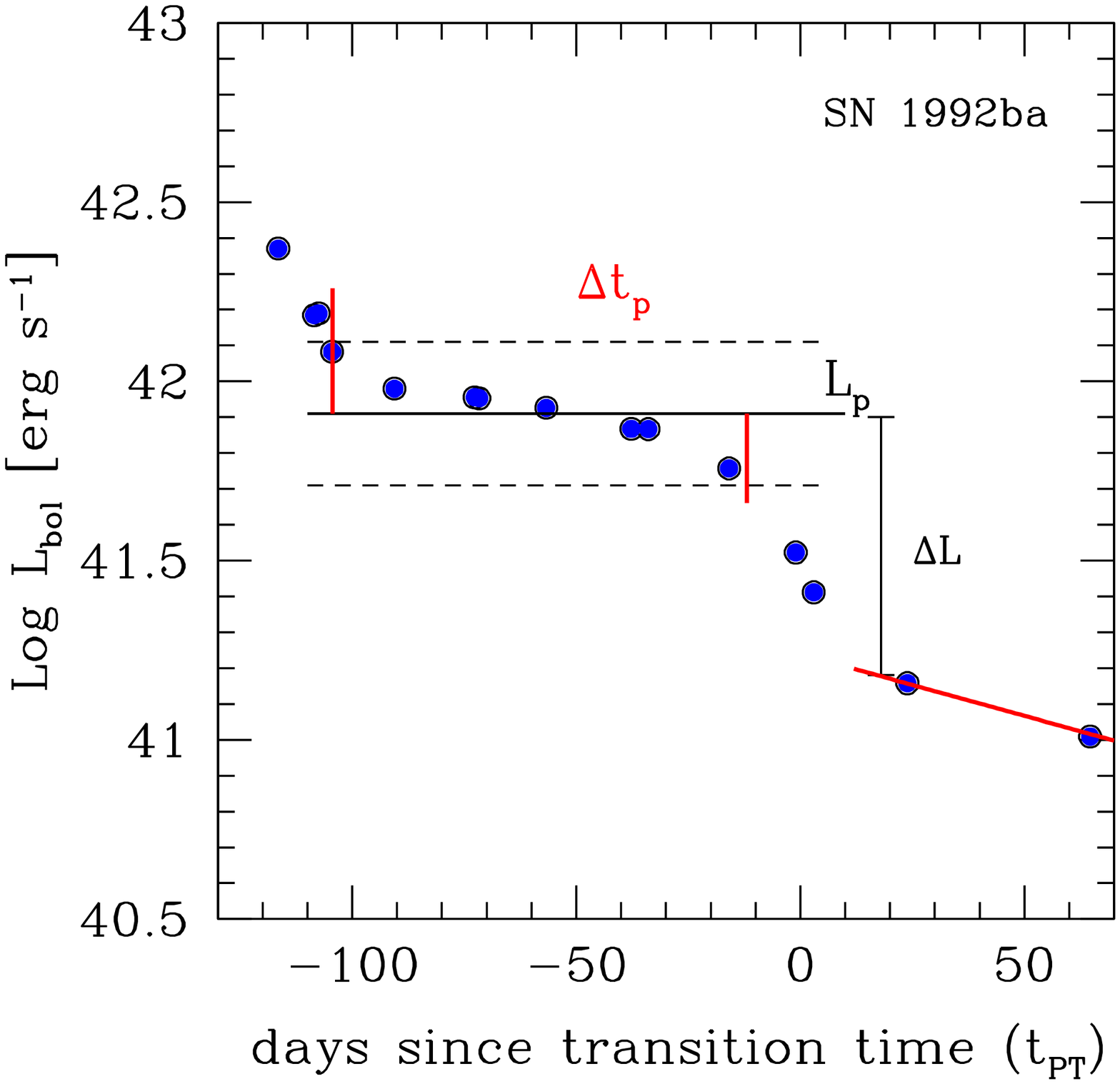}\includegraphics[scale=0.35]{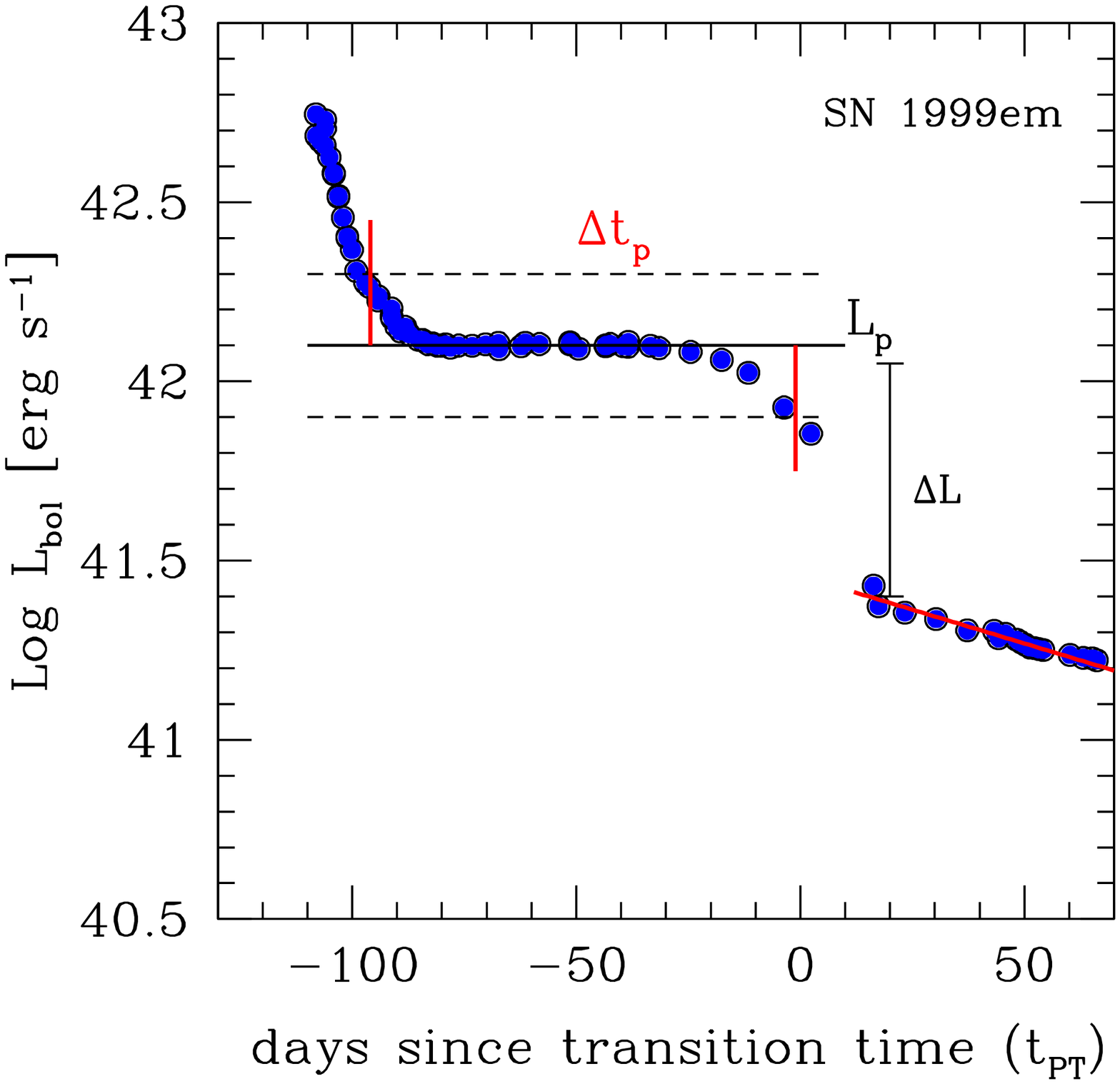}
\includegraphics[scale=0.35]{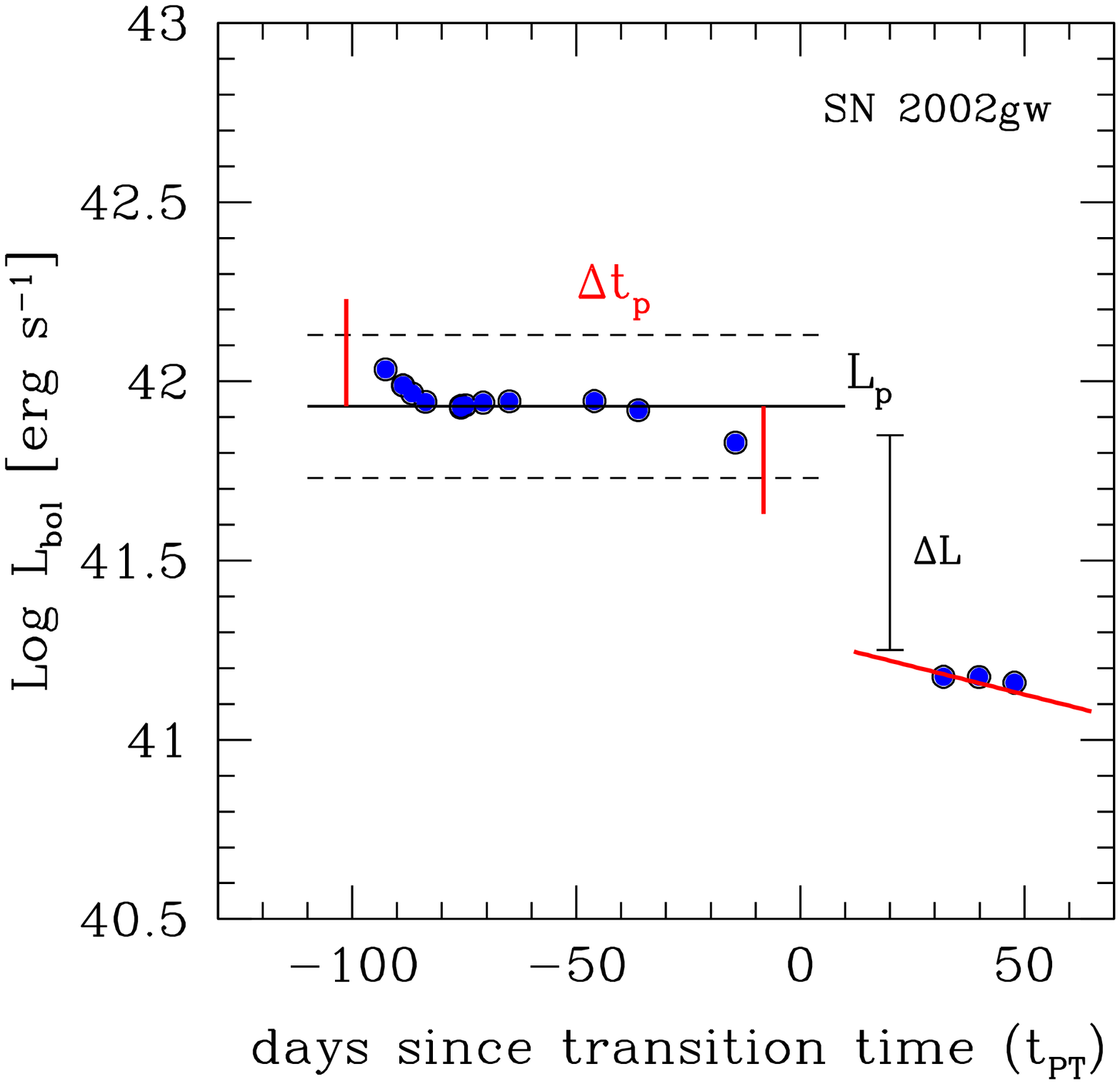}\includegraphics[scale=0.35]{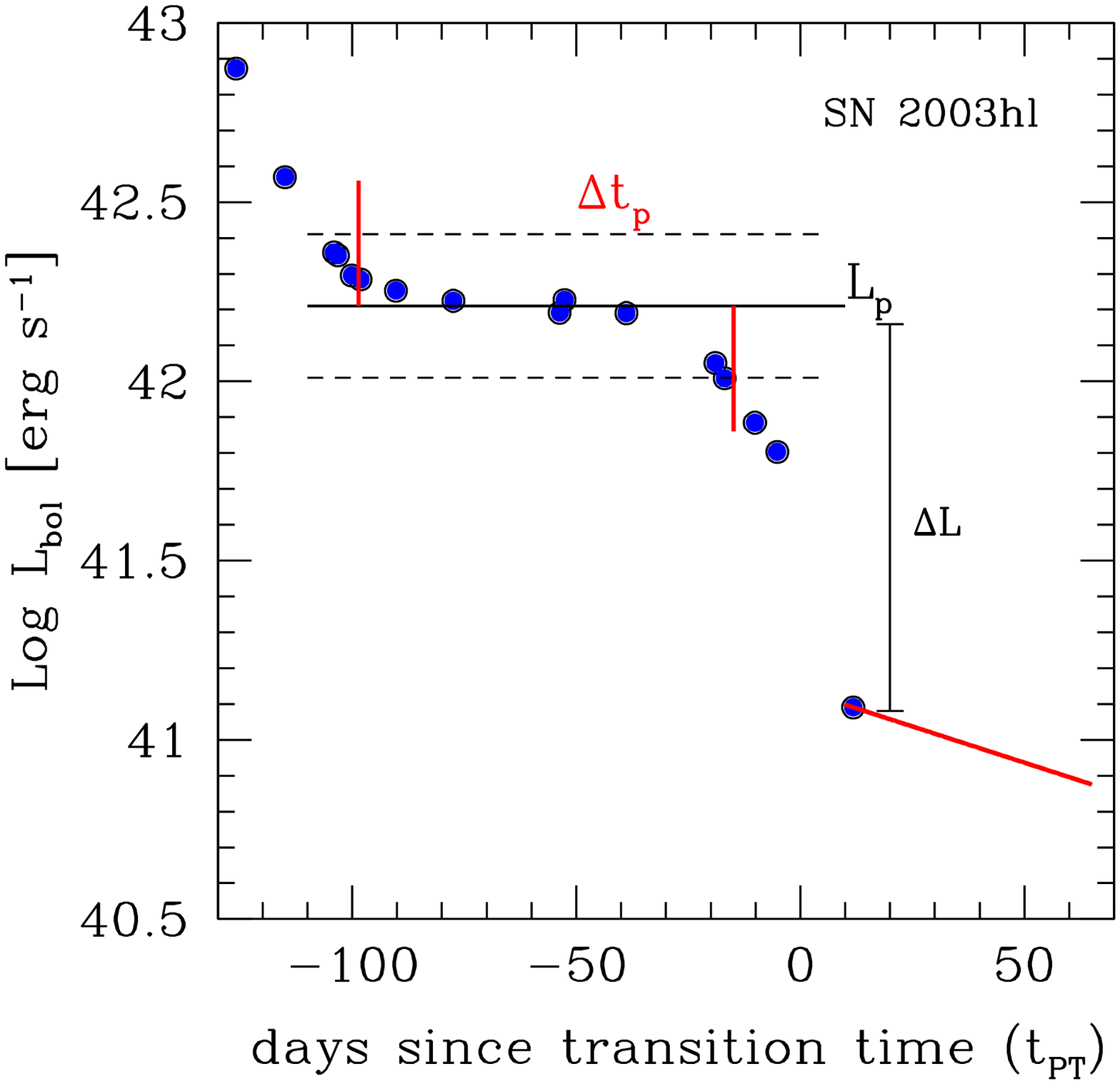}
\caption[Bolometric LC and observed parameters]{Bolometric
  LC and observed parameters ($L_P$,  
    $\Delta t_P$, $\Delta L$ and $M_{\mathrm{Ni}}$) for four SNe in
  our sample (see Table~\ref{parametros}).} 
 \label{fig:allparameters}
\end{center}
\end{figure}

To make a quantitative comparison between different SNe in our sample,
we define a series of parameters that can be measured on the bolometric
LC: (1) plateau luminosity ({\bf $L_P$}), defined as the
mean value between days $-20$ and $-80$ with respect to
$t_{PT}$; (2) plateau duration ({\bf $\Delta t_P$}), defined as
the interval of time (in the rest frame of the SN) during which the
luminosity remains within $\pm 0.2$ dex of the mean value ($L_P$); (3)
post-plateau drop ({\bf $\Delta\log L$}), defined as the
  logarithmic
difference in luminosity  between $L_P$
and the tail luminosity calculated at day 20 with respect to $t_{PT}$;
and (4) $^{56}$Ni mass produced during the explosion ({\bf
  $M_{\mathrm{Ni}}$}). As 
examples we show in 
 Figure~\ref{fig:allparameters} the bolometric LC for SN~1992ba, 
 SN~1999em, SN~2002gw and SN~2003hl together with thier parameters
 (1)--(4). In Table~\ref{parametros} we list 
 the values of the parameters and their uncertainties for for all the SNe in
 our sample. In the following subsection we give a 
 detailed explanation of the calculation of these parameters, and their
 characteristic values and 
 distributions.

\begin{table}[htpp]
\begin{center}
\caption[Observed Parameters for SNe~II-P]{Observed Parameters for SNe~II-P\label{parametros}} 
\scriptsize
\vspace{3mm}
\begin{tabular}{lccccc}
\hline\hline
         & $t_0^{\mathrm{a}}$ & $\Delta t_p$ & $\mathrm{Log} \, L_p$  & $\Delta \mathrm{Log}\, L$ & $M_{\mathrm{Ni}}$\\ 
SN Name  & [JD$-2448000$]   & [days] & [erg s$^{-1}$] & [erg s$^{-1}$] & $M_\odot$\\       
\hline                                    
1991al  & 410.00(30)  & 77.50(30.2)                  & 42.32(0.05)  &  0.701(0.058)  &  0.091(0.011)\\     
1992af  & 736.00(30)  & 97.65(29.9)                  & 42.12(0.04)  &  0.526(0.045)  &  0.093(0.013)\\     
1992am  & 778.10(11)  & 82.54(12.6)                  & 42.71(0.04)  &  0.653(0.057)  &  0.405(0.099)\\   
1992ba  & 883.90(3)   & 92.41(6.4)                   & 41.91(0.05)  &  0.738(0.051)  &  0.041(0.006)\\    
1993A   & 985.50(10)  & 89.44(31.9)                  & 42.02(0.04)  &  $\cdots$      &  $\cdots$     \\ 
1999br  & 3275.6(7.7) & 91.26(0.7)                   & 41.56(0.04)  &  0.763(0.057)  &  0.013(0.003)\\    
1999ca  & 280.00(10)  & 58.36(5.9)                   & 42.38(0.02)  &  0.929(0.131)  &  0.027(0.002) \\  
1999cr  & 221.50(10)  & 74.80(10.5)                  & 42.00(0.03)  &  $\cdots$      &  $\cdots$      \\  
1999em  & 3476.3(1.1) & 94.92(8.3)                   & 42.10(0.02)  &  0.718(0.024)  &  0.055(0.003) \\    
1999gi  & 3517.0(1.2) & 97.30(5.1)                   & 42.06(0.04)  &  0.738(0.047)  &  0.056(0.006) \\    
0210    & $\cdots$    & $>$ 31.40$^{\mathrm{b}}$      & 42.36(0.06)  &  $\cdots$      &  $\cdots$      \\  
2002fa  & $\cdots$    & $>$ 41.61$^{\mathrm{b}}$      & 42.13(0.05)  &  0.350(0.057)  &  $\cdots$      \\    
2002gw  & 4557.9(2.7) & 93.14(38.6)                  & 41.93(0.04)  &  0.713(0.065)  &  0.040(0.004) \\    
2002hj  & $\cdots$    & $>$ 48.56$^{\mathrm{b}}$      & 42.22(0.04)  &  0.972(0.065)  &  0.036(0.009) \\    
2002hx  & $\cdots$    & $>$ 59.37$^{\mathrm{b}}$      & 42.23(0.05)  &  0.489(0.053)  &  0.087(0.008) \\    
2003B   & $\cdots$    & $>$ 58.58$^{\mathrm{b}}$      & 41.94(0.04)  &  0.704(0.043)  &  $\cdots$      \\    
2003E   & $\cdots$    & $>$ 102.70$^{\mathrm{b}}$     & 41.99(0.04)  &  $\cdots$      &  $\cdots$      \\  
2003T   & 4654.2(2.7) & 85.97(56.7)                  & 41.99(0.05)  &  0.717(0.051)  &  0.035(0.006) \\    
2003bl  & 4692.6(2.8) & 75.81(9.8)                   & 41.59(0.04)  &  $\cdots$      &  $\cdots$      \\  
2003bn  & 4693.4(2.7) & 102.34(4.8)                  & 41.96(0.04)  &  0.792(0.057)  &  0.036(0.010) \\    
2003ci  & $\cdots$    & 56.09(47.9)                  & 42.14(0.05)  &  $\cdots$      &  $\cdots$      \\  
2003cn  & $\cdots$    & 58.34(10.3)                  & 41.90(0.04)  &  1.457(0.057)  &  0.005(0.002) \\    
2003cx  & $\cdots$    & $>$ 80.25$^{\mathrm{b}}$      & 42.06(0.04)  &  0.478(0.045)  &  $\cdots$      \\    
2003ef  & 4759.8(4.7) & 75.76(37.4)                  & 42.11(0.17)  &  $\cdots$      &  $\cdots$       \\ 
2003fb  & $\cdots$    & $>$ 65.99$^{\mathrm{b}}$      & 42.14(0.05)  &  0.521(0.064)  &  $\cdots$       \\    
2003gd  & 4716.5(21)  & 93.76(21.0)                  & 41.97(0.03)  &  1.032(0.035)  &  0.020(0.001) \\    
2003hd  & $\cdots$    & 64.10(7.3)                   & 42.12(0.04)  &  0.886(0.063)  &  0.036(0.005)  \\   
2003hg  & $\cdots$    & 98.02(7.8)                   & 42.42(0.06)  &  $\cdots$      &  $\cdots$       \\ 
2003hk  & $\cdots$    & $>$ 45.88$^{\mathrm{b}}$      & 42.18(0.04)  &  1.201(0.051)  &  $\cdots$      \\    
2003hl  & 4872.3(1.7) & 89.75(7.7)                   & 42.21(0.06)  &  1.150(0.057)  &  0.032(0.009) \\    
2003hn  & 4859.5(3.8) & 64.72(12.0)                  & 42.12(0.04)  &  0.882(0.039)  &  0.036(0.003)  \\   
2003ho  & $\cdots$    & $>$ 24.36$^{\mathrm{b}}$      & 42.42(0.11)  &  0.820(0.116)  &  $\cdots$      \\   
2003ip  & $\cdots$    & 47.23(16.8)                  & 42.25(0.05)  &  $\cdots$      &  $\cdots$       \\ 
2003iq  & 4909.6(4.3) & 75.88(10.7)                  & 42.10(0.04)  &  $\cdots$      &  $\cdots$       \\ 
2004dj  & 5167(21)    & 83.61(21.9)                  & 41.90(0.04)  &  0.894(0.038)  &  0.028(0.002)  \\   
2004et  & 5270.5(2)   & 88.01(7.9)                   & 42.10(0.03)  &  $\cdots$      &  $\cdots$      \\  
2005cs  & 5549.9(1)   & 86.71(5.8)                   & 41.72(0.05)  &  0.989(0.065)  &  0.016(0.003)   \\  
\hline
\end{tabular}

\begin{list}{}{}
\item [$^{\mathrm{a}}$]  We generally use $t_0(E96)$ (see Table~\ref{AD} of
  \S~\ref{sec:time}) as the explosion time. In case that this value is
  not available we use  $t_0(D05)$ (see Table~\ref{AD} of
  \S~\ref{sec:time}). If neither value is available we adopt estimations
  of $t_0$ found in the literature (see \S~\ref{sec:time}). Note that
  the explosion time is used to calculate $M_{\mathrm{Ni}}$ and
  both $t_0(E96)$ and $t_0(D05)$ provide consistent
  values for this parameter. 
\item [$^{\mathrm{b}}$] Lower bound for $\Delta t_p$. 
\end{list}

\normalsize
\end{center}
\end{table}

\newpage
\subsection{Plateau Luminosity: $L_p$}
\label{lp}
\begin{figure}[htbp]
\begin{center}
\includegraphics[scale=0.4]{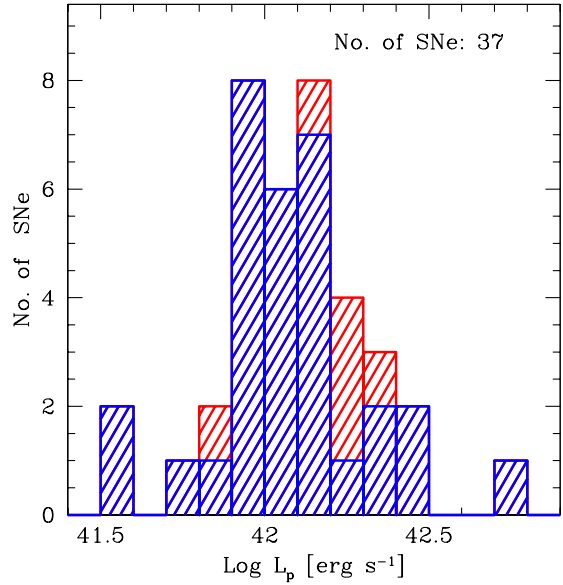}
\caption [The distribution of $L_p$ for SNe~II-P ]{The distribution of
  $L_p$ for SNe~II-P. The SNe considered as intermediate between
  plateau and linear (see \S~\ref{sec:sample}) are marked in red.} 
 \label{fig:Lp_hist}
\end{center}
\end{figure}

In order to estimate the value of this parameter we average the bolometric
luminosity between two well-defined epochs. The choice of
$-20$ and $-80$ days for this was based on the behavior shown by most 
SNe. Column~4 of Table~\ref{parametros} shows the resulting values of this  
parameter and their uncertainties. The estimation of the uncertainty
was done considering the one in 
the bolometric luminosity whose sources are (1)
photometry, (2) extinction (see Table~\ref{AD}), (3)
distance (see Table~\ref{AD}) and (4) calibration of the bolometric
correction. For the latter component, a value of
$0.11$ mag was adopted during the plateau, and of $0.02$ mag during
the tail (see \S~\ref{sec:BCresults}). Figure~\ref{fig:Lp_hist} shows the 
distribution of this parameter among our SNe. We see that the
  intermediate-plateau SNe have plateau luminosities which are
  consistent with the genuine SNe~II-P. The histogram reveals 
that most of the SNe~II-P (22 out of 37) have a characteristic plateau
luminosity between $7.9$ $\times 10^{41}$ erg s$^{-1}$ and $1.6$
$\times 10^{42}$ erg s$^{-1}$. There is only one object, namely
SN~1992am, with $L_p > 3.16$ 
$\times 10^{42}$ erg s$^{-1}$ and two SNe (SN~1999br and SN~2003bl)
with $L_p < 5$ $\times 10^{41}$ erg s$^{-1}$.

\newpage

\begin{figure}[htbp]
\begin{center}
\includegraphics[scale=0.5]{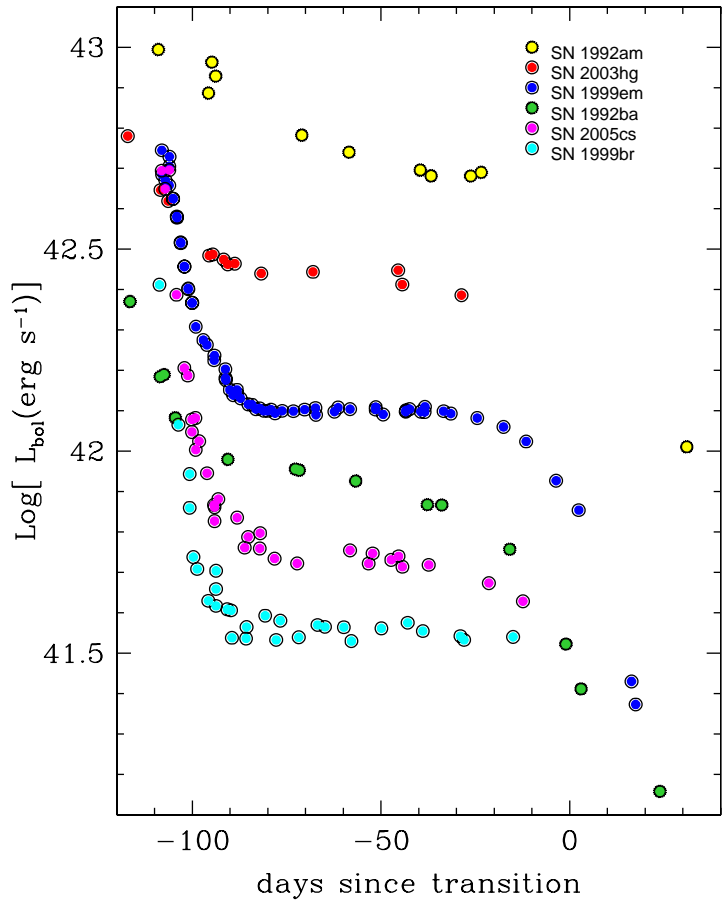}
\caption[Range of the plateau
  luminosities]{Bolometric light curves for five SNe of our sample
  showing the range of variation of plateau
  luminosities.}\label{fig:Lp} 
\end{center}
\end{figure}

In Figure~\ref{fig:Lp} we compare six
exemplary SNe to illustrate that SNe~II-P display  
a wide range of plateau luminosities  similar to the result reported
by \citep{H01} using V-band light curve. The range of luminosities 
encompassed by our sample is $1.15$ dex, equivalent to more than
one order of magnitude of spread in their radiative energy output. The
weighted average value of the characteristic plateau luminosity in
our sample is $<L_P>= 1.26 \pm 0.019 \times 10^{42}$ erg s$^{-1}$. 

\subsection{Plateau duration: $\Delta t_p$}
\label{dtp}
The estimate of this parameter requires the calculation of two
characteristic times: (1) the point where the $L= L_p + \delta L$ 
($L_{+}$),
called $t_{+}$, and (2) the point where the $L= L_p - \delta L$ 
($L_{-}$), called
$t_{-}$. Here, $\delta L$ represents the range of
luminosity adopted to define the plateau
phase. We chose a value of $0.2$ dex which is equivalent to a 
variation of 1 mag. Note 
that $t_{+}$ always occurs before t$_{-}$ given the monotonic behavior of
the LC. Thus by definition, $\Delta t_P =( t_{-} - t_{+} ) / ( 1 +
z)$, where $z$ is the redshift of the
SN. Although it is straight-forward in theory to define $\Delta t_P$ ,
in practice its calculation involves some problems especially for
those SNe which do not have good temporal coverage. We found the following
cases: (a) SNe with data for $L > L_{+}$ and $L < L_{-}$, (b) SNe with data
only for $L < L_{+}$, and (c) SNe with data only for $L > L_{-}$. 

In most of cases (a), we can determine $t_{+}$ and $t_{-}$ and their
uncertainties by interpolating between the nearest points on both
sides of $L_{+}$ and $L_{-}$, respectively. When such points are
separated by more than  30 days from each other an extrapolation is
performed based on the two points right below $L_{+}$ for $t_{+}$,
or above $L_{-}$ for $t_{-}$. In cases (b) and (c) it was only possible to find
$t_+$ or  $t_-$, respectively, by extrapolation. 

In order to compute $\Delta t_P$ for SNe with poorer coverage, we use
data of SNe with the best coverage to derive typical time intervals
which in turn allow to obtain $t_-$ or $t_+$. For the estimation of
$t_-$, we use the typical interval found between this moment and the
transition time, i.e. $\Delta t_{-} = (t_{PT} - t_{-})/( 1 + z)$. A
weighted average, $<\Delta t_{-}>$, and its uncertainty is computed for
the 13 SNe which allow a direct determination of $t_{-}$. We
find $<\Delta t_{-}> = 8.3 \pm 0.5$ days, i.e, $8.3$ days before 
the middle point of the transition between plateau and tail. This
value is then used to obtain $t_{-}$ for other SNe based on their
values of $t_{PT}$.

A similar procedure can be followed for the estimate of
$t{_+}$. In this case its application requires knowledge of the
explosion time ($t_0$). For 13 SNe with both estimates of $t{_+}$ and
$t_0$, we compute the interval between those times, $\Delta t_{+}=
(t_{+} - t_0)/( 1 + z)$, and its 
uncertainty. A weighted average of these values is obtained, 
$<\Delta  t_{+}>= 22.0 \pm 1.5$ days.
This means that, on average, SNe~II-P enter the plateau phase
approximately 22 days after  explosion. 
This result is in very good agreement with predictions from
hydrodynamical models. We are thus able to estimate $t_{+}$ using the
average above only when the 
explosion time is known. 
There are ten additional objects for which we can only estimate a
lower limit of $t_{+}$ and thus of $\Delta t_p$. Column~3 of
Table~\ref{parametros} shows 
the values of $\Delta t_p$ and their uncertainties. The values 
corresponding to lower bounds of $\Delta t_p$ are preceded by
the $>$ symbol. 

\begin{figure}[htbp]
\begin{center}
\includegraphics[scale=0.45]{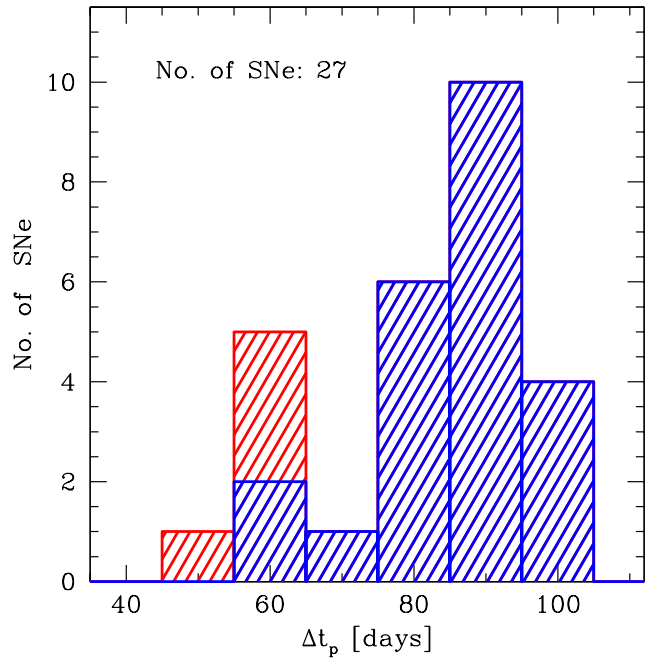}
\caption[Distribution of $\Delta t_p$ for SNe~II-P]{Distribution of plateau
  durations, $\Delta t_p$, values for 
  SNe~II-P. The SNe considered as intermediate between 
  plateau and linear (see \S~\ref{sec:sample}) are marked in red.}
 \label{fig:DLp_hist}
\end{center}
\end{figure}

Figure~\ref{fig:DLp_hist} shows the  
distribution of $\Delta t_p$ (excluding lower limits). The
histogram shows that most  SNe (20 out of 27) have  $\Delta t_p$ values
between 75 and 105 days and the peak is around 90 days. Although
the sample is small, we see hints of a bi-modal 
distribution with a secondary peak lying at about 60
days. However note that all but two of the SNe that form this
  secondary peak are intermediate-plateau SNe.

\begin{figure}[htbp]
\begin{center}
\includegraphics[angle=-90,scale=0.6]{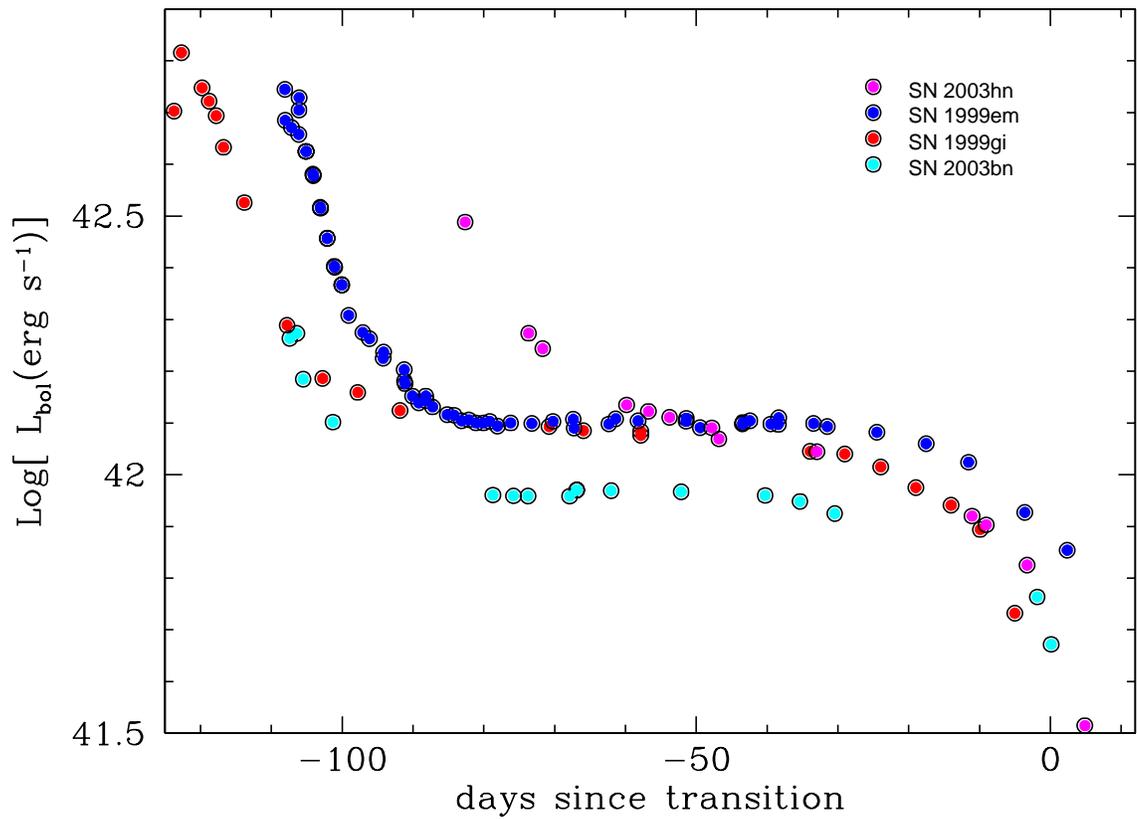}
\caption[Range of variation of the plateau
  lengths]{Bolometric light curves for four SNe in our sample
  showing the range of variation of the plateau
  lengths.}\label{fig:DLp}  
\end{center}
\end{figure}

Figure~\ref{fig:DLp} shows a comparison 
of the light curves of five SNe selected to illustrate the variety of
plateau durations. The range of plateau durations is between 47 and 103
days. However, if we restrict the sample to those SNe with bolometric
LC showing a genuine plateau, the range is reduced to 64 -- 103 days. The
weighted average  value of  the  plateau duration  in our whole
sample is $<\Delta t_p>= 90.47 \pm 1.18$ days.

\subsection{Luminosity drop: $\Delta\log L$}
\label{dlp}
The purpose of this  parameter is to
quantify the luminosity drop between the plateau and the
radioactive phase which is sensitive to the nickel mass and its level
of mixing.  
The calculation of $\Delta\log L$ involves
the use of the plateau luminosity $L_P$ (see~\ref{lp}), 
and an estimation of the tail luminosity ($L_t$) at some specific moment 
during the radioactive tail phase. The luminosity drop is
simply $\Delta\log L= \log L_P -\log L_t$. 

We choose day 20 with respect to $t_{PT}$ \citep{F09} as a convenient moment
for the calculation because at that time
all SNe have fully in the radioactive tail phase. It is
therefore necessary to have at least one data point during the tail
phase. In cases where more than two data points are available, a
straight-line fit to the data is used to provide the value of
$L_t$ and its uncertainty. If only two data points are available, an
interpolation is used. We were able to perform such fits and
interpolations for 20 SNe, which additionally allowed us to determine a
characteristic slope during the tail phase,  $<\mathrm{slope}> =
-0.004$ dex day$^{-1}$ which is in agreement with the slope expected
for  radioactive decay of $^{56}$Co. This slope is 
used to derive 
$L_t$ for those SNe which only have one data point during the
radioactive tail.

\begin{figure}[htbp]
\begin{center}
\includegraphics[scale=0.45]{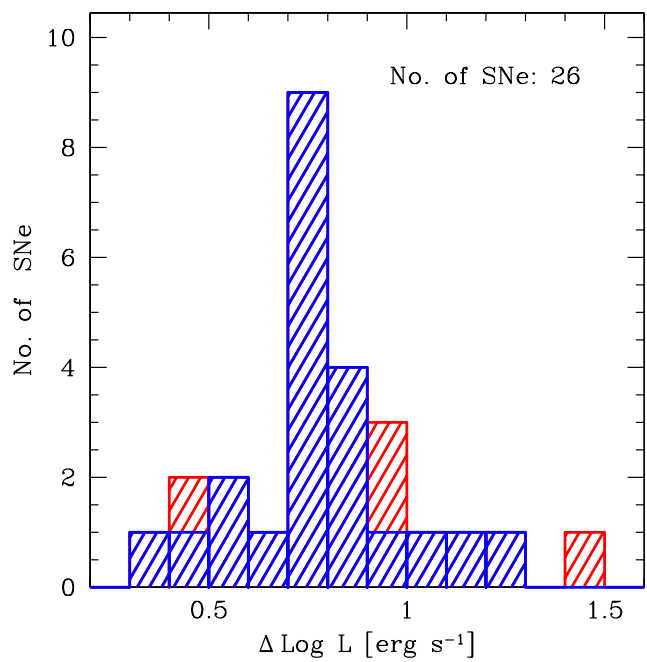}
\caption[Distribution of $\Delta\log L$ for SNe~II-P]{Distribution of
  $\Delta\log L$ for SNe~II-P. The SNe considered as intermediate between  
  plateau and linear (see \S~\ref{sec:sample}) are marked in red.}
 \label{fig:DL_hist}
\end{center}
\end{figure}

The values and uncertainties of $L_t$ are given in column~5 of
Table~\ref{parametros}. Figure~\ref{fig:DL_hist} shows the 
distribution of $\Delta\log L$ with a sharp peak at $\Delta\log L=
0.75$ and a concentration of SNe (18 out of 26) in the range between 0.7 and
1. Note that the  intermediate-plateau SNe have $\Delta\log
  L$ values consistent with the genuine SNe~II-P, with the exception
  of SN~2003cn that shows the largest value of
  $\Delta\log L$ as a consequence of its low estimated $^{56}$Ni mass 
  (see next section). Also note that there is only one object, namely
  SN~2002fa, with $\Delta\log L < 0.4$.  

\begin{figure}[htbp]
\begin{center}
\includegraphics[scale=0.5]{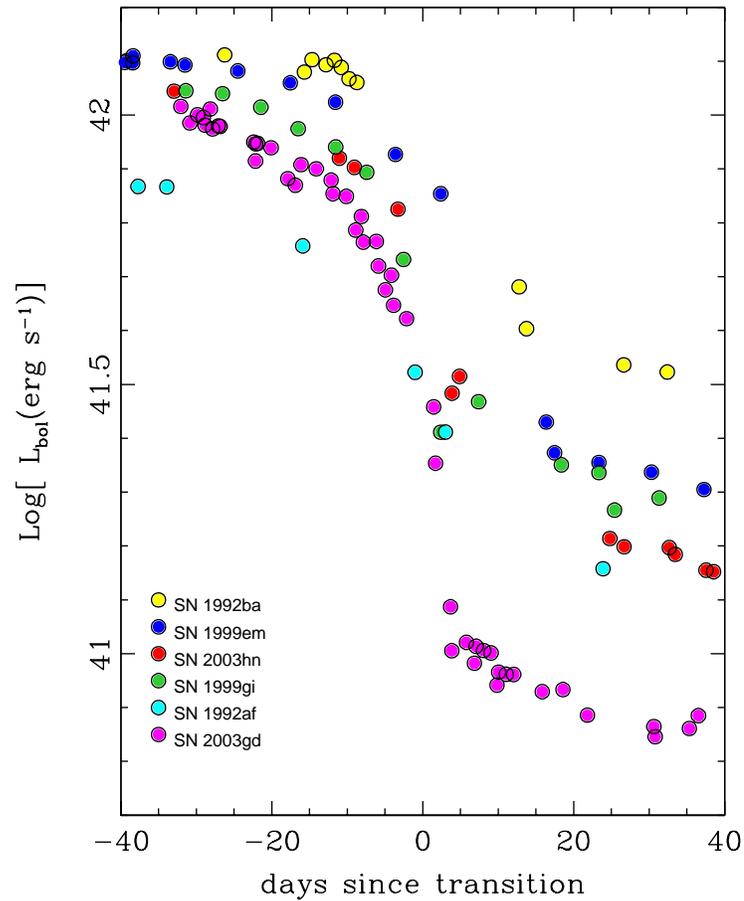}
\caption[Different transition properties]{Bolometric light curves of
  six SNe of the sample with different transition
  properties.}\label{fig:transition}  
\end{center}
\end{figure}

\newpage
In Figure~\ref{fig:transition} we compare the shape of the transition
between the plateau phase and the radioactive tail for SNe having a
well-sampled transition. The behavior of the transition is
related to the $^{56}$Ni mass synthesized in the explosion and its
degree of mixing. More $^{56}$Ni mixing produces a more gradual
transition between the plateau  and the tail
\citep{E94,2007A&A...461..233U}.  The drop for SN~2003gd appears to be
steeper and larger than for the other SNe, which is indicative of
less mixing and a smaller $^{56}$Ni mass. For the whole sample the
luminosity drop is in the range of $0.35$--$1.46$. The weighted
average value of the luminosity drop for the current sample is 
$<\Delta\log L>= 0.783 \pm 0.054$.

\subsection{$^{56}$Ni mass: $M_{\mathrm{Ni}}$}
\label{Mni}
The bolometric luminosity during the radioactive tail phase is a direct
indicator of the mass of $^{56}$Ni synthesized during the explosion if
we assume that all of the $\gamma$-rays resulting from 
$^{56}$Co $\longrightarrow$ $^{56}$Fe are fully
thermalized\footnote{$^{56}$Co is daughter of $^{56}$Ni, which has a
  half-life of only $6.1$ days}. This a reasonable assumption given that
most of the SNe~II-P have a late-time decline rate consistent with
$^{56}$Co $\longrightarrow$ $^{56}$Fe decays (see \S~\ref{dlp}), at least in
the initial 
phase of the nebular era. However, if some $\gamma$-rays escape
before being thermalized, the estimate of $M_{\mathrm{Ni}}$ must be
considered as a lower limit. To calculate $M_{\mathrm{Ni}}$, we
assume that gamma-ray 
deposition (see \S~\ref{sec:gamma}) is equal to 1 in the whole structure
of the object and thus $L_{bol}(t)= M_{\mathrm{Ni}} \, \epsilon_{\rm rad}(t)$
during the tail, with $\epsilon_{\rm rad}(t)$ given by
equation~(\ref{code:erad}) of \S~\ref{sec:gamma}. Therefore,
$M_{\mathrm{Ni}}$ in units of $M_\odot$ is given by

\begin{equation} 
M_{\mathrm{Ni}}[M_\odot] = \frac{L_{bol}(t)} {6.41 \times 10^{43}
  \, \exp \big(\frac{-(t-t_0)}{\tau_{\rm Ni}\,(1 + z)} \big) + 
  1.33 \, \times  10^{43} \,  \exp \big(\frac{-(t-t_0)}
{\tau_{\rm Co}\,( 1 + z)}\big)},
\label{eq:mni} 
\end{equation}

\noindent where $\tau_{\rm Ni} = 8.8$ days, and $\tau_{\rm Co}=113.6$
days are the mean lifetimes of the radioactive isotopes. Note that 
$t_0$ is the explosion time. Based on equation~(\ref{eq:mni})
we can calculate $M_{\mathrm{Ni}}$ for the SNe with known $t_0$ and
tail luminosity. Specifically,  we calculate a value of
$M_{\mathrm{Ni}}$ and its uncertainty for each data point during the
tail. The resulting $M_\mathrm{Ni}$ is the weighted average of these
values. Given that for some SNe we have two estimates of 
$t_0$ ($t_0(E96)$ and $t_0(D05)$ see \S~\ref{sec:time}), two different
values of $M_{\mathrm{Ni}}$ can be calculated. However, in column 6 of
Table~\ref{parametros} we present only one value of
$M_{\mathrm{Ni}}$. This is in general the value corresponding 
to $t_0(E96)$. In cases that this value is not known,
we adopted $t_0(D05)$. And if neither value is known we used estimates
of $t_0$ found in the literature (see \S~\ref{sec:time}). 
We remark that the values of $M_{\mathrm{Ni}}$ calculated using
$t_0(D05)$ are systematically larger but consistent
within the uncertainties with those based on $t_0(E96)$. That is why
we present only one value of $M_{\mathrm{Ni}}$. 

Additionally, we include in our analysis SNe with
$t_+$ (see \S~\ref{dtp}) well determined. Using $t_+$ and 
the characteristic $<\Delta t_+>$ (see \S~\ref{dtp}) we
can estimate the explosion time, $t_0$, and therefore calculate
$M_{\mathrm{Ni}}$ for more SNe in our sample. The uncertainty
associated with this calculation is large because
$M_{\mathrm{Ni}}$ is strongly dependent on the explosion time
and the calculation of the bolometric luminosity (see
  \S~\ref{lp}).

\begin{figure}[htbp]
\begin{center}
\includegraphics[scale=0.45]{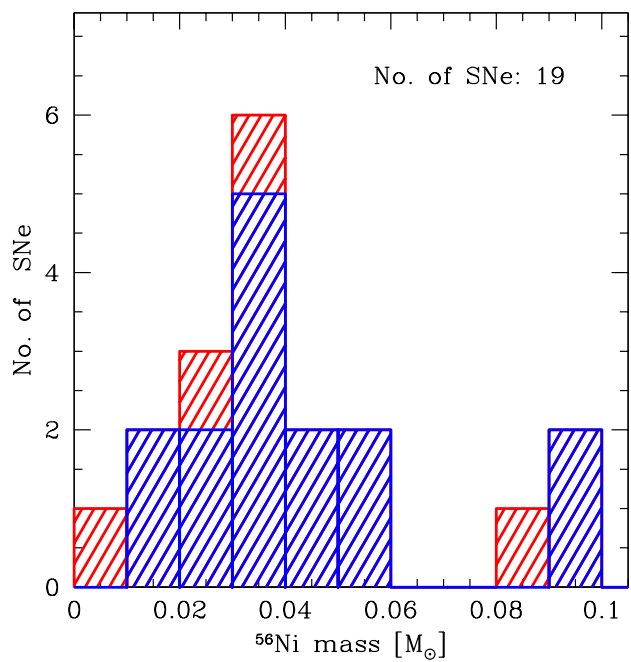}
\caption[Distribution of $^{56}$Ni masses for SNe~II-P]{Distribution
  of $^{56}$Ni masses for SNe~II-P. The SNe 
  considered as intermediate between   
  plateau and linear (see \S~\ref{sec:sample}) are marked in red.}
 \label{fig:Mni_hist}
\end{center}
\end{figure}

Figure~\ref{fig:Mni_hist} shows the distribution of $M_{\mathrm{Ni}}$
for all the SNe for which we can estimate this parameter. We do not
include in this analysis SN~1992am for which we find a
  significantly higher value of the $^{56}$Ni 
  mass as compared with the rest of the sample ($\sim 0.4 M{_\odot}$
  see Table~\ref{parametros}). Considering that there 
  is a wide range of values for the explosion time of this SN  
  in the literature, we checked if this
  high value of $M_{\mathrm{Ni}}$ could be due to an uncertainty in the
  value of $t_0$ assumed. For example, if we use the estimate of $t_0$
  derived from $t_+$ and $<\Delta t_+>$ we find that 
  the $M_{\mathrm{Ni}}$ is reduced to $0.26\pm0.07$
  $M_\odot$. Although smaller, this
  value is still significantly large, which may imply that there are
  SNe~II-P than can produce   
  significant nucleosynthesis of radioactive material. Nevertheless we
  decided to exclude this SNe in the following analysis and leave it
  for a closer scrutiny in the future.
The histogram shows that the SNe in our sample produce less than $0.1
M_\odot$ of $^{56}$Ni, with a peak at $0.035 M_\odot$. There is a
slight tendency to a bi-modal distribution, 
but this should be confirmed with a larger sample of SNe.
With the exception of SN~2003cn that shows the lowest value of 
  $M_\mathrm{Ni}$ in the whole sample, the intermediate-plateau SNe
  have values of $M_{\mathrm{Ni}}< 0.1 M_\odot$ consistent with
  the genuine SNe~II-P. 

\begin{figure}[htbp]
\begin{center}
\includegraphics[scale=0.5]{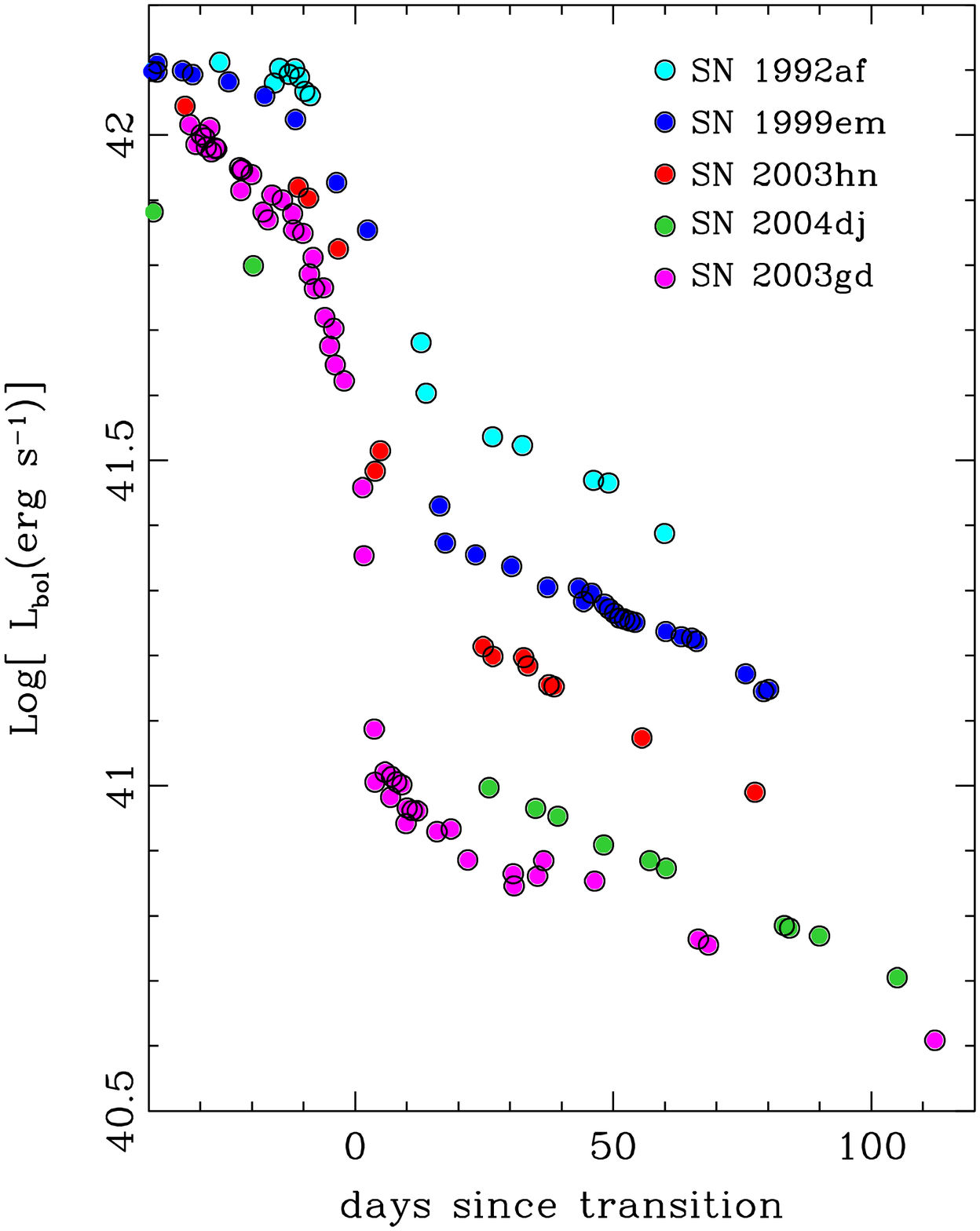}
\caption[Tail luminosity]{Bolometric light curves of
  five SNe of our sample with 
  different tail luminosities.}\label{fig:tail} 
\end{center}
\end{figure}

\noindent In Figure~\ref{fig:tail} we compare the tail luminosity for a subset
of five SNe. Note that the amount of $^{56}$Ni synthesized by the SN
determines the height of the radioactive tail. The weighted average of
$^{56}$Ni masses obtained in our sample of SNe~II-P is
$M_{\mathrm{Ni}}= 0.024\pm0.004 M_\odot$.  

Before proceeding to the next section, we note that with the
  exception of SN~2003cn we did not find
  any peculiarity in the values of the {\bf $L_P$}, {\bf $\Delta \log L$} and
  $M_{\mathrm{Ni}}$  for the intermediate-plateau SNe included in
  the analysis.  This consistency strengthens
  the idea of a connection between linear and plateau subtypes.

\section{Grid of Hydrodynamical Models}
\label{sec:grid}
With the aim of comparing parameters between models and
observations, and of studying correlations between observables and physical
parameters, we calculated a grid of hydrodynamical models for different 
values of the initial mass ($M_0$), initial radius ($R_0$), explosion
energy ($E$), $^{56}$Ni mass ($M_{\mathrm{Ni}}$) and mixing of nickel. Other
parameters like chemical composition and density profile shape were
kept fixed. The ranges of physical parameters used were:
$M_0=$ 10, 15, 20 and 25 $M_\odot$, $E= 0.5$, 1, 2 and 3 foe, $R_0=$ 500,
1000, and 1500 R$_\odot$,   $M_{\mathrm{Ni}}= 0.02$, $0.04$ and $0.07$
$M_\odot$, and mixing of nickel out to fraction of $M_0$ of $0.23$,
$0.5$ and $0.8$.

For each 
model we calculated {\bf $L_P$}, {\bf $\Delta t_P$}, 
and {\bf $\Delta \log L$} using the same definitions as in the 
previous section to perform 
a consistent comparison  between model and observation.  Additionally, we
calculated the photospheric velocity at $-30$ days ({\bf $v_{-30}$})
with respect to $t_{PT}$. 
In Table~\ref{gid:tbl-1} we present
our results  for a total of 46 hydrodynamical models. Although the
table includes all the models, some sets of physical parameters seem
unlikely to happen in nature. For example, 
models with initial masses of 15, 20 and 25 $M_\odot$ and explosion
energy of $0.5$ foe, independently of the values of $R_0$,
$M_{\mathrm{Ni}}$ and nickel mixing, yield values of $\Delta t_P >
110$ days, which is outside the observed range for
our sample of SNe. Thus, we can rule out models and refine the set of physical
parameters that could produce SNe~II-P. 

\begin{figure}
\includegraphics[angle=-90,scale=0.3]{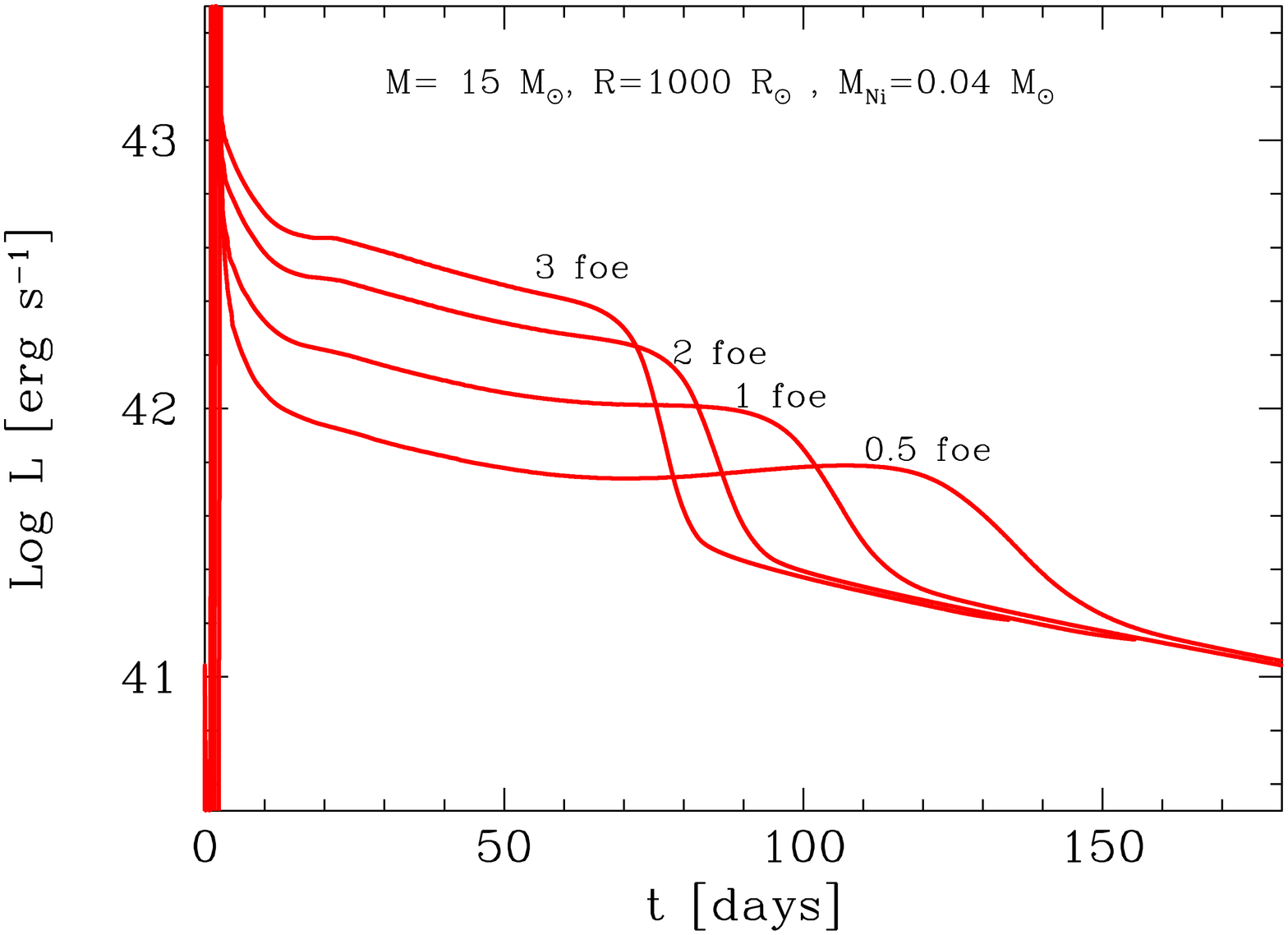}\includegraphics[angle=-90,scale=0.3]{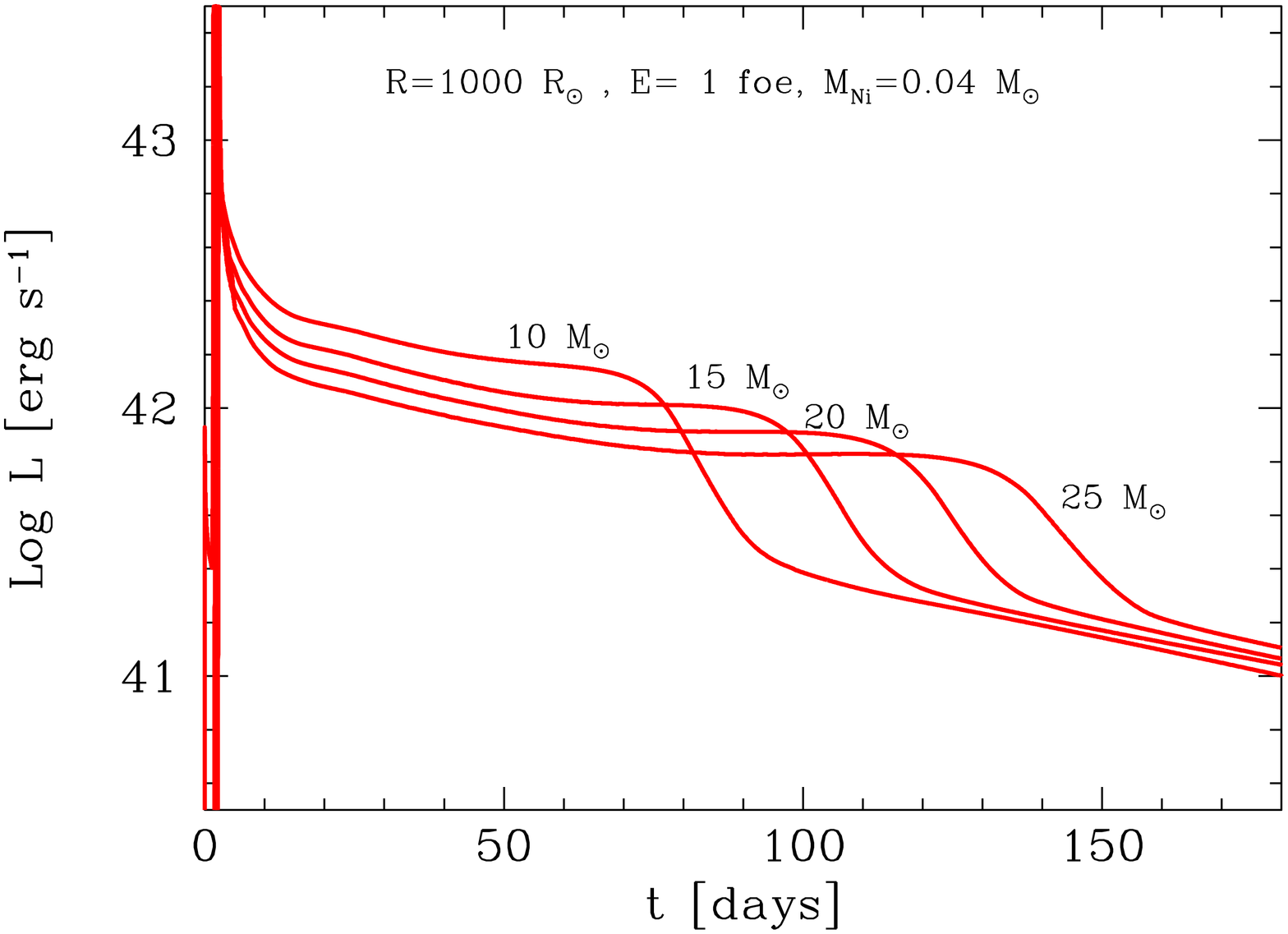}\\
\includegraphics[angle=-90,scale=0.3]{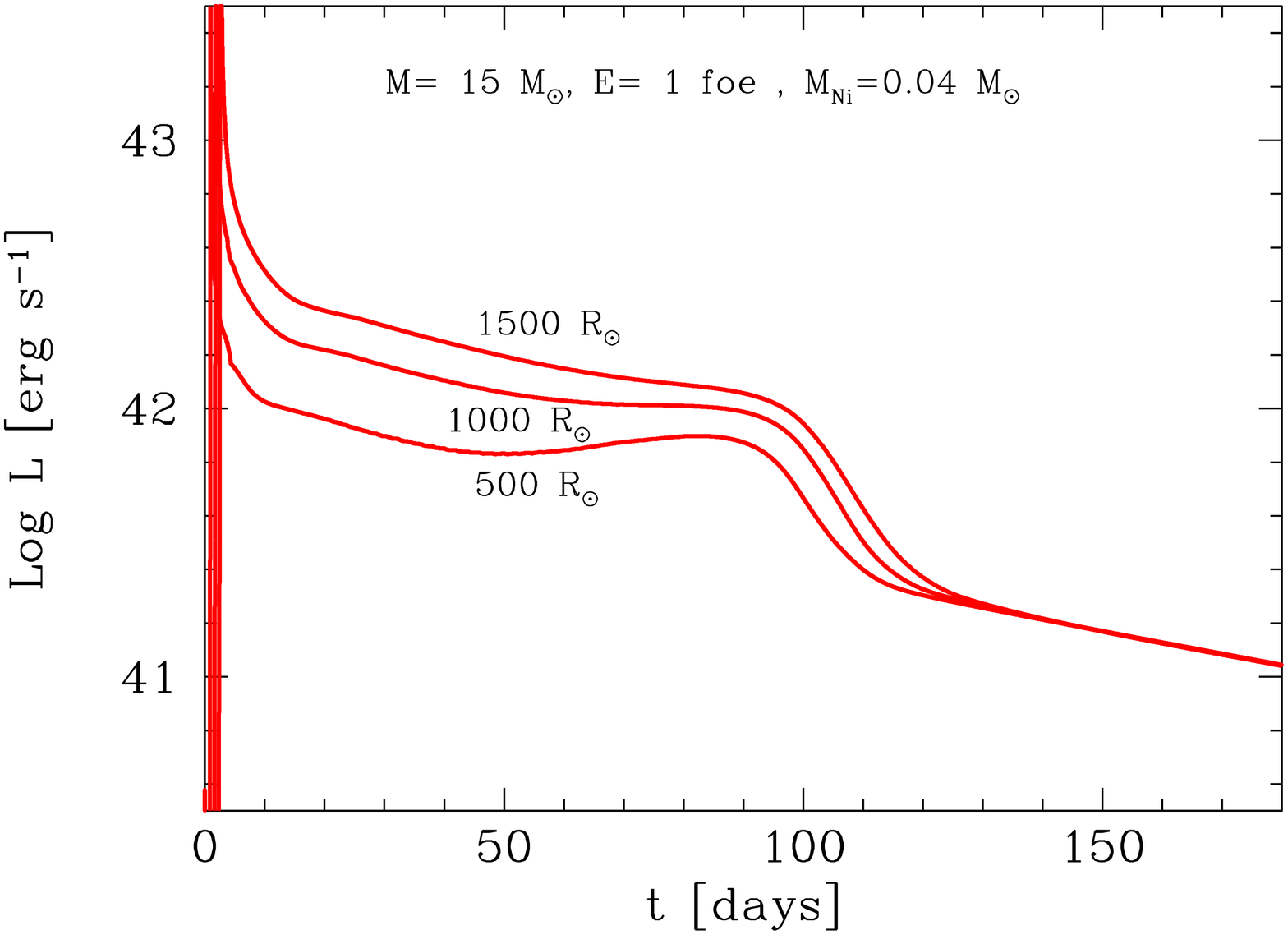}\includegraphics[angle=-90,scale=0.3]{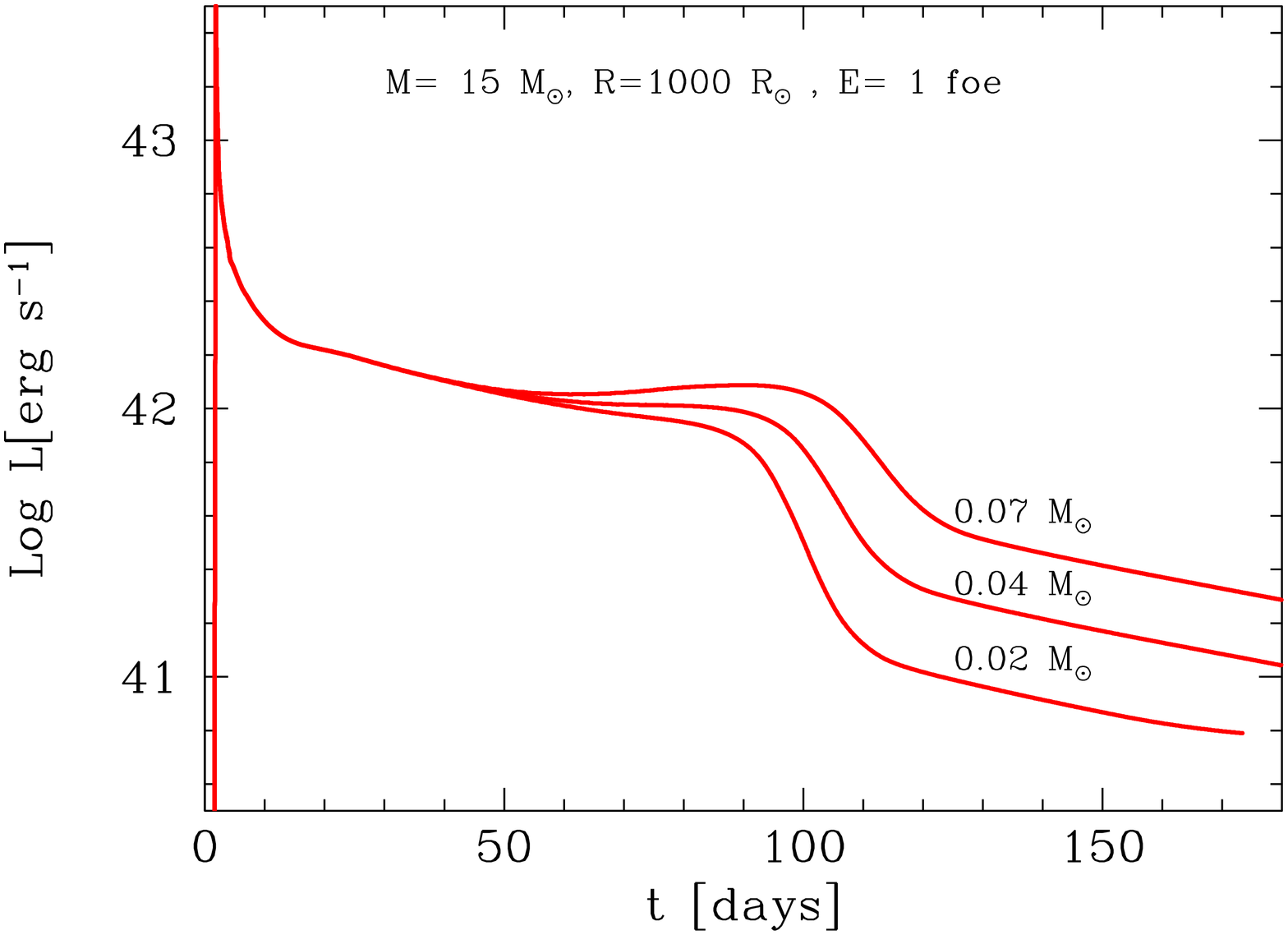}
\caption[Bolometric LC for different physical parameters]{Bolometric
  LC for different {\bf (upper left)} 
  injected energies ($E$), {\bf (lower left)} initial radius ($R_0$),  
    {\bf (upper right)} initial mass ($M_0$), and {\bf (lower right)}
  $^{56}$Ni mass ($M_{\mathrm{Ni}}$). 
   The shape of the LC, being sensitive to the hydrodynamics,
   is a useful tool to infer physical parameter of the SN
   progenitor.}\label{grid:LCV}
\end{figure}

\begin{figure}[htbp]
\begin{center}
\includegraphics[scale=0.5]{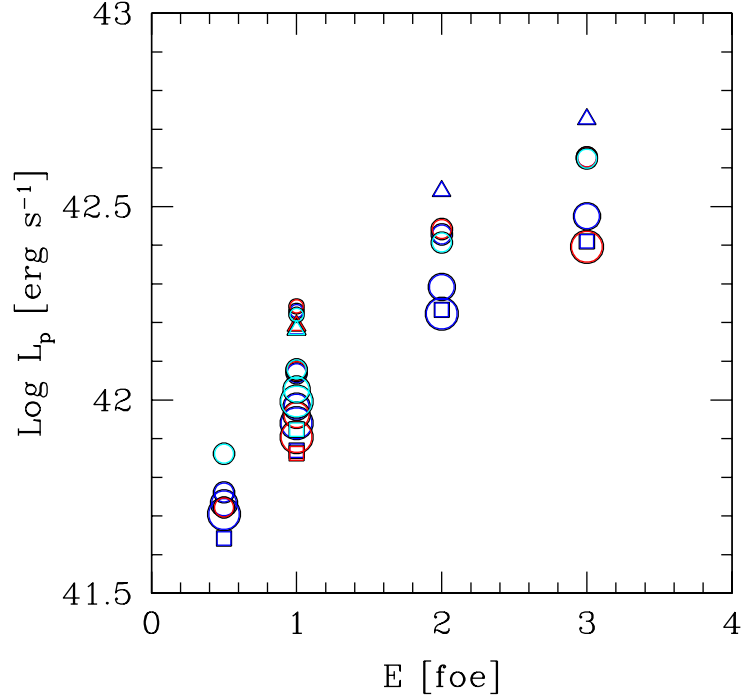}
\caption[Model dependence of $L_P$ on explosion energy]{The
  characteristic plateau luminosity ($L_P$) for all of the
  models presented in Table~\ref{gid:tbl-1} (with $^{56}$Ni mixing
  of 0.5 M$_0$) as a function of  explosion
  energy. The size of the symbols is proportional to the initial mass.
  The shape is related to the initial radius (square symbols for $R_0=
500 R_\odot$, circles for $R_0=1000 R_\odot$, and triangles for $R_0=1500
R_\odot$). The colors indicate different values of $^{56}$Ni mass (red
color for $M_{\mathrm{Ni}}= 0.02 M_\odot$, blue for 
$M_{\mathrm{Ni}}= 0.04 M_\odot$, and cyan for $M_{\mathrm{Ni}}=
0.07 M_\odot$). }\label{LpE}   
\end{center}
\end{figure}

\begin{figure}[htbp]
\begin{center}
\includegraphics[scale=0.5]{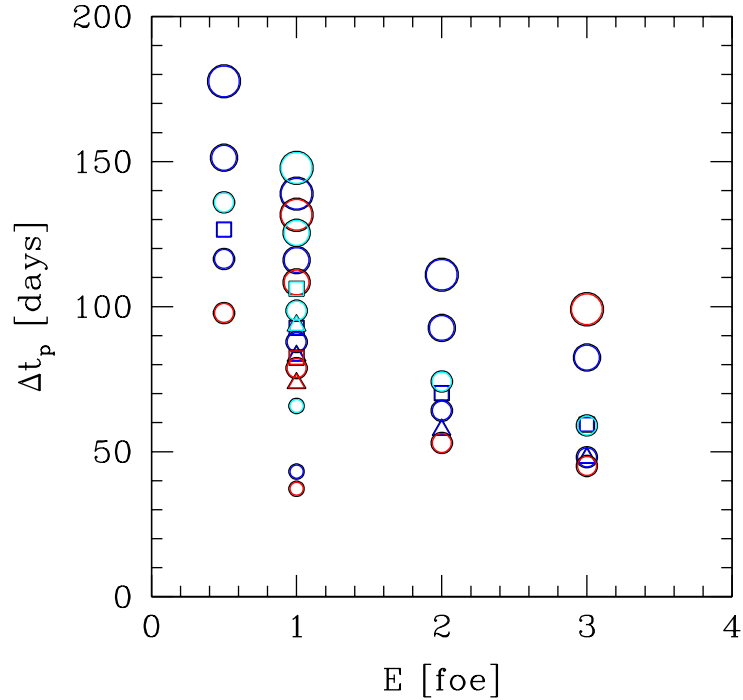}
\caption[Model dependence of $\Delta t_P$  on explosion
  energy]{The plateau duration ({\bf $\Delta t_P$}) for  all of the
  models presented  in Table~\ref{gid:tbl-1} (with $^{56}$Ni mixing
  of 0.5 M$_0$) as a function of explosion
  energy. The size of the symbols is proportional to the initial mass. 
  The shape is related  to the initial radius (square symbols for $R_0=
500 R_\odot$, circles for $R_0=1000 R_\odot$, and triangles for $R_0=1500
R_\odot$). The colors indicate different values of $^{56}$Ni mass (red
color for $M_{\mathrm{Ni}}= 0.02 M_\odot$, blue for
$M_{\mathrm{Ni}}= 0.04 M_\odot$, and cyan for $M_{\mathrm{Ni}}=
0.07 M_\odot$).}
\label{DtpE}  
\end{center}
\end{figure}

\begin{figure}[htbp]
\begin{center}
\includegraphics[scale=0.5]{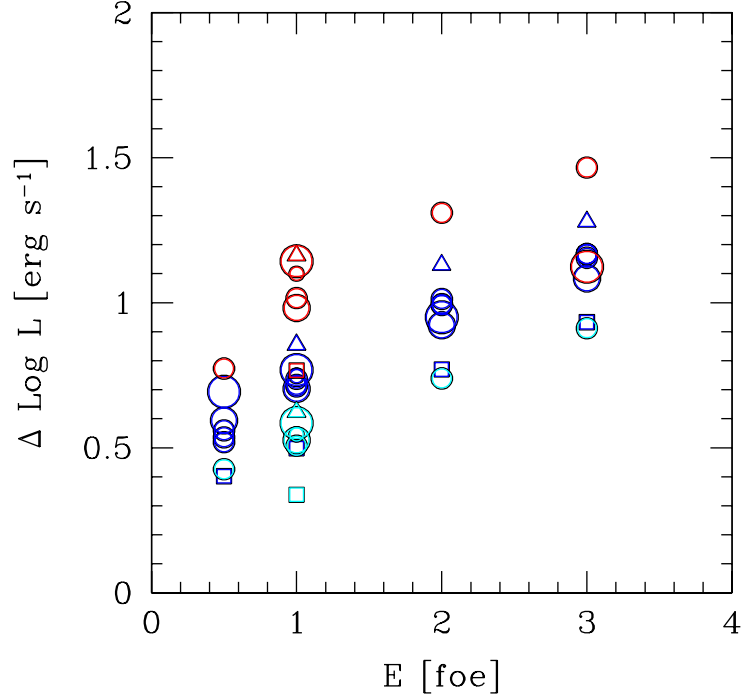}
\caption[Model dependence of $\Delta L$ on explosion
  energy]{The drop in luminosity ({\bf $\Delta L $}) for all of the
  models presented in Table~\ref{gid:tbl-1} (with $^{56}$Ni mixing of
  0.5 M$_0$) as a function of explosion
  energy. The size of the symbols is proportional to the initial mass. 
  The shape is related to the initial radius (square symbols for $R_0=
500 R_\odot$, circles for $R_0=1000 R_\odot$, and triangles for $R_0=1500
R_\odot$). The colors indicate different values of $^{56}$Ni mass
(red color for $M_{\mathrm{Ni}}= 0.02 M_\odot$, blue for
$M_{\mathrm{Ni}}= 0.04 M_\odot$, and cyan for 
$M_{\mathrm{Ni}}=0.07 M_\odot$).}\label{DLE}
\end{center}
\end{figure}

\begin{figure}[htbp]
\begin{center}
\includegraphics[scale=0.5]{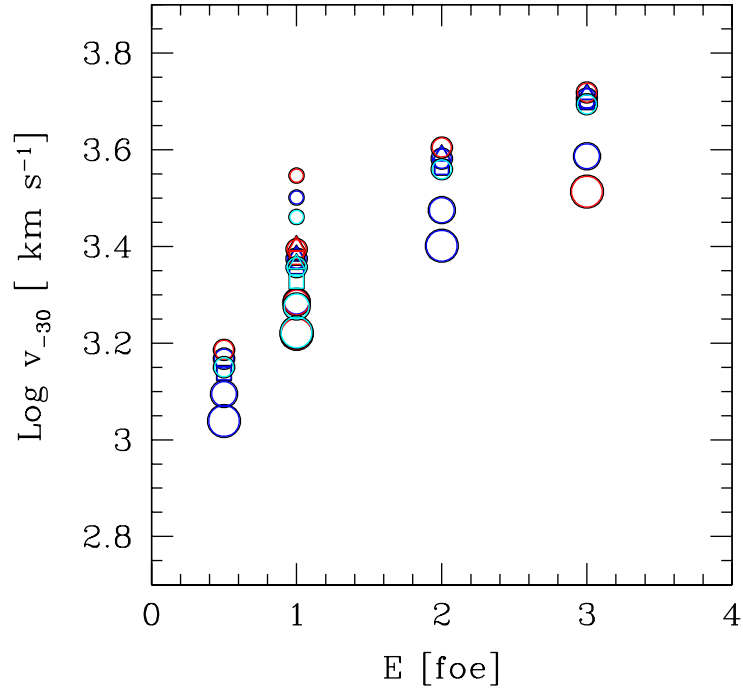} 
\caption[Model dependence of $\mathrm{Log}\, v_{-30}$ on explosion
  energy]{The photospheric velocity at $-30$ days with respect to
  $t_{PT}$ ({\bf $v(-30)$ }) for all of the 
  models presented in Table~\ref{gid:tbl-1} (with  mixing $^{56}$Ni of
  0.5 M$_0$) as a function of the explosion
  energy. The size of the symbols is proportional to the initial mass. 
  The shape is related to the initial radius (square symbols for $R_0=
500 R_\odot$, circles for $R_0=1000 R_\odot$, and triangles for $R_0=1500
R_\odot$). The colors indicate different values of $^{56}$Ni mass
(red color for $M_{\mathrm{Ni}}= 0.02 M_\odot$, blue for
$M_{\mathrm{Ni}}= 0.04 M_\odot$, and cyan for $M_{\mathrm{Ni}}=
0.07 M_\odot$).}
\label{veloE}  
\end{center}
\end{figure}

In Figure~\ref{grid:LCV} we
show the sensitivity of our model light curves to the variation of
physical parameters for some selected models of
Table~\ref{gid:tbl-1}. A glance at this figure indicates that   
(a) increasing the injected energy results
in higher luminosities and shorter plateau durations, (b) more massive
progenitors produce longer plateaus and lower plateau luminosities, (c)
more extended progenitors produce higher plateau luminosities and longer
plateaus, and (d) a greater amount of $^{56}$Ni produces a longer plateau
and a higher tail luminosity. The behavior of our bolometric LC on
variation of physical parameters is consistent with those
shown in previous works \citep[][among
  others]{1983Ap&SS..89...89L,1985SvAL...11..145L,1993ApJ...414..712P}.
Based on the results of our models we study in greater  detail the
dependence of observable  parameters on physical quantities ($E$,
$R_0$, $M_0$ and $M_{\mathrm{Ni}}$). Figures~\ref{LpE}, \ref{DtpE},
\ref{DLE} and 
\ref{veloE} show {\bf $L_P$}, {\bf $\Delta t_P$}, {\bf $\Delta\log L$} 
 and  {\bf $v(-30)$}, respectively, versus explosion
energy. In these figures, the size of the symbols is proportional to
the progenitor mass and the shape 
is related to the initial radius (square symbols for $R_0=
500 R_\odot$, circles for $R_0= 1000 R_\odot$ and triangles
for $R_0= 1500 R_\odot$). The colors indicate different values
of  $^{56}$Ni mass (red color for $M_{\mathrm{Ni}}= 0.02
M_\odot$, blue for  $M_{\mathrm{Ni}}= 0.04 M_\odot$  and cyan for
$M_{\mathrm{Ni}}= 0.07 M_\odot$). In all these 
figures we only show models with a mixing of $^{56}$Ni of $0.5
M_0$. 
\begin{table}[htpp]
\begin{center}

\caption[Parameters of the grid of models]{Model
  Parameters \label{gid:tbl-1}} 
\scriptsize
\vspace{3mm}
\begin{tabular}{lccccccccc}
\hline
\hline
      &  Energy & Mass &  Radius  &  $M_{\mathrm{Ni}}$ &  &$\mathrm{Log} \, L_p$ & $\Delta t_p$  & 
$\Delta \mathrm{Log}\, L$ & $v^{\mathrm{b}}$  \\
 Model  & [foe] &  [$M_\odot$]  &  [$R_\odot$]  & [$M_\odot$]  &
 mixing$^{\mathrm{a}}$ & 
 [erg s$^{-1}$] &  [days] & [erg s$^{-1}$] & [km s$^{-1}$]\\    
\hline

   1  &   0.50  &   15 &  1000 &  0.04 &   0.23  & 41.70  &119.91   & 0.519   & 1213 \\   
   2  &   1.00  &   15 &  1000 &  0.04 &   0.23  & 42.02  & 86.16   & 0.711   & 2093 \\   
   3  &   2.00  &   15 &  1000 &  0.04 &   0.23  & 42.39  & 62.48   & 0.992   & 3531 \\   
   4  &   3.00  &   15 &  1000 &  0.04 &   0.23  & 42.61  & 45.71   & 1.167   & 4809 \\   
   5  &   0.50  &   15 &  1000 &  0.02 &   0.5   & 41.72  & 97.79   & 0.773   & 1534 \\   
   6  &   1.00  &   15 &  1000 &  0.02 &   0.5   & 42.07  & 78.80   & 1.016   & 2479 \\   
   7  &   2.00  &   15 &  1000 &  0.02 &   0.5   & 42.44  & 53.02   & 1.310   & 4023 \\  
   8  &   3.00  &   15 &  1000 &  0.02 &   0.5   & 42.63  & 45.07   & 1.466   & 5226 \\  
   9  &   0.50  &   15 &  1000 &  0.04 &   0.5   & 41.76  &116.43   & 0.536   & 1471 \\  
  10  &   1.00  &   15 &  1000 &  0.04 &   0.5   & 42.07  & 87.83   & 0.717   & 2369 \\  
  11  &   2.00  &   15 &  1000 &  0.04 &   0.5   & 42.43  & 64.13   & 0.995   & 3815  \\ 
  12  &   3.00  &   15 &  1000 &  0.04 &   0.5   & 42.63  & 48.07   & 1.155   & 5080  \\ 
  13  &   0.50  &   15 &  1000 &  0.07 &   0.5   & 41.86  &135.94   & 0.426   & 1413  \\  
  14  &   1.00  &   15 &  1000 &  0.07 &   0.5   & 42.08  & 98.59   & 0.507   & 2276  \\ 
  15  &   2.00  &   15 &  1000 &  0.07 &   0.5   & 42.41  & 74.19   & 0.739   & 3628  \\ 
  16  &   3.00  &   15 &  1000 &  0.07 &   0.5   & 42.62  & 59.03   & 0.912   & 4943  \\ 
  17  &   0.50  &   15 &  1000 &  0.04 &   0.8   & 41.82  &112.78   & 0.560   & 1565  \\ 
  18  &   1.00  &   15 &  1000 &  0.04 &   0.8   & 42.11  & 84.85   & 0.738   & 2543  \\  
  19  &   2.00  &   15 &  1000 &  0.04 &   0.8   & 42.46  & 60.91   & 1.013   & 4026  \\  
  20  &   3.00  &   15 &  1000 &  0.04 &   0.8   & 42.64  & 49.89   & 1.168   & 5384  \\  
  21  &   0.50  &   20 &  1000 &  0.04 &   0.5   & 41.73  &151.32   & 0.594   & 1244  \\  
  22  &   0.50  &   25 &  1000 &  0.04 &   0.5   & 41.70  &177.66   & 0.693   & 1093  \\  
  23  &   1.00  &   10 &  1000 &  0.04 &   0.5   & 42.23  & 43.13   & 0.747   & 3170  \\ 
  24  &   1.00  &   20 &  1000 &  0.04 &   0.5   & 41.98  &116.04   & 0.703   & 1930  \\ 
  25  &   1.00  &   25 &  1000 &  0.04 &   0.5   & 41.94  &138.93   & 0.768   & 1660  \\
  26  &   2.00  &   20 &  1000 &  0.04 &   0.5   & 42.29  & 92.64   & 0.922   & 2985  \\  
  27  &   2.00  &   25 &  1000 &  0.04 &   0.5   & 42.22  &110.99   & 0.950   & 2519  \\ 
  28  &   3.00  &   20 &  1000 &  0.04 &   0.5   & 42.47  & 82.43   & 1.084   & 3860  \\ 
  29  &   3.00  &   25 &  1000 &  0.02 &   0.5   & 42.39  & 99.09   & 1.124   & 3260  \\ 
  30  &   1.00  &   10 &  1000 &  0.02 &   0.5   & 42.24  & 37.21   & 1.100   & 3519  \\  
  31  &   1.00  &   20 &  1000 &  0.02 &   0.5   & 41.96  &108.36   & 0.982   & 1923   \\ 
  32  &   1.00  &   25 &  1000 &  0.02 &   0.5   & 41.90  &131.66   & 1.143   & 1655   \\ 
  33  &   1.00  &   10 &  1000 &  0.07 &   0.5   & 42.22  & 65.77   & 0.546   & 2889   \\ 
  34  &   1.00  &   20 &  1000 &  0.07 &   0.5   & 42.03  &125.44   & 0.526   & 1887   \\ 
  35  &   1.00  &   25 &  1000 &  0.07 &   0.5   & 41.10  &147.83   & 0.585   & 1668   \\ 
  36  &   0.50  &   15 &   500 &  0.04 &   0.5   & 41.64  &126.58   & 0.402   & 1373   \\  
  37  &   1.00  &   15 &   500 &  0.04 &   0.5   & 41.87  & 92.70   & 0.498   & 2280   \\
  38  &   1.00  &   15 &  1500 &  0.04 &   0.5   & 42.18  & 83.05   & 0.855   & 2404   \\ 
  39  &   2.00  &   15 &   500 &  0.04 &   0.5   & 42.23  & 70.20   & 0.770   & 3656    \\
  40  &   2.00  &   15 &  1500 &  0.04 &   0.5   & 42.54  & 57.53   & 1.130   & 3877    \\
  41  &   3.00  &   15 &   500 &  0.04 &   0.5   & 42.41  & 59.36   & 0.933   & 4997    \\
  42  &   3.00  &   15 &  1500 &  0.04 &   0.5   & 42.73  & 47.90   & 1.279   & 5180    \\
  43  &   1.00  &   15 &   500 &  0.02 &   0.5   & 41.86  & 82.41   & 0.766   & 2381    \\
  44  &   1.00  &   15 &  1500 &  0.02 &   0.5   & 42.19  & 73.60   & 1.163   & 2516    \\
  45  &   1.00  &   15 &   500 &  0.07 &   0.5   & 41.92  &106.16   & 0.338   & 2124    \\
  46  &   1.00  &   15 &  1500 &  0.07 &   0.5   & 42.18  & 93.49   & 0.623   & 2309     \\
\hline
\end{tabular}
\begin{list}{}{}
\item[$^{\mathrm{a}}$] The degree of $^{56}$Ni mixing is given as a
    fraction of the initial mass of the model ($M_0$)
\item[$^{\mathrm{b}}$] $v$ is the photospheric velocity at $-30$ days
  with respect to
  the transition between the plateau and radioactive tail phases.
\end{list}
\end{center}
\end{table}

From Figure~\ref{LpE} it is clear that there is a strong 
correlation between  plateau luminosity and explosion
energy. Larger explosion energies 
produce higher plateau luminosities (see also upper-right panel of
Figure~\ref{grid:LCV}). However, there is 
significant dispersion in this relation ($\sim$0.4 dex). Note that for a fixed
value of explosion energy, more massive progenitors produce lower plateau
luminosities and smaller initial radius also produce  
lower values of $L_p$. There is an additional contribution to the
dispersion due to the $^{56}$Ni mass, although it is lower than for
the other parameters.  

The plateau duration also correlates with  explosion energy (see
Figure~\ref{DtpE}) but in this case the dependence is wearer
than in the case of the plateau luminosity. The
progenitor mass appears to be the most important factor 
affecting $\Delta t_P$. Note, for example, that almost
the full range of variation of $\Delta t_P$ is covered for fixed $E=
1$ foe, only by changing the initial mass between 10-25
M$_\odot$. Neverthe less the effect of  $^{56}$Ni mass is not
negligible. Increasing the $^{56}$Ni mass the plateau duration becomes
noticeably larger.  The effect of $^{56}$Ni mass seems to depend
on energy, being  greater  for smaller values of 
explosion energy. The initial radius produces only a minor effect on
$\Delta t_P$. Another  point is that we found several
models with $\Delta t_P < 75$ days, interestingly
 these models tend  to a  show a linear behavior 
consistently with the observations.

From Figure~\ref{DLE} we see that the post-plateau drop
$\Delta L$ is slightly dependent on $E$. The $^{56}$Ni mass appears as
the main factor in the determination of this parameter, as expected
from the large effect of $M_{\mathrm{Ni}}$ on the tail luminosity. 
Lower values of $M_{\mathrm{Ni}}$ lead to reduced brightness in the
tail and therefore to higher values of $\Delta L$, and vice versa. Also note that
$\Delta L$ is not very sensitive to the initial mass but there is some 
dependence on the initial radius.

In Figure~\ref{veloE} we plot the photospheric velocity at $-30$ days
with respect to $t_{PT}$ versus explosion energy. The selection of this
particular epoch to evaluate the photospheric velocity was based on
the work of \citet{F09} which calibrate the standard-candle method 
for estimating  distances using expansion velocities at this epoch.
We found a strong dependence of $v_{-30}$ on the explosion
energy, as expected from the fact that the explosion energy is almost
completely converted into kinetic energy a few days after the
explosion (see \S~\ref{sec:SW}). Also note that the value of $v_{-30}$ 
depends on the progenitor mass in the sense that more massive
progenitors produce lower velocities for a fixed energy. There is also
some dependence on  $^{56}$Ni at least for lower values of the
progenitor mass but we do not find a dependence on initial radius.

The analysis presented above is a qualitative demonstration of the
correlations among  
observable and physical parameters
based on a limited amount of models. Further work is planned to
enlarge the set of models in order to provide quantitative relations.

\section{Observed and Modeled Correlations}
\label{sec:correlatios}
In this section, we use the parameters {\bf $L_P$}, {\bf $\Delta t_P$}, 
{\bf $\Delta\,\mathrm{Log}\,L$}, and {\bf $M_{\mathrm{Ni}}$},
calculated both from the observations of our
SN~II-P sample (see Table~\ref{parametros}) and from our grid of
hydrodynamical models (see Table~\ref{gid:tbl-1}) to analyze correlations
among them. We also include in this analysis the
expansion velocity measured from the Fe~II~$\lambda$5169 line at
$-30$ days with respect to $t_{PT}$ as calculated by \citet{F09} (see Table~4.1
of that work). 

\begin{figure}[htbp]
\begin{center}
\includegraphics[scale=0.5]{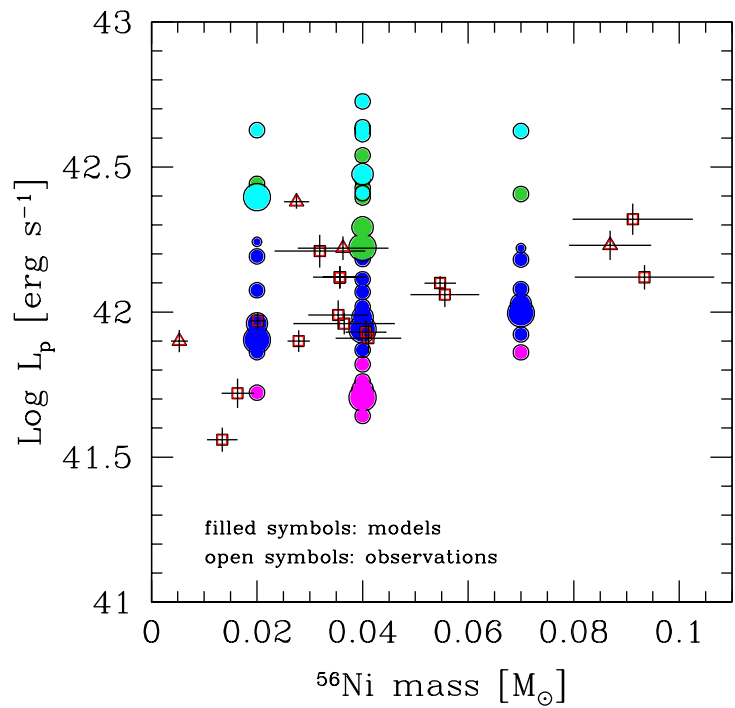}
\caption[Plateau luminosity vs. $^{56}$Ni mass]{The characteristic
  plateau luminosity as a function of $^{56}$Ni mass for models (filled
  symbols) and observations (open symbols). The size of the symbols is
  proportional to the initial mass. Colors indicate the
  explosion energy: magenta for $E= 0.5$ foe,  blue for $E= 1$ foe, green for
   $E=$ 2 foe and cyan for $E= 3$ foe. The
    intermediate-plateau SNe are shown with triangles while 
    squares are used for genuine plateau SNe.}
\label{LpMNi}  
\end{center}
\end{figure}

\begin{figure}[htbp]
\begin{center}
\includegraphics[scale=0.5]{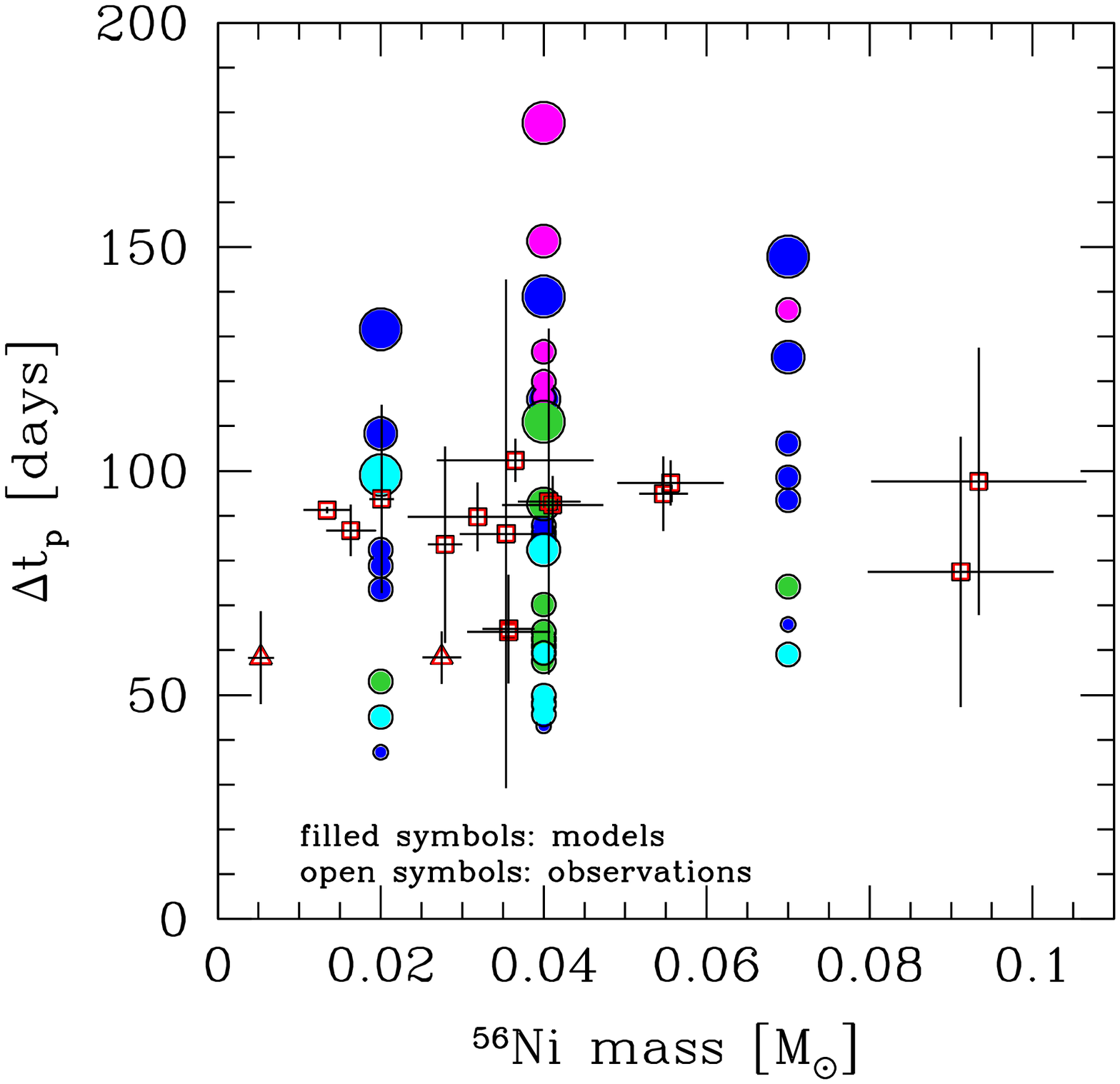}
\caption[Plateau duration vs. $^{56}$Ni mass]{The plateau duration
  as a function of $^{56}$Ni mass for models (filled 
  symbols) and observations (open symbols). The size of the symbols is
  proportional to the initial mass. Colors indicate the
  explosion energy: magenta for $E= 0.5$ foe,  blue for $E= 1$ foe, green for
   $E=$ 2 foe and cyan for $E= 3$ foe. The
    intermediate-plateau SNe are shown with triangles while 
    squares are used for genuine plateau SNe.}
\label{DtpMNi}  
\end{center}
\end{figure}

\begin{figure}[htbp]
\begin{center}
\includegraphics[scale=0.5]{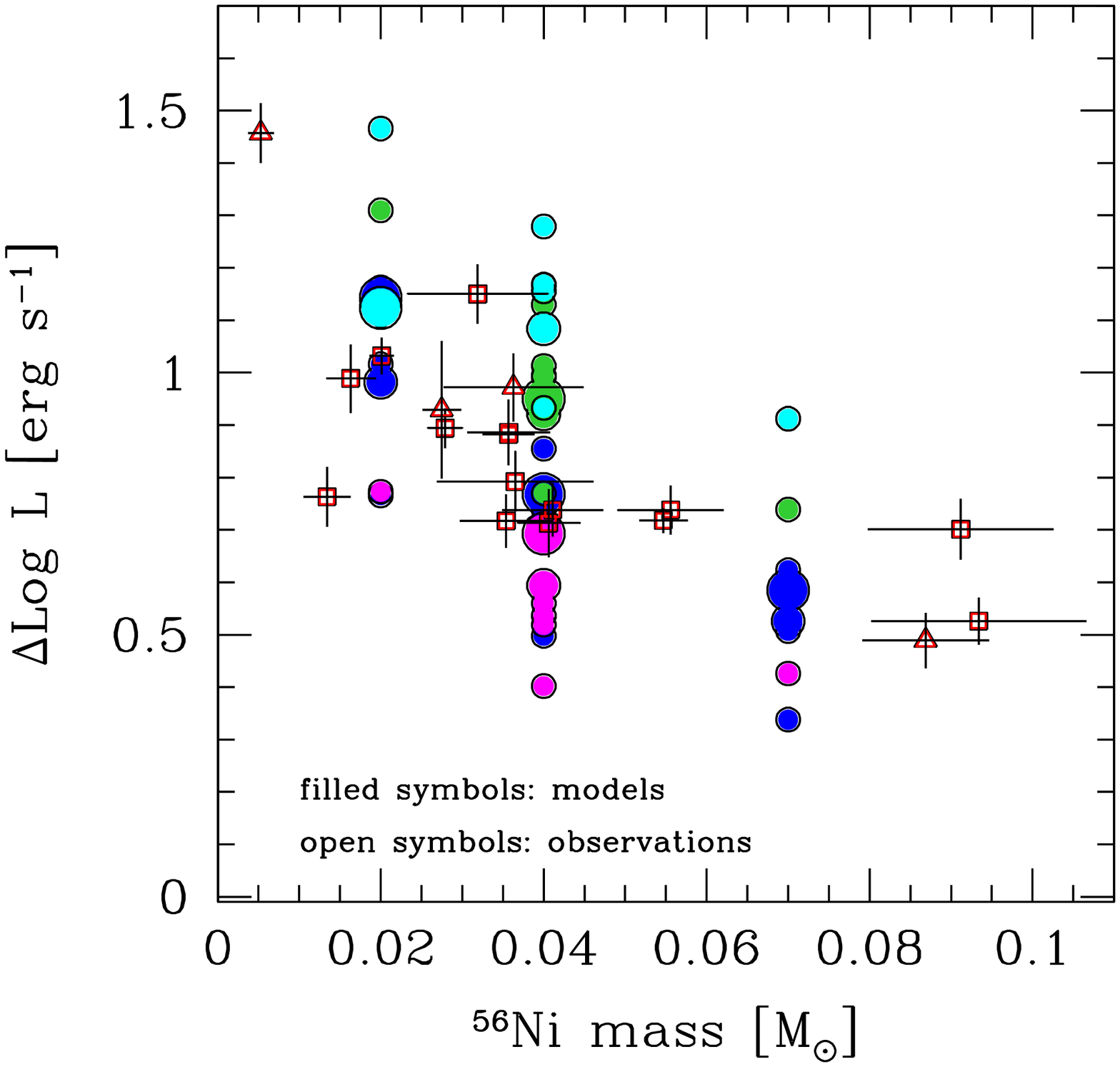}
\caption[The drop in luminosity vs. $^{56}$Ni mass]{The drop in luminosity 
  as a function of $^{56}$Ni mass for models (filled 
  symbols) and observations (open symbols). The size of the symbols  is
  proportional to the initial mass. Colors indicate the
  explosion energy: magenta for $E= 0.5$ foe,  blue for $E= 1$ foe, green for
   $E=$ 2 foe and cyan for $E= 3$ foe. The
    intermediate-plateau SNe are shown with triangles while
    squares are used for genuine plateau SNe.}
\label{DLMNi}  
\end{center}
\end{figure}

\begin{figure}[htbp]
\begin{center}
\includegraphics[scale=0.5]{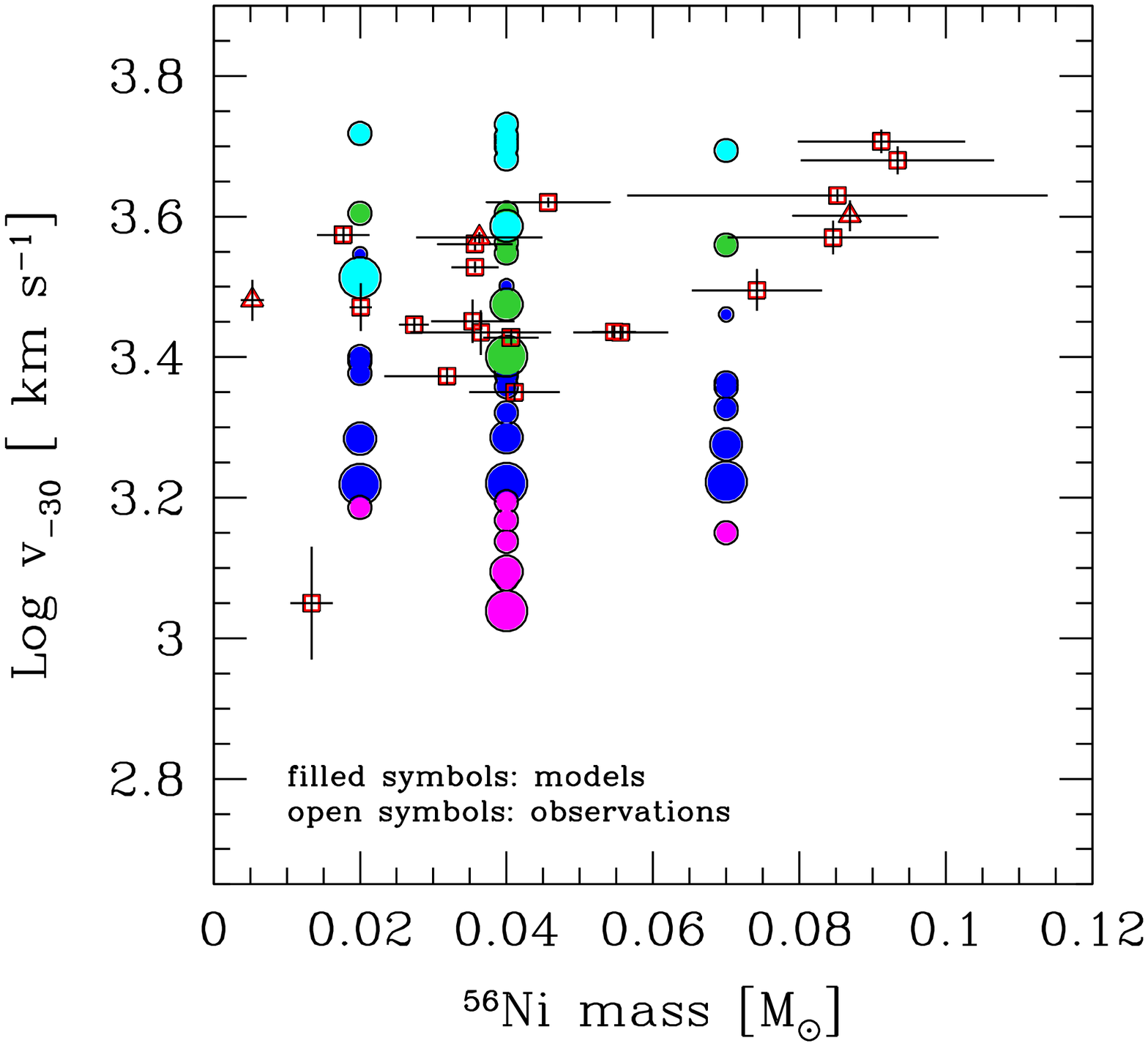}
\caption[The photosperic velocity vs. $^{56}$Ni mass]{The
  photospheric velocity at $-30$ days with respect to $t_{PT}$
  as a function of $^{56}$Ni mass for models (filled 
  symbols) and observations (open symbols). The size of the symbols  is
  proportional to the initial mass. Colors indicate the
  explosion energy: magenta for $E= 0.5$ foe,  blue for $E= 1$ foe, green for
   $E=$ 2 foe and cyan for $E= 3$ foe. The
    intermediate-plateau SNe are shown with triangles while 
    squares are used for genuine plateau SNe.}
\label{vpMNi}  
\end{center}
\end{figure}

We begin by analyzing how   
{\bf $M_{\mathrm{Ni}}$} is related to other observable
parameters. Figures~\ref{LpMNi},  
\ref{DtpMNi}, \ref{DLMNi} and \ref{vpMNi} show {\bf $L_P$},
{\bf$\Delta t_P$},  {\bf $\Delta\,\mathrm{Log}\,L$} and
{\bf $v_{-30}$}, respectively, as a function of the $^{56}$Ni mass. In these
figures, filled symbols represent the models and open 
symbols show the observations. Different colors are used for
identifying models with different explosion energies (magenta color
for $E= 0.5$ foe, blue for $E= 1$ foe, green for $E= 2$ foe, and cyan
for $E= 3$ foe), and the size of the
model symbols is proportional to the progenitor mass. We use
. The intermediate-plateau SNe are
shown with a different symbol (triangles) than genuine plateau SNe
(squares). 

\citet{2003ApJ...582..905H}
found a correlation between plateau luminosity and $^{56}$Ni mass
in the sense that more luminous supernovae produce more
$^{56}$Ni. The observational data in Figure~\ref{LpMNi} shows a slight
trend in agreement with such 
correlation, although less pronounced than the one found by
\citet{2003ApJ...582..905H}. The models show a range of plateau
luminosities for any specific value of $^{56}$Ni mass which is mainly
caused  by the explosion energy. It is
important to note that the 
$^{56}$Ni mass is introduced as a free parameter in our model while it
is actually determined by the nucleosynthesis produced during the 
 explosion and by how much $^{56}$Ni falls back on the
 compact remnant. In any case, we see that the observations are consistent
 with models of explosion energy $\sim$1 foe and different
 progenitors properties (mass an radius). 

As shown in Figure~\ref{DtpMNi}, the data show no correlation between
{\bf$\Delta t_P$} and $^{56}$Ni mass. Thus is consistent with the
models which show only a mild dependence of the plateau
duration with $^{56}$Ni mass.  

Figure~\ref{DLMNi} shows that
 there is a strong correlation between {\bf $\Delta\,\mathrm{Log}\,L$}
 and $^{56}$Ni mass, both in the models and observations. The
 lower the of $^{56}$Ni mass the greater the  luminosity drop. 

Finally,
 Figure~\ref{vpMNi} compares  $^{56}$Ni mass with
 {\bf$v_{-30}$}. The observations suggest a trend where faster SNe produce
 larger amounts of $^{56}$Ni, as was previously pointed out by
 \citet{2003ApJ...582..905H}. Since {\bf$v_{-30}$} is closely related
 with the kinetic energy which is a considerable percentage of the
 explosion energy, this correlation suggests that SNe with higher 
 explosion energy produce more nuclear burning or suffer less fall-back
 of the reprocessed material. From the comparison between
 observations and models, we see that explosion energies around $0.5$
 foe occur, but are not realized very often in Nature.    

\begin{figure}[htbp]
\begin{center}
\includegraphics[scale=0.5]{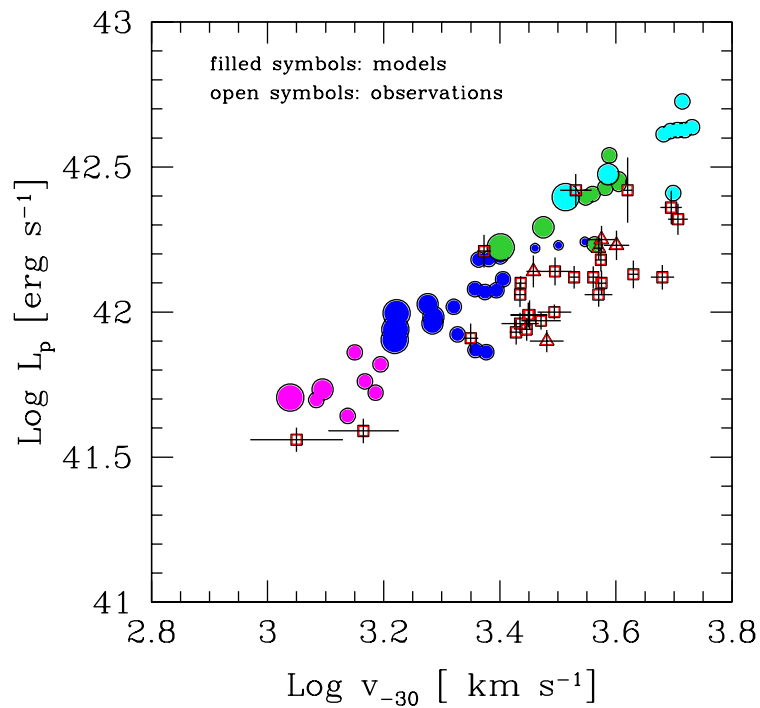}
\caption[Plateau luminosity vs. photosperic velocity]{The characteristic
  plateau luminosity as a function of $v_{-30}$ for models (filled
  symbols) and observations (open symbols). The size of the symbols is
  proportional to the initial mass. Colors indicate the
  explosion energy: magenta for $E= 0.5$ foe,  blue for $E= 1$ foe, green for
   $E=$ 2 foe and cyan for $E= 3$ foe. The
    intermediate-plateau SNe are shown with triangles while 
    squares are used for genuine plateau SNe.}
\label{SCM}  
\end{center}
\end{figure}

\begin{figure}[htbp]
\begin{center}
\includegraphics[scale=0.5]{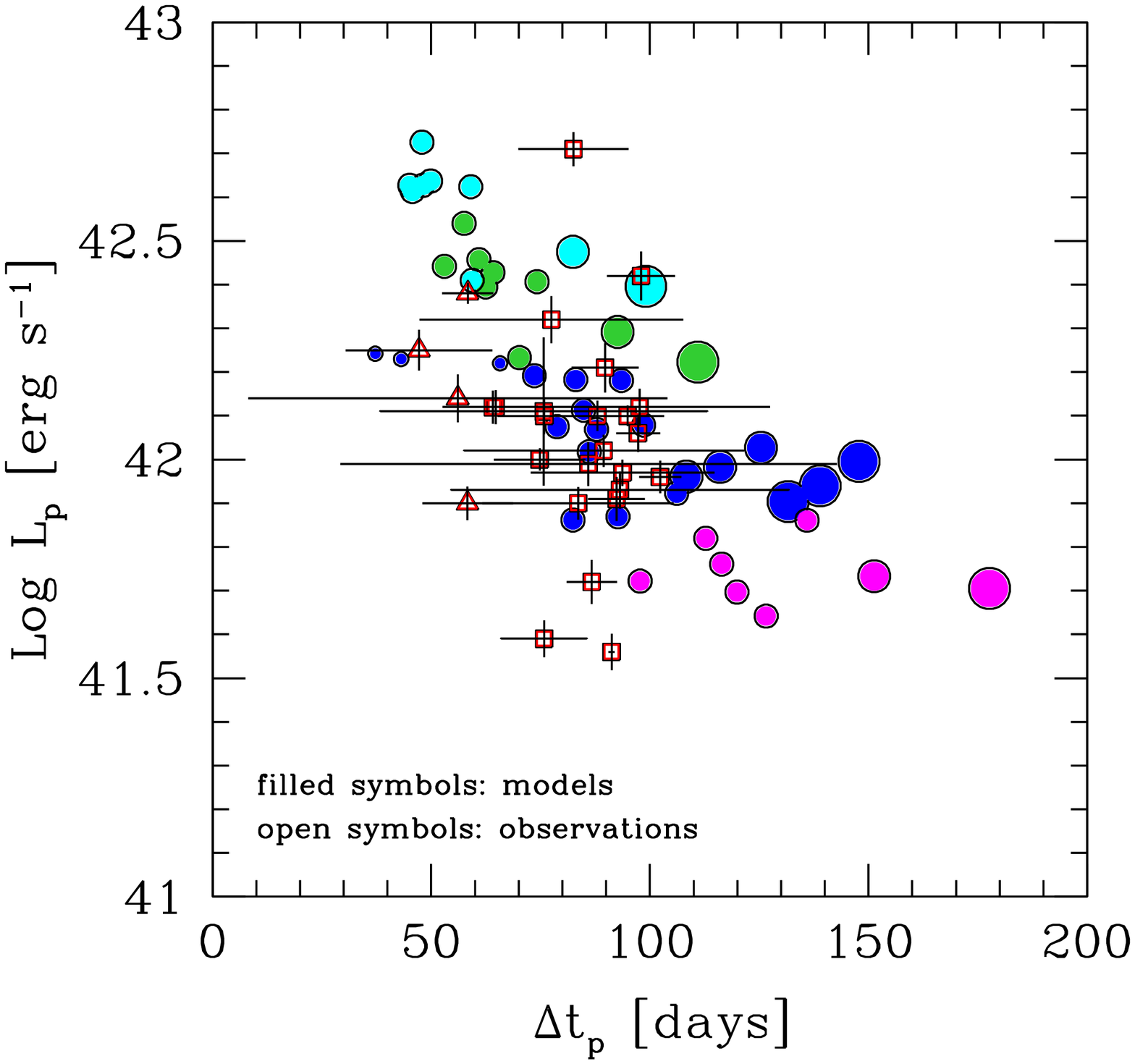}
\caption[Plateau luminosity vs. plateau duration]{The characteristic
  plateau luminosity as a function of plateau duration for models (filled
  symbols) and observations (open symbols). The size of the symbols is
  proportional to the initial mass. Colors indicate the
  explosion energy: magenta for $E= 0.5$ foe,  blue for $E= 1$ foe, green for
   $E=$ 2 foe and cyan for $E= 3$ foe. The
    intermediate-plateau SNe are shown with triangles while 
    squares are used for genuine plateau SNe.}
\label{LpDtp}  
\end{center}
\end{figure}

\begin{figure}[htbp]
\begin{center}
\includegraphics[scale=0.5]{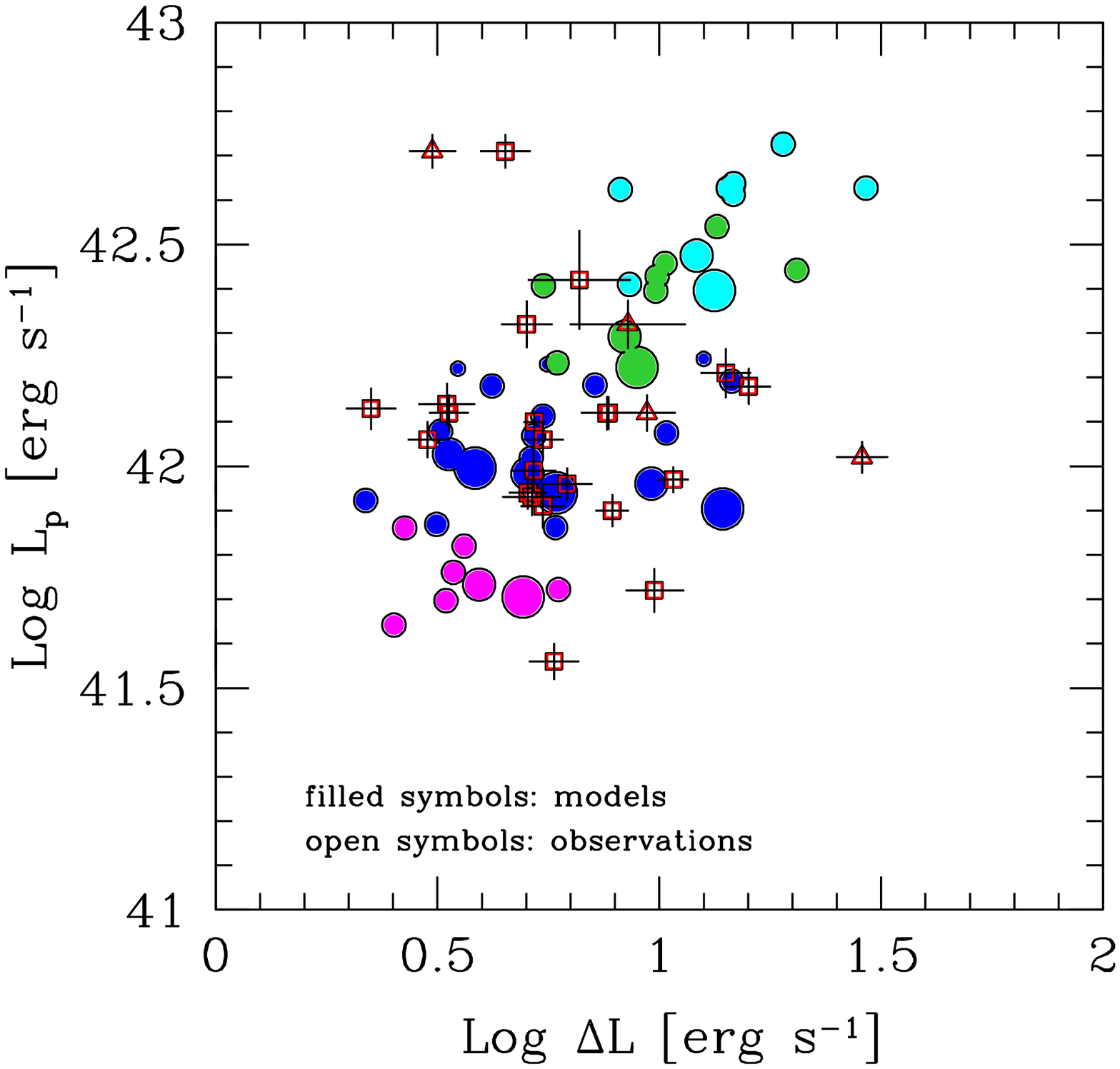}
\caption[Plateau luminosity vs. the drop in luminosity]{The characteristic
  plateau luminosity as a function of the drop in luminosity for models (filled
  symbols) and observations (open symbols). The size of the symbols is
  proportional to the initial mass. Colors indicate the
  explosion energy: magenta for $E= 0.5$ foe,  blue for $E= 1$ foe, green for
   $E=$ 2 foe and cyan for $E= 3$ foe. The
    intermediate-plateau SNe are shown with triangles while 
    squares are used for genuine plateau SNe.}
\label{LpDL}  
\end{center}
\end{figure}

Given that the luminosity is an important parameter in the
determination of distances, we examine the dependence of this
parameter on other 
observables. In Figure~\ref{SCM}, \ref{LpDtp}, and  \ref{LpDL} we  show
its dependence on {\bf$v_{-30}$}, {\bf$\Delta t_P$}, and {\bf
  $\Delta\,\mathrm{Log}\,L$}, respectively. The symbol shapes and
colors have the same meaning as in the previous figures. 

It is known
that there is a strong correlation between 
luminosity and the expansion velocity during the plateau
phase \citep{HP02}. For our sample of SNe this correlation was studied
in detail by \citet{F09} using $-30$ days with respect to $t_{PT}$ as
the epoch where the correlation was calibrated. Figure~\ref{SCM} shows
this correlation for our models and the observational data. From the figure we
see that the models reproduce 
very well the trend of  the 
correlation, and we can identify the explosion energy as the main
driving parameter. In addition, the models show that the dispersion is
related to the progenitor properties. For example, more massive 
progenitors produce lower expansion velocities for approximately the same
luminosity. Although the trend of  the correlation is well 
reproduced by the models we can 
 see that there is a shift in the relationship between
the models and the observations. One reason for the shift
could  be related to luminosity 
 calibration used by \citet{F09} to estimate distances for the SNe
in our sample. The zero point is based on two object, SN~1999em and
SN~2004dj, with known Cepheids distance. These two SNe show a
difference of $\sim$$0.9$~mag in their SCM-calibrated absolute
magnitude, which could lead 
to a shift of $\sim$$0.2$ dex 
in luminosity. Other reason for the discrepancy may
be that we are using photospheric velocities from the models and line
velocities of Fe~II~$\lambda$5169 ($v_{\mathrm{FeII}}$) from the
observations. The right way to 
do the comparison would be to convert $v_{\mathrm{FeII}}$ into
a photospheric velocity using the calibration based on atmospheric
models given by \citet{2009ApJ...696.1176J}. The problem is that this
calibration is only valid for $v \gtrsim 5000$ km s$^{-1}$ (see
\S~\ref{sec:vphoto}) and therefore  
is out of range for most of our data. Another possibility is that our
set of  models does not cover the appropriate range of physical
parameters. In this sense, we note that models with lower
values of the initial mass are more consistent with the
observations. We will explore these issues in greater detail in a future
research.

From the models  shown in Figure~\ref{LpDtp} we see that there is a
relation between plateau duration and plateau luminosity. This
relation  is essentially a consequence of the explosion energy 
but we see an important dispersion associated with the
progenitor properties, especially the progenitor mass. The relationship
is not present in the observational data although the uncertainties in
the estimation of plateau durations are large and the statistics are small.
We also see that models with the lowest values of explosion energy and
hight masses are not consistent with the observations. 

Figure~\ref{LpDL} shows that the values
of {\bf $L_P$} and {\bf $\Delta L$} from  models and observations 
are in very good agreement, especially for models with $E= 1$ and 2
foe. We also see, for the models, a  dependence between
these two parameters which is not observed in the data.

\begin{figure}[htbp]
\begin{center}
\includegraphics[scale=0.5]{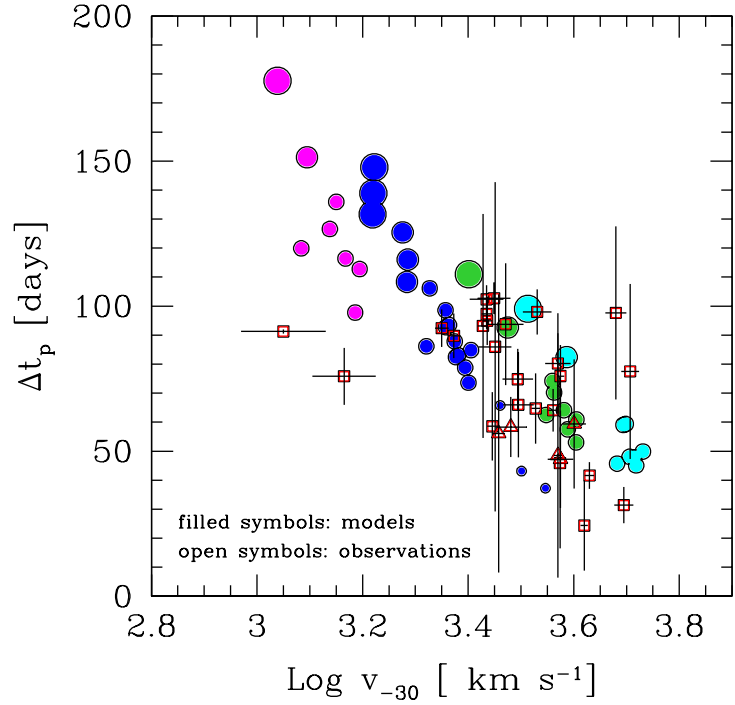}
\caption[Plateau duration vs. photosperic velocity]{The plateau
  duration as a function of $v_{-30}$ for models (filled
  symbols) and observations (open symbols). The size of the symbols is
  proportional to the initial mass. Colors indicate the
  explosion energy: magenta for $E= 0.5$ foe,  blue for $E= 1$ foe, green for
   $E=$ 2 foe and cyan for $E= 3$ foe. The
    intermediate-plateau SNe are shown with triangles while 
    squares are used for genuine plateau SNe.}
\label{vpDtp}  
\end{center}
\end{figure}

Finally, we compare the behavior of $v_{-30}$ with respect to the plateau
duration in Figure~\ref{vpDtp}. We see that models predict a
strong correlation between plateau durations and expansion
velocities. This correlation  
can be understood based on the dependence of the expansion velocity
on explosion energy. It is a well known fact that more energetic
explosions produce  
shorter plateaus and vice versa. The observational data appear to follow this
relation but the dispersion and the errors are large, making it difficult
to draw a definitive conclusion. We note, however, that the
  observations rule out the lowest energy values.

Based on the comparison of different observables between models and
observations we conclude  in general models that
with $E=0.5$ foe do not occur very often
discarded. Also models with $E=3$ foe and $M < 25 M_\odot$ or $E=1$ foe
and $M \geq 20 M_\odot$ are unlikely  to explain the observations of
our sample of  SNe~II-P.

\chapter{Summary}
\label{ch:summ}

The main propouse of this thesis is the development of a one-dimensional,
Lagrangian hydrodynamic code to model bolometric light curves of
SNe~II-P with the ultimate goal of analyzing the physical properties of
an available large sample of SNe with highly-precise, well-sampled
observations. In the first part of the thesis I focused on the
development of the 
hydrodynamic code, the incorporation of the required micro-physics 
and the construction of the progenitors models. 

Since our code
produces bolometric light curves and effective temperatures, 
it was also necessary to  calculate such quantities from the observed
photometry before performing a comparison between models and
observations. Hence, the second (Chapter~\ref{ch:model_obs}) of the thesis was
devoted to deriving reliable  
calibrations for bolometric corrections (BC) and effective temperatures
based on $B V I$ photometry and applicable to SNe~II-P. In this analysis I
used data of three well-observed SNe~1987A, 1999em,  and 2003hn; and
two series of atmosphere models by Eastman et al. (1996) and Dessart $\&$
Hillier (2005b). I found a tight correlation betewen BC and
photometric colors which allows one to estimate bolometric luminosity
with an uncertainty of $0.05$ dex during the plateau phase. On the
radioactive tail I found that the BC was independent of color with a
value of $-0.70$~mag and a scatter of $0.02$~mag based only on the
behavior of SN~1987A. I noticed the importance of including $L$- and
$M$-band photometry in order to derive an accurate BC during this
phase.

Then, I studied in detail the performance of the code via
its application to one of the best observed SNe~II-P, 
namely SN~1999em, obtaining a very good agreement with the
observations using the following physical parameters $E=1.25$ foe, $M=
19 \,M_\odot$, $R= 800 
\,R_\odot$, and $M_{\mathrm {Ni}}=0.056 \,M_\odot$. In this analysis,
I
found that an extended mixing of $^{56}$Ni is needed in order to
reproduce a plateau as flat as that shown by the observations of
SN~1999em. I note that the plateau phase, at least in the case of
this SN, is powered by the energy deposited by the shock wave and
released by the recombination process, plus some extra energy
deposited by the radioactive material. I compared our rusults 
  with two previous hydrodynamical studies
of SN~1999em given by \citet{2005AstL...31..429B} and
\citet{2007A&A...461..233U}. Even if these studies are based on
a more sofisticaded prescription of the radiative
transfer, there is a very good agreement among the results obtained by
the three works. In particular,  the physical  parameters obtained in
our calculations are  intermediate between those estimated by
\citet{2005AstL...31..429B} and \citet{2007A&A...461..233U}.

The remarkably good agreement of our model with the observations of
SN~1999em, and the 
consistency found with previous hydrodynamical studies of this SN
using more sophisticated code give us confidence in our attempt
to model SNe~II-P at least during the plateau phase and early
evolution of the radioactive tail phase, in spite of the
simplifications assumed in our calculation.

The last part of this thesis was focused on the study of the intrinsic
properties of our sample of SNe. First, I used the calibrations of BC
and effective temperature to 
derive bolometric luminosities and effective temperatures for 
our complete sample of SNe~II-P. In order to make a quantitative
comparison among the objects, I calculated a set of parameters that
characterize the bolometric LC: (1) plateau luminosity ({\bf $L_P$}), 
(2) plateau duration ({\bf $\Delta t_P$}), (3) the drop in luminosity
between plateau and radioactive tail ({\bf $\Delta \log L$}) and (4)
the nickel mass ($M_{\mathrm{Ni}}$). 

I found that the 
SN sample shows a range of $1.15$ dex in plateau luminosity with a
weighted average value of $<L_P>= 1.26 \pm 0.019 \times 10^{42}$ erg 
 s$^{-1}$, and plateau durations between 64 and 103 days with a
weighted average value of $<\Delta t_p>= 90.47 \pm 1.18$ days.
Comparing the shape of the transition between the plateau and the 
radioactive tail, I found that the size of the drop in
  luminosity to be in the range of $0.35$ to $1.46$~dex with a weighted
average of $<\Delta \log L>=0.783 \pm
0.054$~dex. The radioactive nickel masses calculated for all of the SNe in
our sample are below $0.1 M_\odot$ with a weighted average
of $0.024\pm0.004 M_\odot$. However, it is important to note that I
have excluded the extreme case of SN~1992am from the analysis. For this
  SN I found  $M_{\mathrm{Ni}}$  $> 0.26\pm0.07$ $M_\odot$ implying
  that there are SNe~II-P which produce significant 
  nucleosynthesis during the
  explosion. We leave a detailed analysis of this SN for a future
  work.   SNe with
a more linearly declining  
  behavior during the ``plateau'' phase show consistent values of 
 {\bf $L_P$}, {\bf $\Delta \log L$}, and $M_{\mathrm{Ni}}$, which strengthens
  the idea of a connection between linear and plateau SNe~II. In
  addition, our models show a strong dependence of the {\bf $\Delta t_P$}
on the progenitor mass in the sense that progenitors with lower masses
produce shorter {\bf $\Delta t_P$}. This could mean that linear
SNe in general have less massive progenitors, as has been previously proposed
(see \citet{2003LNP...598...21T}).

Then I proceeded to calculate a grid of 46 hydrodynamical models for a range of
 initial masses ($M_0$), radii ($R_0$), explosion energies ($E$),
 nickel masses ($M_{\mathrm{Ni}}$), and nickel mixing. For each model
 I calculated {\bf $L_P$}, {\bf $\Delta t_P$}, 
 {\bf $\Delta \log L$} and the photospheric velocity ($v_{ph}$), and I
 studied the dependence of these observable parameters on physical
 quantities ($E$, $R_0$, $M_0$ and $M_{\mathrm{Ni}}$). 

 From this
 analysis I found that the explosion energy is the main factor in
 the determination of {\bf $L_P$} and $v_{ph}$, with a smaller dependence
 on $M_0$, and on $R_0$ for the case of {\bf $L_P$}. I
 found that $M_{\mathrm{Ni}}$ has no noticeable effect on the
 those parameters. In the case  of {\bf $\Delta t_P$}, I identify the 
 progenitor mass as the most important parameter but $E$  and
 $M_{\mathrm{Ni}}$ also produce an important effect. Finally, I found
 that {\bf $\Delta \log L$} largely depends on $M_{\mathrm{Ni}}$, with
 some dependence on $E$ and $R_0$ but not on $M_0$. 

I studied correlations among the observable parameters, {\bf $L_P$},
{\bf $\Delta t_P$}, {\bf $\Delta \log L$}, $M_{\mathrm{Ni}}$, and
$v_{ph}$ calculated both from the observations and hydrodynamical
models. The most important results of this analysis are: (1) our
models reproduce the strong correlations shown by the observations
between luminosity and expansion velocity during the plateau. The
slope of the  relation is consistent with the observations, but there
is a general shift between observational and model data. The shift is
possibly related with the zero point of the distance 
calibration used to calculate distances for the SNe in our sample
\citep[see][]{F09}, which could lead to a shift of $\sim$$0.2$ dex in
luminosity. I identified the explosion energy as the main
driving parameter of this correlation. (2) I did not find a clear
correlation between $M_{\mathrm{Ni}}$ and other observables 
with the exception of {\bf $\Delta \log L$} where I found that lower
values of $M_{\mathrm{Ni}}$  correspond to higher {\bf $\Delta \log
  L$} and vice versa. (3) Our models show a certain relationship
between {\bf $L_P$}  and {\bf $\Delta t_P$} in the sense that more
luminous SNe have shorter  {\bf $\Delta t_P$}, but this is not seen in the
observations. It is intresting to note that this correlation was
  also found in a recent work by \citet{2009ApJ...703.2205K}  using
  their hydrodynamical code and initial models from stellar-evolution
  calculations.  (4) The models show a relationship between 
{\bf $\Delta  t_P$}  and  the expansion velocity with  smaller values
of {\bf $\Delta  t_P$} associated to SNe with greater expansion
velocities. The observations show a similar trend 
but the scatter is large and the statistics is too poor to provide a
definitive conclusion. 

Finally, based on the comparison of different observables
between models and observational data we conclude  in general models
with (a) $E=0.5$ foe and $M \geq 10 M_\odot$, (b) $E=3$ foe
and $M < 25 M_\odot$, and (c) $E=1$ foe and $M \geq 20 M_\odot$
do not occur very often.

\appendix
\chapter{Finite Difference Approximations}\label{ch:anexoA}
\section{Computational grid}
\label{anexoA:grid}
A finite difference representation of a system of time dependent
equations, such as the one presented in \S~\ref{sec:equations}
(equations \ref{sec:equations:eq1}--\ref{sec:equations:eq5}),
consists of defining a space-time grid upon which finite 
difference approximations are performed to evaluate the functions and
their derivatives which constitute the partial differential equations
under study. As an 
example, Figure~\ref{anexoA:grid} shows a schematic  
space-time grid which consists of a set of fixed coordinate positions
$x_k$ defined at discrete times $t^n$. Throughout
this discussion, the  
subscript $k$ represents the spatial index and $n$, the temporal
index. The temporal variable
changes in a discrete fashion such that $t^{n+1}= t^n + \Delta t^{n+1/2}$,  
where $\Delta t^{n+1/2}$ is the time step between $t^{n+1}$ and
$t^n$. If the spatial coordinates are fixed, as is the case of
Lagrangian coordinates, they do not carry a temporal
index. The interval  between two spatial coordinates,
$x_{k+1}$ and $x_k$, is denoted by $\Delta x_{k+1/2}$ and it is
commonly referred to as grid cell or zone. Note that 
$\Delta x_{k+1/2}$ and  $\Delta t^{n+1/2}$  need not be constant
throughout the grid.

Dependent variables are defined on this space-time grid and the
evolution equations take them from one spatial level to the
next. However, there are different ways to represent a dependent
variable on the grid. For example, given the function $u(x,t)$ this
can be defined at cell corners denoted
by $u_k^n= u(x_k,t^n)$, at cell edges denoted by $u_{k-1}^{n+1/2}=
u(x_{k-1},t^{n+1/2})$ with $t^{n+1/2}= t^n + 0.5 \, \Delta t^{n+1/2}$,
or at cell centers denoted by $u_{k+1/2}^{n+1/2}=u(x_{k+1/2}, t^{n+1/2})$ with
$x_{k+1/2}= x_k + 0.5 \, \Delta x_{k+1/2}$ (see
Figure~\ref{anexoA:grid}). The choice of the representation is
based on the physical meaning of each variable. 

\begin{figure}[htbp]
\begin{center}
\setlength{\unitlength}{2.cm}
\begin{picture}(4.5,4.5)
\thicklines
\put(0,0){\vector(1,0){5}}
\put(5.3,-0.1){$x$}
\put(0.95,-0.23){$k-1$}
\put(2.35,-0.23){$k$}
\put(3.35,-0.23){$k+1$}
\thicklines
\put(0,0){\vector(0,1){4}}
\put(-0.3,3.8){{$t$}}
\put(-0.8,1){$n-1$}
\put(-0.5,2){$n$}
\put(-0.8,3){$n+1$}

\linethickness{0.1mm}
\multiput(0,0)(1.2,0){4}%
{\line(0,1){3.8}}
\multiput(0,0)(0,1){4}
{\line(1,0){4.5}}

\multiput(0.2,0.5)(0.2,0){22}
{\line(1,0){0.11}}
\multiput(0.2,1.5)(0.2,0){22}
{\line(1,0){0.11}}
\multiput(0.2,2.5)(0.2,0){22}
{\line(1,0){0.11}}

\multiput(1.8,0.1)(0,0.2){18}
{\line(0,1){0.11}}
\multiput(3,0.1)(0,0.2){18}
{\line(0,1){0.11}}

\linethickness{.2mm}
\put( 0.2,0.5){\line(1,0){.1}}
\put(2.4,1){\circle*{0.15}}
\put(2.5,1.15){$u_k^{n-1}$}
\put(2.4,2){\circle*{0.15}}
\put(2.5,2.15){$u_k^{n}$}
\put(1.2,1.5){\circle*{0.15}}
\put(1.21,1.7){$u_{k-1}^{n-1/2}$}
\put(1.8,2.5){\circle*{0.15}}
\put(1.82,2.65){$u_{k-1/2}^{n+1/2}$}
\put(3.,1.5){\circle*{0.15}}
\put(3.05,1.65){$u_{k+1/2}^{n-1/2}$}
\put(3.6,1){\circle*{0.15}}
\put(3.75,1.15){$u_{k+1}^{n-1}$}
\put(3.6,2.5){\circle*{0.15}}
\put(3.78,2.65){$u_{k+1}^{n+1/2}$}
\end{picture}

\vspace{0.5cm} 
\caption[Space-time grid]{Space-time grid. The values of the
  discretized function $u(x,t)$ can be defined either at cell corners
  (e.g., $u_k^n$), cell edges (e.g., $u_{k-1}^{n-1/2}$)  or cell
  centers (e.g., $u_{k+1/2}^{n-1/2}$). \label{anexoA:grid} }
\end{center}
\end{figure}
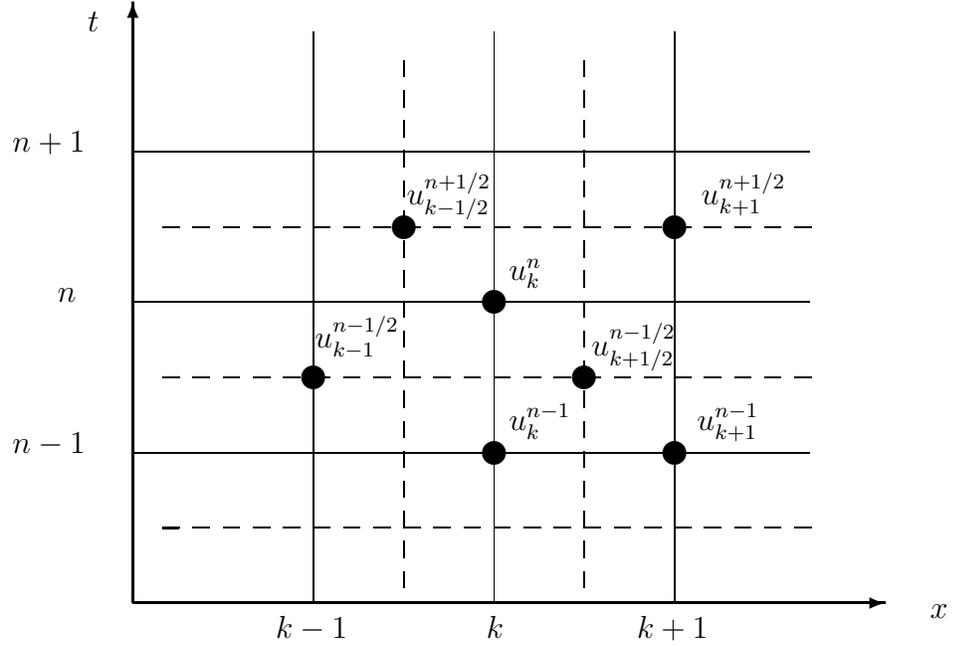

\section{The Difference Equations}

In the Lagrangian prescription of the hydrodynamic equations in spherical
symmetric approximation
(equations~\ref{sec:equations:eq1}--\ref{sec:equations:eq3} of 
Section~\ref{sec:method}), the object is represented by  
concentric spherical shells (zones) of mass $\Delta m_{k+1/2}$ which
represent the spatial coordinates of the system (analogous to 
$\Delta x_{k+1/2}$ in the previous discussion). The radii $r_k^n$ are
defined at cell corners, the velocities $v_k^{n+1/2}$ are defined at
the centers of vertical  cell edges and the intensive quantities 
($\rho$, $P$ and $E$) are defined at the centers of the horizontal cell
edges; for example the pressure is represented as $P_{k+1/2}^{n}$. Two time
steps are also required: one to advance the velocity $\Delta t^n$
defined by $\Delta t^n$= $0.5 (\Delta t^{n-1/2} + \Delta t^{n+1/2})$, and
another to advance the material state variables, $\Delta t^{n+1/2}=
t^{n+1}-t^n$.

The calculation procedure is to take known radii and velocities at
$t^n$ and determine the new values $v_k^{n+1/2}$, $r_k^{n+1}$ and
$V_{k+1/2}^{n+1}$ for all values $k$ using the following explicit 
difference representation of the hydrodynamics equations

\begin{equation}
v_k^{n+1/2}= v_k^{n-1/2} - 4 \pi (r_k^n )^2 [ P_{k+1/2}^n + Q_{k+1/2}^n
  - (P_{k-1/2}^n + Q_{k-1/2}^n) ] \frac{\Delta t^n}{\Delta m_k} -
\frac{G m_k} {(r_k^n)^2} \Delta t^n
\end{equation}
 
\begin{equation}
r_k^{n+1}= r_k^{n} + v_k^{n+1/2} \Delta t^{n+1/2}
\end{equation}

\begin{equation}
V_{k+1/2}^{n+1} \equiv (\rho_{k+1/2}^{n+1})^{-1}= \frac{4 \pi}{3} \frac{[(r_{k+1}^{n+1})^3 - (r_{k}^{n+1})^3 ]}{\Delta m_{k+1/2}}
\end{equation}

\noindent where $Q$ is the {\em artificial viscosity}  term
   included in the
equation to take into account numerical discontinuities which are
produced by the shock wave. For this term I have adopted the following
expression given by \citet{VN50} 

\begin{displaymath}
Q_{k+1/2}^{n+1/2}= \left\{ \begin{array}{ll}
C_q \, ( v_{k+1/2}^{n+1/2} - v_k^{n+1/2})^2 \rho_{k+1/2}^{n+1/2} &
\textrm{if  $v_{k+1}^{n+1/2} < v_{k}^{n+1/2}$}\\
0 & \textrm{otherwise}
\end{array} \right.
\end{displaymath}

\noindent where $C_q$ is a constant set to a value of 2 in 
our calculations. The time step $\Delta
t^{n+1/2}$ is chosen  subject  to assure
stability and accuracy. The stability of the explicit numerical 
scheme is obtained if the time step is constrained to the Courant
condition, i.e.,

\begin{equation}
\Delta t^{n+3/2}= \min \{ \Delta m_{k+1/2}^{n+1}/ v_{s,k+1/2}^{n+1}\},
\end{equation}

\noindent where $v_{s,k+1/2}^{n+1}$ is the sound speed which varies from
zone to zone. The time step is taken as a fraction of the Courant
condition for all zones. Additional
conditions on the time step are imposed to assure the accuracy of the results:
it is required that changes in temperature, density and flux over
one time step be less than 5\%. This constraint added
  to that on  the
time step is  a consequence of  the iterative
procedure used to solve the 
radiative transfer (see following discussion). Note that  the
iterations usually do not converge if $\Delta t$ is too large. Even if we do
not need to iterate, the size of $\Delta t$ in the implicit scheme is
limited by the error growth rate. To keep the error low, changes
in the physical quantities over one time step must be less than, for
example,  5\%. So, the time step is limited, by reason of accuracy,
to the physical time scale on which quantities change. However, in
general, the physical time scale is much longer than the Courant time.

The energy and radiation transfer equations
(equation~\ref{sec:equations:eq4} and 
~\ref{sec:equations:eq5} of Section~\ref{sec:method}) are solved using an
iterative procedure.  The discretization of the energy equation 
in the diffusion limit for a semi-implicit scheme is written as 

\begin{eqnarray}
\lefteqn{ E_{k+1/2}^{n+1} - E_{k+1/2}^{n} + \frac{1}{2} ( P_{k+1/2}^{n+1} +
P_{k+1/2}^{n} + Q_{k+1/2}^{n+1}+Q_{k+1/2}^{n}) \, \Delta V_{k+1/2}}\nonumber\\  
& & = - [\theta (L_{k+1}^{n+1}- L_{k}^{n+1} ) + (1 - \theta)
  (L_{k+1}^{n}- L_{k}^{n})] \frac{\Delta
  t^{n+1/2} }{\Delta m_{k+1/2}} \label{anexoA:ener}
\end{eqnarray}

\noindent where $\Delta V_{k+1/2}= V_{k+1/2}^{n+1}- V_{k+1/2}^{n}$ is
written by numerical convenience as

\begin{equation*}
 \Delta V_{k+1/2}= \frac{4 \pi}{3}\frac{(\Delta r_{k+1}^3 -\Delta
  r_k^3)}{\Delta m_{k+1/2}}  \hspace{0.5cm} \textrm{with} \hspace{0.5cm}
 \Delta r_{k}^3= \Delta r_{k} \,( 3 \,r_{k}^n \,r_{k}^{n+1} + \Delta r_k^{n+1}).
\end{equation*}

\noindent Here $\theta$ is a parameter with reflects the
time-centering: $\theta=1$ fully implicit scheme, and $\theta= 0.5$ for
time-centering. This latter value for $\theta$ is the one I usually
adopted in the calculations. The luminosity is written as

\begin{equation}
L_k^n= -[4\, \pi (r_k^n)^2]^2 \,\, \frac{a c \lambda_k^n}{ 3
  \overline{\kappa}}\,\,  \frac{(T_{k+1/2}^n)^4
  -(T_{k-1/2}^n)^4}{\Delta m_k},
\label{anexoA:lum} 
\end{equation}

\noindent where $\lambda_k^n=\frac{ 6 + 3 R_k^n}{6 + 3 R_k^n +
  (R_k^n)^2}$ is the ``flux-limiter'', and $R_k^n$ is written as

\begin{equation*}
R_k^n= \frac{4 \pi (r_k^n)^2}{0.5 [(T_{k+1/2}^n)^4
  +(T_{k-1/2}^n)^4]} \frac{ \mid(T_{k+1/2}^n)^4-
  (T_{k-1/2}^n)^4\mid} {\overline{\kappa}\, \Delta m_k}.   
\end{equation*}

\noindent Here, $\overline{\kappa}$, which also appears in
equation~\ref{anexoA:lum}, is 
the Rosseland mean opacity  evaluated at  $m_k$ and $t^n$, i.e.,
$=\kappa_k^n=\kappa(T_k^n,\rho_k^n)$. Note, however, that the
intensive quantities are 
known at the middle points of the grid. Therefore, it is needed to
average the value of 
$\kappa$ between the points $m_{k-1/2}$ and $m_{k+1/2}$.  Special care
must be taken to 
do this average in order to prevent numerical noise which appears when
the radiation 
flow is propagated through  a steep front (such as a shock-wave front)
where $T$ and  
$P$ change abruptly. Following \citet{Christy67},  instead of
$\overline{\kappa}$, I have used an effective opacity  defined by 

\begin{equation}
\left( \frac{1}{\kappa} \right) _{eff}= \frac{T_{k+1/2}^4 / \kappa_{k+1/2}
  + T_{k-1/2}^4 / \kappa_{k-1/2}} { T_{k+1/2}^4 + T_{k-1/2}^4}.
\end{equation}

\noindent Note that knowing the values of the temperature
$T_{k-1/2}^{n+1}$, $T_{k+1/2}^{n+1}$ and $T_{k+3/2}^{n+1}$  for all values
$k$, the transport problem is solved. However,
equation~(\ref{anexoA:ener})  is nonlinear with respect to the new
temperatures at $n+1$, according to the dependence of $E^{n+1},
P^{n+1}$ and $L^{n+1}$ on  these new values. Defining
an extrapolated temperature as 
$T^{n+1}_{k+1/2}= T^{n}_{k+1/2} + \delta T_{k+1/2}$, one can 
linearize the energy equation in corrections $\delta T_{k+1/2}$, which
leads to a system of $N$ (equal to the number of points grid) coupled
linear equations. Each equation takes the form

\begin{equation}
A_{k+1/2} \, \delta T_{k+3/2} + B_{k+1/2} \, \delta T_{k+1/2} + C_{k+1/2}\,
\delta T_{k-1/2}= D_{k+1/2},
\label{anexoA:linear}
\end{equation}

\noindent where $D_{k+1/2}= G(T_{k+1/2},
T_{k+1/2},T_{k+1/2})|_{t=t^n}$, is a function of the temperature
evaluated at only three spatial points and at
the previous time step $t^n$. The other constants in
equation~\ref{anexoA:linear} are derivatives of this function ``$G$''
with respect to $\delta T$ at the three different spatial
points evaluated at $t^n$, i.e.,  

\begin{equation*}
A_{k+1/2}= \frac{\partial G}{\partial (\delta T_{k+3/2})}
\bigg|_{t=t^n}, \hspace{0.5cm} B_{k+1/2}= \frac{\partial G}{\partial
  (\delta T_{k+1/2})}\bigg|_{t=t^n} \textrm{and} \hspace{0.5cm} C_{k+1/2}=
\frac{\partial G}{\partial (\delta T_{k-1/2})}\bigg|_{t=t^n}.
\end{equation*}

\noindent Because each equation of the system depends
on the temperature at only three adjacent mass points (see
equation~\ref{anexoA:linear}), the system of  
algebraic equations takes the from of a tridiagonal matrix. This
tridiagonal system is solved for all the correction
  temperatures $\delta T$ (inverting the tridiagonal matrix). The zone  
temperatures are corrected and the process is repeated until the corrections
are reduced to a sufficiently small amount. The condition that is
usually adopted in the code for the convergence of the iterative
process is that max$(\frac{\delta T}{T}) < 10^{-5}$, where
max$(\frac{\delta T}{T})$ is the maximum value of the $\frac{\delta
  T}{T}$ in all the grid points. 

\chapter{Solution of Gamma-ray transfer}\label{ch:anexoB}

In this appendix I describe the solution of the $\gamma$-ray transfer
in gray approximation assuming that the  $\gamma$-ray opacity is
purely absorptive and independent of the energy, and allowing a
spherically symmetric distribution for the sources of the $\gamma$
photons ---i.e., the radioactive material. From this solution, the
energy deposited by $\gamma$-rays can be calculated becuase, by
definition \citep[see][]{1983psen.book.....C}, the total energy
absorbed by matter per unit of time and mass is  

\begin{equation*}
\frac{dE}{dm}= \kappa_\gamma \int I d\Omega,
\end{equation*}

\noindent where the integral is done in all solid angles, $I$ is the
specific intensity and $\kappa_\gamma$ is the opacity of the
$\gamma$-rays for which we adopted the value $\kappa_\gamma= 0.06 \,
y_e$ cm$^{2}$ g$^{-1}$ \citep{1984ApJ...280..282S}
and $y_e$ is the number of electrons per baryon. Although our code
solves the equations in spherical symmetry, in order to correctly take
into account the energy deposited in a specific point, it is necessary to 
consider rays which come from different directions. Nevertheless, the
axial  symmetry is preserved. Therefore, the previous integral in
spherical coordinates and assuming axial symmetry can 
be written as

\begin{equation}
\frac{dE}{dm}= 2 \pi \kappa_\gamma
\int_{\theta_{min}}^{\theta_{max}}I(r,\theta) \, sin \theta \, d\theta.
\label{eq:depo}
\end{equation}

\noindent  As can be seen in the above expression, to calculate
the energy deposited by 
$\gamma$-rays one should find the specific intensity in each point
of the object by solving the transport of $\gamma$-rays.
The radiative transfer equation in gray approximation is

\begin{equation}
\frac{dI}{d\tau}= - I + S,
\label{eq:transt}
\end{equation}

\noindent where $I$ and $S$ are the specific intensity and the
source function integrated in frequency, and $ d\tau= \kappa_\gamma
\,\rho\,ds$ is the differential  
optical depth of $\gamma$-rays along the direction of the beam,
$\hat{s}$. The formal solution of this equation is

\begin{equation}
I(\tau)= I_0  \, e^{-\tau} + \int^{\tau}_0  S(\tau) \,
e^{-(\tau-\tau')} d\tau'.
\label{eq:integroeq}
\end{equation}

\noindent The source function, $S$, is a known function in our
problem and it depends on the distribution of the radioactive
material. Specifically, it can be written as
  
\begin{equation}
  S= \epsilon_{rad}(t) \, \xi(r) / \kappa_\gamma,
\label{eq:source}
\end{equation}

\noindent where $\epsilon_{rad}(t)$ denotes the time-dependent rate
of energy released per gram by Ni--Co--Fe decay, 
\begin{equation*}
\epsilon_{\rm rad} = \, 3.9 \times 10^{10} \,\, \exp(- t/\tau_{\rm Ni}) +
\, 6.78 \times 10^{9} [\exp(- t/\tau_{\rm Co}) -  \exp(- t/\tau_{\rm
    Ni})] \;{\rm erg\; g}^{-1} \,{\rm s}^{-1},
\end{equation*}

\noindent and $\xi(r)$ of equation~\ref{eq:source} is the initial
distribution of 
$^{56}$Ni\footnote{Since $^{56}$Co is a decay product of $^{56}$Ni,
  its spatial distribution is described by the same function
  $\xi(r)$.}. Note that this function 


\end{document}